\documentstyle[12pt,wuthesis,wurevtex,epsf]{report}

\input epsf
\epsfverbosetrue
\newcommand {\epsfxy}{\epsfxsize=4.5in \epsfysize=3.18in}
\newcommand {\erf}{\hbox{erf}}
\newcommand{\mathrm}{\rm}

\newcommand{\Sizinf}{\sum_{i=0}^{\infty}}
\newcommand{\Sioinf}{\sum_{i=1}^{\infty}}

\newcommand\gthth{ g_{\theta\theta} }
\newcommand\gphiphi{ g_{\phi\phi} }
\newcommand\sinsq{ \sin^2\theta }

\newcommand\dAlem{{\vcenter{\hrule height.5pt
                \hbox{\vrule width.5pt height8pt \kern8pt
                        \vrule width.5pt}
                \hrule height.5pt}}}

\newcommand{\pdmu}{\phi_{;\mu}}
\newcommand{\pdnu}{\phi_{;\nu}}

\newcommand{\eupv}{e^{u+v}}

\newcommand{\evmu}{e^{v-u}}

\newcommand{\dtphi}{\frac{\partial\phi}{\partial t}}

\newcommand{\be}{\begin{equation}}
\newcommand{\ee}{\end{equation}}
\newcommand{\ben}{\begin{eqnarray}}
\newcommand{\een}{\end{eqnarray}}
\newcommand{\bc}{\begin{center}}
\newcommand{\ec}{\end{center}}

\newcommand{\pst}{\varphi}
\newcommand{\psia}{\phi}
\newcommand{\varphia}{{\varphi}}

\newcommand{\domst}{\frac{d\omega}{d\varphi}}
\newcommand{\omsts}{\frac{\omega}{\varphi^2}}
\newcommand{\gumunu}{g^{\mu\nu}}
\newcommand{\gdmunu}{g_{\mu\nu}}

\newcommand{\ddmu}{\partial_{\mu}}
\newcommand{\ddnu}{\partial_{\nu}}

\newcommand{\pte}{\frac{8\,\pi}{\varphi}\,T}
\newcommand{\pstl}{\varphi_{;\lambda}^{;\lambda}}

\newcommand{\msun}{M_{\odot}}
\newcommand\dal{{\vcenter{\hrule height.5pt
                \hbox{\vrule width.5pt height8pt \kern8pt
                        \vrule width.5pt}
                \hrule height.5pt}}}

\newcommand{\s}{\sigma}

\newcommand{\fpi}{4\,\pi}

\newcommand{\del}{\nabla}

\newcommand{\bosa}{{\psi_1}}
\newcommand{\bosb}{{\psi_2}}
\tighten
\begin{document}

\title{A Numerical Study of Boson Stars:\\
     Einstein Equations with a \\
       Matter Source}
\author{Jayashree Balakrishna}
\dept{Physics}
\submitdate{August 1999}
\principaladvisor{Wai-Mo Suen/Matt Visser} 
\firstreader{Clifford M. Will}
\secondreader{Carl M. Bender}

\beforepreface
\signaturepage
\tablespagefalse

\prefacesection{Abstract}

The study of the properties and dynamics of self-gravitating bosonic objects in Einstein gravity
was conducted. Bosons are promising candidates for dark matter. They can form compact
objects through a Jeans instability mechanism. We studied boson stars made up of self-gravitating
scalar fields, with and without nonlinear self-couplings. These are non-topological solutions of
the coupled Einstein--Klein--Gordon equations. We studied the stability of boson stars in the
ground and excited states, and determined the quasinormal mode 
($QNM$) frequencies of stable boson stars
in spherical symmetry. The study was carried out in the
standard Einstein theory of General Relativity and in Brans--Dicke theory.
We also studied the formation of these objects in Brans--Dicke theory
showing that they can form from the self-gravitation of bosonic matter. We also studied the
possibility of a bosonic halo surrounding galaxies. These halo models predict the observed flatness
of galactic curves. We studied their formation and stability.

After an extensive study in spherical symmetry we carried out numerical studies of boson star
dynamics in full 3+1 dimension. One focus of the 3 spatial dimension (3D) study was on the validation
of the numerical code constructed to solve Einstein equations with matter sources. The
use of the scalar field has unique advantages: Boson Stars do not suffer from difficulties
associated with hydrodynamic sources (like shock waves or the surface problems of neutron stars).
They also do not suffer from the difficulties related to the singularities of black holes.
The code was first tested with spherical perturbations and
compared with the
spherical results. We determined the coordinate conditions needed to provide
stable evolutions under radial perturbations. We then went on to study their
behavior under non-spherical perturbations. Both scalar and gravitational radiation produced
under these perturbations were studied. We reproduced the $QNM$
frequencies of the stars, as determined by perturbation studies carried
out by other groups. The energy generated by the perturbation was
studied with different radiation indicators. We also observed the
collapse to black holes of unstable boson-star configurations. We
simulated the collision of two boson stars. This is of interest as the
two body problem is as yet unresolved in general relativity.

\prefacesection{Acknowledgements}
I want to thank my advisor Wai-Mo Suen and Edward Seidel for their
pioneering work on Boson Stars in Numerical Relativity. 
My $1D$ work involved three problems that were done using a code
developed originally by Wai-Mo Suen and Ed Seidel and which was subsequently
modified by me to study my related problems.
Hisa-aki Shinkai and Greg Comer provided the equilibrium boson
star model in Brans--Dicke theory and Hisa-aki and I subsequently
collaborated to work on the evolution problem. Franz Schunck
provided a model for the massless scalar field halo that he
fit to the rotation curves of galaxies. We collaborated
on the work on bosonic halos. 

In $3D$ the work was done using the ``G'' code. The ``G'' code
was developed by members of the NCSA/Wash.\ U. Numerical
Relativity Group. In addition to Wai-Mo and Ed, members of the group who
have contributed to the code include Peter Anninos,
Greg Daues, Steve Brandt, Karen Camarda,
Joan Mass\'o, Malcolm Tobias, John Towns, Paul Walker,
and others.

I would especially like to thank Greg Daues for vital
collaboration in my $3D$ project. The coordinate drifting
problems were solved due to vital collaborations with
Greg Daues and Malcolm Tobias. I would like to thank Greg Daues
for his work on the elliptic initial value solver. The implementation
of the Apparent Horizon Boundary Conditions ({\em AHBC}) in my
code to study the boson star collapse problem,
was possible due to his own research work in the field ({\em AHBC}).

The numerical work in $1D$ was carried out
with the C90 at the Pittsburgh Supercomputing center and local
machines at Washington University. The $3D$ work was carried
out on
the CM5 of the National Center for Supercomputing
Applications. The research was supported in part by
NSF grants 94-04788 and 96-00507.

I would {\bf especially} like to thank Matt Visser for his
thorough reading of the thesis and suggestions. Without him this work
would not have been completed. I thank Pranoat Suntharothak-Preiesmeyer
for all her assistance with graphs and tremendous emotional
support through tough times. I would like
to thank my parents for their patience and my husband
Greg for his forbearance. Buster Balakrishna was as much
a part of this as anyone else, sitting patiently by me
as I worked. I thank him for his companionship.

Last but not least I would like to thank Stefan,
Aziz, Kim, Raffa, MJ, Steph,
Liliana, K.C. Leung,
Hisa-aki, Laura, Malcolm, Nils,
and Ed Wang for
their warm friendship. I would
like to thank Julia Hamilton, Beej, Valerie and all the other members of the staff and faculty
as well as Henrietta for making my stay memorable.

\afterpreface

\chapter{Introduction}

The study of bosonic compact objects is motivated by several considerations.
From a purely aesthetic point of view, the existence of compact objects
made up of fermionic subparts (like neutron stars), leads one to postulate
the existence of bosonic counterparts made up of particles satisfying
Bose-Einstein statistics. This was the motivation for the original works of Kaup
~\cite{kaupgeon}, and Ruffini and Bonazzola~\cite{ruf1}. 

Although not observed thus far, these purely theoretical objects nevertheless
can provide many useful physical insights.
They can be a source
of scalar radiation and gravitational radiation, and can under certain
conditions collapse to black holes. Studying the evolution of
the resultant black holes then adds a new dimension to black hole
studies since now matter is present in the spacetime.

Recently, Bosonic objects have 
been suggested as strong candidates for dark-matter
in the universe. This has changed their status from being
purely theoretical toy models, to possibly being
extremely real physical objects.

They also occupy a significant place in the field of numerical relativity.
For the first time one can consider the full Einstein system of equations
with matter sources present.
Before the advent of this field, boson stars could only be studied under
infinitesimal perturbations using perturbative
techniques~\cite{leepert,lee2}. 

In one-dimensional numerical relativity
we have observed the behavior of boson stars under larger and more general spherical perturbations. In $3D$,
for the first time, a fully relativistic problem
with matter present was considered. Boson stars provide a model which does not
suffer from
the problematic singularities of black-hole spacetimes.
In addition the presence of a scalar field with an exponential fall-off
that, in principle, stretches to spatial infinity, prevents the formation of
sharp outer surfaces that are characteristic of neutron stars.

The $3D$ code could be tested against the well studied 1D codes by
considering spherical perturbations. The gravitational wave signals of
non-spherical configurations in the linear regime were compared 
against the results of linear 
non-spherical perturbation studies. A comparison of the signals
calculated by different methods provided a further successful code test.
In order to stabilize the code, various numerical techniques were implemented
and studied. With a stable code one could then study the signals
that came out of large non-spherical perturbations. These physical
studies were precursors to a planned series of future neutron star studies.
The collapse of an unstable boson star to a black hole, followed by
a ``turning on'' of apparent horizon boundary conditions, was a new aspect
of black hole studies that boson stars provided. In addition we began simulations 
of the collision of two boson stars. This is of importance
in the generalized two-body problem since the resulting dynamics
may not be sensitively dependent on the constituent objects.

A study of compact bosonic objects touched upon numerous physical
issues. In one dimension, the evolution of boson stars in the
ground and excited states, with and without self-coupling,
and the evolution and formation of boson-stars in Brans-Dicke theory
were investigated.
The stability and formation of objects formed from massless scalar fields
was also studied in a one-dimensional halo model.
In 3D the study of gravitational
waves from the non-spherical perturbation of a boson star was followed by
a study of black hole evolutions of the black holes formed from collapsed 
boson stars. 

We begin by introducing the Einstein equations with general matter
sources, followed by a
historical perspective on the nature of boson stars and a summary of
earlier work in the field.
The notion of dark matter is then introduced,
what makes one believe it exists, and why bosons are good candidates for
it. This is followed by the basics of a Jeans instability mechanism to
give an idea of how boson stars could actually form. 

The final topics in the introduction are about boson stars without spherical
symmetry.
%in higher
%dimensions.
Since they can be a source of gravitational waves,
we first review how Einstein's equations predict
gravitational radiation in a non-spherically symmetric setting. 
In this regard we also
introduce the idea of waveform extraction through
the Zerilli, Bel-Robinson, and Newman-Penrose functions
which we use in our code. We discuss some of
the other work in the field of Boson stars in higher dimensions as
well as the results of $3D$ perturbation studies to which we compare our
results.
The penultimate topic in the introduction is about the
$3+1$ $ADM$ split of space-time and how the Einstein equations
can be separated into constraint and evolution equations. 
Since our work is numerical, we also
present an overview of the kinds of difficulties that we face in a general
numerical code. 
The actual details of the code and numerics are explained
in the chapters that follow.

\section{Einstein Equations with Matter: Variational Approach}
\label{section:variat}
%\begin{document}

The central tenet of {\em GR} is that matter must respond to geometry 
by moving in a certain way, and
geometry must in turn respond to matter by curving. 
We briefly outline the various quantities that we encounter in
the equations that we will have to solve, and then use a variational
approach to derive the equations which a fully relativistic system
with matter sources must obey~\cite{mtw}.

\begin{itemize}

\item{Metric:}
The invariant measure of proper length or proper time between neighboring
events in spacetime is given by
\be
ds^2 = g_{\mu\nu}\;dx^{\mu}\;dx^{\nu}= -d\tau^2,
\ee
where the metric $g_{\mu\nu}$ in a Minkowski spacetime is
the diagonal metric $\eta_{ij}=diag(-1,1,1,1)$.

\item{$\Gamma$:}
In curved space variations in tensors are affected by the fact that the
basis vectors themselves are twisting and turning from spacetime point to
spacetime point.
The {\em Connection Coefficient} $\Gamma$ measures the changes in the
basis vectors through the relation
\be
{\bf \nabla}_{\nu}{\bf e}_\mu=\Gamma^{\delta}{}_{\mu\nu}\;{\bf e}_\delta.
\ee
Here ${\bf \nabla}_{\nu}$ measures changes along the basis vector ${\bf e}_\nu$.
In any coordinate basis ($ [{\bf e}_\mu, {\bf e}_\nu] =0$ ) the
connection coefficients are given by
\be
\Gamma^{\alpha}{}_{\mu\nu} = \frac{1}{2}\,g^{\alpha\delta}\left(
g_{\delta \mu,\nu} + g_{\delta \nu,\mu}-g_{\mu\nu,\delta}\right)
= \Gamma^{\alpha}{}_{\nu\mu}.
\ee

\item{Geodesic Equation:}
The path of particles between spacetime points ${\cal P}$ and ${\cal Q}$
is one that
extremizes $\int_{\cal P}^{\cal Q} \,d\tau$ leading to
the geodesic equation
\be
\frac{d^2 x^\beta}{{d\lambda}^2} + \Gamma^{\beta}{}_{\mu\nu} \frac{dx^{\mu}}{d\lambda}\frac{dx^{\nu}}{d\lambda} = 0,
\label{gammon}
\ee
where $\lambda$ is a parameter that parametrizes the path and agrees
with the proper time $\tau$ within $\lambda =a\tau+b$, with
arbitrary constants $a$ and $b$.

\item{Riemann tensor:}
Spacetime curvature results in geodesics deviating from nearby geodesics,
with test particles accelerating relative to one another.
Measuring the deviation of one geodesic from another involves quantifying
this curvature through the Riemann curvature tensor.
In flat space (zero curvature) it must vanish.
The Riemann curvature tensor components in terms of the connection coefficients
are
\be
R^{\lambda}{}_{\alpha\mu\beta} = \frac{\partial \Gamma^{\lambda}{}_{\alpha\beta}}
{\partial x^{\mu}}-
\frac{\partial \Gamma^{\lambda}{}_{\alpha\mu}}
{\partial x^{\beta}}+
\Gamma^{\lambda}{}_{\sigma\mu}\Gamma^{\sigma}{}_{\alpha\beta}-
\Gamma^{\lambda}{}_{\sigma\beta}\Gamma^{\sigma}{}_{\alpha\mu}.
\label{Reqn}
\ee
Physically in a curved spacetime parallel transport of a vector
around a closed loop brings it back to the original point visibly changed.
The Riemann tensor is a measure of
the change. A 4 vector $A$ transported around a closed curve formed by
vectors ${\bf u}$ and ${\bf v}$ (and their commutator to close the gap)
suffers a change
\be
\delta  A^{\alpha} + R^{\alpha}{}_{\beta \gamma \delta}\; A^{\beta}\;u^{\gamma}\;v^{\delta} =0.
\ee
The Ricci tensor and scalar follow from
$R_{\mu\nu} = R^{\alpha}{}_{\mu\alpha\nu}$ and $R= g^{\mu\nu}R_{\mu\nu}$.
\end{itemize}

In contrast to the geometrical quantities above, the stress energy
tensor ${\bf T}$ contains information about the 
matter sources of gravity. The conditions of energy and
momentum conservation are contained in the relation ${\bf \nabla \cdot T}=0$.
The components of $T_{\mu\nu}$ are symmetric.

The equations that we must rely on to tell us how the system behaves are
the Einstein equations
\be
{\bf G }=8\,\pi\,{\bf T},
\label{gee}
\ee
where the geometry of spacetime that tells matter how to move is on the
left hand while the matter sources telling spacetime how
to curve are on the right hand. Thus ${\bf G}$ must only
be geometry dependent and so must be entirely made from the Riemann tensor
and the metric. For simplicity it is made linear in the Riemann tensor
and since it is an indication of curvature it must vanish in flat space.
Like the right hand side, it must be a dual tensor that is symmetric. 
The only second rank tensor we can build fulfilling these criteria
is (up to an overall constant)
\be
G_{\mu\nu} = R_{\mu\nu} -\frac{1}{2}R\;g_{\mu\nu}.
\ee
The factor $8\,\pi$ in (\ref{gee}) follows from comparisons to
well known situations in Newtonian
gravity ($g_{\mu\nu} \sim \eta_{\mu\nu}$). 

It can be shown that we can associate the $tt$ component of the
stress-energy
tensor $T^{00}$ with the density of mass energy, while $T^{j0}$
is the density of the $j$ component of the momentum.
The $00$ component of the Einstein equation is, therefore, called the
Hamiltonian constraint and the $0i$ components are the momentum
constraints.

The Einstein curvature tensor
${\bf G}$ has to have
zero divergence and from this the divergence of $T_{\mu\nu}$ follows.
Why this is so can be seen from the fact that a given spacetime
hypersurface has a metric given by
\be
ds^2=g_{\mu\nu}\;dx^{\mu}\;dx^{\nu}.
\label{lineadm}
\ee
Using the metric symmetries we get ten independent components of $g_{\mu\nu}$.
When the geometry evolves then, if we had ten independent
equations
\be
G_{\mu\nu}=8\,\pi\, T_{\mu\nu},
\label{Geq}
\ee
we could determine all ten $g_{\mu\nu}$ on a slice. This is not
possible since we have the freedom to choose our coordinates:
\be
x^{\mu} \rightarrow x^{'\mu},
\ee
and this transformation would keep the line element (\ref{lineadm}) unchanged,
but lead to the transformation  $g_{\mu\nu}\rightarrow {g'}_{\mu\nu}$.
Therefore four of the ten conditions must be eliminated so that we have only
six independent Einstein equations. Here is where the 
vanishing divergence of the
geometry independent
$T_{\mu\nu}$ plays a part. The equation
\be
T^{\mu\nu}{}_{;\nu} =0,
\ee
gives us the required four conditions that are subtracted from the ten equations leaving
six independent geometrical equations. 

In order to actually derive equation (\ref{Geq}), a variational
principle is applied with action written in the form
\be
I= \int\,{\cal{L}}\;d^4x = \int\,L\,(-g)^{\frac{1}{2}}\;d^4x.
\ee
 The Lagrangian density
is separated into a geometrical and a field part
\ben
{\cal{L}} = {\cal{L}}_{\mathrm geom} + {\cal{L}}_{\mathrm field} =
(-g)^{\frac{1}{2}}\,L,\\ \nonumber
L =L_{\mathrm geom}+ L_{\mathrm field}.
\een
One then extremizes this action. 
In the $ADM$ formalism,
one finds it convenient to think of the dynamics as
representing the evolution from the specified 3-geometries of an
initial spacelike to a final spacelike slice. The action integral is
to be extremized with regard to choice of spacetime between these faces:
\be
S \equiv S({}^3\xi),
\ee
with the action changing with 3-geometry according to
\be
\delta S = \int\,\pi^{ij}\;\delta g_{ij}\;d^3x.
\ee
Here $\pi^{ij} = \frac{\delta S}{\delta g_{ij}}$
is the field momentum conjugate to $g_{ij}$. Consider the simplest 
action principle one can imagine:
\be
L_{geom} = \left(\frac{1}{16\pi}\right)\;{}^4\!R.
\ee
Here ${}^4\!R$ is the four-dimensional scalar invariant. To be an invariant,
$L_{geom}$ must be independent of the choice of coordinate systems. In the
neighborhood of a point one can always choose coordinates such that the
first derivatives of $g_{\mu\nu}$ vanish. Therefore, apart from a constant,
no scalar invariants can be built homogeneously out of the metric coefficients and their
first derivatives. Unlike the case of mechanics one must, therefore,
consider second derivatives of the metric coefficients. One then has
20 distinct components of the curvature tensor with six parameters of
a local Lorentz transformation giving 14 distinct choices. Only, one of these,
${}^4\!R$, is linear in second derivatives of metric components.
\be
{}^4\!R =g^{\alpha \beta}\,R_{\alpha\beta},
\ee
where
\be
R_{\alpha\beta} = R^{\lambda}{}_{\alpha \lambda \beta}.
\ee
Every $\Gamma$ should be symmetric in its two lower indices giving a total
of $4 \times (3*2 +4) = 40$ independent components. In order for
the action to be an extremum, variations due to changes in
$g^{\mu\nu}$, as well as the $\Gamma$s' should vanish.
\be
0 =\delta I = \frac{1}{16\pi} \int \,\delta \left(
g^{\alpha\beta}\,R_{\alpha\beta}\,\sqrt{-g}\right)\;d^4x + \int\,\delta
\left[L_{\mathrm field}\,\sqrt{-g}\right]\; d^4x.
\label{geqn}
\ee
The $\Gamma$s themselves do not transform like tensors although their variations
$\delta \Gamma$ do.
\be
\Gamma^{\bar \gamma}{}_{{\bar \alpha}{\bar \beta}} = \left[
\Gamma^{\lambda}{}_{\sigma\tau}\;\frac{\partial x^{\sigma}}{\partial x^{\bar \alpha}}\;
\frac{\partial x^{\tau}}{\partial x^{\bar \beta
}} + \frac{\partial^2x^{\lambda}}{\partial x^{\bar \alpha}\partial
x^{\bar \beta}}\right]\frac{\partial x^{\bar \gamma}}{\partial x^{\lambda}}.
\ee
The last term destroys the tensor character of the $\Gamma$s but cancels
in $\delta \Gamma^{\lambda}_{\sigma\,\tau}$. From the fact that tensors
can be studied in any coordinate system, we conveniently choose
one in which the $\Gamma$s vanish at the point in question and so from
(\ref{Reqn}) (after replacing commas by semicolons for correct formulas in
any coordinate system)
\be
\delta R^{\lambda}{}_{\alpha\mu\beta} = \delta \Gamma^{\lambda}{}_{\alpha\beta;\mu}
-\delta \Gamma^{\lambda}{}_{\alpha\mu;\beta},
\ee
and
\be
\delta R_{\alpha\beta} =  \delta \Gamma^{\lambda}{}_{\alpha\beta;\lambda}
-\delta \Gamma^{\lambda}{}_{\alpha\lambda;\beta}.
\ee
Also
\be
\delta(-g)^{\frac{1}{2}} =\frac{1}{2}\,(-g)^{-\frac{1}{2}}\;(-g)\;\frac{\delta g_{\mu\nu}}{ g_{\mu\nu} }
= -\frac{1}{2}\,(-g)^{\frac{1}{2}}\;g_{\mu\nu}\;\delta g^{\mu\nu}.
\ee
Thus from (\ref{geqn}) we see that
\ben
\frac{1}{16\,\pi} && \int
\Bigg\{
g^{\alpha\beta}\,R_{\alpha\beta}\left(-\frac{1}{2}\, \sqrt{-g}\,g_{\mu\nu}\,\delta g^{\mu\nu}\right)
+ g^{\alpha\beta}\,\delta R_{\alpha\beta}\,\sqrt{-g}
\\ \nonumber
 &+&R_{\alpha\beta }\,\sqrt{-g}\,\delta g^{\alpha\beta}\Bigg\}\,d^4x
+ \int\,\delta \left(L_{\mathrm field}\,\sqrt{-g}\right)\;d^4 x,\\ \nonumber
&=&
\frac{1}{16\, \pi}\,\left[\int\,
\left(-\frac{1}{2}\,g_{\alpha\beta}\,R +R_{\alpha\beta}
+\left(\delta \Gamma^{\lambda}{}_{\alpha\beta;\lambda}-\delta
\Gamma^{\lambda}{}_{\alpha\lambda;\beta} \right)\right)
\sqrt{-g}\;\delta g^{\alpha
\beta} \; d^4 x\right]\\ \nonumber
&&\qquad+\int\,\delta\left(L_{field}\,\sqrt{-g}\right)\;d^4 x.
\een
The $\delta \Gamma$ term is evaluated as follows:
\ben
\frac{1}{16\pi}\, \int && g^{\alpha\beta} \,\left(\delta \Gamma^{\lambda}{}_{\alpha\beta;\lambda}-\delta
\Gamma^{\lambda}{}_{\alpha\lambda;\beta} \right)\,\sqrt{-g}\;d^4 x \\
\nonumber
&=& \frac{1}{16\, \pi} \, \int\,\Bigg(
+\left\{{ g}^{\alpha\beta}\,
\delta \Gamma^{\lambda}{}_{\alpha \beta} \right \}_{;\lambda}\,
-{g}^{\alpha\beta}{}_{;\lambda}\,\delta \Gamma^{\lambda}{}_{\alpha \beta} 
\\ \nonumber
 &-& \left\{{ g}^{\alpha\beta}\,
\delta \Gamma^{\lambda}
{}_{\alpha \lambda}\right\}_{;\beta}
+ { g}^{\alpha\beta}{}_{;\beta}\,\delta \Gamma^{\lambda}{}_{\alpha \lambda}
\Bigg)\,\sqrt{-g}\;d^4 x.
\een
Using the fact that the metric is covariantly constant
($g^{\alpha\beta}{}_{;\alpha}=0$) we get
\ben
\frac{1}{16 \, \pi}&& \int \, g^{\alpha\beta} \,\left(\delta \Gamma^{\lambda}{}_{\alpha\beta;\lambda}-\delta
\Gamma^{\lambda}{}_{\alpha\lambda;\beta} \right)\,\sqrt{-g}\;d^4 x \\
\nonumber
&=& \frac{1}{16 \, \pi}\, \int\,\left(
\left\{{ g}^{\alpha\beta} \,
\delta \Gamma^{\lambda}{}_{\alpha \beta} \right \}_{;\lambda}\,
- \left\{{ g}^{\alpha\beta} \,
\delta \Gamma^{\lambda}
{}_{\alpha \lambda}\right\}_{;\beta}
\right)\,\sqrt{-g}\;d^4 x.
\een
Using Gauss' theorem we get an integral over the surface. Since
variations vanish on the boundary this integral identically vanishes
leaving
\be
\frac{1}{16\,\pi}\, \left[\int\,
\left(-\frac{1}{2}g_{\alpha\beta}\,R +R_{\alpha\beta}
\right)\,\sqrt{-g}\,\delta g^{\alpha
\beta}\;d^4x\right] +
\int\,\delta\left(L_{\mathrm field}\,\sqrt{-g}\right)\;d^4 x =0.
\label{evalvar}
\ee
The stress-energy tensor is defined to be
\be
T_{\alpha\beta} \equiv 
\frac{2}{(-g)^{\frac{1}{2}}}\,\left(\frac{
{\delta {\cal L}_{\mathrm field
}}}{\delta g^{\alpha\beta}}\right)=
-\frac{2}{(-g)^{\frac{1}{2}}}\,\left(
\frac{ {\delta {\sqrt{-g}\, L_{\mathrm field
}}}
}{\delta g^{\alpha\beta}}
\right).
\label{variatmunu}
\ee
Using this definition 
\be
\int\,\delta
\left(L_{\mathrm field}\,\sqrt{-g}\right)\;d^4 x =
\int \, \delta 
g^{\alpha \beta} 
\frac{ {\delta ({\sqrt{-g}\, L_{\mathrm field
}})}
}{\delta g^{\alpha\beta}}\;d^4 x = -\frac{1}{2} \,\int\,
T_{\alpha\beta} \, \sqrt{-g} \, \delta g^{\alpha \beta}\;d^4x,
\ee
which gives the equations of motion
\be
G_{\alpha\beta} = 8\, \pi \, T_{\alpha\beta}.
\ee
If we now restrict attention to a matter Lagrangian that
depends on the metric but not metric derivatives
(this assumption holds for usual matter fields such as spinors, gauge bosons
and minimally coupled scalars but fails for non-minimally coupled
scalars), then
\be
G_{\alpha\beta} = 8\,\pi 
\left(g_{\alpha\beta}\,L_{\mathrm field} -2\, \frac{\partial L_{\mathrm field}}{\partial g^{\alpha\beta}}\right).
\ee

This then sets up the equations for general metrics and matter
sources in {\em GR}. We then specialize them to particular
metrics, for example, spherically symmetric spacetimes, and
a particular matter source that describes bosonic spin zero objects.
Since most of the work has been on boson stars we start by describing
what those are.

\section{What are Boson Stars?}

Boson stars are compact self-gravitating objects made up of scalars that
are held together by the balance between the attractive force of
gravity and the dispersive effects of a wave equation. The
scalar field is complex and satisfies the Klein--Gordon equation, and the system has 
field solutions with different numbers of nodes. The
lowest energy state, or ground state, is characterized by a field distribution
that has no nodes. Unlike the case of fermions, where one can invoke
the perfect fluid approximation and write an equation of state, a system
of bosons has an anisotropic distribution of stress near its center,
making such a concept inadmissible. Nevertheless, they
have many similar properties to neutron stars. 
For a review, see~\cite{balarev}.

\subsection{An Historical Perspective}

From an historical point of view the study of boson stars began with
the work of Kaup~\cite{kaupgeon}. Electromagnetic geons, as introduced by
Wheeler~\cite{wheeler} were gravitational-electromagnetic objects satisfying
the electromagnetic and gravitational field equations. Their spherical
configurations were found to be in unstable equilibrium.
Compelled by this concept,
Kaup introduced a Klein--Gordon geon, which satisfied the coupled
{\em KG} and Einstein equations (the {\em KGE} equations) in a spherically symmetric
metric. 
From the Lagrangian
\begin{equation}
L=R+ \gumunu \; \phi^{\ast}{}_{;\mu}\,\pdnu -\frac{m^2}{\hbar^2}\,\phi^{\ast}\,\phi,
\end{equation}
with scalar field $\phi$
and a spherically symmetric metric
\begin{equation}
d\tau^2=b^2dt^2
-r^2d\Omega^2
-a^2dr^2,
\end{equation}
is derived a scalar field equation (in appropriately defined units)
\begin{equation}
\Psi''+\left(\frac{2}{r}+\nu'-\lambda'\right)\Psi'+a^2\left(\frac{E^2}{b^2}-1\right)\Psi=0,
\label{bc1}
\end{equation}
where $\phi(r,t)=e^{i\,E\,t}\Psi(r)$, $a=e^{\lambda}$ and $b=e^{\nu}$.

Transforming to a new radial coordinate $s$ where
\begin{equation}
s=\int_{0}^{r}\,\frac{a}{b}\;dr,
\end{equation}
so that $ds/dr=a/b$, we get
\begin{equation}
\frac{d^2\Psi}{{ds}^2}+\frac{2}{r}\frac{b}{a}\frac{d\Psi}{ds}+\left(E^2-b^2\right)\Psi=0.
\end{equation}
This equation is similar to the radial Schr$\ddot{\text {o}}$dinger equation for a potential
$b^2$. In the
asymptotic region $b^2\sim 1-2\,M/r$ so the potential is similar to the
hydrogen atom case in that it has a $1/r$ dependence. This suggests countably infinite localized solutions for $\Psi$ with different numbers of nodes.
The conserved current associated with the $U(1)$ symmetry of the problem
is
\begin{equation}
J_{\nu}=\frac{1}{2}\,i\,(\Phi^{\ast}{}_{;\nu}\,\Phi-\Phi_{;\nu}\,\Phi^{\ast}),
\end{equation}
With the particle number times $m$ the mass of a single boson
\begin{equation}
N \,m = -i \int \,\sqrt{-g}\;J_0 = \int\, r^2\,a\,b\,J^4\;dr.
\end{equation}
The ground state solutions are then numerically obtained and a plot of
mass
and particle number profiles for different central densities is made.
The nature of the plot is shown in Fig.~2 (chapter 2).
%FFig Put mass profile figure here of yours. 
The mass of the bosonic geon
increases
to a maximum and then starts to decrease as does the particle number with
its maximum at the same value of central density. However $Nm$
is greater than the mass of the bosonic geon for central densities up to
a point beyond the maximum. 

This result is very interesting since the mass profile as a function
of central field density is very similar to that of neutron stars
and white dwarfs when plotted as a function of density $\rho$.
This is so, as pointed out in this paper, despite
the absence of an equation of state in the case of a the {\em KGE}
system where the radial pressure $-T^r_r$ is different from
$-T^{\theta}_{\theta} = -T^{\phi}_{\phi}$. Not only
are the profiles similar, but remarkably, so also are
the stability properties,
although the usual perfect fluid approach
to stellar stability can no longer be applied.

A first-order perturbation approach in spherical symmetry is
developed in Kaup's paper, whose
essential features are that the metric components are perturbed as
$\nu \rightarrow \nu+\delta \nu$ and $\lambda \rightarrow \lambda
+ \delta \lambda$ so that $\delta g_{rr}=-2\,a^2\delta\lambda$ where $\delta
\nu$ and $\delta \lambda$ are
 functions of
 $r$ and $t$. The scalar field perturbation is of the form
 \be
 \delta\Phi=e^{i\,E\,t}\,\left[R(r,t)+i\,I(r,t)\right].
 \ee

All time dependences of the perturbations are taken to be of the
form $e^{i \omega t}$ and solutions with imaginary $\omega$ or
negative $\omega^2$ are obviously unstable. A variational approach follows,
to establish the ground state $\omega$ and this gives an erroneous
result that all ground state solutions are stable. 

A Newtonian approach to the problem is presented in~\cite{ruf1}.
The nonrelativistic Schr$\ddot {\text{o}}$dinger equation with
a potential $V$ satisfying the Poisson equation $\nabla^2 V = -4\pi G \rho$ is used. A mass profile plot as a function of central density yields
a maximum mass of $1.653 \, {M_{Pl}^2}/{m}$, far greater
than the relativistic value of $0.633 \, {M_{Pl}^2}/{m}$ because of
the continued use of the Newtonian approach to dense configurations no
longer in the Newtonian regime. In~\cite{based1} an $N$-body Hamiltonian
for an $N$ boson configuration interacting via Newtonian gravity
is written. The hydrogen-like resulting Hamiltonian then yields 
a ground state expectation value by analogy.
A lower bound on the ground state energy (better by a factor of almost 3
than that of~\cite{ruf1})
is obtained by separating out the center of mass kinetic energy (which
corresponds to the global motion of the system and should not contribute
to the ground state energy). This is now much lower than
the relativistic result.
A semi-relativistic calculation with a 
Hamiltonian of the form
\begin{equation}
H=\sum_{i=1}^{N}\,{\left({\bf p_i^2}+m^2\right)}^{1/2}
 - \sum_{i<j=1}^{N}\,\frac{G\,m^2}{r_{ij}},
 \end{equation}
yields an answer of $1.51\, {M_{Pl}^2}/{m}$, lower than~\cite{ruf1}
but higher than the relativistic values. 

A Hartree approximation~\cite{membrado}
to the problem of $N$ identical bosons interacting
through two-body attractive Yukawa forces with force range $\frac{1}{\mu}$
(the gravitational case corresponding to $\mu=0$) shows bound
state solutions for $\frac{\mu}{N} <$ some critical value. In the Hartree
approximation the wavefunction of the $N$ particle ground state is
in terms of single particle minimum energy wave functions $|f>$ 
\begin{equation}
|\Psi_0> =|f>_1|f>_2.....|f>_N,
\end{equation}
with the Hamiltonian
\begin{eqnarray}
{\bf H} &=& {\bf T} + {\bf V}, \nonumber \\
{\bf T} &=& -\frac{{\hbar }^2}{2\,m}\,\sum_{i=1}^{N} \, {\bf \nabla}_i^2,
\nonumber
 \\
 {\bf V} &=& -g^2 \sum_{i>j=1}^{N}\, \frac{e^{-\mu|r_i-r_j|}}{|r_i-r_j|}.
 \end{eqnarray}
 This paper~\cite{membrado} posits a variational
 problem in the particle density $n(r)=N[f^\ast(r)f(r)]$ with energy
 minimized as a function of $n(r)$ under the normalization constraint
$\int\,d ^3{\bf r}\,n(r) = N$. For the gravitational case, the results
of~\cite{ruf1} for bound state energy are recovered.

With the particle number density decaying
slowly to infinity,
the velocity
profile of a test particle (calculated from $v(r) = \sqrt{ G M(r)/r}
= \sqrt{G \, m \,\int_{0}^{r} \, n(r')\, \fpi\, {r'}^2 \,dr'/r}$ by equating
centripetal
and gravitational forces) as a function of radius does not fall off as
steeply as the finite boundary fermion case.
One of the suggestions of this paper was, therefore, to
consider boson halos around luminous objects (such as galaxies) as a possibility. 

In the literature, there are also treatments of finite temperature boson
stars in the classical and quantum regimes~\cite{meraf1,ing1}
but our work has involved systems of a condensate of zero temperature bosons all
of which occupy the same state.

By far the most thorough analyses on the stability of boson stars
from a perturbative point of view have been by T.D. Lee {\em et.~al}. While
the previous papers have dealt only with boson stars in the ground
state T.D. Lee {\em et.~al}.~discuss excited state configurations as well.
In~\cite{friedlee} they discuss the cusp structure of the boson
star mass and particle number profile in great detail. This cusp structure
has important stability consequences. The equilibrium configurations
have a field with an underlying characteristic frequency $\omega$
so that $\phi = \sigma(\rho)\, e^{i\omega t}$.
The action for the system consists of a gravitational part and
a matter part given by:
\ben
A(g)&=&\int_{-\infty}^{+\infty}\,dt\,L(g),\\ \nonumber
A(m)&=&-\int\,\left[\phi^{\ast;\mu}\,\pdmu+U(\phi^{\ast}\,\phi)\right]\;d\tau,
\een
with volume element $d\tau=\sqrt{|g|}dt\,d\rho\,d\theta\,d\phi$. The
metric is taken to be of the form $ds^2=-e^{2u}dt^2
+{\rho}^2d\Omega^2
+e^{2v}d{\rho}^2$
with matter Lagrangian
\be
L(m)=4\pi\int\,\left(-U-V+W\right)\,e^{u+v}\,\rho^2\;d\rho.
\label{frieb}
\ee
The total Lagrangian $L ={\cal K}-{\cal V}$ has components
\be
{\cal K}=4\,\pi\int_{0}^{\infty}\,\eupv\,\rho^2\,W\;d\rho
=2\,\pi\int_0^\infty\,\evmu\,\rho^2\,\dtphi^{\ast}\,\dtphi\;d\rho,
\label{fridama}
\ee
where spherical symmetry ensures no kinetic energy is carried off by
the gravitational field and
\ben
{\cal V}&=&V(m)+V(g)
=4\,\pi\,\int_0^\infty\,\eupv\,\rho^2\,\left(U+V\right)\;d\rho
-L(g)
\\ \nonumber
\text{with}\quad U&=&\frac{1}{2}m^2\sigma^2+\frac{1}{4}f^2\sigma^4 \quad\\
\nonumber
\text{and} \quad V&=&\frac{1}{2}e^{-2\,v}{(\frac{d\sigma}{d\rho})}^2 .
\een
The Hamiltonian is then
\be
H={\cal K}+{\cal V}={\cal K}+V(m)-L(g)=2\,{\cal K}-L(m)-L(g)=2\,{\cal K}-L.
\ee

The particle
number  $N=\int\,j^0{|g|}^{\frac{1}{2}}dx_1dx_2dx_3$ is derived
from the conserved current discussed 
earlier in the context of~\cite{kaupgeon}( 
$j^0 = -\frac{i}{2}\left(\phi^{\ast}
\dtphi - \dtphi^{\ast} \phi\right) = \omega \sigma^2$)
giving
\ben
2\,{\cal K}=N\,\omega \quad \text{and} \quad E=N\,\omega-L
\\ \nonumber
\text{and hence}\quad \frac{dM}{dN} = \left[\frac{\partial E}{\partial N}
\right]_{u,v,\sigma}
= \omega.
\een
The equilibrium configurations are determined numerically from
the Einstein equations, with mass determined from the Schwarzschild nature
of the asymptotic region: namely $GM = \lim_{\rho \to \infty}\rho\,u$. A
plot of the mass versus particle number has slope $\omega$. For
any number of nodes $n$ it increases up to a point $(M(n,1),N(n,1))$ where
it has a cusp and abruptly changes course till $M(n,2),N(n,2)$ and then
another cusp causes it to change course again till a third cusp and so on.
The sign of $dN/d\omega$ changes when a cusp is passed and the function
$M(N)$ has a cusp therefore at
$\frac{dN}{d\omega}=\frac{d^2}{d\omega^2}L=0$.
% Since $N=\frac{dL}{d\omega}$
At the cusp, in fact,
\be
\frac{dN}{d\omega}=\frac{dM}{d\omega}=0.  
\ee
This change of sign at the cusp is an important part of stability
discussions in~\cite{leepert} where a perturbation technique is used.

In their stability analysis the
perturbed
wave function is expressed in terms of the radial part of the
unperturbed wave function as
\be
\phi =\frac{1}{\sqrt{2}}e^{-i\theta(t)}\left[\sigma(\rho)+
\chi_R(\rho,t)+i\chi_I(\rho,t)\right],
\ee
where the unperturbed field function satisfies the Klein--Gordon equation
written for brevity here in operator form ${\bf h}\sigma=0$.
The $\chi$ functions are expanded in terms of a complete set of orthonormal
radial basis functions $f_i$ such that
\be
\chi_R(\rho,t)=\Sizinf \, R_i(t)\;f_i(\rho)\quad \text{and}\quad
\chi_I(\rho,t)=\Sioinf\, I_i(t)\;f_i(\rho).
\ee
The perturbed Hamiltonian 
and Hamilton's equations
are written down.
The nature of the eigenvalues of $R$ and $I$ hold the key to stability.
For an $n$ node state it is shown that of the eigenvalues $\lambda_a$ of the
basis functions $f_a$ under operator ${\bf h} $,
$n$ of them are negative using the Sturm-Liouville theorem. For
a ground state wave function, therefore, all these are positive. The sign of these
eigenvalues determines the nature of the squared eigenvalues of $R$ and
$I$, being real if $\lambda$ is positive and imaginary if not. For
a ground state star, therefore, the squared eigenvalues
of $R$ and $I$ are real. To
set about finding out whether the eigenvalues themselves
are real (if the squared eigenvalues are positive) or imaginary indicating stability or instability respectively,
a simplifying trick is used. The change of sign only occurs at the
cusp hence, if any configuration on the first branch is stable, all
configurations on that branch must be so. Thus a Newtonian configuration
on that branch is considered, thereby simplifying the equations
and it is shown that first branch of ground state boson stars
are stable. For excited states though, all those configurations are
unstable. This analysis for stability was done without including any
self-coupling terms in the matter Lagrangian.

The need to incorporate the self-coupling term is mostly motivated by
astrophysical considerations. A single boson star, also called a mini-soliton
star, is not very large. Although, there are no limits on how many of them
there could be it would be nice to be dealing with large objects.
The radius of the star held up from collapse by the Heisenberg uncertainity
principle is $p \sim 1/R$ which for a relativistic star gives
$R \sim 1/m$ where  $m$ is the mass of a boson. The mass of the star is
of order $M_{Pl}^2/m$, much smaller than a fermion star of mass of order
$M_{Pl}^3/m_F^2$. To get a physical dimension on this, a $1 GeV$ boson mass
gives us a boson star mass of $ M \sim 10^{-19} \msun $.
The importance of the self-interacting term is measured by the
ratio $V(\phi)/m^2|\phi|^2=\lambda | \phi |^2/m^2$.  The energy density of a
system without self coupling is given by $\rho \sim M/R^3\sim M_{Pl}^2m^2$.
Since the energy density of non-interacting bosons is given by
$\rho\sim m^2|\phi|^2$ this gives $|\phi| \sim M_{Pl}$ and hence the
self interaction may be ignored if $\lambda << m^2/M_{Pl}^2$. The
mass of the system with self-coupling is of the order
$M \sim \sqrt{\lambda}M_{Ch}$ so that even a small self-coupling parameter
can significantly increase the sizes of these objects as discussed in~\cite{colp1}. 

In this paper~\cite{colp1}, the equilibrium configurations are set up (with
self-coupling term) in a spherically symmetric spacetime. The
usual mass profile (as a function of central field density) for
ground-state boson stars with different values of the dimensionless
self-coupling parameter $\Lambda = \frac{\lambda}{m^2 \fpi\, G}$
is plotted. The maximum mass of these profiles steadily increases with
increase in $\Lambda$. In the limit of large $\Lambda$ it is
shown that one can write an effective equation of state, and the
perfect fluid theorems can be used to study their stability, indicating
that the branch to the left of the maximum in the mass profile
are stable and those to the right are unstable. 

This effective equation of state, and vanishing of anisotropies, makes these
high $\Lambda$ boson stars very much like neutron stars. They have a
well defined radius with their wave character suppressed. This suggests
that one might be able to use the self-coupling parameter as
an anisotropy control parameter~\cite{gleianisot}. In realistic systems,
one
expects some anisotropies to be present in systems.
These anisotropies could affect surface red-shifts and
critical masses~\cite{bow1}. Usually one puts them in by hand in neutron star
models in an ad-hoc manner. With self-coupled boson stars this can be done
more naturally. Interestingly the fractional anisotropy $(T^{r}_{r}-
T^{\theta}_{\theta})/T^r_r$ at the ``radius'' of the star
\be
R \equiv  \frac{\fpi\int_{0}^{\infty} \, r\,\rho \, r^2 \; dr}{\fpi\int_{0}^{\infty} \,
\rho \, r^2 \; dr}= \frac{\fpi}{M}\int_{0}^{\infty} \, r\,\rho \, r^2 \; dr,
\ee
is roughly the same for all central densities.

\subsection{Boson Stars as Dark Matter Candidates}

In this section we discuss why boson stars, although they have never been seen, might
indeed really exist. To do so we introduce the notion of dark matter and the
evidence for this kind of matter. We then show why
bosons themselves might be good candidates for this kind
of matter. We then describe by what mechanism these bosons could
form compact self-gravitating objects.

\subsubsection{Evidence of Dark Matter}

Almost all the information about the Universe has been obtained through photons
(radio photons from neutral hydrogen gas, X-ray photons from ionized gas,
optical photons from stars etc.)~\cite{scottrem,peebles}.
What about matter that does not emit
detectable radiation? For example, the star formation process could result in
extremely dim main sequence stars with masses below the lower limit for hydrogen burning 
on the main sequence ($0.08 \msun$) which could only be detected at 
distances of less than $1 pc$ with present instruments.

Even within a given type of astronomical object, there is no
reason to expect mass and luminosity to be perfectly correlated.
Stars fainter than the sun make up about $75 \%$ of the mass, while
stars brighter than the sun account for $95 \%$ of the luminosity in
the solar neighborhood. It is found that most astronomical
systems have extremely high mass to light ratios.

Dark matter is defined to be matter whose existence we know of only
because of gravitational effects. Objects
like white dwarfs in the solar neighborhood
are not dark matter candidates as their existence is inferred from present densities
of visible white dwarfs, theories on stellar evolution, and the history of
star formation rates in the solar neighborhood.

Zwicky in 1933 was the first to suggest that dark matter existed. His work,
based on the measurement of radial velocities of 7 galaxies in the Coma cluster,
showed that individual galaxies had radial velocities differing from the mean
velocity of the cluster, with an RMS dispersion of $700 km s^{-1}$. This 
dispersion was taken to be a measure of the kinetic energy per unit mass
for the galaxies in the cluster and a crude estimate of the radius
of the cluster then provided a measure of the mass of the cluster through
the virial theorem. The mass to luminosity ratio based on this
model was found to be almost a factor of $400$ times larger than the
mass to luminosity ratio calculated from the measurement of rotational
curves of the nearby spirals. He concluded that virtually all the cluster mass was in the form of some invisible or dark matter undetectable except by its
gravitational force. Although based on uncertain cluster radius and distance
scales with minimal statistics, it was nevertheless a fairly accurate result.
The mass to light ratio of the Coma cluster as a whole exceeds the mass to
light ratio of luminous parts of typical galaxies in the cluster by more 
than a factor of 30.

Ostriker and Einasto {\em et al.} proposed in 1974 that even isolated galaxies had
large amounts of dark matter around them with spiral galaxies having dark halos
several times the radius of luminous matter. We have studied one
such halo model made up of bosonic matter. 

Since boson models have been used to fit rotation curves,
and because scalar field cosmology is much discussed, we present a few
dark matter scenarios in these contexts.

\begin{itemize}
\item{Galactic Rotation Curves}
\end{itemize}

From emission lines in HII regions and the 21 $cm$ emission line of neutral
hydrogen, the rotation curves of galaxies are optically traced. Using the fact that
the gravitational force provides the centripetal force
\be
\frac{v^2}{r}= \frac{GM(r)}{r^2},
\ee
one sees that in the inner region of roughly constant density, where
$M(r) = \rho \, \frac{4}{3} \pi r^3$, the speed $v(r)$ should rise linearly with distance
$r$ from the center. An intermediate region should exist, where the
speed reaches a maximum and starts to decline, until
the outer {\em Keplerian} region where the system acts like a point mass concentrated at
the center, so $v \propto r^{-1/2}$. 

However, most of the rotation curves of the over 70 spiral galaxies 
studied are either flat or slowly rising up until the last point measured.
The few that have falling rotation curves either fall off at a lower than
Keplerian rate, or have neighbors that may perturb their velocity profiles.
This absence of the Keplerian region indicates the absence of well determined
masses of galaxies, even for those galaxies whose rotation curves extend to large enough
radii to contain all the light. Thus there are no spiral galaxies
with accurately determined total mass. This suggests the presence of a massive
dark halo extending beyond optical radii. A flat rotation curve (constant
velocity) would indicate halo masses increasing linearly with radius out
to radii beyond the last observed point.
\begin{itemize}
\item{Groups of Galaxies}
\end{itemize}

Collections of galaxies with separation distances much smaller than
typical intergalactic separations would seem to indicate a gravitational
binding between component galaxies.

If a large number of stars are in a potential $\phi({\bf x},t)$ at a given time  $t$ the number of stars in a given volume $d^3x$ with velocities in
the range $d^3{\bf v}$ centered on ${\bf v}$ is given by
\be
f({\bf x},{\bf v},t)\; d^3{\bf x}\; d^3{\bf v}
\ee
where $f$ is the distribution function.
The density of stars must satisfy a continuity equation since the drift
of stars must be star conserving. Hence
\be
\frac{\partial f}{\partial t} + \sum_{i=1}^{3}\, \left(v_i\, \frac{\partial f}{
\partial x_i} + \dot{v_i}\,\frac{\partial f}{\partial v_i}\right).
\label{jean}
\ee
On integrating the above equation over all velocities 
(after multiplying them by $v_j$ and replacing
acceleration with the negative gradient of the potential) we get
\be
\int \,\frac{\partial f}{\partial t}\, v_j \; d^3 {\bf v} +
\int \,v_i\, v_j\, \frac{\partial f}{\partial x_i}\; d^3 {\bf v} -
\frac{\partial \phi}{\partial x_i} \int \,v_j \, \frac{\partial f}{\partial v_i}\; d^3 {\bf v} = 0.
\ee
Using the divergence theorem, and the fact that no stars are moving infinitely
fast, the last integral in the above equation when integrated by
parts can be modified to
\be
\int \,v_j \, \frac{\partial f}{\partial v_i}\, d
^3 {\bf v} = -\int\,\frac{\partial v_j}{\partial v_i}\, f\, d^3{\bf v} 
= -\int\,\delta_{ij}\, f\; d^3{\bf v}.
\ee

 Also, since the
velocity and the coordinate are independent, the partial $x$ derivative
in the second term can be brought outside the integral. This gives
\be
\frac{\partial \rho\, \bar {v_j}}{\partial t} + \frac{\partial (\rho \,
{\overline {v_i\,v_j}})}{\partial x_i} + \rho\, \frac{\partial \phi}{\partial x_j}
=0,
\ee
where $\rho = \int\,fd^3{\bf v}$ with $\bar {v_i} =
\frac{1}{\rho}\int\,fv_id^3{\bf v}$.
Multiplying the above equation by $x_k$ and integrating over all space variables, we get
\be
\int\,x_k\frac{\partial \rho \bar {v_j}}{\partial t}d^3{\bf x} + 
\int\, x_k\frac{\partial (\rho \,
{\overline {v_i\,v_j}})}{\partial x_i}\;d^3{\bf x} +
\int\, \rho\,x_k \, \frac{\partial \phi}{\partial x_j}\,d^3{\bf x}
=0,
\label{vir1}
\ee
In a steady state, the first term vanishes,
and one gets the tensor virial
theorem
\be
2\,K_{jk} +W_{jk} = 0,
\ee
where the kinetic energy tensor $K$ is 
\be
K_{jk} =  \frac{1}{2}\int\, \rho \, {\overline {v_j\, v_k}}\; d^3{\bf x}.
\ee

The kinetic term was obtained by integrating the second integral of
(\ref{vir1}) by parts.
The potential energy tensor is given by the third
integral in (\ref{vir1}). In order to see this consider the $W$ tensor
\be
W_{jk} = -\int \, \rho (\bf x) \, x_k \, \frac{\partial \phi}{\partial x_j}
\; d^3{\bf x}.
\ee
Substituting for $\phi$ in terms of $\rho$, we get
\be
W_{jk} = G\,\int \, \rho (\bf x) \, x_k \, \frac{\partial }{\partial x_j}
\,\int \, \frac{\rho({\bf x'})}{{\bf |x-x'|}} \; d^3 {\bf x'} \; d^3{\bf x}.
\ee
Differentiating through the second integral ($\rho$ is a function of the
prime coordinate and is unaffected by a derivative in the unprimed one),
we get
\be
W_{jk} = G\,\int \, \int\, \rho (\bf x) 
\, \rho({\bf x'})\,
\frac{x_k \, (x_j-{x'}_j)}{{\bf |x-x'|^3}} \; d^3 {\bf x'} \; d^3{\bf x}.
\ee
Exchanging the dummy coordinates (${\bf x}$ and ${\bf x'}$) and adding the
result to the above integral, we get
\be
W_{jk} = -\frac{1}{2}\,G\,\int \, \int\, \rho (\bf x) 
\, \rho({\bf x'})\frac{({x'}_k -x_k)\, ({x'}_j -x_j)}{{\bf |x-x'|^3}} 
\; d^3 {\bf x'} \; d^3{\bf x}.
\ee
Taking the trace of both sides of this equation, we get
\ben
W &=& -\frac{1}{2} \, G \, \int \, \rho ({\bf x}) \, \int \, \frac
{\rho (\bf x')}{{\bf |x-x'|}} \; d^3 {\bf x}\; d^3 {\bf x}
\\ \nonumber
  &=& \frac{1}{2} \, \int \, \rho({\bf x})\, \phi({\bf x})\; d^3 {\bf x}.
\een
By considering the change in potential energy of a system, whose
density and potential are $\rho({\bf x})$ and $\phi(x)$,
when we bring
in a small mass from spatial infinity to position ${\bf x}$, we can show that
the above expression for $W$ is one of many alternative expressions for the potential
energy of the system. So we have a scalar virial theorem 
\be
2K + W = 0.
\ee

Consider a group of $N$ galaxies with masses $M_i$. Then by
the virial theorem
\be
\sum_{i=1}^{N} \, M_i\,<v_i^2>_t \,=\, \sum_{i=1}^{N}\sum_{j<i}\,
G\,M_i\, M_j \left<\frac{1}{|{\bf r_i}-{\bf r_j}|}\right>_t,
\ee
where averages are time averages. Assuming that the mass to light
ratios of each member galaxy is the same $\gamma = M_i/L_i$ and
writing the above equation in terms of luminosity, we get
\be
\gamma = \frac{3\,\pi}{2\,G} \,\frac{\sum_{i=1}^{N}\, L_i \,<v_{p,i}^2>_{t,\Omega}}
{\sum_{i=1}^{N}\sum_{j<i}\,
L_i \,L_j \,\left<\frac{1}{|{\bf R_i}-{\bf R_j}|}\right>_{t,\Omega}},
\ee
where we have replaced $v_i^2$ by $3<v_{p,i}^2>_\Omega$, where $v_p$ is the
line of sight component (which should have the same mean square value
as the orthogonal components). Also, ${\bf R_i}$ is the
projection of ${\bf r_i}$ onto the plane of the sky, and one
can  write the average $\frac{1}{|{\bf r_i}-{\bf r_j}|}$ as $2/\pi$
times the average $\frac{1}{|{\bf R_i}-{\bf R_j}|}$ by averaging over
the angles. Although all the temporal and angular quantities cannot
be measured, if one has a large number $N$ of galaxies, then for
galaxy orbits of random phases and orientations the observable quantity
\be
\gamma_{est} = \frac{3\,\pi}{2\,G} \,\frac{\sum_{i=1}^{N}\, L_i\, v_{p,i}^2}
{\sum_{i=1}^{N}\sum_{j<i}\,
L_i\, L_j \,\frac{1}{|{\bf R_i}-{\bf R_j}|}},
\ee
should approach $\gamma$ (and should be a good estimator for $\gamma$).
If there is dark matter present then this estimate
would be lower than the actual value. By estimating
$\gamma$ using the virial theorem, taking into account a dark
matter distribution of the same order as the galactic spatial distribution,
(note that this would still be smaller than the actual matter distribution
if there were a more extensive dark matter distribution), Huchra and Geller found
a median value of $260h\msun$ in the visible band
for their groups. The mass to light
ratios for groups was seemingly
much larger than values seen in the luminous parts
of galaxies. This suggests again the presence of large amounts of
dark matter.
\pagebreak
\begin{itemize}
\item{Dark Matter Cosmology}
\end{itemize}
%FFig fillin
%k <1 open universe should be discussed when u do our halo model.

Estimating a lower limit on the density of the universe using the Einstein cosmological
model, and comparing it to the measured luminosity density, indicates
there is far more dark matter than visible matter.

The universe on a large scale appears to be very homogeneous and isotropic,
so much so that the small scale anisotropies might be considered as
perturbations on a homogeneous background. In the idealized version,
considering total homogeneity and isotropy of spatial geometry
in Einstein's theory  of gravity
$G_{\mu\nu} = 8\pi T_{\mu\nu}$, if we assume a boundary condition of closure
a $three$ $sphere$ would satisfy all conditions.
The spatial metric of a three sphere can be built up step by step starting
from a $1$ $sphere$. Visualized as embedded in an Euclidean space of one
higher dimension, $S^1$ satisfies $x^2+y^2 = a^2$, which in polar coordinates
transforms into $x=a \cos \phi$ and $y=a\sin\phi$, giving
the metric $d\sigma^2 = a^2 d\phi^2$. A 2 sphere $S^2$,
$x^2 +y^2 +z^2 = a^2$, under the transformation $ x= a\sin\theta\cos \phi$,
$y =a \sin\theta \sin\phi$, and $z= a \cos \theta$ gives a spatial metric
$d\sigma^2 = a^2(d\theta^2 + \sin^2\theta d\phi^2)$. A 3 sphere
$S^3$, given by $ x^2 + y^2 + z^2 + w^2= a^2$, under the
transformation $x= a\sin \chi \sin\theta \cos \phi$, $y = a
\sin \chi \sin\theta \sin\phi$, $z= a \sin \chi \cos \theta $, and
$w = a \cos \chi$, therefore, gives $d\sigma^2 = a^2 (d\chi^2
+ \sin^2 \chi (d\theta^2 + \sin^2\theta d\phi^2))$. Hence, the 
spacetime geometry is described by
\be
ds^2 = -dt^2 + a^2 \,\left(d\chi^2
+ \sin^2 \chi \,(d\theta^2 + \sin^2\theta \,d\phi^2)\right).
\ee
Calculating the connection coefficient
symbols $\Gamma$ as described above, and the 
Riemann tensor from them, one can calculate
Ricci tensor and Ricci scalars. The $tt$ component of the Einstein equation
is
then
\be
\frac{3}{a^2} \,\left(\frac{da}{dt}\right)^2 + \frac{3}{a^2} = 8 \,\pi\, \rho.
\label{expand}
\ee
Where $T_{tt} = \rho$. 
From red shift measurements and distance measurements
one can calculate the ratio of the velocity of recession of a
galaxy, to the distance to the galaxy, which should equal the ratio of the
rate of increase in the radius of the universe to the radius of the 
universe. This is the Hubble parameter today:
\be
H_0 = \frac{1}{a} \,\frac{da}{dt}{\Bigg|_{\mathrm now}}.
\ee
From this and (\ref{expand})
\be
\rho >  \frac{3}{8\,\pi\, a^2} \,\left(\frac{da}{dt}\right)^2 = \frac{3}{8\,\pi}
\,{H_0}^2.
\ee
From a Hubble time today of $\frac{1}{H_0}$ (Hubble expansion rate
$H_0 \sim 55 km/sec/Mpc$) one gets a lower limit on
the density of $\rho_c = 5\times 10^{-30} g/cm^3$ as compared to an observed
luminosity density of $\rho_l \sim  2\times 10^{-31} g/cm^3$. A large
amount of dark matter must be present.

\subsubsection{Bosons as Dark Matter Candidates}

There are suggestions that dark matter is mostly of a non-baryonic nature.
About $10^{-4}$ secs after the Big Bang, at a black body temperature
of $10^{12}K$, nuclear reactions like $n+p\leftrightarrow d +\gamma$ were in equilibrium.
With many high energy photons present, the deuterons were as likely to be
disassociated as formed, keeping the net deuterium density low. Once
the temperature cooled to about $10^{10}K$, the photon energy ($kT$)
was no longer high enough to disassociate the deuterons. The primordial
deuterium densities were ``frozen in'' and therefore, became a measure of baryon density then and now (baryon number conservation). However, whatever
deuterons formed rapidly burned into {\em He}${}^3$, {\em H}, 
{\em He}${}^4$, and {\em Li}${}^7$;
with {\em He}${}^4$, having the most binding energy, dominating. The higher the density
of baryons, the faster the deuteron producing reactions; and a higher
density of deuterium at the epoch of nucleosynthesis would cause more of it
to burn off to helium. The observed abundances of
deuterium (which is really a lower limit on primordial abundances)
therefore puts a limit on the contribution of baryons to the mass density
of the universe.
%From abundances of primordial deuterium, $He^4$, $He^3$ and $Li^7$ 
The present
mass density in baryons calculated comes out to less than $6\%$ of
$\rho_c$. 

As a candidate for structure formation in the universe, bosons are in many
ways suitable candidates as suggested by~\cite{madpower}.
Spontaneous fluctuations in the pre-inflationary
epoch could have been greatly magnified by inflation, producing regions
slightly denser than their surroundings which were amplified by gravity to
set up the coalescence to present day structures. These fluctuations, or
matter density variations over the range of clusters of galaxies, can be
measured. The square of the density fluctuation strength multiplied by
the volume over which they are sampled provides a power spectrum.
If one plots the power spectrum against the length scale over which
fluctuations are detected for models dominated by cold dark matter and
hot dark matter, one finds the former is unable to explain large scale
structure
while the latter is unable to explain small scale structure. Hot dark 
matter candidates have low masses and large random velocities. Cold dark
matter candidates have large masses and low velocities.
However, low mass bosons which follow Bose-Einstein statistics
also have a large number of particles with low velocities
unlike neutrinos which follow Fermi-Dirac statistics. These are capable
of keeping the power peak for large scale structures as well as having enough
power on small scales.

Besides this, many particle physics and cosmological models rely on
the presence of scalars. These scalars have never been seen experimentally
and are strong
dark matter candidates. The actual formation of a boson star out of these
scalars
must rely on a Jeans instability mechanism described in the next subsection.
There are various dark halo models made up bosonic objects that have been
used to fit rotation curves. We have investigated the stability and
formation of one such model in the context of a  cosmological
model universe that is not closed.

\subsubsection{Formation of Compact Objects---The Jeans Instability Mechanism}

Although many particle physics and cosmological models predict the 
existence of scalars, the question arises as to how these
scalars could come together to form compact objects like boson stars.
In this regard, we discuss the Jeans instability mechanism.
From a homogeneous background of
matter, local fluctuations can grow in
time and cause clustering of matter.

Unlike plasmas that have both positive and negative charges, so as to
be neutral on large scales, an always attractive gravitational force
prevents gravitational systems from being in static homogeneous equilibrium.
In principle an infinite homogeneous gravitational system in equilibrium is
impossible.
If density and pressure are constant and mean velocity ${\bf v_0}$ is
zero, then Euler's equation
\begin{equation}
\frac{\partial {\bf v}}{\partial t} + {\bf \nabla}\cdot{\bf v} =
-\frac{1}{\rho}\, {\bf \nabla}p - {\bf \nabla}\phi,
\end{equation}
leads to ${\bf \nabla} \phi_0 =0$. However Poisson's equation gives
${\bf \nabla^2}\phi_0= \fpi\,G\,\rho_0$ clearly in contradiction unless
$\rho_0=0$. In a homogeneous gravitational system, there
are no pressure gradients to balance the existing gravitational attraction.
In order to construct an infinite homogeneous gravitational system 
the ``Jeans effect'' is used.

The conditions of the Jeans effect invoke Poisson's equation only
when the perturbed density and potential are involved~\cite{scottrem,peebles},
while the
unperturbed potential is assumed to be zero ($\phi_0=0$; $\nabla^2 \phi=\fpi
\,G\,(\rho-\rho_0)$). In uniformly rotating
homogeneous systems, where one has centrifugal forces in place of
pressure gradients to balance the equilibrium gravitational field, no
Jeans effect is necessary. So a homogeneous system can be in static equilibrium in a rotating frame.
\begin{itemize}
\item{Physical Basis of the Jeans Instability}
\end{itemize}
Consider an infinite homogeneous fluid of density $\rho_0$ and pressure $p_0$
with no internal motions so that ${\bf v_f}=0$. Now draw a sphere of radius
$r$ around any point and compress this region by reducing the volume from
$V$ to $V-\alpha\,V$. Thus the density $\propto 1/V$ is perturbed by an amount
$\rho_1 \sim \alpha\,\rho_0$, and as a result there is a pressure perturbation
$p_1 = (dp/d\rho)_0\rho_1 = v_s^2\alpha\rho_0$. 
The pressure force per unit mass is ${\bf F}_p = -{\bf \del}p/\rho$,
and so there is an extra pressure force of magnitude
$|{\bf F}_{p_1}| = |{\bf \del} p_1/\rho_0| \sim p_1/(\rho_0\,r)\sim
\alpha v_s^2/r$ (assuming $p_1$ is of the form 
$ r^n$ then ${\bf \del} p_1$ goes as $p_1/r$). Also because of the increased
density there is an extra gravitational force 
${{\bf F}_G}_1 =
-{\bf \del}\phi_1$. Here $\phi_1 = G\,M/(r-\delta r) -G\,M/r\sim G\,M \,\delta r/r^2$. Since $\delta V =  -\alpha V $ implies $\delta r \sim -\alpha r$
therefore ${\bf F}_{G_1} \sim G\,M\,\alpha/r^2$ 
with $M = \frac{4\,\pi}{3}\rho_0\,r^3$.
Hence the net force is ${\bf F}_{p_1} +
{\bf F}_{G_1}$ which if outward, implies the compressed fluid reexpands and
the perturbation is stable. If the fluid continues contracting then the
perturbation is unstable with the gravitational force larger than the
pressure force. This happens if
\be
G\,\rho_0 \,r\,\alpha > \alpha\,v_s^2/r,
\ee
or
\be
r > \sqrt{\frac{v_s^2}{G\,\rho_0}}.
\ee
Perturbations on a scale larger than this are unstable.

In the case of stellar systems stability can be discussed in the context
of this mechanism.
The density of states $f$ satisfies
\be
\frac{\partial f_1}{\partial t} + {\bf v}\cdot\frac{\partial f_1}{\partial {\bf x}}
-{\bf \del}\phi_1 \cdot \frac{\partial f_0}{\partial {\bf v}} =0.
\ee
Here we have made use of the time independence and homogeneity of
the unperturbed density $f_0({\bf x},{\bf v},t) = f_0({\bf v})$ as well
as the Jeans effect $\phi_0=0$.
Poisson's equation gives
\be
\nabla^2 \phi_1 =\fpi\,G\,\int\,f_1\;d^3{\bf v}.
\ee
The solution is $f_1({\bf x}, {\bf v}, t) = f_a ({\bf v})\,\exp[i
({\bf k\cdot x}-\omega t)],\quad {\bf \phi_1}({\bf x},t)= \phi_a
\exp[i({\bf k\,\cdot \,x}-\omega t)]$ where 
\ben
({\bf k\cdot v}-\omega)\,f_a - \phi_a\,{\bf k}\,\cdot\,\frac{\delta f_0}{\delta {\bf v}}=0\\
-k^2\,\phi_a =\fpi\,G\,\int\,f_a\;d^3{\bf v}.
\een
Noting that $\phi_1$ is only a function of ${\bf x},t$ and not ${\bf v}$
and substituting for $f_a$ from above into the equation for $\phi_a$, we
get
\be
-k^2 = \fpi\,G\,\int\,
\frac{{\bf k}\,\cdot\,\frac{\partial f_0}{\partial{\bf v}}}{{\bf k\,\cdot\,v}-\omega}\;
d^3{\bf v}.
\label{jeank}
\ee

Consider $f_0$ to have a Maxwellian velocity distribution
\be
f_0({\bf v}) =\frac{\rho_0}{(2\,\pi\,\sigma^2)^{\frac{3}{2}}}\;e^{-\frac{v^2}{2
\sigma^2}}.
\ee
Taking the x direction to be the direction of ${\bf k}$, we get
\be
1-\frac{2\sqrt{2\pi}\,G\,\rho_0}{k\,\sigma^3}\,\int_{-\infty}^{\infty}
\,\frac{v_x\,e^{-\frac{1}{2}v_x^2/\sigma^2}}{k\,v_x-\omega}\;dv_x =0.
\label{jeank2}
\ee
where $\rho_0$ is the density. For $\omega =0$ equation (\ref{jeank}) becomes
\be
k^2(\omega=0) = k_J^2= \frac{\fpi\,G\,\rho_0}{\sigma^2}.
\ee

Instability corresponds to imaginary $\omega$, as
can be seen from the form of $f_1$.
Set $\omega = i \gamma$ where $\gamma$ is real and positive.
After multiplying and dividing by $k\,v_x + i\,\gamma$,
(\ref{jeank2}) becomes
\be
1-\frac{2\,\sqrt{2\,\pi}\,G\,\rho_0}{k\,\sigma^3}\,\int_{-\infty}^{\infty}
\,\frac{k\,v_x^2\,e^{-\frac{1}{2}v_x^2/\sigma^2}}{k^2\,v_x^2+\gamma^2}\;dv_x =0,
\ee
where the term with $i \,\gamma$ in the numerator is odd and
hence vanishes (and so has not been written).
Using
$\int_{0}^{\infty}\,x^2\exp(-x^2)\,dx/(x^2+\beta^2) =
\sqrt{\pi} -\pi\,\beta\,\exp(\beta^2)[1-\erf(\beta)]$ gives
\be
k^2 = k_J^2 \,\left\{ 1 -
\frac{\sqrt{\pi}\,\gamma}{\sqrt{2}\,k\,\sigma}\,\exp{\left(\frac{\gamma^2}{2\,k^2\,\sigma^2}
\right)}\left[1-\erf\left(\frac{\gamma}{\sqrt{2}\,k\,\sigma}\right)\right]\right\}.
\ee
If one plots this one sees that this instability ($\omega^2 <0$)
corresponds to
$k^2 < k_J^2$  or $\lambda=2\,\pi/k > \lambda_J=2\,\pi/k_J$.
%FFig 5-1  of Binney and Tremaine.
$\lambda_J$ is called the Jeans length which sets the scale for instabilities.

The validity of the homogeneity assumption and
Jeans effect is legitimate on small scales and so stationary 
stellar systems are generally stable on small scales. The clumping
of material that begins when $\lambda > \lambda_J$  can sometimes be arrested
due to some nonlinear effects. 

The Jeans analysis, and significance of the Jeans length, can also be
used to analyse homogeneous collapsing and expanding systems. The
perturbed gravitational field merely accelerates collapse or decelerates expansion
and no Jeans effect is invoked.
%In this case $\lambda >> \lambda_J$ corresponds to $t^{2/3}$ power
%growth of instability rather than exponential. 
This is the kind
of instability that could have led to galaxy formation out
of the initial homogeneity of the early expanding universe, and
so also a compact self-gravitating object from a homogeneous
soup of bosons.

\section{Boson Stars: Nonspherically Symmetric Configurations}

We have set up boson star configurations in a general $3D$ code, and used them as a  test
of this code by comparisons to spherically symmetric results obtained
using a 1D code. In the process of stabilizing the code to achieve this a study of
coordinate conditions was needed. Once stability was achieved the code
could be used to study the behavior of boson stars under non-spherical
perturbations, which we obviously could not do in $1D$.
When perturbed in $1D$ a boson star emits scalar radiation and loses
mass. In $3D$ we can study another kind of radiation, namely gravitational
radiation.

We now give a brief review on Einstein's equations as a source of
gravitational radiation and then
outline the other works in the field of higher dimensional
boson star studies. Then we set the tone for our own work in higher
dimensions by a description of the $3 + 1$ $ADM$ split of space-time
and introduce the concept of extrinsic curvature. 
A brief description of the tools of numerical relativity and the
underlying numerical errors that one has to deal with follows.
Details of our work, and resolution of these problems in the
context of our own model, are described in the fourth chapter of the thesis.

\subsection{Einstein's Equations as a Source of Gravitational Radiation}

Maxwell's equations have radiative solutions that predict
the existence of electromagnetic waves. Likewise
Einstein's equations have such solutions, giving rise
to gravitational waves. The construction of laser interferometric
detectors like LIGO makes the detection of this kind
of radiation a real possibility. The theory of gravitational
waves is complicated due to the nonlinearity of Einstein's equations.
In order to simplify the system, and yet get insight into the
nature of gravitational radiation consider the weak field limit, with
the metric just slightly perturbed from the Minkowski metric~\cite{sweinberg},
$g_{\mu\nu}= \eta_{\mu\nu} + h_{\mu\nu}$. Since the derivative terms
only come from the $h_{\mu\nu}$ the Ricci tensor is
\be
R_{\mu\nu} = \Gamma^\lambda{}_{\lambda\mu,\nu} - \Gamma^\lambda {}_{\mu\nu,
\lambda} + O(h^2),
\ee
giving
\be
R_{\mu\nu} =\frac{1}{2}\, \eta^{\lambda \sigma}\,\left[h_{\sigma\lambda,\mu\nu}
+h_{\sigma\mu,\lambda\nu} -h_{\lambda\mu,\sigma\nu}\right]-
\frac{1}{2}\,\eta^{\lambda\sigma}\,\left[h_{\sigma\mu,\nu\lambda}
+h_{\sigma\nu,\mu\lambda} - h_{\mu\nu,\sigma\lambda}\right] + O(h^2).
\ee
Thus
\be
R_{\mu\nu} = \frac{1}{2}\,\left[\frac{\partial^2}{\partial x^\mu\,\partial x^\nu}
h^\sigma {}_\sigma -h^\lambda {}_{\mu,\nu\lambda} - h^\lambda {}_{\nu,\mu\lambda}
+ \dal h_{\mu\nu}\right].
\ee
From the trace of Einstein's equations $R =-8\pi GT^\lambda {}_\lambda$.
The Einstein equations
to order $h$ are then
\be
\frac{\partial^2}{\partial x^\mu\,\partial x^\nu}
h^\sigma {}_\sigma -h^\lambda {}_{\mu,\nu\lambda} - h^\lambda {}_{\nu,\mu\lambda} 
+ \dal h_{\mu\nu} = 16\,\pi\, G\, S_{\mu\nu},
\label{graveqn}
\ee
where 
\be
S_{\mu\nu} = T_{\mu\nu} -\frac{1}{2}\,\eta_{\mu\nu}\,T^\lambda {}_\lambda.
\ee
Exploiting the gauge invariance of the theory to choose a convenient
one
\be
g^{\mu\nu} \,\Gamma^\lambda {}_{\mu\nu} =0, \qquad \text{(harmonic coordinate
system)}
\ee
gives to order $h$:
\ben
0&=&\frac{\eta^{\lambda\delta}}{2}\,\left[ \eta^{\mu\nu}\,h_{\delta\mu,\nu} +
\eta^{\mu\nu}\,h_{\delta\nu,\mu} - \eta^{\mu\nu}\,h_{\mu\nu,\delta}\right]
\\ \nonumber
&=& \frac{\eta^{\lambda\delta}}{2}\,\left[2\,h^{\nu} {}_{\delta,\nu}-
\frac{\partial}{\partial x^{\delta}} h^{\mu} {}_{\mu}\right].
\een
This means $\frac{\partial}{\partial x^{\mu}}h^{\mu} {}_{\nu} = \frac{1}{2}
\frac{\partial}{\partial x^{\nu}} h^{\mu} {}_{\mu}$ making
(\ref{graveqn}) a wave equation of the form
\be
\dal h_{\mu\nu} = 16\,\pi \, G \, S_{\mu\nu}.
\ee

In electromagnetism, a localized source has no charge flowing in or
out of it due to charge conservation. The monopole part
of the potential of a localized source is static, and fields
with a harmonic time dependence $e^{i\omega t}$ have no
monopole terms. Hence, electromagnetic radiation is dipolar. Similarly,
energy conservation ensures the absence of monopolar gravitational
radiation. In addition, the power output of dipole radiation is
related to the second derivative of dipole moment, which for
gravitational radiation is zero because the first derivative of
a mass dipole is momentum and the law of conservation of momentum
makes its derivative zero. That is to say, gravitational radiation
is quadrupolar in nature. 
In spherically symmetric spacetimes with spherically symmetric sources,
there are no gravitational waves.

This can explicitly be shown. Consider the source free or homogeneous
wave equation for gravitational waves and write out its
plane wave solution
\be
h_{\mu\nu}(x) = e_{\mu\nu}\,\exp(i k_\lambda x^\lambda) +
c.c.
\ee
The symmetric polarization tensor should have $3\times 2 +4 =10$
independent components. However, the gauge invariance introduces 4
conditions on it giving six independent components. By a suitable
coordinate transformation one can actually show that only two
are physically meaningful and the remaining components vanish.
By subjecting the
coordinate system to a rotation, one can show that the physically
meaningful components have helicity $\pm 2$ while the others which can be
set to zero have helicities zero and one. 
This is analogous to electromagnetism where
Maxwell's equations give four components
of the polarization tensor and using the Lorentz gauge this
is reduced to three independent components. Without
leaving the Lorentz gauge a transformation of the vector potential components
in terms of the derivative of a scalar field allows one to reduce the
polarization vector components to two physical components. A rotational
transformation shows
these have helicity $\pm 1$. 
See~\cite{sweinberg}
for details.

\subsubsection{Gravitational Waveform Extraction}

In order to find out how much energy is
carried by gravitational radiation, and what the nature of this 
radiation is, one needs to extract the gravitational
waveform. In addition, one needs different methods in order
to compare the energies from the different measurements to check
that the code is giving reasonable results.
We list below the various functions that portray these waves in our $3D$ numerical code:

\begin{itemize}
\item{Zerilli Function}
\end{itemize}

The first step to waveform extraction using Zerilli functions is
to write the total metric as a sum of a spherically symmetric background
and a perturbed metric $h_{\mu\nu}$. Following Regge--Wheeler
the perturbed metric is written in terms of spherical harmonics by
associating the components with 
scalars $(h_{00}, h_{r0},h_{rr})$,
vectors $(h_{0\theta}, \, h_{0\phi};\, h_{r\theta},\, h_{r\phi})$ and
tensors. The details for spherical harmonic
expansions can be found in~\cite{rege}.
\be
g_{\alpha\beta} = ({g_{\alpha\beta}})_s + h_{\alpha \beta},
\ee
where
\be
{g_{\alpha\beta}}_s=
\left(
\begin{array}{cccc}
-N^2 & 0 &0 &0 \\
0 & g^2 &0 &0 \\
0 & 0 & R^2 &0 \\
0 & 0 & 0& R^2\,\sin^2\theta
\end{array}\right).
\ee

In terms of spherical harmonics, (actually $m=0$ spherical harmonics since
all the perturbations we consider are axisymmetric) the
perturbed metric $h_{\alpha\beta}$ can be written as the matrix
\be
\left(
\begin{array}{cccc}
-N^2\,H_0\,Y_{l0} & H_1\,Y_{l0} &h_0\,Y_{l0,\theta} &0 \\
H_1\,Y_{l0} & g^2\, H_2\, Y_{l0} &h_1\,Y_{l0,\theta} &0 \\
h_0\,Y_{l0,\theta} & h_1\,Y_{l0,\theta} & R^2\left(K+G\frac{\partial^2}{\partial
\theta^2}\right)Y_{l0}&0 \\
0 & 0 & 0& R^2\,\left(K\sin\theta
+G\cos\theta\frac{\partial}{\partial\theta}\right)Y_{l0} \sin\theta
\end{array}\right),
\ee
where the Regge--Wheeler perturbation functions ($H_0,\,H_1,\, H_2,\,
h_0,\,h_1,\,K,\,G$) are all functions of the radial and
time coordinates only~\cite{Abrahams92a}.
Any information about the gravitational wave content
is contained in these perturbed metric functions  so
that these must be determined.
Using the
orthonormality relations of spherical harmonics these functions can be
found from the full metric. For example, 
\begin{equation}
H_0^l= \frac{2\pi}{N^2}
\int_{0}^{\pi}\, g_{tt} \,Y_{l0}\, \sin\theta\;d\theta,
\end{equation}
since the $l=0$
mode corresponds to the spherical metric $N^2= -\frac{1}{2}\int_{0}^{\pi}
\,g_{tt}\sin\theta\,d\theta$. Similarly, the functions $H_1$ and $H_2$ are
extracted. We also get
\begin{equation}
h_1^l =\frac{2\pi}{3}\int_0^\pi\,
Y_{l0,\theta}\,g_{r\theta}\,
\sin\theta\;d\theta,
\end{equation}
and $h_0$ follows analogously. 
In order to separate $G$ and $K$ consider
\begin{equation}
\frac{1}{R^2}(h_{\theta\theta}\,\sin^2\theta -h_{\phi\phi}) =
G\,\left(Y_{l0,\theta\theta} + \sin\theta\,\cos\theta\, Y_{l0,\theta}\right).
\end{equation}
This is not composed of orthonormal spherical harmonics. Therefore one
must change basis so the $\theta \theta$ and $\phi\phi$
part of the metric can then be written in terms
of orthonormal tensor spherical harmonics. (See~\cite{Abrahams92a}
for details.) We find
\be
G^l =\frac{2\,\pi}{R^2}\int_0^\pi\,
\frac{
\left(\gthth \, \sinsq -\gphiphi\right)\left(-\cos\theta \, Y_{l0,\theta}
+\sin\theta \, Y_{l0,\theta\theta}\right)}{\left[l\,(l+1)\,(l+2)\,(l-1)\right]\,
\sinsq},
\ee
and
\be
K^l = \frac{l\,(l+1)}{2}\,G^l + \frac{\pi}{R^2} \int_0^\pi \,
\left(\gthth +\frac{\gphiphi}{\sinsq}\right)\, \sin\theta \, Y_{l0} \; d\theta.
\ee
Once one has these perturbed metric components one has all the
information about the gravitational wave output of the system. However
these metric components are dependent on gauge choices and, in
this form, cannot yield the actual wave perturbations. Following~\cite{moncrief} one needs to construct gauge invariant quantities
from the perturbed metric components. When the background metric
is Schwarzschild, the Zerilli function can be constructed, describing
the propagation of even parity waves. This is why, when we use
this method, we place detectors at the exterior region of our system.
One defines the Zerilli function~\cite{zer}:
\be
\psi_z= \sqrt{\frac{2\,(l-1)\,(l+2)}{l\,(l+1)}} \left(\frac {4\,r\, (1-\frac{2M}{r})^2 \,
k_2 + l\,(l+1)\,r\,k_1}{\left(l\,(l+1) -2 + \frac{6\,M}{r}\right)}\right),
\ee
where
\ben
k_1 &=& K+ {r}{(1-\frac{2\,M}{r})}G_{,r} - 2\frac{1-\frac{2\,M}{r}}{r}h_1,
\\ \nonumber
k_2 &=& \frac{H_2}{2(1-\frac{2\,M}{r})} -
\frac{1}{2\,\sqrt{1-\frac{2\,M}{r}}}\frac{\partial}{\partial r}
\left(r\,K (1-\frac{2\,M}{r})^{-\frac{1}{2}}\right).
\een
The quantity $\psi_z$ satisfies the wave equation
\be
\frac{\partial^2}{\partial t^2}\psi_z -
\frac{\partial^2}{\partial r_\ast ^2} \psi_z + V_z(r)\psi_z =0,
\label{zereqn}
\ee
where the tortoise coordinate $r_{\ast} = r+ 2M \ln
(\frac{r}{2M} -1)$ has been used. The scattering potential is given by
\ben
V_z(r) = \frac{\left(1-\frac{2\,M}{r}\right)}{\left(l(l+1) -2 +
\frac{6M}{r}\right)}\Bigg\{\frac{1}{\left(l\,(l+1) -2 + \frac{6\,M}{r}\right)}
\label{zerpot}
\\ \nonumber
\left[\frac{72 \,M^3}{r^5} -\frac{12\,M}{r^3} \,(l-1)\,(l+2)\left(1-
\frac{3\,M}{r}\right)
\right]
\\ \nonumber
+ \frac{l\,(l-1)\,(l+2)\,(l+1)}{r^2}\Bigg \}.
\een

\begin{itemize}
\item{Newman-Penrose Spin Coefficients}
\end{itemize}

Consider a four dimensional Riemannian space on which a tetrad system of null vectors
$l_\mu$, $m_\mu$, ${\bar m_\mu}$ and $n_\mu$ is introduced.
The orthogonal null vectors $l$ and $n$ are made by
adding and subtracting a space-like unit vector from a time-like unit
vector. The orthogonal complex null vectors $m$ and ${\bar m}$ are 
complex conjugates of each other. 
Far from the source the Newman--Penrose scalar 
\be
\Psi_4 = R_{\alpha\beta\gamma\delta} \;n^\alpha \; {\bar m}^\beta \;n^{\gamma} \;{\bar m}^\delta,
\ee
represents an outward propagating wave. For details see~\cite{newman,smr79,anhobi}.

\begin{itemize}
\item{Bel--Robinson vector}
\end{itemize}

Constructed in a manner analogous to the Poynting vector in electromagnetism, the
Bel--Robinson spatial vector is
\be
p^\gamma = E_{\alpha\beta}\, \epsilon^{\beta\gamma\delta} \, B_\delta {}^\alpha.
\ee
It is effective in tracking gravitational radiation. Here $E_{\alpha\beta}$ and
$B_{\alpha\beta}$ are the ``electric" and ``magnetic" components
of the Riemann tensor. The normal time-like vector $n^\alpha$ (see
$ADM$ section)
splits the vacuum four-dimensional tensor $R_{\alpha\beta\gamma\delta}$
into
\ben
E_{\alpha\beta} =n^{\gamma}\,R_{\gamma\alpha\beta\delta}\, n^{\delta},
\\ \nonumber
B_{\alpha\beta} = \frac{1}{2} \,\epsilon_\alpha {}^{\gamma\mu}\, n^\epsilon\,
(g^\nu {}_\beta +n^\nu\, n_\beta)\,R_{\epsilon\nu\gamma\mu},
\\ \nonumber
E_{\alpha\beta}\,n^\beta = B_{\alpha\beta}\, n^\beta =0.
\een
The details of this method are in~\cite{smr79,bel}.

\subsection{Axisymmetric Calculations}

Static or equilibrium solutions in the case of spherically symmetric boson
stars have been described in an earlier subsection. Static solutions also exist for
axisymmetric configurations (no azimuthal dependence) as shown in~\cite{yoshistat}. Again, the field itself can have a time dependence which does not affect
the static nature of the configuration. The field is of the form $\phi=
\phi(r,\theta)e^{i\omega t}$, with
a metric of the form
\be
ds^2=-e^{2\nu}\,dt^2
+e^{2\alpha}\,\left(dr^2+ r^2 \,{d\theta}^2\right)
+e^{2\beta}\,r^2\,\sin^2\theta\,{d\varphi}^2.
\label{yoshmeta}
\ee
Here $\nu$, $\alpha$ and $\beta$ are functions of
$r$ and $\theta$. The three relevant Einstein and scalar field equations are
set up, and this set of equations define an eigenvalue problem. The basic
partial differential
equations now are of the elliptic kind, and boundary conditions are more
easily
enforced by writing them in integral form using a Green's function approach.
That is, if for an operator ${\bf O}$ and functions $F$ and $H$
\be
{\bf O} F(r,\theta) = H(r,\theta),
\ee
then $F = \int{\bf dV'} \; G(r,,\theta,r',\theta')\, H(r',\theta')$ where
${\bf O}  G = {\bf \delta(r-r')}$. The Green's function is then expanded
in terms of the Legendre polynomials $P_{2n}$ with equatorial symmetry ensuring that only the even
ones appear. The particle number (and mass) versus radial
derivative of the central field shows marked similarity in profile with
the spherical case. The maximum mass is $1.05\,M_{Pl}^2/\mu$ as opposed
to $0.633$ in the spherical case. 

Using this Green's function technique, the analysis is extended to the case
of rotating boson stars in~\cite{yoshrot}. In~\cite{kobarot} a perturbation
approach, where perturbations were axisymmetric over a spherically symmetric
configuration, was considered and it was shown that no rotation was
possible. However the assumption of axisymmetry was also extended to the
perturbed boson field. In~\cite{schurot} this limitation was removed.
The scalar
field was itself allowed to have an azimuthal dependence without affecting
the axisymmetry of spacetime. In~\cite{yoshrot} this idea was again
implemented but without the drawbacks of the previous attempt.
Namely, this time several equilibrium configurations were
found, not just one or two, for different values of $m$. In addition
non-Newtonian configurations were considered as opposed to
just Newtonian ones in the former case. Also a ``spikelike'' solution in the
energy distribution in the former is believed to be caused by
inappropriate boundary conditions.  This was explained in~\cite{yoshrot}.
The metric is written in the form
\be
ds^2=-e^{2\nu}\,dt^2
+e^{2\alpha}\,\left(dr^2+ r^2 \,{d\theta^2}\right)
+e^{2\beta}\,r^2\,\sin^2\theta\,\left(d\varphi-\omega\,dt\right)^2.
\label{yoshmet}
\ee
Here $\nu$, $\alpha$, $\beta$ and
$\omega$ are functions of $r$ and $\theta$. As in~\cite{schurot}
the scalar field is taken to be
\be
\phi=\phi_0(r,\theta)\,e^{-i(\s\,t-m\,\varphi)},
\label{yoshphi}
\ee
where single valuedness as $\varphi \rightarrow \varphi + 2\,\pi $
dictates that $m$ is an integer. This time the Green's function associated
with the scalar field equation
has to be expanded in terms of the associated Legendre functions due to
this azimuthal dependence in the boson field. The expansions are
done till order $2n=12$ and numerically the system is iterated using estimated
values that step by step converge to the correct value. 

The angular momentum 
\be
J=\int\,T^0_3\,\sqrt{-g}\;dr\,d\theta\,d\varphi = m\,N,
\ee
is related to the particle number $N=j^0$ (${\bf j}=conserved
\,\,\,\,current$) through the azimuthal number $m$. $J=0$ for the
$m=0$ case as it should be. This means there is 
a conservation law in operation with
specific angular momentum
$\tilde j =T^0_3/j^0=m$
being constant for these configurations.
This corresponds to the $\tilde {j}$ constant law of perfect fluid stars
except that the constant in this case is an integer.
The
maximum mass was $M_{max}=1.05$, $1.31$, and $> 2.22$ for
$m=0,$ $1$, and $2$ respectively (in units of
$M_{Pl}^2/\mu$). 
The mass-energy density $-T^0_0$ for the $m=1$ and $m=2$
case shows a
clearly non spherical almost toroidal 
distribution.
The main difference for $m=1$ and $m=2$ is the nonvanishing of this function
near the origin on the symmetry axis in the former case although it
vanishes in the latter.

A rotating boson star configuration with large self-coupling is described
in~\cite{ryanspin}. This is with a view to having a structure emitting
enough gravitational radiation as to be capable of being detected by
future gravitational wave detectors. It comes with the added bonus
of a simplified set of equations in which field gradients other than
those with respect to the azimuthal angle $\varphi$ can be ignored. As
a result of the large self-coupling assumption, one can write an effective
equation of state. In the tail region, where gradients cannot be ignored, the
field is itself vanishingly small. The Einstein equations are set up and
a Green's function method is again employed to find solutions. On the left
hand side, terms for which the flat space Green's function are known are
kept, while the right hand side consists of the other terms.
As a result the
function on the left is the integral of the Green's function times the terms
on the right side. The Green's function is then expanded in Legendre
polynomials as described previously. The mass profile and toroidal geometry
of the density are extracted as per the earlier discussion. 
The structure of the star is
determined by the azimuthal quantum number $m$, the eigenvalue $\sigma$ and the value of
$\sqrt{\lambda}/m^2$ where $\lambda$ is the self-coupling parameter. As the
mass of the star increases its radius slightly decreases and the strongest
gravity region would be a configuration near the maximum mass. These
stars would give the largest gravity wave signals. From the output gravitational
waves, the multipole moment structure of the star is revealed. If, from the
waves, the mass, spin, mass quadrupole moment, and spin octopole moment
are determined, then the object could be confirmed to be a particular
configuration (three quantities completely parametrize it) of a boson star. The maximum
mass as a function of the star's spin angular momentum is plotted. The
multipole moments are encoded in the asymptotic form of the metric
coefficients and can be determined from comparisons with the same
order terms in the previously described series expansions.  Plots
are made of the mass quadrupole moment and spin octopole moment as functions
of mass and spin. If the gravitational wave measurements give four
moments, the mass $M$, spin $S_1$, quadrupole moment $M_2$, and spin
octopole moment $S_3$, one could determine from these plots the
value of the self-coupling parameter and identify the boson star.
Plots can also be made of the energy of a test particle orbitting and entering
a boson star for a particular mass boson star, as a function of orbital
frequency. One can also make such plots against other boson parameters,
thereby providing the gravitational energy per frequency band. This could
serve as a map of the interior of a boson star if there was some
way of calculating the total energy emitted by measurements from a single
detector. However, this would involve knowing the exact angular pattern of
emitted waves.

\subsection{Boson stars in $3D$: Perturbation Studies}
\label{subsection:bs}

The nonradial quasinormal modes of a boson star are described in~\cite{yoshinonrad}.
The even-parity axisymmetric perturbations of the metric are
written using the Regge--Wheeler gauge~\cite{rege}, with the perturbed
metric
functions and components 
of the complex boson field having frequency $\sigma$, that is, a time dependence of
the form $e^{-i \sigma t}$. The perturbation equations are then written
down for the perturbed Einstein and scalar field equations. Outside
the star, the background spacetime becomes almost Schwarzschild and the
scalar fields perturbations decouple from those of the metric.
The metric perturbation equations then reduce to a
system of perturbation equations for a black hole, and one can then
calculate the Zerilli equation as discussed above. Since the time dependence is
of the form $e^{-i \sigma t}$, the second time derivative in the
Zerilli equation (\ref{zereqn}) gives a $-\sigma^2 \psi_z$ contribution and
one gets the radial part of this function $Z$ in the 
far zone to be of the form
\be
Z = A_{in} e^{-i \sigma r_{\ast}} + A_{out}e^{i \sigma r_{\ast}}
+ O\left(\frac{1}{r}\right),
\ee
since the Zerilli potential $V_z$ is of order $\frac{1}{r^2}$ according to
(\ref{zerpot}). The scalar field equations for the two fields reduce
to Schr$\ddot {\text {o}}$dinger type wave equations and
and the quasinormal modes for the star are determined by numerically solving
the perturbed equation for the $l=2$ or quadrupole modes. 

The most important features of the quasinormal modes are that
the imaginary parts of the frequencies are large and the difference in
real parts of the frequencies between nearest modes is almost constant.
The large imaginary parts of the frequencies means a damping time
scale which is very small compared to other relativistic stars. This is
of consequence to our own $3D$ studies described in chapter 4 and so
we provide an analysis of the reason for these differences.

The reason for the differences from other stars comes from the fact that
the scalar field
extends to
infinity, as opposed to ordinary stars which have definite surfaces. The fact that
perfect fluid stars have at least two families of $QNM$ is explained
by a ``two string model''~\cite{kokkostring}. The star and the spacetime
are described by a finite and a semi-infinite string respectively.

For small coupling constants of the connecting string, there are two kinds
of normal modes. One has a small imaginary part to its eigenfrequency, and
for
this mode the amplitude of the finite string is very large compared to
the semi-infinite string. This mode is weakly damped. The other mode is
strongly damped and has an amplitude that is large compared to the
finite string.

\begin{figure}
\hspace{-36pt}
\vspace{-130pt}
%\epsfbox[0 -100 350 500]{stringport.ps}
\epsfbox[100 -100 650 500]{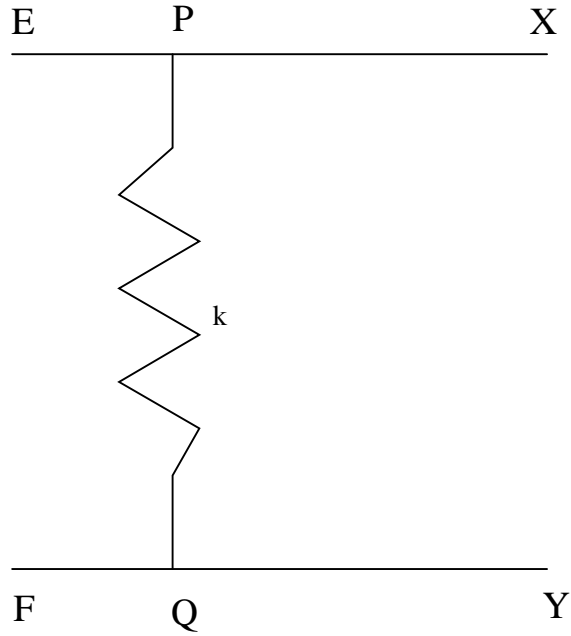}
\caption{A schematic diagram of the two string model for a boson star
is shown. The scalar field, extending to spatial $\infty$, is represented
by a semi-infinite string. It is coupled
to the spacetime (also a semi-infinite string) through a spring
of coupling constant k. As a result there is no place to store the 
energy (as in a finite string) and the modes of the
system are strongly damped.
}
\label{fig_string}
\end{figure}

Here on the other hand, we have two semi-infinite strings as the
scalar field also
extends to infinity (refer to Fig.~\ref{fig_string}).
%\vspace{3.5 in}

The displacement in each segment can be expressed as
\be
y=A\,\exp(i\,\omega(t+x/c))+B\,\exp(i\,\omega(t-x/c)).
\ee
The complex amplitudes $A$ and $B$ characterize the solutions in each
segment.
At $x=0$ strings are fastened so that
\be
y_1(0)=y_2(0)=0.
\ee
Continuity at the position of the spring gives
\ben
y_{EP}(l) &=& y_{PX}(l), \\ \nonumber
y_{FQ}(l) &=& y_{QY}(l).
\een

There are no incoming waves at $\infty$ so $A_{PX}=A_{QY}=0$. In terms
of the spring constant $k$ the tension $T$ is given by
\ben
T\left[(\frac{\partial y_{EP}}{\partial x})\Bigg|_{x=l}-(\frac{\partial y_{PX}}
{\partial x})\Bigg|_{x=l}\right]&=&-k\,(y_P-y_Q),
 \label{yoshitension} \\ \nonumber
 T\left[(\frac{\partial y_{FQ}}{\partial x})\Bigg|_{x=l}-(\frac{\partial y_{QY}}
 {\partial x})\Bigg|_{x=l}\right]&=& k\,(y_P-y_Q),
 \een
 where the $y$s are displacements of points and segments of the string.
 Using the boundary conditions, and the fact that $A$ and $B$ are
 constant in each segment yields
 \ben
 B_{EP}&=&-A_{EP},\\ \nonumber
 B_{FQ}&=&-A_{FQ}, \\ \nonumber
 B_{{PX}/{QY}}&=&(e^{2z}-1)A_{EP/FQ},
 \een
 where $z=i\,\omega\,l/c$.

 If you divide the two equations of (\ref{yoshitension}) by each other
 you get $A_{FQ}=-A_{EP}$ and putting this into either of the equations
 gives
 \be
 ze^{z}=K(e^{-z}-e^z),
 \ee
 where $K=\frac{k\,l}{T}$ is the coupling constant. So for small coupling
 constants, and small imaginary parts for the frequencies, we have
 $ze^z=0$. This has the trivial solution $z=0$, meaning there
 are no normal modes for the weak damping case. On the other hand,
 for perfect fluid stars the corresponding expression is~\cite{kokkostring}
 \be
 z(e^z+e^{-z})=K\,(e^{-z}-e^z)\,\frac{(2+e^{-2z})}{2},
 \ee
 and there are non trivial solutions in the weak damping case.
 For the strongly damped case ($e^z$ subdominant to
 $e^{-z}$)
 both cases yield
 \be
 z\sim K\,e^{-2z}.
 \label{yoshz}
 \ee

 If $e^{-2z}$ is dominant then $z$ must have a real part which is negative.
 So $Re(z)=Ke^{-2Re(z)}\cos(Im(2z))<0$ or $\cos(Im(2z))<0$. Hence
 \be
 z \sim -a+i\,(2n+1)\frac{\pi}{2}+i\, b
 \ee
 where $a$ is large and positive while $b$ is very small. Writing
 $z$ in terms of $\omega$ ($z=i\omega l/c$)
 the eigenfrequencies are approximately
 \be
 \omega_n=\frac{c}{l}\,\left(\frac{(2n+1)\,\pi}{2}+b\right)+i\,\frac{a\,c}{l}.
 \ee
 By equating the real and imaginary parts of (\ref{yoshz}) we see that
 $a=Ke^{2a}$ (since $b$ is small). The presence of a coupling constant,
 however small, is necessary to ensure that we have a nontrivial solution.
 The smaller the coupling constant the larger the damping part of
 the frequency ($a$). The imaginary part of the eigenfrequencies
 are equidistantly spaced. These are like the $w$ modes of a
 perfect fluid star. In the perfect fluid case we have a finite
 string that has no mechanism to radiate its energy except through
 a spring coupling to a semi-infinite string, however weak the coupling may
 be.
 On the other hand, the two semi-infinite string case allows each string to
 radiate energy excited by an oscillation to infinity without
 coupling to the other string. Thus, there is no place to
 store energy like in a finite string. Therefore, radiation must be rapid and
 there are no weakly damped modes.

\subsection{Boson Stars in Full {\em GR}: Numerical Relativity}

Eventually what one wants is the solution of
the Einstein equations with matter terms present.
This is the crux of our work. Of course, Einstein's equations are far
too non-linear and complicated to be solved analytically and so we
must rely on numerical techniques. 

We look at spacetime from the point of view of a Cauchy problem. That
is, we view the evolution as being from one spacelike hypersurface to
the next. In order to construct the gravitational field, one solves the
initial-value
problem and then integrates the dynamical equations along trajectories of a
prescribed reference system. 

Four of Einstein's equations, $G_{0\mu} = 8\,\pi \, T_{0\mu}$, do not contain
second time derivatives of the metric. The solutions of these equations
constitute the initial-value problem. The most natural geometrical variables
on the hypersurfaces are the metric of the hypersurface, which is denoted
$\gamma_{ij}$ or ${}^3g_{ij}$, and the extrinsic curvature $K_{ij}$.

\subsection{ADM}
\label{subsection:ADM}

In order to develop the equations that describe the system we must
understand the spacetime that we are studying. In the first place, we
must revise our concept of time. In Newtonian theory it is absolutely
defined. In special relativity time is somewhat ambiguous in that the
concept of simultaneity is not universal, but by specification of an
inertial frame the concept is made precise. On the other hand, in {\em GR}
we have to replace the concept of time of an event by the notion of
a spacelike hypersurface. The entire spacetime is divided into these
spacelike hypersurfaces, or rather sliced into them, and the parameter
separating one hypersurface from the other is called ``time". At each event
on a given hypersurface, a local Lorentz frame exists whose surface of
simultaneity coincides locally with the hypersurface. This Lorentz frame is
one with its 4-velocity orthogonal to the hypersurface. 

Thus to explore spacetime conveniently it is
divided into a series of spacelike hypersurfaces that are parametrized
by successive values of a time parameter $t$. The 3-geometry on two
faces of a spacetime sandwich are connected by a 4-geometry in between that
must extremize the action described in \ref{section:variat}. 

This $3\,+\, 1$ split of spacetime, with a hypersurface parametrized
by time coordinate $t= constant$ followed by one parametrized by $t+dt =
constant$, is convenient but this parameter $t$ is not the ``proper time''.
The 3-geometry of the lower hypersurface is given by
\be
\gamma_{ij}(t,x,y,z)\,dx^i\,dx^j,
\ee
and that of the upper one by
\be
\gamma_{ij}(t+dt,x,y,z)\,dx^i\,dx^j,
\ee
with the lapse of proper time between the two being related through a lapse function $N$. We have
\be
\text{lapse of proper time} = N(t,x,y,z)\,dt.
\ee
The spatial coordinates are shifted by a shift vector from one hypersurface to another
\be
x^i_{high}=x^i_{low} + N^i(t,x,y,z)\,dt.
\ee
This is part of the gauge or coordinate feedom we have in the theory.

The 4-geometry connects a point $(t,x^i)$ to $(t+dt,x^i+dx^i)$ with
proper interval
\be
ds^2= \gamma_{ij}\,(dx^i+N^i\,dt)\,(dx^j+N^j\,dt) - (N\,dt)^2 = {}^4 g_{\alpha\beta}\,dx^{\alpha}\,
dx^{\beta}.
\label{admmetric}
\ee
This gives the 4-metric in terms of the lapse, the shift, and
the 3-metric.

A nongeometrical way of looking at this is to introduce the notion of time through a coordinate
$t=t(x^\mu),$ $ \mu = 0,1,2,3$ and a time flow vector 
$t^{\mu}$. The time flow vector is normalized
\be
t^{\mu}\,\nabla_{\mu}t=1.
\ee
Since this vector of time flow is in general not going to be normal to the spacelike hypersurfaces, the normal is given in terms of a lapse function
\be
n_{\mu} = -N\,\nabla_{\mu}t.
\ee
Thus,
\be
t^{\mu}\,n_{\mu} = -N.
\ee
This normal to the spacelike hypersurfaces must be timelike and it is normalized to unity so that
\be
n^{\mu}\,n_{\mu}=-1.
\ee
Hence
\be
N=-t^{\mu}\,n_{\mu}=(n^\mu\,\nabla_\mu t)^{-1} =
\frac{1}{\sqrt{-g^{\mu\nu} \, \nabla_\nu t \, \nabla_\nu t}}.
\ee
We now separate the time vector into a spacelike and a timelike part as
follows
\be
t^{\mu} = P^{\mu}{}_{\nu}t^{\nu} - n^{\mu}\,n_{\nu}\,t^{\nu}=
\left[ \left(\delta^\mu {}_{\nu} + n^{\mu}\,n_{\nu} -n^{\mu}\,n_{\nu}\right)
\right]\,t^{\nu}
= N^{\mu} + N\,n^{\mu}.
\label{definep}
\ee
Here the first term, $(\delta^\mu_{\nu} + n^{\mu}n_{\nu})t^{\nu}= N^{\mu}$,
defines the shift vector. It is perpendicular to the timelike
component $Nn^{\mu}$, as can be seen by the fact that we have subtracted off the
timelike component from the whole term to get the first term, and also
by simply contracting to get
\be
n_{\mu}N^{\mu} = (\delta^{\mu} {}_{\nu} \, n_{\mu} \, t^{\nu} +
n_{\mu}\, n^{\mu} \,n_{\nu} \,t^{\nu}) = n_{\mu}\,t^{\mu} (1+(-1))=0.
\ee
Since all physics must be independent of coordinates, a system $x^{\mu} =
(x^0,x^i)$ is chosen with the spacelike hypersurfaces described by
$x^{i}$ and the time by $t=x^0$. The components $t^\mu$ are then
\be
t^\mu = (1,0,0,0).
\ee
The lapse function is 
\be
N=-t^\mu \,n_{\mu} = -n_0,
\ee
and the shift
\be
n_i = -N \,\nabla_i t =0.
\ee
Also, $n_\mu n^{\mu} =-1$ gives 
\be
n^0 = \frac{1}{N}.
\ee
Similarly,
\be
n_{\mu}\,N^{\mu} =0,
\ee
implies $N^0=0$ and
\be
N_{\mu}\,n^{\mu} =0,
\ee
implies
\be
\frac{N_0}{N} + N_i \,n^i=0.
\ee
Thus,
\be
n^i =\frac{a^i}{N} \quad \text{and}\quad N_0 = N_i b^i, \quad
\text{where } \qquad
a^i = - b^i
\ee
Choosing $b^i =N^i$ gives
\be
n^\mu = (N^{-1}, -N^{-1}\,N^i),
\ee
and
\be
N_{\mu} = (N_i\,N^i, N_i).
\ee
Finally, we can extract the metric components as follows. From
\ben
g_{\mu\nu}\,n^{\mu} &=& n_{\nu}\\ \nonumber
&\Rightarrow&
g_{00}\,n^0 + g_{i0}\,n^i = n_0 \\ \nonumber
&\Rightarrow&
g_{00} -g_{i0}\,N^i = -N^2,
\een
and
\ben
g_{\mu\nu}\,N^{\nu} &=& N_{\mu} \\ \nonumber
&\Rightarrow&
g_{00}\,N^0 + g_{i0}\,N^i = N_0 = N_i\,N^i \\ \nonumber
&\Rightarrow&
0+g_{i0}\,N^i = N_i\,N^i,
\een
one gets
\be
ds^2=g_{\mu\nu}\,dx^{\mu}\,dx^{\nu} = -(N^2 - N^i\,N_i)\,dt^2 +
2\,N_i \,dx^i\, dt + \gamma_{ij}\,dx^i\,dx^j,
\ee
where $\gamma_{ij}$ is the three metric on the hypersurface.
In matrix form
\be
g_{\mu\nu}=
\left(
\begin{array}{cc}
-(N^2-N^i\, N_i)& N_i \\ 
N_i & \gamma_{ij} 
\end{array}\right),
\label{downmet}
\ee
and
from $g_{\mu\nu}g^{\mu\epsilon} = \delta^{\epsilon}_{\nu}$ one gets the
inverse matrix
\be
g^{\mu\nu} = 
\left(
\begin{array}{cc}
-N^{-2}& N^{-2}\,N^i \\
N^{-2}N^i & \gamma_{ij} -N^{-2}\,N^i\, N^j
\end{array}\right).
\label{upmet}
\ee

\begin{itemize}
\item{Intrinsic and Extrinsic curvature and Connection Coefficients}
\end{itemize}

Consider a vector ${\bf A}$ lying in the hypersurface
\be
{\bf A} = {\bf e}_i\,A^i.
\ee
If we parallel transport this vector along a route in the hypersurface and
compare
${\bf A}$ at the transported point to the parallel transported ${\bf A}$
\be
{}^4 {\bf \nabla}_i{\bf A} = {}^4 {\bf \nabla}_i({\bf e}_j\, A^j)
= {\bf e_j} \,\frac{\partial A^j}{\partial x^i} + {}^4\Gamma^{\mu}{}_{ij}\,{\bf
e}_\mu \,
A^j.
\ee
Here the connection coefficient $\Gamma$ is the usual measure of the variation 
in the basis ${\bf e}$ itself. We started out on the hypersurface but
now have a component outside the hypersurface $A^j {}^4\Gamma^0_{ij}{\bf
e}_0{\bf \cdot n}$. If one projects ${}^4 {\bf \nabla}{\bf A}$ orthogonally
onto the hypersurface so as to get rid of the component outside the
hypersurface then we are only dealing with the 3-geometry intrinsic to
the hypersurface. Writing the 3-derivative,
of some vector
${\bf A}$ that lies on a spacelike hypersurface,
intrinsic to the surface 
and
then taking the projection in a particular diection gives for covariant
components

\be
{\bf e_j}\cdot {}^3{\bf \nabla}_i {\bf A} = \frac{\partial A_j}{\partial x^i}-
A^m \,\Gamma_{mji}.
\ee
Here the connection coefficients are those in three dimensions
\be
{}^3\Gamma_{mji} = \Gamma_{mji} = {\bf e}_m {\bf \cdot}{\bf \nabla}_i {\bf e}_j.
\ee

The concept of extrinsic curvature deals with the embedding of the
3-geometry in an enveloping spacetime. Consider the ``normal''
vector that stands at a point on the hypersurface
(${\cal P}+d{\cal P}$), and compare it to the one parallel tansported to this
point from a neighboring point ${\cal P}$. This can be regarded as the
limiting concept of the 1-form ${\bf dn}$. Hence the 1-form surface lies
perpendicular to the vector, and so lies along the hypersurface.
\be
{\bf dn} = -{\bf K}({\bf d}{\cal P}).
\ee
The components of the extrinsic curvature $K_i^j$ are obtained by
displacements along
a particular coordinate direction ${\bf e}_i$, so that
\be
{}^4{\bf \nabla}_i{\bf n} = -{\bf K} ({\bf e}_i) = -K_i{}^j \,{\bf e_j}.
\ee
Then
\ben
K_{ik} &=& K_i{}^j \, g_{jk} = K_i{}^j\, ({\bf e}_j \cdot {\bf e}_k) = -({}^4{\bf
\nabla}_i{\bf n})\cdot {\bf e}_k= -{}^4{\bf\nabla}_i({\bf n}\cdot {\bf e}_k)
+{\bf n}\,\cdot {}^4{\bf\nabla}_i({\bf e}_k) 
\\ \nonumber
&=& 0+ {\bf n}\, \cdot {}^4{\bf\nabla}_i({\bf
e}_k)
=({\bf n} \cdot {\bf e_\mu})\, {}^4\Gamma^{\mu}{}_{ki}
= ({\bf n} \cdot {\bf e_0})\,{}^4\Gamma^0{}_{ki} = {\bf n}\, \cdot {}^4{\bf \nabla}_k{\bf e_i}
=K_{ki}.
\een

The change of a vector ${\bf e_m}$ can be written in terms of its component
along the normal and along the basis vectors ${\bf e_i}\quad i=1,2,3$
and so
\be
{}^4{\bf \nabla}_i{\bf e}_j = ({}^4{\bf \nabla}_i{\bf e}_j )\cdot {\bf {\hat n}}
\; {\bf {\hat n}}
+ ({}^4{\bf \nabla}_i{\bf e}_j )\cdot {\bf e}_k \; {\bf e}_k \\ \nonumber
= K_{ij} \; \frac{{\bf n}}{{\bf n \cdot n}} + {}^3\Gamma^k {}_{ij}\; {\bf e}_k.
\ee

Similarly, defining $A^j_{|i}= {\bf e}_j \cdot {}^3{\bf \nabla}_i{\bf A}$,
\be
{}^4{\bf \nabla}_i{\bf A} = A^j {}_{|i}\,{\bf e}_j +K_{ij}\,A^j \frac{\bf n}{
({\bf n\cdot n})}.
\ee

The extrinsic curvature components in terms of the connection coefficients
and the ADM metric
then are
\be
K_{ij} = -n_{i;j} = -n_{\mu}\, {}^4\Gamma^{\mu}{}_{ij} = -n_0\, {}^4\Gamma^{0}{}_{ij}
= -N\, \Gamma^{0}_{ij}= -N
\left({}^4g^{00}\, \Gamma_{0ij}+{}^4g^{0k}{}^4 \, \Gamma_{kij})\right).
\ee
Thus
\ben
K_{ij} &=& \frac{1}{2\,N}\left[
g_{0i,j}+g_{0j,i} -g_{ij,0} -2\,\Gamma_{kij} \, N^k\right]
\label{extrink}
\\ \nonumber
&=&\frac{1}{2\,N}\left[\frac{\partial N_i}{\partial x^j} +
\frac{\partial N_j}{\partial x^i} -
\frac{\partial g_{ij}}{\partial t} -
2\,\Gamma_{kij} \, N^k\right]
\\ \nonumber
&=& \frac{1}{2\,N}\left[N_{i|j}+N_{j|i} - \frac{\partial g_{ij}}{\partial t}
\right].
\een

\begin{itemize}
\item{The $3+1$ Gauge Freedom}
\end{itemize}

One sees, therefore, that the spacetime metric $g$ naturally separates into
the six $\gamma_{ij}$ (symmetric), the three shift terms $N^i$, and the lapse
$N$ relative to a given foliation and choice of time vector. The only
second-time-derivative terms in the Einstein equations are those of
the $\gamma_{ij}$'s and these are the dynamical degrees of freedom.
However the initial-value problem, that we have hitherto not discussed
and which we shall address in chapter 4, shows that the $\gamma_{ij}$ need only be
known up to an overall conformal factor $\phi^4$, with $\phi$ determined
by constraints. Since 
\ben
\gamma_{ij} = (\phi^6)^{2/3}\; {\tilde \gamma}_{ij} &=&
\left(
\frac{\sqrt \gamma}{\sqrt {\tilde \gamma}}\right)^{2/3}\,
\tilde \gamma_{ij},
\\ \nonumber
&=& (\gamma)^{1/3} \, \tilde {\gamma_{ij}},\quad\text{  $\tilde \gamma=1$},
\een
one can regard the separation of the configuration coordinates $\gamma_{ij}$
into the square root of its determinant $\sqrt{\gamma}$ and the conformal
metric $\tilde \gamma_{ij}$. Since $tr\,K$ turns out to be a canonically
conjugate variable to $\sqrt{\gamma}$ one can as well regard the six
$(
\tilde \gamma_{ij}, tr\,K)$ as configuration variables. However, we have four
constraint equations and hence only two dynamical degrees of freedom.
Assuming the constraints have been satisfied we then have four
gauge degrees of freedom that we impose as four conditions on the velocities
$\partial_t (tr K)$ and $\partial_t \tilde \gamma_{ij}$. These lead
to equations on the lapse and the shift~\cite{York79}.
\pagebreak
\begin{itemize}
\item{The $3+1$ Equations}
\end{itemize}

Consider the projection $n^\mu n^\nu T_{\mu\nu}= n_\mu n_\nu T^{\mu\nu}$, 
where $n$ is the
normal vector defined in the earlier subsection. 
This is
$n_{\mu}n_{0}\, T^{0}_{0}\,g^{0\mu} = n^0n_0 \,T^{0}_{0}$ which is the energy density $\rho=-T^0_0$. The {\bf Hamiltonian
constraint equation} is determined from 
\be
n_\mu \,n_\nu\; G^{\mu\nu}=8\,\pi\, \rho.
\ee
The projection $-n^\nu P^\sigma {}_{i}T_{\nu \sigma}$,
where (\ref{definep}) defines the projection operator $P_i$, is
$-n^\nu \delta_i {}^\sigma T_{\nu \sigma} 
- n^\nu n_i \, n^\sigma T_{\nu \sigma} $. Since $n_i=0$
it simplifies to $-n^\nu T_{\nu i} = - n^\nu T_{i} {}^{0}g_{0\nu}  = - n_0 T_{i} {}^{0}$,
which is just the momentum density ${\cal J}_i$. This gives three
{\bf momentum constraint equations}
\be
-n^\nu\, P_i {}^\sigma\, G_{\nu \sigma} = 8\,\pi\, {\cal J}_i.
\ee
Finally, the projection $P_i {}^\mu P_j {}^\nu T_{\mu\nu} =T_{ij}=S_{ij}$
provides six evolution equations.
The extrinsic curvature relation (\ref{extrink})
in the $3+1$ approach, guides the
evolution of the metric through six more equations. 

We can write $G_{\mu\nu}$ in terms of the Ricci tensor $R_{\mu\nu}$ and Ricci scalar  
$R$, which we can write in terms of the $\Gamma$s'
themselves. These are then written in terms of the extrinsic curvature, to get the
constraint and evolution equations:
\be
\frac{1}{2}\left[{}^3R-K_j^i\,K_i^j +K^2\right] = 8\,\pi\, \rho, \qquad
\text{Hamiltonian Constraint}
\label{hamilform}
\ee
\be
K_i{}^j|_j -K|_i =8\,\pi \, J_i, \qquad
\text{Momentum Constraint}
\ee
and evolution equations
\ben
-\frac{1}{N}\left[
\frac{d}{dt} K_j {}^i \right. &-& \left.
N^{i}{}_{,k}\, K_j{}^k + N^k{}_{,j}\, K_k{}^i +N^{|i}{}_{|j}\right]+
{}^3 R_j{}^i + K\,K_j{}^i 
\label{kevolve} 
\\ \nonumber
&=& 8\,\pi\,\left[S_j {}^i - \frac{1}{2}\,(S-\rho)\,\delta_j {}^i
\right],
\een

and
\be
\frac{d \,\gamma_{ij}}{dt} + N^k {}_{,i} \,\gamma_{kj} +  N^k {}_{,j} \,\gamma_{ki}
=-2\,N\,K_{ij},
\ee
where $S =S_i^i$ and $K =K_i^i$.
Instead of six second-order evolution equations for $\gamma_{ij}$, by
defining extrinsic curvature in terms of the first time derivative of
the metric we have 12 first-order equations. By letting $j=i$
and summing over all $i$ in (\ref{kevolve}) we get an evolution equation
for $tr\, K$
\be
\frac{d \,tr \,K}{dt} = -N^{|i}{}_{|i} +N\,\left[R +K^2 -\fpi\, (3\,\rho-S)\right].
\label{trkevolv}
\ee

Data on the initial hypersurface satisfying the 4 constraint equations
evolves to the next hypersurface,
according to the 12 evolution equations. On the next hypersurface, but for the shortcomings
of numerical approximation schemes, the data should consistently satisfy the constraint
equations. 

\begin{itemize}
\item{Coordinate Conditions}
\end{itemize}

Part of the gauge freedom we have is that we can impose four conditions on
$\partial_t tr\,K$  and
$\partial_t tr\,\tilde \gamma_{ij}$ as discussed before. This gives us equations for the lapse and the shift.
We will discuss the various coordinate conditions we use on our lapse and shift, and the
specific advantages of these in chapters 2, 3, and 4 as is pertinent to the
problem being considered.

\begin{itemize}
\item{Finite Differencing Errors}
\end{itemize}

There are three kinds of partial differential equations that one
has to deal with. Our evolution equations are typically
hyperbolic equations (or wave equations) and the constraint equations are
elliptic equations. In addition, there are parabolic equations, and parabolic
terms may be added to our equations as part of developing a numerical
algorithm.
Elliptic equations represent equilibrium and the constraint equations are
of this form. 
In addition, when we use maximal
slicing (described in chapter 4) we have an elliptic equation for
the lapse that has to be solved on each time slice. This is in fact the
most time consuming part of the numerics. The constraint equations
are solved on the initial timestep and then monitored as
the code evolves, on each time step, as indicators of the accuracy of
the code. We circumvent the $3D$ constraint problem, for
spherical configurations, by getting data from the $1D$ code
and putting it on a $3D$ grid. We are dealing with highly
nonlinear equations but to get an idea of the nature of
these equations, we write them down in the linear case.
\ben
{\partial_t}{\partial_t}\phi - c^2 \,{\partial_x}{\partial_x}\phi &=& 0
 \qquad \text{hyperbolic (wave) equation},
 \\ \nonumber
 {\partial_y}{\partial_y}\phi + {\partial_x}{\partial_x}\phi &=& 0
  \qquad \text{elliptic equation},
  \\ \nonumber
 {\partial_t}\phi-D {\partial_x}{\partial_x}\phi&=& 0
 \qquad \text{parabolic (diffusion) equation}. 
 \een
 The factor $D$ is a diffusion coefficient. 

We have already shown how the linearized Einstein's equations give rise
to the
gravitational wave equations. In fact the full evolution equations are
like the wave equation~\cite{Smarrfd}.
By defining the extrinsic curvature in terms
of the time derivative of the metric we evolve a set of
first order equations. So a prototypical hyperbolic equation to consider
is of the form 
\be
\partial_t \phi + v\,\partial_x \phi =0.
\ee
We use finite differencing techniques to solve these
equations.

The ``spacetime'' for the numerical algorithm consists of a discrete
set of points and not a continuum; the grid spacing, and timesteps,
being dictated by the resources of memory and real time available to us.
The derivatives in the $PDE$ are replaced by finite difference
approximations. In doing so one introduces spurious solutions of the
finite difference equations that have no bearing to the
solutions of the actual equation we are solving. There could
be several different ways to finite difference the problem, each with its
own solution. Choosing a finite differencing scheme is, therefore, of
great importance. 

One obvious kind of error is the {\em Truncation Error}. 
We typically use second order accurate
schemes for finite differencing in the $3D$ code and up to 6th order
in the $1D$ code. There can also be {\em  Propagation Errors}. For example
if one writes the prototypical hyperbolic equation as
\be
\phi^{n+1} = G \phi^n,
\ee
where $G$ is an operator connecting the function on time slice $n$
with that on time slice $n+1$, one can have propagation errors.
If the solution of the finite differenced equation exhibits a wave behavior
that
is not shared by the $PDE$ it represents, the solution
can grow without bound if $|G| >1$ and it will dissipate away if $|G| <1$.
If $G=1$ then the wave amplitude is propagated without change as
the physical solution. For example, a forward-time centered-space derivative scheme can become unstable. We discuss this
in chapter 3 under our leapfrog method discussion for the
numerical code. In the staggered leapfrog method, one uses second-order accuracy in space and time and $|G|=1$. 

In nonlinear hyperbolic
equations, we often see grid point to grid point instabilities due
to our using a  leapfrog scheme. In a leapfrog scheme, one uses centered
time and space derivatives. As a result the time levels in the
time derivative terms ``leapfrog'' over time levels in space derivative
terms. For example, in our prototypical hyperbolic equation
\be
u_j^{n+1} - u_j^{n-1} = \frac{\Delta t}{\Delta x} \left(u_{j+1}^n
-u_{j-1}^n\right).
\ee
This kind of method for a non-linear equation can result in odd and even
mesh points totally decoupling, and results in mesh drifting with
odd and even grid points evolving in their own separate ways.
For large gradients, this can result in grid point to grid point
instabilities
that can invalidate the code. To prevent this, one needs to connect odd and
even grid points through a viscosity term. This is a term of the form
$D \left(u_{j+1}^n -2u_j^n + u_{j-1}^n\right)$ where $D$ is a small parameter.
We have had  to use a very tiny term of this form in our $3D$ code.
Putting this kind of term everywhere usually invalidates our code at our
boundaries
so
that we have to exponentially damp it out in the exterior of the
star. The steep gradients occur only near the center of the star, so
we never need diffusion anywhere else anyway. Typically we add diffusion
in the central $1/3rd$ of the grid and then damp it out.
 
\chapter{Equilibrium Bosonic Objects in Spherical Symmetry}

We now go on to continue the analysis of the Einstein equations with
matter sources in a spherically symmetric environment.
The general spherically symmetric metric is of the form
\be
ds^2 = -(N^2-N^r\,N_r)\,dt^2 + 2\,N_r \,dt \,dr + g^2\, dr^2 + r^2\, (d\theta^2+
\sin^2\theta\,
d\phi^2),
\ee
where the metric functions $N^2$ and $g^2$, and the shift function $N_r$, are
only functions of $r$ and $t$.
The angular metric functions are $g_{\theta\theta} =r^2$ and
$g_{\phi\phi}=r^2\sin^2\theta$.

We study the stability and dynamics of such a system using a numerical
code.
The success of a numerical code relies on the choice of coordinates. A
general code with general metric functions and $x,y,z$ coordinates would
have
to perform far too many algebraic operations to calculate all
the Ricci components.
A good coordinate system takes advantage of the symmetries of the system and
avoids physical singularities by slowing down evolution in the spatial
region near the singularity. Also, it should itself not develop any
coordinate singularities.

The lapse function determines the proper time
separation between nearby spacelike hypersurfaces and determines the time
coordinate. The choice of lapse is often times determined by a condition on
the extrinsic curvature (refer to \ref{subsection:ADM}).
Our $1D$ code uses a polar slicing condition
$K^\theta_\theta+ K^\phi_\phi =0$~\cite{bard}.
In a spherically symmetric spacetime
with no $N^{\theta}$ and $N^{\phi}$ terms this
translates to
\ben
0&=&
\frac{1}{N}\,\left[
\Gamma^r{}_{\theta\theta}\, g^{\theta\theta} +
\Gamma^r{}_{\phi\phi} \, g^{\phi\phi}\right] N_r  \\ \nonumber
&=& -\frac{1}{N}\,\left[g^{rr}\,  g^{\theta\theta}\, g_{\theta\theta,r}+
g^{rr} \, g^{\phi\phi}\, g_{\phi\phi,r}\right]\,N_r.
\een
{\em Thus the shift $N_r$ is zero.} 

The polar slicing condition in a spherically symmetric spacetime is very
singularity avoiding, matching up with Schwarzschild coordinates.
In the spherical collapse problem (in this case when a boson star
is collapsing to a black hole), the evolution slows down so that not
even an apparent horizon forms as the lapse collapses ($N\rightarrow 1-2M/2M=0$). Thus when an apparent horizon is approached, the
numerical code breaks down as the lapse collapses and the radial metric rises
sharply. This is because sharpening gradients invalidate the finite difference
approximations.
We now outline the features of the various studies we have conducted in
spherical symmetry.

We first calculate the connection coefficients from (\ref{gammon}).
%\be
%\Gamma^{\mu}_{\nu\delta} = \frac{1}{2}g^{\mu\lambda} \left(
%g_{\lambda \nu,\delta} +
%g_{\lambda \delta,\nu} - g_{\nu\delta,\lambda}\right).
%\ee
\ben
\Gamma^{0}{}_{00} &=& 
\frac{1}{2}\,g^{00}\,g_{00,0} = \frac{\dot N}{N}, \\ \nonumber
\Gamma^{0}{}_{0r} &=& \frac{1}{2}\, g^{00}\,g_{00,r} = \frac{N'}{N}, \\
\nonumber
\Gamma^{0}{}_{rr} &=& -\frac{1}{2}\,g^{00}\,g_{rr,0} = -\frac{g\,\dot g}{N^2},\\
\nonumber
\Gamma^{r}{}_{00} &=& -\frac{1}{2}\, g^{rr}\,g_{00,r} = -\frac{N N'}{g^2},
\\ \nonumber
\Gamma^{r}{}_{0r} &=& \frac{1}{2}\, g^{rr}\,g_{rr,0} = \frac{\dot g}{g},
\\
\nonumber
\Gamma^{r}{}_{rr} &=& \frac{1}{2}\, g^{rr}\,g_{rr,r} = \frac{g'}{g},
\\ 
\nonumber
\Gamma^{r}{}_{\theta\theta} &=& -\frac{1}{2}\, g^{rr}\, g_{\theta\theta,r}=
-\frac{r}{g^2},
\\
\nonumber
\Gamma^{r}_{}{\phi\phi} &=& -\frac{1}{2}\, g^{rr}\,g_{\phi\phi,r}=
-\frac{r\,\sin^2\theta}{g^2},\\
\nonumber
\Gamma^{\theta}{}_{\theta r} &=& \frac{1}{2}\, g^{\theta\theta}\,g_{\theta\theta,r}
                          = \frac{1}{r}, \\
\nonumber
\Gamma^{\phi}{}_{\phi r} &=& \frac{1}{2}\, g^{\phi\phi}\,g_{\phi\phi,r}
= \frac{1}{r},\\
\nonumber
\Gamma^{\phi}{}_{\phi \theta} &=& \frac{1}{2}\, g^{\phi\phi}\,g_{\phi\phi,\theta}
= \frac{\cos\theta}{\sin\theta}\\
\nonumber
\een
Here $prime$ refers to a derivative with respect to $r$ and $dot$ to
a derivative with respect to $t$.

The rest of the coefficients are zero (other than the symmetric 
exchange of the lower two indices)
as there is no variation with
$\theta$ or $\phi$.
These are the connection coefficients we must now use to calculate
the Ricci tensor and the Ricci scalar,
$R_{\mu\nu}$ and $R$.
\be
R_{\mu\nu} = R^{\lambda}{}_{\mu\lambda\nu}
= \Gamma^{\lambda}{}_{\mu\nu,\lambda} - \Gamma^{\lambda}{}_{\mu\lambda,\nu}
+ \Gamma^{\lambda}{}_{\lambda\rho}\,\Gamma^{\rho}{}_{\mu\nu}
-\Gamma^{\lambda}{}_{\nu\rho}\,\Gamma^{\rho}{}_{\mu\lambda}.
\ee

From the Riemann tensor we calculate the Ricci tensor and Ricci scalar
to give us
\be
G_{\mu\nu} = R_{\mu\nu} - \frac{1}{2}\, g^{\mu\nu}\,R.
\ee

This gives
\be
G_{00} = \frac{N^2}{g\, r^2}\left(2\,r\,g' +g\,(g^2-1)\right),
\label{good}
\ee
\be
G_{rr} = \frac{1}{N \, r^2}\left[ 2 \,N'\, r - N \,g^2 +N\right],
\ee
and
\be
G_{tr} = \frac{2\,\dot g}{g\, r}.
\ee
Since $G_{00}= 8\pi T_{00}$ according to Einstein's equations,
we get an equation for $g'$
\be
g' = \frac{1}{2}\,\left[\frac{r\,g}{N^2} \,(8\,\pi\, G) \,T_{00} -\frac{g}{r}\,(g^2-1)
\right].
\label{gprima}
\ee
The $rr$ component of the Einstein equation is $G_{rr}=8\pi T_{rr}$.
This gives us an equation for $N'$ which is
\be
N' =\frac{1}{2} \,N\, \left[ 8\,\pi \, G \,T_{rr}\, r + \frac{g^2-1}{r}\right].
\label{lapsa}
\ee
Similarly from the $tr$ component of Einstein's equations
we get
\be
\dot g = 4\,\pi \, G \,r\,g\, T_{tr}.
\label{gdot}
\ee
These are the equations in {\em GR}. Equation (\ref{gprima}) is
the Hamiltonian constraint equation and is used to
monitor the accuracy of the code. Equation (\ref{lapsa})
is used to calculate the lapse on each time step and equation
(\ref{gdot}) is an evolution equation for the radial metric which
we use to get the radial metric on a given time step from its value
on a previous time step.
In Brans--Dicke theory, which
we will discuss in the next section,
the form of the field equations changes and
the $8\pi T_{\mu\nu}$ terms in the above equations
would have to be replaced with
those plus additional terms. 

\section{Alternative Theories of Gravity}

We have also studied boson stars in Brans--Dicke theory, and now give an
overview of the history of these theories. We then derive
the equations for spherically symmetric boson stars in a Brans--Dicke theory.
By doing this we are also able to get the equations in {\em GR} by
setting the {\em BD} parameter $\omega =\infty$ in these equations.

Alternative theories of gravity, like Brans--Dicke and Scalar--Tensor theories,
were first propounded to explain the coincidence in values of large numbers
made from dimensionless combinations of physical and cosmological constants.
The idea as originated by Dirac and outlined by Jordan~\cite{jord1}  was as follows: 

There are six constants that determine the large scale structure of the
universe namely
\begin{itemize}
\item{1) } The velocity of light $c$.
\item{2) } The Gravitational constant $ k = \frac{8\pi G}{c^2}$.
\item{3) } The age of the universe $A$.
\item{4) } The Hubble effect red shift
to order 1/r, is, $H_0 = \frac{c\Delta \lambda}{\lambda r}$. 
Here the observed nebula is at distance $r$ from the observer and
its observed red shift is $\Delta \lambda$.
\item{5) } The mass density throughout the universe $\mu$.
\item{6) } the radius of the universe $R$.
\end{itemize}

Amazingly all the dimensionless quantities $H_0 A$, $\frac{R}{c A}$,
and $ k \mu c^2 A^2$ are of order unity. The first of these
relations on substituting for $H_0$ shows that when the age of the
universe was small the expansion must have been rapid. The second
relation suggests the radius of the universe is increasing
proportional to its age.
The last relation in the context of a homogeneous and
isotropic closed universe can be  written as
\be
k \, \mu \, c^2 \, A^2 \sim k \, \frac{M}{R^3}\, R^2 = \frac{k\, M}{R} \sim 1.
\ee
Since the radius of the universe is not fixed it means that either $k$,
$M$ or both are varying.

Turning now to constants derived from microphysics: consider the
ratio of the radius $R$ of the universe to the elementary length $l$
(the electronic radius $e^2/m_0 c^2$). This is a number of the
same order of magnitude ($10^{40}$) as the ratio of the electromagnetic
to the gravitational force, and since $R/l$ increases proportional to the
age of the universe, it suggests that the gravitational constant might not be constant but might
be inversely proportional to the age of the universe.

Assuming one accepts that dimensionless quantities made from physical
and cosmological constants are of the same order, then
the gravitational constant is cosmologically not after all a constant.
Under this premise a modified relativistic theory
of gravitation
compatible with Mach's principle was developed~\cite{brans1}.
In Newtonian and earlier theories space itself is portrayed as having its
own physical structure and properties. Mach's view on the other hand was that
the physical properties of space were intimately connected with the matter
therein. The only meaningful motion of a particle would then be
motion relative to another. {\em GR} partially accomodates to this point of view, with
spatial geometries being affected by mass distributions. However they are
not uniquely specified by those distributions. In order to fully incorporate
Mach's principle, boundary conditions would have to be imposed on the
field equations, eliminating solutions with no mass present.

As seen earlier, the gravitational constant, under the premises of
the large numbers hypothesis, is dependent on mass distribution
in a uniformly expanding universe through
\be
\frac{G\,M}{R\,c^2}\sim 1.
\ee
Here $M$ is the finite mass of the visible universe, assumed to be
of finite radius $R$. This relation suggests that
either the $M$ to $R$ ratio is constant or the locally observed gravitational
radius ($G/c^2$) is determined by the mass distribution about the point in question.
The former might come out of some boundary conditions on the
{\em GR} field equations. The latter however is not compatible with the
``strong equivalence principle" and {\em GR}.

As per Mach's principle, the motion of a test particle is only
meaningful in relation to the rest of the matter in the universe, and
a laboratory accelerated relative to distant matter must experience inertial
forces equivalent to a laboratory that is fixed due to the existence of 
distant accelerated matter. 
If the inertial reactions are interpreted as gravitational forces due to distant accelerated matter then locally observed values of a particle's
inertial mass could depend upon the matter distribution around it. However
there is no way to compare the mass of a particle at one point with the mass of
another particle at a different point.
Mass {\em  ratios} can be compared at two different points
but not masses. Using the gravitational constant we get a 
characteristic mass the Planck mass,
\be
\left(\frac{\hbar\,c}{G}\right)^{\frac{1}{2}} = 2.16\times 10^{-5}g,
\ee
with 
\be
m_e\,\left(\frac{\hbar\,c}{G}\right)^{\frac{1}{2}} \approx 5\times 10^{-23},
\ee
providing a way to compare an electron's mass at different spacetime points.
Likewise the notion of $\hbar$ and $c$ being the same at all spacetime
points is meaningless. However $\hbar$ and $c$ can be defined
to be constant without ambiguity if $m$ or $G$ or both are allowed
to vary from position to position. (Earlier we discussed the possibility
of variation in time.) The formal structure of the theory would be
different depending on whether the mass changes or not. 

In {\em GR} the representation uses units so that atoms have physical properties
independent of location. Assuming such a choice is possible the theory
is simpler if the gravitational constant is the one allowed to vary with
inertial masses of elementary particles being constant.
In {\em GR}, the strong equivalence principle states that the laws of physics,
and their numerical content (like dimensionless
physical constants) are independent of the location of
the laboratory in spacetime. Mach's principle as outlined above is only
compatible with the ``weak equivalence principle''.
This regards local equivalence of gravitational
forces and accelerations and is the only one experimentally supported by the
E$\ddot {\text{o}}$tv$\ddot {\text{o}}$s experiment.

The Brans--Dicke theory has its foundation on the above arguments.
A generalization of {\em GR} is made in which the variation of the gravitational
``constant'' is effected by writing it as a function of a scalar field
which is coupled to the mass density of the universe. 
The contracted metric tensor is a constant and is uninteresting. The scalar curvature and other scalars formed from the curvature tensor contain gradients of the metric
tensor components and fall off faster than $1/r$ from  a mass source. Hence 
these are determined by nearby rather than distant matter. Some new scalar
rather than those provided by {\em GR} must therefore be
introduced which would determine
the local value of the gravitational field.
This scalar field satisfies a scalar field equation of the form
\be
\dal\, \varphi = \fpi \lambda \,{T^{\mu}{}_{\mu}}_{{}M}
\ee
where $\lambda$ is the coupling constant tying the field to matter sources.
The average value of the field can be roughly estimated by taking the
universe to be a uniform sphere of density $\rho \sim 10^{-29}g/cm^3$
and radius $R \sim 10^{28}cm$. This gives an average value of the field
($\sim \lambda \rho R^2$)
of roughly $1/G$  (units $c=1$)
if $\lambda\sim1$. These considerations suggest an
equation of the form
\be
G_{\mu\nu} = -\frac{8\pi}{\varphi}\left[{T_{\mu\nu}}_M + {T_{\mu\nu}}_\varphi
\right].
\ee
However one does not want to give up the successes of the
equivalence principle, namely the equality
of gravitational and inertial masses, so the {\em BD} parameter
should not enter into the equations of motion of the particles leaving 
$T^{\mu}{}_{\mu;\nu} =0$ and of the same form as before.

One starts from the usual variational principle of {\em GR}
\be
0=\delta\int\,\left[R+\frac{16\,\pi\,G}{c^4}L_M\right]\sqrt{-g}\;d^4x,
\ee
(with scalar curvature $R$ and matter Lagrangian $L_M$ including all
nongravitational fields). Dividing through by $G$, and replacing
the inverse of $G$ by a scalar field $\varphi$ gives
a scalar field
equation: 
\be
0 =\delta\int\,\left[
\varphi \,R + \frac{16\,\pi\,}{c^4}L_M - \omega\,(\varphi_{,\mu}\,\varphi^{,\mu}/\varphi)
\right](-g)^{\frac{1}{2}}\; d^4x.
\label{varbd}
\ee
Note that a Lagrangian density for the scalar field has also been added.
In the original Brans--Dicke theory $\omega$ is constant. Writing $\varphi^{,\mu}
= g^{\mu\nu}\varphi_{,\nu}$  and finding the variations as we did in the
Introduction we get, using the traditional $c=1$ units
\ben
R_{\mu\nu}-\frac{1}{2}\,g_{\mu\nu}\, R = 8\,\pi\,\varphi^{-1}\,T_{\mu\nu} &+&
\frac{\omega}{\varphi^2} \,\left(
\varphi_{,\mu}\,\varphi_{,\nu} - \frac{1}{2}\,g_{\mu\nu}\,\varphi_{,\gamma}\,\varphi^{,\gamma}
\right) \\ \nonumber
&+& \varphi^{-1}\,\left(\varphi_{,\mu;\nu} -g_{\mu\nu}\,\Box \varphi\right).
\een

Here $T_{\mu\nu}$ is the usual source term from (\ref{variatmunu}). All
the right hand terms have been multiplied by the factor of $G$ or
$\varphi^{-1}$ in this case. 
%FFig check this:The last term on the right hand side 
%comes from the $R\delta \varphi$ part of the calculation which would
%have been absent in the case of constant $G$. The integral is
%done by parts and the second derivatives of the metric present in $R$ are
%eliminated giving the term.

\subsection{Boson Stars in Brans--Dicke Theory: Spherical Symmetry}
\label{subsection:sectionbd}

We now turn to the description of Boson Stars in Brans--Dicke theory
and derive the equations that describe the system. We start out
with this model because Boson Stars in {\em GR} can be derived as a 
special case of these and the equations for the case of massless
scalar field halos follow as a further specialization. These are the
three spherically symmetric models that we have studied using our $1D$ code.
In fact we start with a general scalar tensor case where the parameter
$\omega$ is not constant but may in fact be a function of the scalar
and/or tensor
fields.

The matter Lagrangian in this case is one from which the relativistic
scalar field or Klein-Gordon equation is derived and is of the form
\be
L_M=-\frac{1}{2}\,g^{\mu\,\nu}\,\partial_{\mu}\phi^{\ast}\,\partial_{\nu}\phi
-\frac{1}{2}\,m^2\,{|\phi|}^2-\frac{1}{4}\,\lambda\,{|\phi|}^4,
\end{equation}
where $\phi$ is the boson field and the $\lambda\,{|\phi|}^4$ term is
a self-coupling term. This is the simplest way to include a self-interaction
while ensuring stability as well as renormalizability. 

The action for the system is given by
\be
S_J
=\frac{1}{16\,\pi}\,\int\,d^4x\;\sqrt{-g_J^{}}\, \left({\varphi}_J^{}\,R
-\frac{\omega(\phi)\,{{g^{\mu\,\nu}}_J}}{{\varphi}_J^{}}\,\partial_{\mu}
{\varphi}_J^{}\,
\partial_{\nu}{\varphi}_J^{}\right)
+ \int\,d^4x\;\sqrt{-g_J^{}}\,L_M,
\label{actionjordan}
\end{equation}
where the subscript $J$ refers to the fact that we are in the physical
or Jordan frame. 

By a ``physical frame'' we mean a frame in which all fields give
rise to positive--definite energy density. In the {\em Jordan Frame}
gravity is entirely described by the metric $g_{\mu\nu}$. There
is also another frame, related to this one by a conformal transformation,
the {\em Einstein Frame}. In this frame the scalar tensor field acts like
an external matter field which is the source for the metric. Hence, in
the {\em Einstein Frame}, where the equations are like those in
{\em GR}, gravity is not represented by the metric alone but by the
scalar--tensor field as well. Thus scalar--tensor theories and {\em GR}
may be mathematically similar, but are physically dissimilar. The field
variables $g_{\mu\,\nu}$ and $\varphi_J$ define the {\em Jordan Frame}.
By assumption the matter couples to the metric and does not couple to
the gravitational scalar field. Thus the {\em Jordan Frame} is
assumed to be the physical frame~\cite{magnano1}.

Varying the action with respect to the metric gives, as
seen in the previous subsection, the gravitational equation
\ben
G_{\mu\nu} = \frac{8\,\pi}{{\varphi}_J^{}}\,{{T}_{\mu\nu\,J}} +
\frac{\omega}{{{\varphi}_J^{}}^2} \left(\partial_{\mu}{\varphi}_J ^{}\,
\partial_{\nu}
{\varphi}_J^{} 
- \frac{
{{g_{\mu\nu}}_J}
}{2}\,\partial_{\lambda}
{\varphi}_J^{} \, \partial^{\lambda}
{\varphi}_J\right) 
\label{gmunubda}
\\ \nonumber
+ \frac{1}{
{\varphi}_J^{}
}\left(
{
{{\varphi}_{;\mu;\nu}}_J
}-
{{g_{\mu\nu}}_J }\,
{{\varphi_{;\lambda}}^{;\lambda}}_J\right).
\een
Varying the action with respect to the boson field, gives us,
according to the Euler-Lagrange equations
\be
{\partial_\lambda} \frac{\partial L_M}{\partial 
(\partial_{\lambda}\phi )} = \frac{\partial L_M}{\partial \phi},
\ee
the scalar field equation
\be
{g^{\mu\,\nu}}_J\,\partial_{\lambda}\partial_{\nu}\phi \; \delta^{\lambda}{}_{\mu}
-m^2\,\phi-\lambda \,|\phi|^2 \,\phi^{\ast}=0,
\ee
or
\be
\phi_{;\lambda}{}^{;\lambda}-m^2\,\phi - \lambda \, |\phi|^2 \,\phi^{\ast}=0.
\ee
Varying the action with respect to the $ST$ field $\varphi$ gives
\be
\frac{2\,\omega}{{\pst}_J}{{\varphi_J}^{;\lambda}{}_{;\lambda}}
-\omsts \,\partial^{\lambda}{\pst}_J \,
\partial_{\lambda}{\pst}_J + R_J +\domst \, \frac{g_J^{\mu\nu}}{{\pst}_J
}\,\ddmu{\pst}_J \,\ddnu{\pst}_J
=0.
\label{torgundphist}
\ee
Taking the trace of (\ref{gmunubda}) gives
\begin{equation}
R=-\pte+\omsts \,\partial_{\lambda}\varphi \, \partial^{\lambda}\varphi +
\frac{3\,\pstl}{\pst}.
\end{equation}
Also, from the form of the matter Lagrangian, we have that the stress energy
tensor in this frame is
\be
T_{\mu\nu}=\frac{1}{2}\,\left(\ddmu\phi^{\ast}\,\ddnu\phi +c.c\right)
-\frac{1}{2}\,{g^{\mu\nu}}_J\,\left({g^{\rho\kappa}}_J\,\partial_{\rho}\phi^{\ast}\,
\partial_{\kappa}\phi + m^2\,|\phi|^2 + \frac{1}{2}\,\lambda \,|\phi|^4
\right).
\ee

However for convenience we make a conformal transformation to the
{\em Einstein Frame} keeping in mind that when we solve the equations
we must revert back to the {\em Jordan Frame} when interpreting the results
physically.

The transformation from one frame to the other is given by
\begin{equation}
{\gdmunu}_J = e^{2\,a(\pst)}\,\gdmunu.
\end{equation}
We define a constant $G_{\ast}$ through
\begin{equation}
{\phi}_J^{-1}=G_{\ast}\,e^{2\,a(\pst)}.
\end{equation}
Also
\begin{equation}
\alpha(\pst)=\frac{\partial\,a}{\partial\,\pst},
\end{equation}
where
\begin{equation}
\alpha^2=\frac{1}{2\,\omega+3}.
\end{equation}
Here the quantities without subscript $J$ refer to quantities in the Einstein frame.

Going back to the action in the Jordan frame (\ref{actionjordan})
we get from
\be
R_J = e^{-2a}\,\left[R-6\,g^{\mu\nu}\, \nabla_{\mu}\nabla_{\nu} a -
6 \, g^{\mu\nu}\,(\nabla_{\mu} a)\,(\nabla_{\nu} a)\right],
\ee
and the other transformations
\ben
S_J = &&\frac{1}{16\,\pi}\,\int\,d^4x \; e^{4a}\,\sqrt{-g}\,
\Bigg(
e^{-2a}\, e^{-2a} 
\\ \nonumber
&& \qquad\qquad\qquad
\left[
 R-6\,g^{\mu\nu} \, \nabla_{\mu}\nabla_{\nu} a 
- 6\, g^{\mu\nu}\,(\nabla_{\mu} a)\,(\nabla_{\nu} a)\right]
\\ \nonumber
&& \qquad\qquad\qquad 
\left.
-\left(\frac{1}{2\,\alpha^2}-\frac{3}{2}\right)\,e^{2a}\, e^{-2a}\, g^{\mu\nu}\,
{\partial_\mu}
e^{-2a} \,
{\partial_\nu} e^{-2a}\right) \\ \nonumber
&-&\int\,d^4x \;
\sqrt{-g}\,\Bigg[
e^{2a}\,\left(
+\frac{1}{2}\, g^{\mu\,\nu}
\partial_{\mu}\phi^{\ast}\, \partial_{\nu}\phi\right)
+ e^{4a}\, \left( 
-\frac{1}{2}\, m^2\, {|\phi|}^2-\frac{1}{4}\, \lambda\,{|\phi|}^4\right)
\Bigg].
\een
Thus
\ben
S_J = \frac{1}{16\pi}&&\, \int\,d^4x\; \sqrt{-g}\Bigg(
R-6\,g^{\mu\nu}\, \nabla_{\mu}\nabla_{\nu} a 
 \\ \nonumber
&& \qquad\qquad \qquad -6 \, g^{\mu\nu}(\nabla_{\mu} a)\,(\nabla_{\nu} a)
-\left(\frac{1}{2\,\alpha^2}-\frac{3}{2}\right)\,g^{\mu\nu} \,4\, \nabla_{\mu} a\, \nabla_{\nu} a 
\Bigg) \\ \nonumber
&-& \int\,d^4x \;
\sqrt{-g}\,\left[
e^{2a}\,\left(
+\frac{1}{2}\,g^{\mu\,\nu}\,
\partial_{\mu}\phi^{\ast}\, \partial_{\nu}\phi\right)
+ e^{4a} \, \left( 
-\frac{1}{2}\, m^2\, {|\phi|}^2-\frac{1}{4}\,\lambda\,{|\phi|}^4\right)\right].
\een
or
\ben
S_J &=& \frac{1}{16\,\pi}\int\,d^4x \; \sqrt{-g}\,\Bigg(
R -6\,g^{\mu\nu} \, \nabla_{\mu} \,\nabla_{\nu} a - \frac{2}{\alpha^2}
\,g^{\mu\nu} \,
\nabla_{\mu} a \, \nabla_{\nu} a \bigg)
\\ \nonumber
&-& \int\,d^4x \;
\sqrt{-g} \, \left[
e^{2a}\,\left(
+\frac{1}{2} \, g^{\mu\,\nu} \,
\partial_{\mu}\phi^{\ast} \, \partial_{\nu}\phi\right)
+ e^{4a} \, \left( 
-\frac{1}{2} \, m^2 \, {|\phi|}^2-\frac{1}{4}\,
\lambda\,{|\phi|}^4\right)\right].
\een
Since $\nabla_{\mu}\nabla_{\nu} a$ is a total divergence the integral
vanishes and we get the action in the Einstein frame
\ben
S_E &=& \frac{1}{16\, \pi} \, \int \,d^4x \; \sqrt{-g} \, \left(
R - 2 \, g^{\mu\nu} \, \nabla_{\mu} \pst \, \nabla_{\nu} \pst \right) 
\\ \nonumber
&-&\int\,d^4x \;
\sqrt{-g} \, \left[
e^{2a}\,\left(
+\frac{1}{2} \, g^{\mu\,\nu} \,
\partial_{\mu}\phi^{\ast} \, \partial_{\nu}\phi\right)
+ e^{4a}  \, \left(
-\frac{1}{2} \, m^2 \, {|\phi|}^2-\frac{1}{4} \,
\lambda\,{|\phi|}^4\right)\right].
\een
Again using the variational principle described in the earlier subsection we
get
\be
G_{\mu\nu}  = 8 \, \pi \, G_{*} \, T_{\mu \nu} +
\partial_{\mu}\partial^{\nu} \varphia -  \partial_{\tau} \varphia \,
	       \partial^{\tau} \varphia \, g_{\mu \nu} \label{2.8},
\end{equation}
We get the stress energy tensor from the matter Lagrangian term
by variation with respect to the metric as seen in the previous subsection
so that
\begin{eqnarray}
  T_{\mu \nu} &=& {1 \over 2} \,e^{2 a(\varphia)} \, \left(\partial_{\mu}
 \psia^{\ast} \,  \partial_{\nu} \psia + \partial_{\nu}
  \psia^{\ast} \, \partial_{\mu} \psia\right)
  \label{2.7}
 \\ \nonumber
 &&\qquad - {1 \over 2}\, 
g \, e^{2 a(\varphia)} \, \left(
	  \partial_{\tau}
		  \psia^{\ast} \, 
	  \partial^{\tau} \psia +
  e^{2 a(\varphia)}\,
 \left[m^2 \,
  \psia^{\ast} \, \psia + 2 \, 
  V(\psia^{\ast}
 \psia)\right]\right)
	  g_{\mu \nu}  ,
 \end{eqnarray}
 where the first set of terms comes from variations of $L_M$ and
 the second set from variations of $\sqrt {-g}$ with respect to the
 metric.
 Variation with respect to the Brans--Dicke field gives from the
 $E$-$L$ equation
 \ben
 & &
 -2\, g^{\mu\nu}\, {{\nabla}_\lambda} \delta^{\lambda}_{\mu}
 {\nabla_\nu}\varphi 
 -2\, g^{\mu\nu} {{\nabla}_\lambda} \, \delta^{\lambda}_{\nu} \,
  {\nabla_\mu}\varphi 
  \\ \nonumber
&& \qquad = -16\,\pi\, e^{2a}\, 2\, \alpha \, g^{\mu\nu}\,\left(\frac{1}{2}\,
\partial_{\mu}\phi \,\partial_{\nu}\phi^{\ast} \right)
+ 16\, \pi \, e^{4a} \, 4 \alpha \, \left(\frac{1}{2}\,
m^2 \, {|\phi|}^2-\frac{1}{4} \, \lambda\,{|\phi|}^4\right),
\een
So that
\be
{{\nabla}_\lambda} {\nabla^\lambda}\varphi
= -4 \, \pi \, \alpha \, \left[-
g^{\mu\nu} \, \partial_{\mu}\phi \, \partial_{\nu}\phi^{\ast}
- 4 \, \left(\frac{1}{2}\,
m^2\,{|\phi|}^2-\frac{1}{4} \, \lambda\,{|\phi|}^4\right)\right],
\label{phisteq}
\ee
which is nothing but
\be
{{\nabla}_\lambda} {\nabla^\lambda}\varphi
=  -4 \, \pi \, \alpha \, T,
\ee
where $T$ is the trace of the stress-energy tensor.
For the scalar field the $E$-$L$ equations gives
\be
\nabla_{\mu} \left(e^{2a} \,  \nabla^{\mu} \phi\right)
= e^{4a} \, \left(-
m^2\, \phi -\lambda\,{|\phi|}^2 \, \phi\right),
\ee
giving 
\be
\nabla_{\mu}\nabla^{\mu}\phi +
2\, \alpha \, {\partial_\lambda} \phi \, {\partial^\lambda} \pst \,
e^{2a} \, \left(-
m^2 \, \phi -\lambda\,{|\phi|}^2 \, \phi\right).
\ee

\subsection{{\em BD} Stress Energy Tensor: Equilibrium
Conditions}

From (\ref{2.7}) we see that the relevant components of the
stress-energy tensor for the spherically symmetric problem are given by
\begin{eqnarray}
T_{00}&=&\frac{1}{2} \,
e^{2a} \, \left[{N^2 \over g^2} \, |{\partial_r} \phi|^2 + |\dot \phi|^2
 + N^2 \, e^{2a} \, \left(m^2 \, |\phi|^2 + \frac{\lambda}{2}| \, \phi|^4\right)
 \right],
 \\ \nonumber
 T_{rr}&=& \frac{1}{2} \,
 e^{2a} \, \left[ | {\partial_r} \phi |^2
 +{g^2 \over N^2} \, |\dot \phi|^2 - g^2 \, e^{2a}\,
 \left(m^2 \, |\phi|^2 + \frac{\lambda}{2} \, |\phi|^4\right)
 \right] \\ \nonumber
 T_{0r} &=& \frac{1}{2} \, \left( \dot \phi \, {\partial_r} {\phi^{\ast}}
 + \dot {\phi^{\ast}} \, \dot \phi\right).
 \end{eqnarray}

Note that we had been using units with $\hbar=c=1$ so that distance and
time were in units of $1/mass$. 
The boson field has an
underlying oscillation frequency $\omega_E$ for an equilibrium
boson star. If we look at the form of the stress energy tensor
we notice that every term with the boson field or its derivative
is multiplied by $\phi^{\ast}$ or its derivative, indicating that a time
dependence in the field of the form $e^{i \omega_E t}$ does not introduce
any time dependence in any physical quantity. Hence even though the
field has this time dependence the system can still be in equilibrium
provided the metric is not time dependent. This underlying frequency is
crucial for a boson star configuration to be held together. Gravity, ever
attractive, is trying to cause the system to collapse and to prevent this
collapse the dispersion of the wave equation must be used to
balance it. For a given
configuration a very specific frequency of the wave will exactly balance the
gravitational effects. We have used the symbol $\omega_E$ to distinguish
it from the scalar tensor parameter $\omega$.

We now make a change of coordinates to a set of dimensionless ones:

\begin{equation}
r= m \, r_{\mathrm old}, \quad t=\omega_E \, t_{\mathrm old},\quad \sigma=\sqrt{4\, \pi \,  G_{\ast}} \, \Phi, \quad
N= N_{\mathrm old} \, \frac{m}{\omega_E},\quad \Lambda=\frac{\lambda}{m^2\, 4
\,\pi \,  G_{\ast}}.
\label{dimensionless}
\end{equation}
We define $\phi(r,t) = \Phi (r) e^{i \omega_E t}$.

We now use equations (\ref{good})-(\ref{gdot}), with the modification that
in the {\em BD} case all factors of $8\pi G T_{\mu\nu}$ be replaced by
the right hand side of (\ref{gmunubda}). For the equilibrium configurations
we take the {\em BD} field to have no time dependence. So in
dimensionless coordinates we have the equations for an equilibrium boson
star, namely the scalar field equation for the {\em BD} field ((\ref{phisteq}) with
$\varphi$ replaced by $\varphi_{BD}$), plus the scalar field equation
for the boson field,
and the $tt$ and the $rr$ components of the gravitational field equation
\begin{eqnarray}
{\partial_r}{\partial_r}\varphi&=&
\left[-{g^2 +1 \over r} + 2\,  e^{4a}\, r \, g^2 \, V(\sigma)\right]
{\partial_r}\varphi \nonumber \\&&
+ g^2 \, \left\{ 2 \, \alpha \, e^{2a} \,
 \left[{1 \over 2} \, \nabla_\sigma \sigma  \, \nabla^\sigma \sigma
 +2 \,e^{2a} \, V(\sigma) \right ] \right\}, \label{field1-slv} \\
 {\partial_r}{\partial_r} \sigma&=&
 \left[-{g^2 +1 \over r}  +  2 \, e^{4a} \, r \, g^2 \, V(\sigma)\right]
 {\partial_r} \sigma
  -{g^2 \over N^2} \, \sigma  \nonumber \\&&
  + 2 \, g^2 \, \left[e^{2a} \,  {d V(\sigma) \over d \sigma} - \alpha \,
  \nabla_\sigma \sigma  \, \nabla^\sigma \varphi \right], \label{field2-slv} \\
  {\partial_r} (g^2)&=& {g^2} \, \left({g^2 \, r \over N^2} \, 2\, T_{00} +r \,
  (\partial_r \varphi_{BD})^2
  - {g^2 -1 \over r} \right),  \label{field3-slv} \\
  {\partial_r} (N^2)&=& {N^2} \, \left(2\, r \,  T_{rr} + 
  +r\, (\partial_r \varphi_{BD})^2 - {g^2 -1 \over r} \right),
  \label{field4-slv}
  \end{eqnarray}
  where
  \be
  V(\sigma) = \frac{1}{2}\, \left(\sigma^2 + \frac{\Lambda}{2} \, \sigma^4
  \right).
  \ee

We have just derived the equations of the equilibrium
{\em BD}  boson star system and we will now discuss the details of the configurations and code.
We note that for the case of boson stars in Einstein Gravity
we just need to set $a =0$ in the above equation as there is no {\em BD} field 
involved. Finally, we make the further modification $m=0$ to get the
equations for configurations with a massless boson field.

\subsection{Constraints on the value of $\omega$}

The perihelion shift of Mercury provided the first constraint on $\omega$~\cite{sweinberg}.
The perihelion shift is a measure of orbital precession of a planet
that does not follow a closed ellipse. If it did the change in azimuthal
angle at the end of each revolution as it went from its perihelion position
to aphelion and back, would be $2\pi$. The deviation from
$2\pi$ is a measure of the perihelion shift. 

An isotropic static metric 
($ds^2=g^{\mu\nu}dx^\mu dx^\nu$ time independent and all spatial
terms made
from scalar products of ${\bf x}$ and ${\bf dx}$)
can be chosen to be written in the standard form
$ds^2 = - B(r) dt^2 +A(r)dr^2 +r^2 d\Omega^2$. For a static spherically
symmetric body like the sun, we expand the metric in factors of $MG/r$
\be
ds^2 = -\left( 1 + \alpha \, \frac{M\, G}{r} + \beta \, \frac{M^2\, G^2}{r^2}+...\right)\, dt^2 +
\left( 1 + \gamma \, \frac{M \, G}{r}+ ...\right) \, dr^2 + r^2 \, d\Omega^2.
\label{dsschwar}
\ee

For the gravitational case in the weak field limit Einstein's equations
deliver the Schwarzschild metric with
$\alpha =-2$, $\beta=0$ and $\gamma=2$. For the Brans--Dicke case
the equations give $\alpha =-2$, $\beta = 2-\gamma$ and $\gamma =
2\frac{\omega+1}{\omega+2}$. 
Once the form of the metric is determined the geodesic equations can
be written out for a particle in this metric, as, for example
a planet in the
Sun's metric. Since the field is isotropic we can choose to remove $\theta$
dependences and confine calculations to the $\theta=\frac{\pi}{2}$ plane.
We thus arrive at a formula for $\frac{dr}{d\phi}$ with two constants
representing energy and angular momentum. 
At the major and minor axes of the ellipse, where $\frac{dr}{d\phi}=0$,
one can solve the equation to determine the two constants.
Integrating from the minor to the major axis and multiplying by 2 gives the
change in $\phi$ in one revolution and subtracting $2\pi$ provides
the perihelion shift. This effect is maximum for Mercury, the 
closest planet to the Sun. The {\em GR} prediction is a perihelion shift of about
$43.03$ seconds per century. This is in close agreement with observation. 
However one of the assumptions of the calculation was that the Sun was a 
perfect sphere, while in fact it is slightly oblate.
Dicke and Goldenberg~\cite{dickeoblate} photoelectrically scanned the
solar disk and reported that the Sun's polar diameter was shorter than its
equatorial diameter by some $(0.005 \pm 0.0007)\%$. Assuming the
assymetry persisted throughout the Sun, the implied quadrupole
moment could account for an extra precession of $3.4$ seconds per century.
Thus only only $39.6$ seconds could be a general relativistic effect.
Thus Dicke and Goldberg claimed that the {\em GR} prediction was
really in $8\%$ disagreement with observation.
For {\em BD} theory the value is $(3\omega+4)/(3\omega+6)$ of
the {\em GR} value so a bound on $\omega$ would be given by,
\be
\frac{3\, \omega+4}{3\,\omega+6} \, (108/100) > 1
\ee
This put a lower limit $\omega \ge 6.4$. 
Historically,
this was the crucial experiment that brought the Brans--Dicke theory
into prominence.

Recent experiments~\cite{VIKING,VLBI} place a much tighter
constraint $\omega_{BD} > 500$. Note
the $\omega=\infty$ limit of the {\em BD} theory is {\em GR}.

\section{Boson Halos as Dark Matter}
\label{section:halo}

One of the boson models we have studied in $1D$ is made from a massless
complex scalar field. We have studied the evolution and the
formation of such a system~\cite{balamsls}. The model had
been proposed as a dark halo model~\cite{sch1halo,sch2halo}. 

As discussed earlier, the flatness of the galactic rotation curves
indicates the presence of a dark halo. If there was no dark halo,
then beyond edge of the luminous core, the rotation curves should
follow a Keplerian $v \sim \sqrt{\frac{1}{r}}$ fall off.
Instead they are flat, that is
$v \sim constant$ which implies $M(r) = v_{limit}^2 r$. This suggests
a linear relation between mass functions of galaxies with distance
and a density distribution $\rho \sim 1/r^2$.
A model for the mass of this functional form, but that does not fall inward
to the center of the galaxy and form black holes, is needed.

Attempts to fit the galactic rotation curve
using the boson star model have been made in~\cite{sin,jaewonleehalo}. While the ground
state boson stars have a mass falloff which is too fast to
explain rotation curves, for more than three nodes 
the rotation curves resemble observed ones. 
Since excited states of boson stars are inherently
unstable, one justifies these models on the grounds that the time scale of
instability is large.
There is a limit on the number of nodes, since the mass of the system
increases with number of nodes, and $\Omega >1$ if more than
six nodes were present. The best fits are then with four or five nodes.
The similarity in ``quantum numbers'' for all galaxies is explained on the
basis that on an astronomical time scale all galaxies were formed
simultaneously, the dark matter having collapsed from some more dispersed
state of higher quantum number to the present one. The absence of lower
quantum numbers is speculated to be due to higher energy level states being
closer together while lower energy levels are more separated. Since
the energy dissipation mechanism is inefficient evolution to lower
states would be harder.
The calculated density profile is $\rho \approx r^{-1.6}$. This is
slightly increasing compared to the flat curve prediction of
$\rho \approx 1/r^2 $. By including a repulsive self-coupling term one
can increase the time scale of stability. Also one has a wider range
of mass and self-coupling to fit curves with. However
regardless of the value of
self-coupling, the rotation curve slightly increases compared
to the prediction with
$\rho \approx r^{-1.7}$. These attempts to fit
unstable configurations of boson stars to explain galactic rotation curves
could be considered rather fanciful. A more natural choice seems to
involve massless scalar fields.

The massless scalar field model is governed by
equations that can be derived from the {\em BD} equations
above by setting $m=0$ and $a=0$. To get an idea of the
nature of the solutions we consider the $M \rightarrow 0$ limit
Newtonian system described in~\cite{sch1halo,sch2halo}.
The scalar field equation is then of the form
\be
\frac{\partial^2}{\partial t^2} \psi - \nabla^2\psi=0.
\ee
For a field $\psi =\sigma(r)e^{i \omega t}$ this equation 
has a  Bessel solution 
\be
\sigma (r) = A\, \frac{\sin (\omega \,r)}{\omega \,r}.
\ee
Regularity at the origin rules out the ``cosine'' solution.
(This provides an approximate solution to the low density
system. We have studied the evolution of configurations of this form
in our evolution code to see if they settle down to stable configurations.
This is described in the next chapter.)

The density
$\rho =-T^{0}_{0}$ is for this case given by
\be
\rho = \frac{A^2}{\omega^2 \, r^2}\left[ 1- \frac{\sin(2\,\omega \,r)}{\omega \,r}
+ \frac{\sin^2(\omega \, r)}{\omega^2 \, r^2}\right].
\ee
For small $r$ this behaves like a constant and for large $r$ has a
$1/r^2$ dependence making it a viable halo model.
The rotational velocity profile $v^2_{Newt} = M(r)/r \sim A^2$.
The mass scales as $\frac{1}{\omega}$ while the energy density scales as
$\omega^2$. The central amplitude of the scalar field,
namely the parameter $A$ of the Newtonian solutions, determines the orbital
velocity while
the scalar field frequency varies the mass and density near the galactic
center. The best fits to the rotation curves are found if an almost dominating
scalar field halo is added to low luminosity galaxies. The energy
density decreases in leading order as $A^2/r^2\omega^2$
and must eventually attain an equilibrium with some other
matter at
the same pressure. Alternatively the radius of these entities
may be able to estimate the size of the cosmological constant $\Lambda$.

We have investigated the nature of these solutions in full {\em GR}
as well as studied the formation of these entities. We will
describe the equilibrium configurations
in the next section. The results of our investigation of
their stability and formation are detailed
in chapter three of this thesis. The
main question that remains is the basis for truncation of
this halo at some outermost value of radius.

\section{Equilibrium Configurations}

We now describe the results of numerically integrating the equilibrium equations for the three
spherically symmetric cases, Boson stars in {\em GR} and {\em BD}, and Boson Halos. We show the
mass profiles determined from our code and in the next chapter we discuss the results of dynamical
evolutions. In all cases the equilibrium profiles have static metrics although the fields
themselves have an $e^{i \omega_Et}$ dependence which leaves all physical quantities
time independent.

\subsection{{\em GR} Case}

The equations in the dimensionless coordinates defined in (\ref{dimensionless}) for
a self-coupled equilibrium boson star are given by (\ref{field2-slv})--(\ref{field4-slv}) with
parameter $a$ set to zero (and no {\em BD} scalar field equation).
Note that there is a self-coupling term which we
have already motivated in the introduction. For
very high values of the self-coupling parameter, it has
been shown in~\cite{colp1} that the field varies very slowly
over a very large radius. As a result of this one can ignore the field
derivative terms in the field and metric equations as compared to
the field itself. In fact in the $\Lambda \rightarrow \infty$
limit one can write an equation of
state with the field effectively vanishing at the boundary. Thus
the perfect fluid theorems
of stellar stability might be used for very high values of $\Lambda$~\cite{colp1} where an approximate equation of state might be written.
These theorems, however,
cannot apply for smaller values of self-coupling. We choose to study
the stability of these stars numerically~\cite{seinum,balalam}.

The system of equations describing the equilibrium configurations (static
metrics although the scalar field itself has an $e^{i\omega_Et}$
dependence), form an eigenvalue problem that we have to solve
in order to get the equilibrium
boson star solution. This is because we have an overdetermined system. We
specify the field at the origin. The radial metric at the origin is by
regularity required to be unity. The field derivative and derivatives of the
metrics must be zero at the origin by conditions of spherical symmetry and
absence of divergences at the origin. This gives five
conditions with four first order equations. As shown before we have the
freedom to rescale the lapse with the underlying frequency of the
boson field. Thus the eigenvalue giving the value of the lapse at the
origin is physically a way to calculate the
frequency of the field such that the system has the exact amount of
dispersion
required to prevent gravitational collapse. For a given value of
central field density $\sigma(0)$ the different eigenvalues correspond
to different numbers of nodes. 

We also have a boundary condition on the boson field to vanish at $\infty$. 
We therefore use a shooting method to determine the eigenvalues.
This is done by making a guess of the $\omega$
rescaled lapse at the origin,
and then using this guess to integrate the equations. If one is looking
for a ground state solution one knows the field must have no nodes.
It also should be monotonically decreasing and must vanish at $\infty$.
So at some stage with this guess the field will either go negative
or will develop a positive slope. The trick is to bisect between two values
one of which gives a positive slope and the other of which causes the
field to become negative and to continue this process as far as the
machine permits so that the field falls off giving us the solution to
greater and greater radii until the machine limit is reached. In the one
node case a guess value will be used until the field crosses zero a second
time, or if the slope changes sign a second time and so on.
A fourth order {\em Runge Kutta} technique~\cite{Press} is used to
integrate the equations.

\begin{figure}[t]

\begin{center}
\leavevmode
\makebox{\epsfysize=15cm\epsfbox{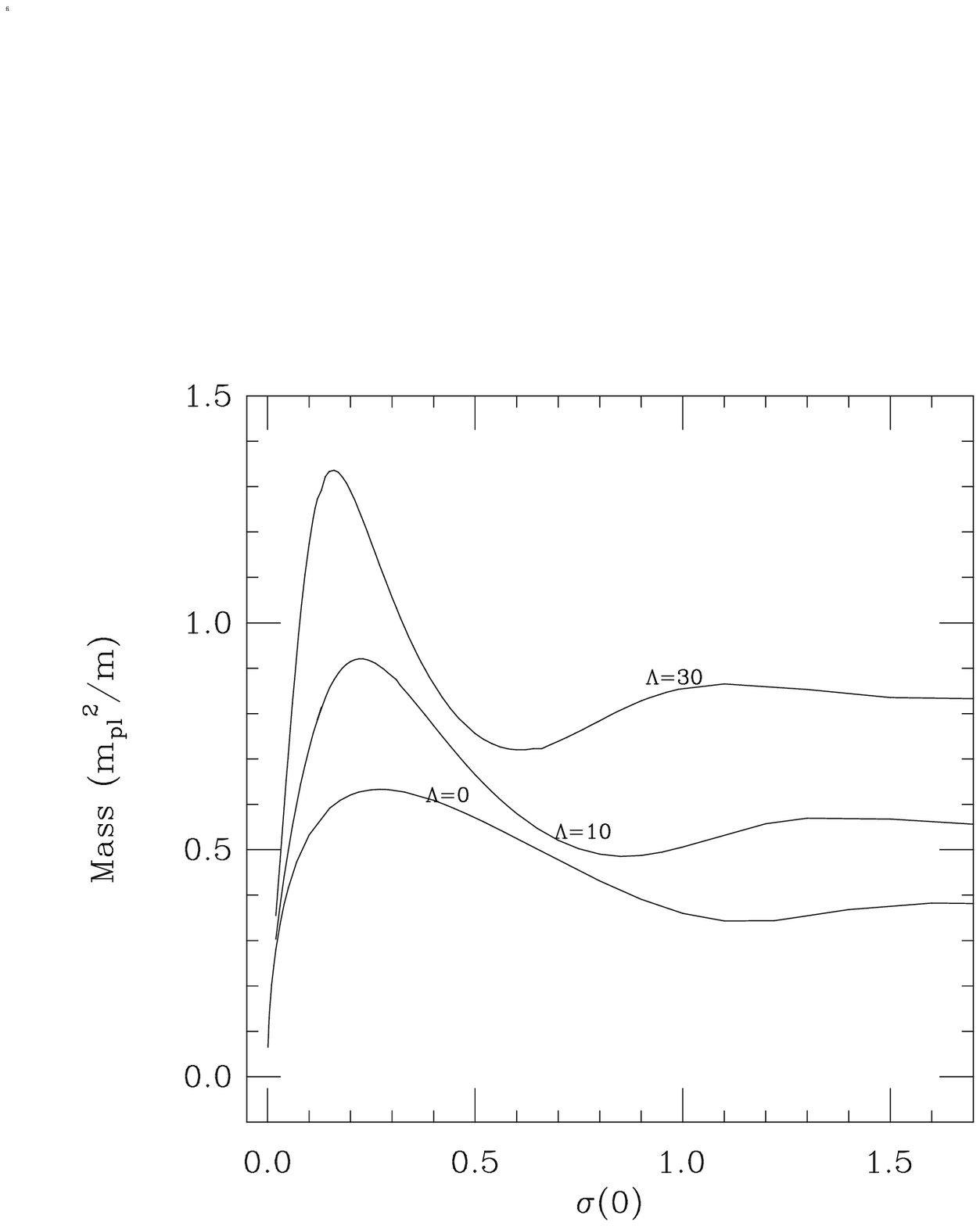}}
\end{center}
%%\begin{figure}
%%\hspace{-36pt}
%%\vspace{-130pt}
%%\epsfbox[-50 -150 350 500]{figs/masslame.ps}
%newstuff
%\begin{figure}[t]
%\begin{center}
%\leavevmode
%\makebox{\epsfysize=15cm\epsfbox{newfigs/masslam.ps}}
%\end{center}
%endnewstuff
\caption{
Mass profiles of ground state boson stars for different values
of the self-coupling $\Lambda$ are shown. The increase in mass with
$\Lambda$ is clear although
the profiles are very similar.}
\label{fig_masslam}
\end{figure}

The mass profile of boson stars is a much discussed feature
which we have already commented on. However, for the sake of
completeness, we show the ground state mass profiles as determined from our equilibrium code
in Fig.\ref{fig_masslam}. Clearly the mass increases with increased
self-coupling. 

\begin{figure}[t]

\begin{center}
\leavevmode
\makebox{\epsfysize=15cm\epsfbox{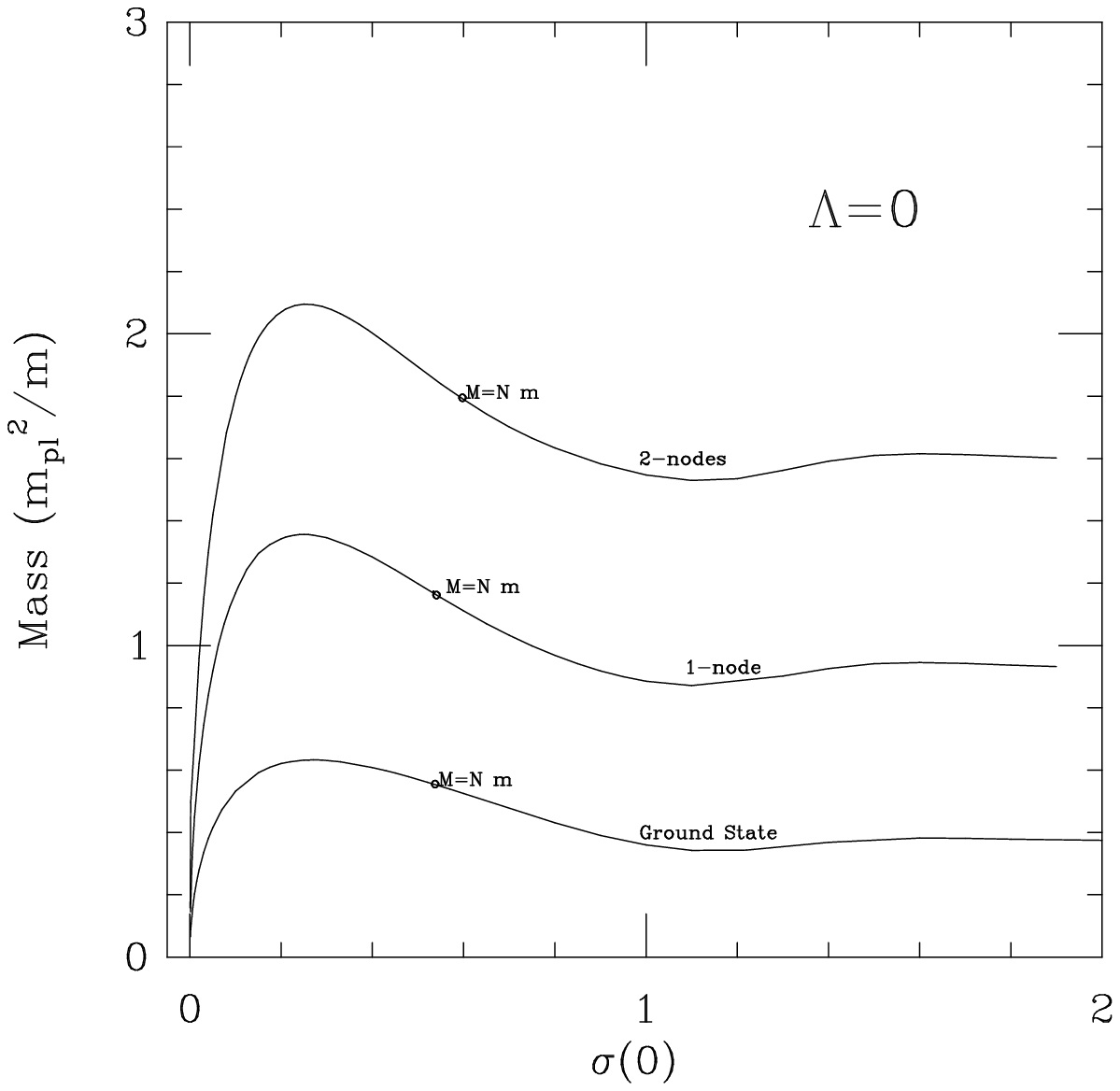}}
\end{center}
%%\begin{figure}
%%\hspace{-36pt}
%%\vspace{-130pt}
%\epsfbox[0 -250 450 500]{figs/lamexcmass.ps}
%%\epsfbox[0 -100 350 500]{figs/lamexcmasse.ps}
\caption{
Masses of zero-node, one-node and two-node boson stars
without self-coupling are plotted as a function of central
density. The maximum mass of one-node stars is $1.356 \,M_{Pl}^2/m$ while the
maximum mass for two-node stars is (as expected) greater at $2.095 \, M_{Pl}^2/m$.
The profiles are deceptively similar to their ground state counterparts.
Excited state stars are
inherently unstable irrespective of the branch they lie on, unlike ground
state stars that can be termed stable or unstable depending on whether they
lie on the branch to the left of the maximum mass or to the right
respectively.
}
\label{fig_lamexcmass}
\end{figure}

The mass profiles of excited state boson stars are
similar to ground state stars.
Fig.\ref{fig_lamexcmass} shows the mass versus central density curves 
for ground, first, and second
excited states of boson stars without self coupling. The maximum mass increases
with the number of
nodes as expected.

\begin{itemize}
\item{{\bf Runge Kutta ({\em RK}) Method}}
\end{itemize}

Whenever we have ordinary differential equations of any order we
are able to rewrite them as a series of first order equations, as
we have done above for the boson eigenvalue problem, by redefining the
field derivative as a new variable, and with this definition serving as a
a new equation. In each of the other equations the second derivative of the
field is replaced by the first derivative of this new variable, and the
first derivative of the field is replaced by the new variable itself
\cite{Press}.
In general
we then have to satisfy $N$ equations of the form
\be
\frac{dy_i(x)}{dx} = f_i(x,y_1,...,y_N), i=1,2,...,N.
\ee
For simplicity consider $N=1$.
This can be solved by
\be
y_{n+1} = y_n + \Delta x\,f(x_n,y_n) + O({\Delta x}^2).
\label{yneuler}
\ee
which moves the solution from $x=x_n$ to $x=x_n+\Delta x$. This is the Euler
method. To increase accuracy
and stability we expand the second term in a Taylor's series about
$f(x_n+\Delta x/2, y_n + \Delta y/2)$ (where $\Delta y = \Delta x f(x_n,y_n)$)
giving
on writing the $O({\Delta x}^2)$ term as well
\ben
y_{n+1} &=& y_n + \Delta x\,f(x_n+\Delta x/2, y_n + \Delta y/2) -\frac{{\Delta
x}^2
}{2} f'(x_n+\Delta x/2, y_n + \Delta y/2) \\ \nonumber
&+& \frac{{\Delta x}^2}{2} f'(x_n+\Delta x/2, y_n + \Delta y/2)/2
+ O({\Delta x})^3.
\een
where the $+ \frac{1}{2}{\Delta x}^2 f'(x_n\Delta x/2, y_n + \Delta y/2)$
term comes from the $O({\Delta x}^2)$
term of (\ref{yneuler}). This sets up
a cancellation in the
first order terms and we have a second order accurate method
\be
y_{n+1} = y_n + \Delta x\,f(x_n\Delta x/2, y_n + \Delta y/2) + O({\Delta x})^3.
\ee
This then is the spirit of the second-order {\em RK} scheme.
The code uses a fourth order {\em RK} scheme. This uses derivatives at the
initial point, at the half step, and at a full step, with weights of
$1/6$, $2/3$ and $1/6$ respectively as can again be confirmed by
Taylor expansions. Guesses for the $y$ value at the $1/2$ step
are made using the derivative at the original point and the derivative
at the previously guessed point and these two are then averaged. For
the full step the derivative used is that at a $y$ value calculated using
the second of the just described half step derivatives.

\subsection{{\em BD} Case}

Equilibrium configurations of boson stars in {\em BD} theory have been discussed in~\cite{CS97}.
They used a modified version of the code we used to get equilibrium 
{\em GR} configurations. They added the {\em BD} terms to the equations as well 
as adding the {\em BD} scalar field. The equilibrium configurations in this 
case were those for which the boson field has the
$e^{i \omega_E t}$ dependence in time as described before, while the
metrics and the {\em BD} field are static. The value of the central {\em BD} field was not a
free parameter but depended on its specified value at $\infty$.
This was chosen to be zero to
some tolerance. The equations were integrated for different values of the central lapse
value for a given central boson field value and guessed central {\em BD} field
value. The value of the Brans--Dicke field at the
edge of the grid was checked against its
expected value ($\sim \varphi_{\infty} + C/r$) to the predetermined
tolerance. If it did not match the value of the central {\em BD} field was changed
and the process repeated for different central lapse values until a match was
found. Since these configurations were quite similar 
to the {\em GR} case the central lapse values were close to
the {\em GR} values. For all self-couplings the {\em BD}  particle number
was very slightly smaller than the {\em GR} case while the mass was slightly lower
for small values of self-coupling and slightly larger for higher values of 
self-coupling. This was true for both the ground state and excited state
configurations.

\subsection{Boson Halos}

\begin{figure}
\centering
\leavevmode\epsfysize=7.5cm \epsfbox{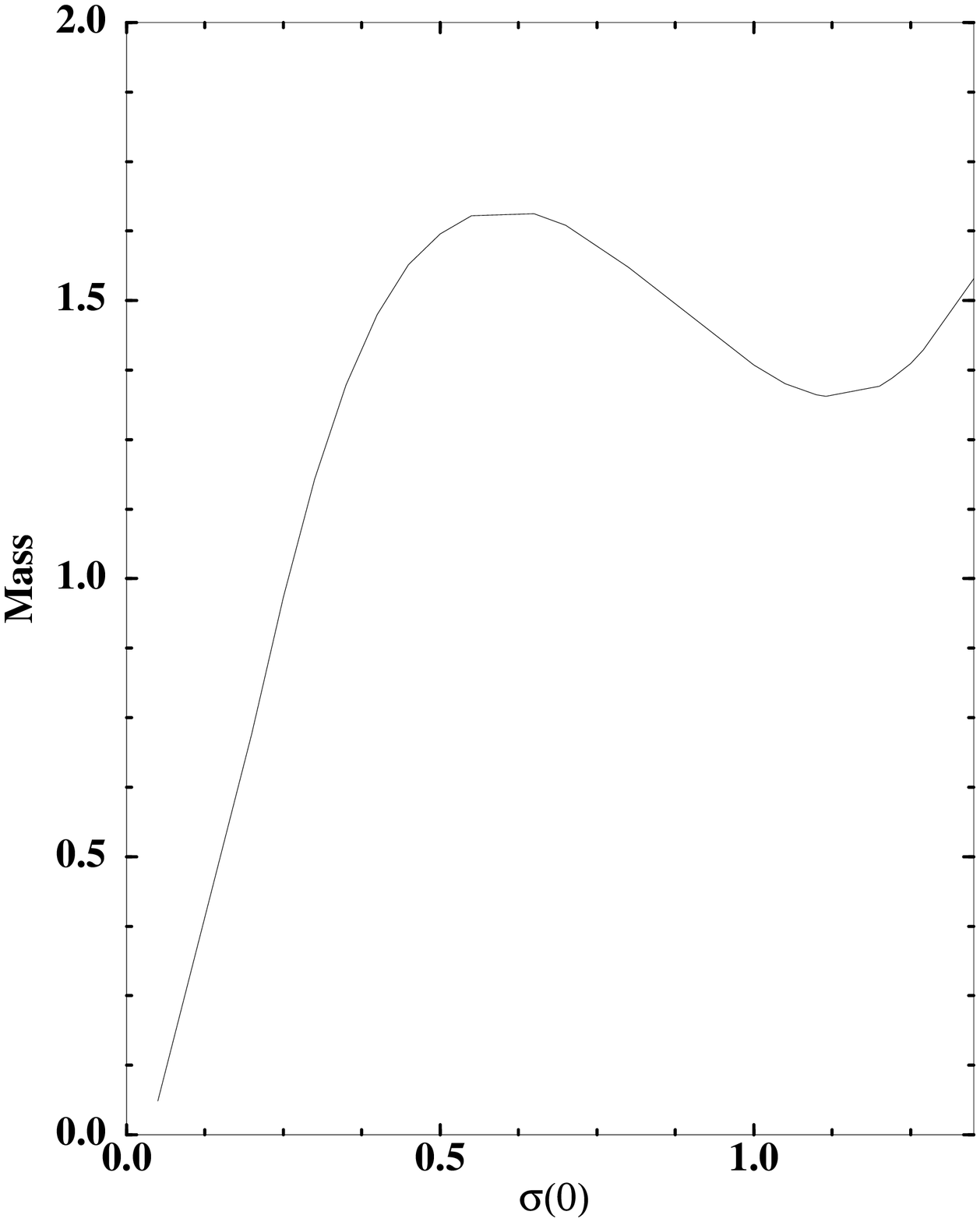}\hskip0.5cm
\leavevmode\epsfysize=7.5cm \epsfbox{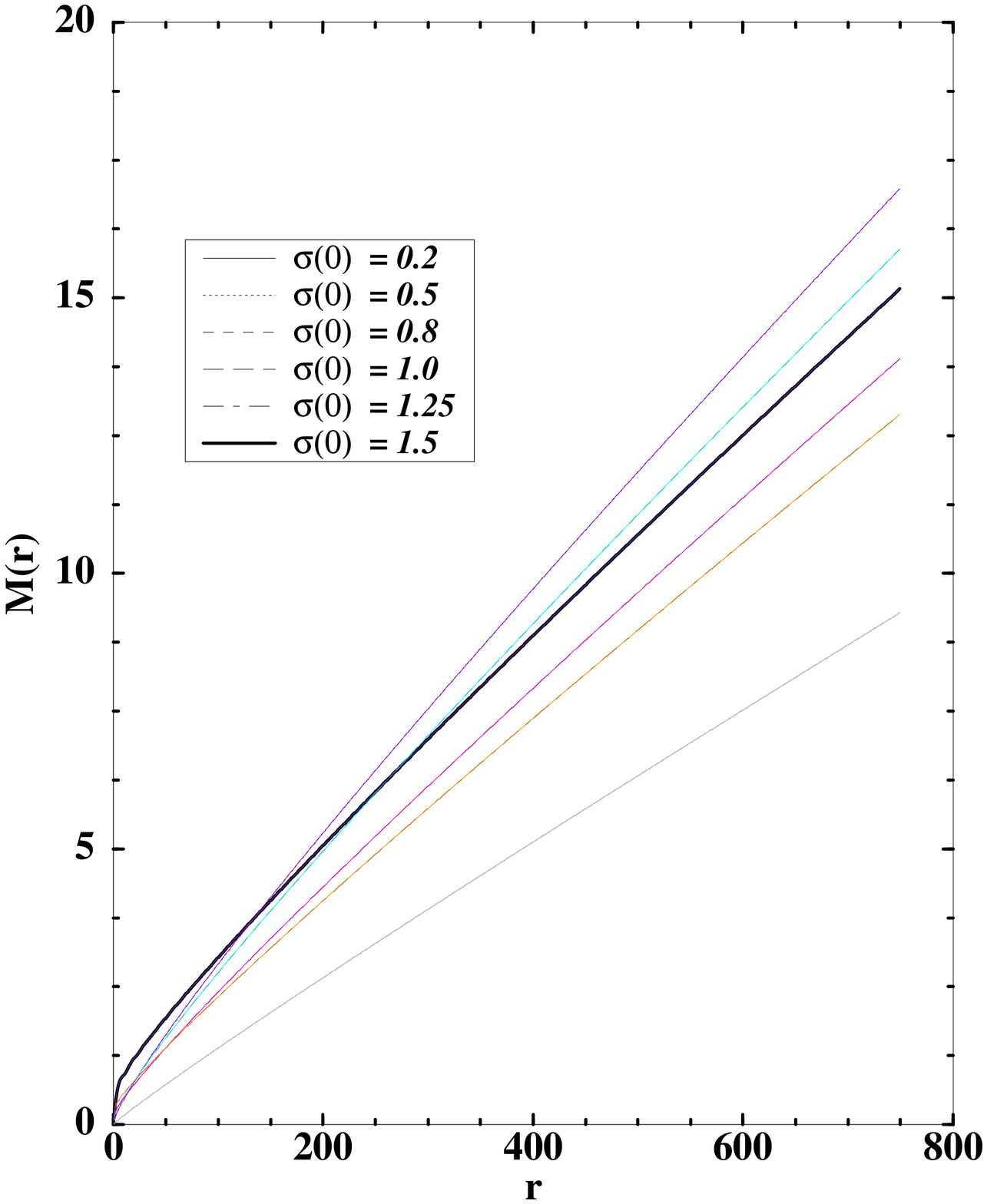}\hskip0.5cm
\caption[Mass Profile of Massless Boson Field Halo Configuration]
{(a) Left: The mass profile of equilibrium configurations of massless scalar
fields, are interestingly
similar to the profiles of massive scalar fields and neutron stars.
The mass for a given radius
increases to a maximum with central density before it decreases with
further increase in
central density. However by calculating the particle number, we see
no division into stable and unstable
configurations at the peak. The particle number depends very sensitively
on the initial value of the lapse function $N$, so that the clear cusp
structure seen in the mass-particle-number diagram for boson stars
cannot be derived here.
(b) Right: The profiles are largely independent of radius although the mass
itself increases with $r$.
}
\label{msls1}
\end{figure}

The equilibrium configurations are obtained
({\em BD} equations with $m=a=0$), from the equations
\begin{equation}
\sigma' = \chi \; , \label{chii}
\end{equation}
\begin{equation}
\chi'= -\left[\frac{1}{r}+\frac{g^2}{r}\right]\chi -
\frac{\sigma g^2}{N^2} \; ,
\end{equation}
\begin{equation}
g'=\frac{1}{2} \, \left[\frac{g}{r}-\frac{g^3}{r}+\frac{\sigma^2 \,r \, g^3}{N^2}
+ r \, g \, \chi^2 \right] \; ,
\end{equation}
\begin{equation}
N' =\frac{1}{2} \,\left[-\frac{N}{r}+\frac{N \, g^2}{r}+\frac{r \, g^2 \, \sigma^2}{N}
+ r \, N \, \chi^2 \right] \; . \label{chif}
\end{equation}
Here we use the dimensionless variables $r=\omega {\bf r}$,
$t=\omega {\bf t}$ and $\sigma = \sqrt{4\pi G\, } \phi$.
These equilibrium configurations are characterized by static metrics
although the fields themselves again have an $e^{i\omega t}$ dependence.
Regularity at the center implies $g(r=0)=1$.
The equilibrium configurations are characterized by saddle points in the
density $\rho$, which might
have some significance in stability issues. These systems are characterized
by a two parameter family
of solutions. Firstly, we can vary $\sigma(0)$, the central density and
for each $\sigma(0)$ we can vary $N(0)$. This is effectively changing the
frequency and for each value of the frequency of the system there
is a configuration for which the dispersive effects cancel the gravitational
effects. We have noticed an interesting feature of the mass function,
\begin{equation}
M(r)=\frac{r}{2} \left ( 1 - \frac{1}{g^2} \right ) \; ,
\end{equation}
where $g^2$ is the radial metric. If plotted at a given value of $r$ the
profile is very similar to boson stars as shown in figure
\ref{msls1}a. We also see the increase in mass linearly
with radius for large $r$ in the relativistic case. Figure
\ref{msls1}b shows mass versus radius for different central densities
and a given value of $N(0)$.

\chapter{Numerical Evolution of Bosonic Objects in Spherical Symmetry}

In the previous chapter we developed (from the action for a {\em BD} system with
a matter Lagrangian that produces a scalar field equation for bosons,
and for a spherically symmetric metric) the basic equations that govern the
structure of an equilibrium boson star (see \ref{subsection:sectionbd}). The 
{\em GR} case was a special case of these with the parameter $a$ set to zero
and discarding the {\em BD} field
equation. (The {\em BD} field decouples from the physics). We now write down the evolution equations for the {\em BD} case. These of course include the
time derivatives of the metric and the {\em BD} field. 
For convenience, so that we have no second time derivatives, we introduce
new variables
\begin{eqnarray}
\Pi_\varphi&=& {g \over N} \, \partial_t (r \, \varphi)
\equiv {1\over \beta} \, \partial_t (r \, \varphi), \\
\Pi_\Psi&=& {g \over N} \, \partial_t (r \, \Psi) \equiv {1\over \beta}
 \partial_t (r \, \Psi),
  \end{eqnarray}
  where $\Psi = \sqrt{4\pi G_*}\phi$.

  We thus get the field equations (for $\phi$, $\Psi$)
  \begin{eqnarray}
  \partial_t (r \, \varphi) &=&  % {N \over g } \Pi_\varphi =
  \beta \, \Pi_\varphi,  \label{dyn-1} \\
  \partial_t \Pi_\varphi
  &=& (\partial_r\beta) \, \partial_r(r \, \varphi)
  +\beta \, \partial_r\partial_r(r \, \varphi)
   -(\partial_r\beta)\, 
   (r \, \varphi)\, {1\over r} \nonumber \\&&
  - N\,g \, r~
2\, \alpha \, e^{2a}
   \left[{1 \over 2} \, \nabla_\sigma \Psi \, \nabla^\sigma \Psi^\dagger
 +2\, e^{2a} \,  V(\Psi\Psi^\dagger) \right], \label{dyn-2}  \\
 \partial_t (r \, \Psi) &=& % {N \over g } \Pi_\Psi =
 \beta \, \Pi_\Psi,  \label{dyn-3} \\
\partial_t \Pi_\Psi
&=& (\partial_r\beta) \, \partial_r(r \, \Psi) + \beta \,
\partial_r\partial_r(r\, \Psi)
-(\partial_r\beta)\, (r\Psi) \, {1\over r}  \nonumber \\&&
 - 2\, N \, g \,r \left[ e^{2a} {d V(\Psi\Psi^\dagger) \over d \Psi^\dagger} -
\alpha \,
\nabla_\sigma \Psi \, \nabla^\sigma \varphi\right] \label{dyn-4}.
\end{eqnarray}
where there are really two components for the second equation above
(since we have a complex boson field). On each time slice 
we solve for the lapse $N$ by using the $rr$ component of the Einstein equation.
However, instead of using the $00$ component to get the radial metric
on each time slice it is more convenient to use this component to record the
  time development of the radial metric through the $0r$ equation 
  (the momentum constraint equation). The momentum constraint and 
  $G_{rr}$ equations are respectively
  \begin{eqnarray}
  \partial_t g &=&
  N \, [ \Pi_\varphi \, \partial_r\varphi +  e^{2a}\,  {1\over 2} \, 
  (\Pi_\Psi^\dagger \, \partial_r\Psi+\Pi_\Psi \,
  \partial_r\Psi^\dagger)],
  \label{momco} \\
  \partial_rN &=& {N\over 2 \, r} \,(g^2-1) + {N \, r \over 2} \left(
  (\partial_r\varphi)^2 + \Pi_\varphi^2 \, {1\over r^2}
  \right.  \nonumber \\&& \left.
  + e^{2a} \,  \left[(\partial_r\Psi) \, (\partial_r\Psi^\dagger)
  +\Pi_\Psi \, \Pi_\Psi^\dagger \, {1\over r^2} -2 \, g^2 \, e^{2a}\,
  V(\Psi\Psi^\dagger) \right]
   \right). \label{branlap}
    \end{eqnarray}

Since $T_{00}$ is the energy density the $G_{00}$ equation is
the Hamiltonian constraint equation. 
It is given by
%\begin{eqnarray}
 \be
%&&
     2 \, {\partial_r g \over g\, r} +{g^2-1 \over r^2}=
     {\Pi_\varphi^2 \over r^2} +(\partial_r\varphi)^2
     %\nonumber \\&&
     +e^{2a}\, \left[ {\Pi_\Psi \, \Pi_\Psi^\dagger \over r^2}
      +(\partial_r\Psi) \, (\partial_r\Psi^\dagger)
	+2 \, g^2 \, e^{2a} \, V(\Psi\Psi^\dagger) \right].
	  \label{branham}
	    \ee
This is used to monitor the code by keeping track of the 
accuracy with which it is satisfied throughout
the evolution. Orders of $10^{-5}$ or better are typical at the end of
our runs.

\section{Evolution Equations and Code}

The equations that describe the evolution of the boson star in {\em GR} are
given by those in (\ref{dyn-3})--(\ref{branlap})
with the parameter $a$ set to zero,
and again discarding the wave equation for the {\em BD} field. In
order to finite-difference the derivatives, one must rely on well
studied methods used in the study of linear differential equations,
although we are in fact dealing with very non-linear systems. Our
equations are of the hyperbolic form (wave equation) and we use a
leapfrog method for evolution~\cite{Press}.

\begin{itemize}
\item{{\bf Leapfrog Method}}
\begin{itemize}

Consider a simple outgoing wave equation of the form
\be
\frac{\partial u}{\partial t} = -v \frac{\partial u}{\partial x}.
\ee
Now take a one-sided differencing scheme for the time-derivative 
($n$ representing the timestep and $j$ the position of a space coordinate)
\ben
u_j^{n} = u_j^{n+1} - \frac{\partial u}{\partial t} \Bigg|_{j,n} \Delta t 
+ O((\Delta t)^2)\\ \nonumber
{\text {so } }
\frac{\partial u}{\partial t}\Bigg|_{j,n} = \frac{u_j^{n+1} - u_j^{n}}{\Delta t}+
O(\Delta t).
\een

For spatial derivatives consider a second-order representation
\ben
u_j^n &=& u_{j+1}^n - \Delta x \,u'|_{j,n} + (\Delta x)^2 u'' + O((\Delta x)^3)
\\ \nonumber
u_j^n &=& u_{j-1}^n + \Delta x \,u'|_{j,n} + (\Delta x)^2 u'' + O((\Delta x)^3)
\een
so that on subtracting the two equations
\be
u'|_{j,n} = \frac{u_{j+1}^n - u_{j-1}^n}{2\Delta x} + O((\Delta x)^2).
\ee

Thus this `Forward-Time Centered-Space' scheme gives
\be
\frac{u_j^{n+1} - u_j^{n}}{\Delta t} = -v \left(\frac{u_{j+1}^n -
u_{j-1}^n}{2\Delta x}\right).
\ee

This scheme is unstable as shown by Von Neumann stability analysis. Consider
the coefficient $v$ to be slowly varying enough to
be considered constant. Now consider eigenmodes of the difference
equation of the form $u_j^n = \xi^n e^{i kj\Delta x}$ so that
\be
\xi -1 = -\frac{v\Delta t}{\Delta x} \,  i \, \sin (k\Delta x).
\ee
The time dependence of the eigenmodes grows like $\xi^n$ and since
$|\xi|= \sqrt{1+ (\frac{v\Delta t}{\Delta x})^2 \sin^2 (k\Delta x)} > 1$ 
this instability grows and renders the scheme
unstable.

So instead one uses a $staggered$ $leapfrog$ method using centered-time and
centered-space derivatives.
\be
\frac{u_j^{n+1} - u_j^{n-1}}{\Delta t} = -v \left(\frac{u_{j+1}^n -
u_{j-1}^n}{\Delta x}\right).
\ee
Now we get a quadratic equation for $\xi$ with a solution $|\xi|=1$ for
any $\frac{v\Delta t}{\Delta x}<1$. Thus there is no growth of instability or
amplitude dissipation ~\cite{Press,Smarrfd}.
It is called a leapfrog scheme because the
time derivative terms are on odd time slices while the time levels
in the space derivative terms are on even time slices. 

When one has second-order time derivatives (as we do) it is more appropriate
to redefine the first time derivative in terms of another variable
($\pi =\frac{g}{N}\frac{\partial (r\psi)}{\partial t}$) that is then given
on the $n+\frac{1}{2}$ steps. 
\end{itemize}
\end{itemize}

The two components of the equilibrium scalar field have components
$\Psi_1$ with time dependence as $\cos(\omega_Et)$ and 
$\Psi_2$ with time dependence as $\sin(\omega_Et)$. Thus quite generally
at $t=0$ we set the field $\Psi_2=0$. Our equilibrium
code calculates the initial field configuration $\Psi_1$ and the metric components
as functions of $r$. This is the initial data for the evolution code.
The first spatial derivatives of the metrics
and fields are then determined by using $f'\Delta x= 
a\left(f(x+3\Delta x)-f(x-3\Delta x)\right)+
b\left(f(x+2\Delta x)-f(x-2\Delta x)\right)+
c\left(f(x+\Delta x)-f(x-\Delta x)\right)$. The coefficients $a$, $b$ and $c$ 
are determined ($a=1/60$,$b=-9/60$, $c=45/60$)
from the conditions that the coefficient of the first derivative of the 
Taylor expansion be $\Delta x$ and  the third and fifth derivative coefficients be zero.
This makes $f'$ sixth order accurate since the even derivatives automatically
vanish.
In order to get
sixth order accuracy in the second derivatives the form above is applied
for $f''(\Delta x)^2$ this time with every $f(x-s)$ and $f(x+s)$ added rather
than subtracted so odd derivatives vanish. The coefficients
are then determined ($a=2/180$, $b =-27/180$ ,
$c=490/180$).
At the last but one grid point though one only has terms to second order accuracy. 

The metrics and fields have vanishing first derivatives at the origin. From
the
equations it is clear their parities are conserved. A pair
of fictitious points corresponding to negative $r$ are set up.
The code evolves $r\psi$ and $\pi$ which are antisymmetric about the origin.
The derivative of $r\psi$ at the origin is $\psi$ and that is how we
determine it. The radial metric at the origin is unity but the lapse is not
prescribed there. Instead we choose to fix the lapse at the outer boundary
and integrate inwards on each slice using a fourth order {\em RK} method. 
Rather than integrate the Hamiltonian constraint equation on each
slice to determine $g$ we monitor its evolution from time step to
time step. We only determine the boundary value of $g$ from the Hamiltonian
constraint equation. 

The outer region of the system is the low field region and the mass of
the system can be determined using the value of the radial metric there.
\be
M = \lim_{r\rightarrow \infty} \frac{r}{2} \, \left[1-\frac {1}{g^2(r)}\right].
\label{massaha}
\ee

The outer boundary condition on the fields is an outgoing wave condition.
If in the asymptotic region the field
is of the form $\sim e^{i({\bf k.r} -\omega
t)}$, then this results in the condition (from the {\em KG} equation)
\ben
\frac{k^2}{g^2} &=& \frac{\omega^2}{N^2} -m^2,
\\ \nonumber
{\mbox giving} \quad
k &=&  \frac{g \, \omega}{N} (1-\frac{1}{2} \, m^2 \, \frac{N^2}{\omega^2}+...),
\een
which can be written (in terms of the fields and dimensionless units of the
code) as
\be
\dot \pi =- \beta \pi' -\frac{N^2}{2} r \psi,
\ee
where $\beta = \frac{N}{g}$. This eliminates first order reflections off
the grid. In order to remove higher-order reflections, we add a ``sponge''
in the outer region of the grid to absorb reflections. To the $\dot \Pi$
equation (\ref{dyn-4}) we add a damping term at the edge of the grid.
This ``sponge'' is effectively like adding a potential term that is proportional to
$k+\omega$ for an incoming wave and proportional to $\omega-k$ for
an outgoing wave. Thus it forms a potential barrier against incoming radiation.
This sponge is smoothly phased in from some point $r(j)$ out to the edge of the
grid. The width of the sponge is of the order of the wavelength of
the scalar radiation.

\section{Nature of the Perturbations}
We study the dynamical properties of the boson stars
by perturbing the equilibrium field
distribution. The
accretion or annihilation of scalar particles is simulated by the addition
of
field in the outer regions of the star or by decreasing it in denser regions
of
the star respectively. Another type of perturbation that has been  studied
is
changing the $\dot\psi_1$ and $\dot\psi_2$ of the equilibrium
configuration. This perturbation changes the kinetic energy
density distribution. In either case the changes in the metric functions
$g$ and $N$ are determined by the
constraint equations
and the polar-slicing condition. That is, we reintegrate the
Hamiltonian constraint and lapse equations of (\ref{branlap}) and
(\ref{branham}) on the initial time slice.
The magnitude and
the length scale of the perturbations can be chosen arbitrarily. The
perturbations
are always spherically symmetric.

\section{Results and Conclusions for 1D Boson Stars with Self-Coupling}
\begin{itemize}
\item{Ground  State Results}
\end{itemize}

An equilibrium configuration of boson stars when perturbed reacts to
these perturbations in a myriad of ways. As discussed before, in
the works of Pang and Lee~\cite{leepert}, ground state
boson stars without self-coupling on the $S$ branch have very
specific quasinormal modes of oscillation under infinitesimal perturbations.
In ~\cite{kus} the mass function is plotted against radius for a given
particle number. The curve is seen to have a ``well'' with a minimum at the
exact equilibrium mass corresponding to that particle number. For
any other given mass at that particle number there are oscillating regimes
from a minimum to a maximum radius. In a numerical scenario we are in a
position to test more general and realistic perturbations.
While we do study the effects of infinitesimal perturbations (by applying 
perturbations for which the mass change from equilibrium is less than $0.1 \%$.)
larger perturbations are also analyzed. The effects of larger perturbations
cannot be studied analytically using perturbative methods.
From the former we get
quasinormal modes of oscillations, while from the latter we get scalar radiation
that results in mass losses in these stars followed by their eventual
settling to stable configurations of lower mass, their collapse to black holes,
or their total dispersal. 
Once perturbed the metric equations are reintegrated to get
consistent metrics for those new initial configurations. Thereafter the
system is allowed to evolve under the evolution equations.
\begin{itemize}
\item{$S$ Branch Stars}
\end{itemize}

\begin{figure}[t]

\begin{center}
\leavevmode
\makebox{\epsfysize=15cm\epsfbox{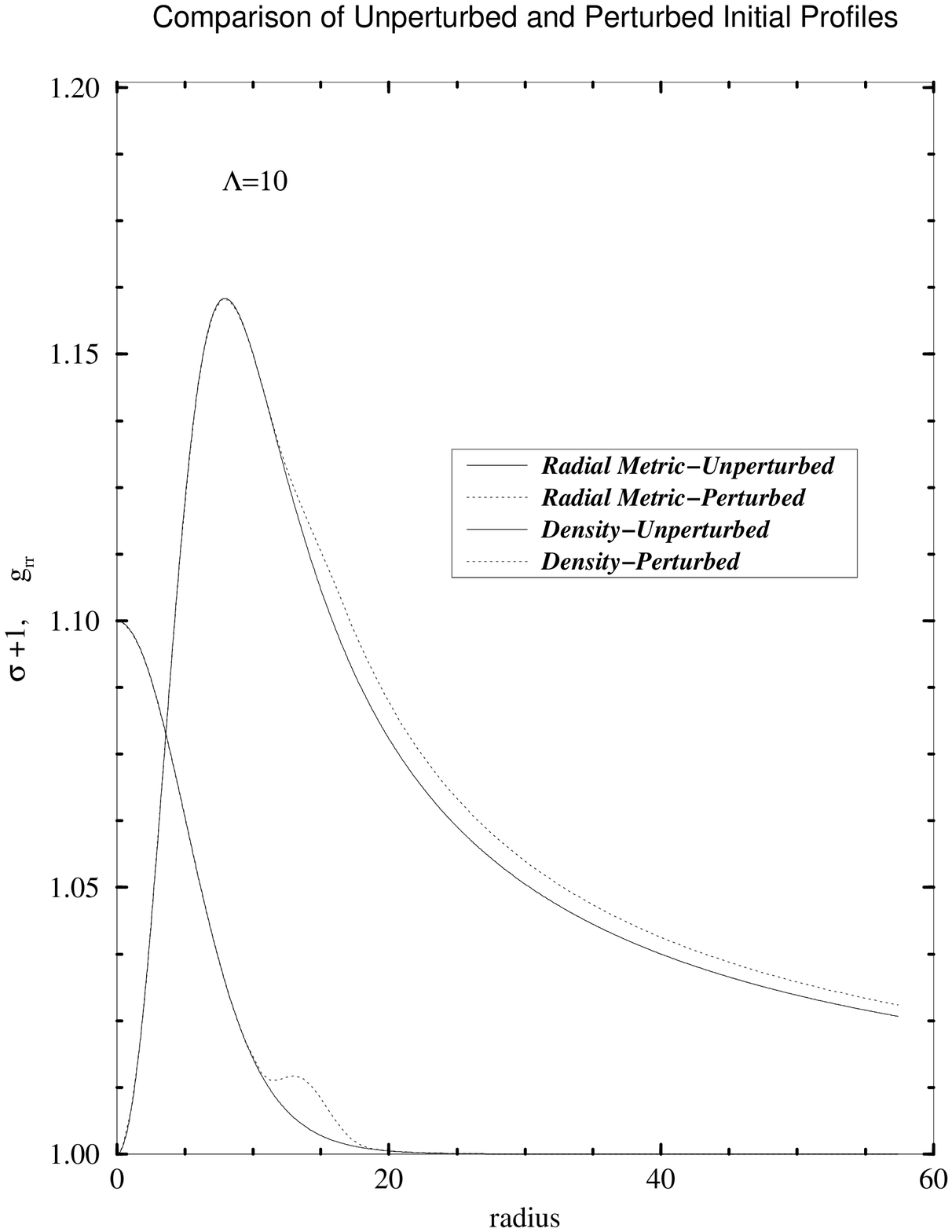}}
\end{center}

%%\begin{figure}
%\hspace{-106pt}
%\vspace{-230pt}
%%\hspace{-0pt}
%%\vspace{-130pt}
%%\epsfbox[0 -130 450 620]{figs/lampert.ps}
%\epsfbox[100 -130 350 500]{figs/lamperte.ps}
\caption{
Comparison of a strongly perturbed
ground state $\Lambda=10$, $S$-branch star
[$ M=0.781 \,M_{Pl}^2/m, \sigma (0)=0.1 $] to the unperturbed
configuration [$ M=0.722 \, M_{Pl}^2/m$].
The solid lines correspond to the unperturbed configuration
and the dashed ones to the perturbed star.
The perturbation shown corresponds to
the addition of scalar field $\sigma$ at $t=0$.}
\label{fig_lampert}
\end{figure}

In the case of a free field ($\Lambda=0$ case) a perturbed $S$-branch star oscillates with a definite frequency,
losing mass through bursts of scalar radiation at each expansion. It
finally settles to a new $S$-branch configuration of lower mass.
Here the effect of a self-interaction term
on this behavior is examined. In Fig.~\ref{fig_lampert}
a typical example of the perturbed field configuration, and
radial metric, of a star with $\Lambda = 10$ and a central density
of $\sigma (0)=0.1$ is shown. This star has been perturbed by
accretion of scalar particles in a region of
lower density. Its evolution is detailed below.

\begin{figure}[t]

\begin{center}
\leavevmode
\makebox{\epsfysize=15cm\epsfbox{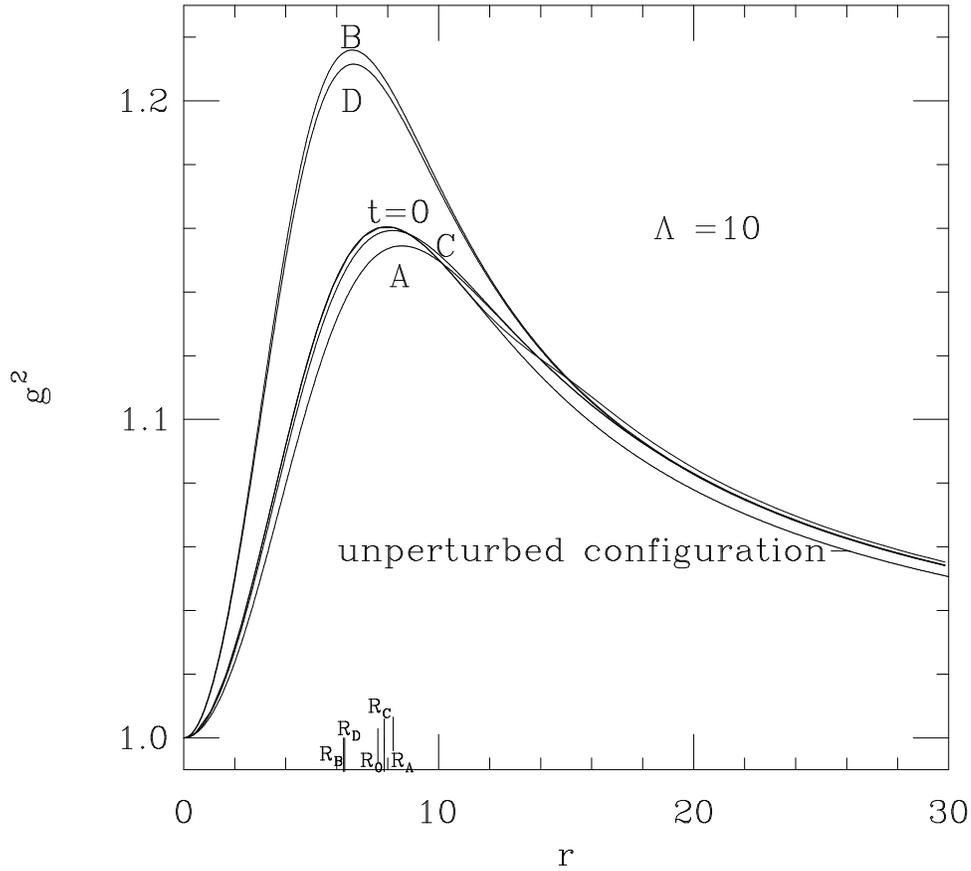}}
\end{center}
%%\begin{figure}
%%\hspace{-106pt}
%%\vspace{-130pt}
%\epsfbox[0 -150 450 250]{figs/lamgrra.ps}
%%\epsfbox[0 -70 350 500]{figs/lamgrrae.ps}
\caption{
The evolution of the radial metric $g_{rr}=g^2$ for
the configuration shown in the previous figure.
The initial perturbed configuration 
labeled $t=0$ and unperturbed configuration are shown. 
$A,\,\,B,\,\,C,$ and $D$ correspond to times $t=192,\,\,306,\,\,391,$ and $505$ respectively. The
positions of the peaks for these times are labeled $R_A,\,\,R_B,\,\,R_C,$ and $R_D$.}
\label{fig_lamgrraa}
\end{figure}

\begin{figure}[t]

\begin{center}
\leavevmode
\makebox{\epsfysize=15cm\epsfbox{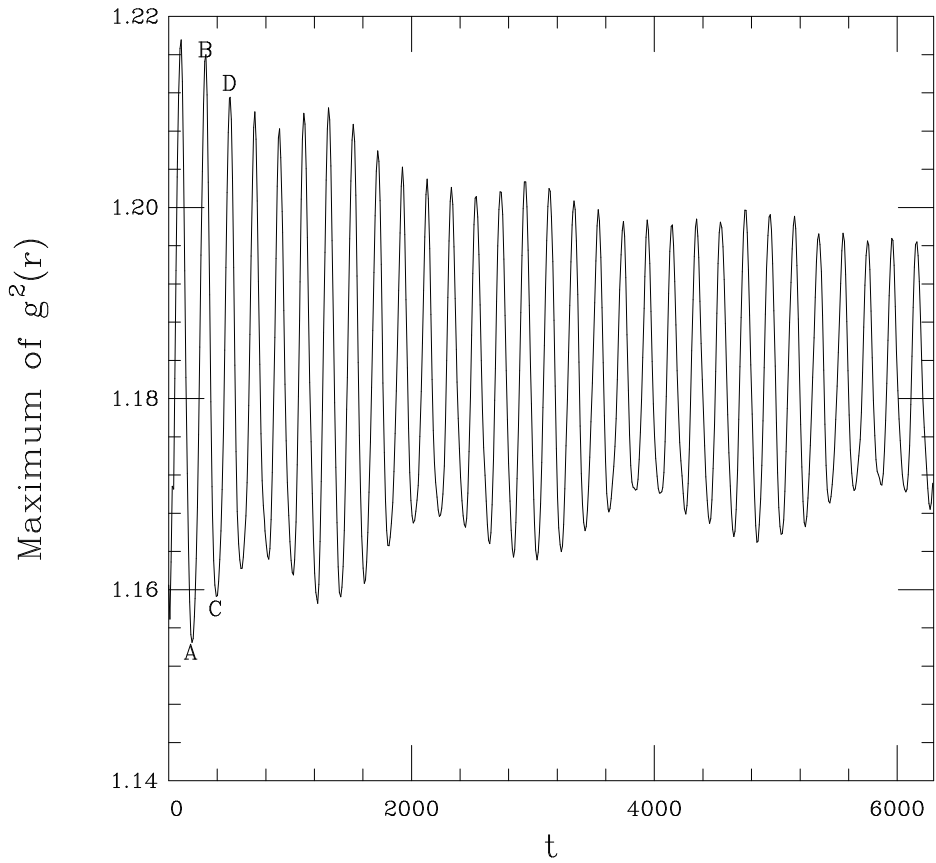}}
\end{center}
%%\begin{figure}
%%\hspace{-36pt}
%%\vspace{-130pt}
%\epsfbox[0 -130 450 620]{figs/lamgrrb.ps}
%%\epsfbox[0 -150 350 500]{figs/lamgrrbe.ps}
\caption{
The peak value of the perturbed radial metric is plotted over a
long time ($6000\, \hbar/mc$). The points labeled $A$, $B$, $C$,
and $D$ correspond to the same labels in
the previous figure. The oscillations decay in time.}
\label{fig_lamgrrab}
\end{figure}

\begin{figure}[t]

\begin{center}
\leavevmode
\makebox{\epsfysize=15cm\epsfbox{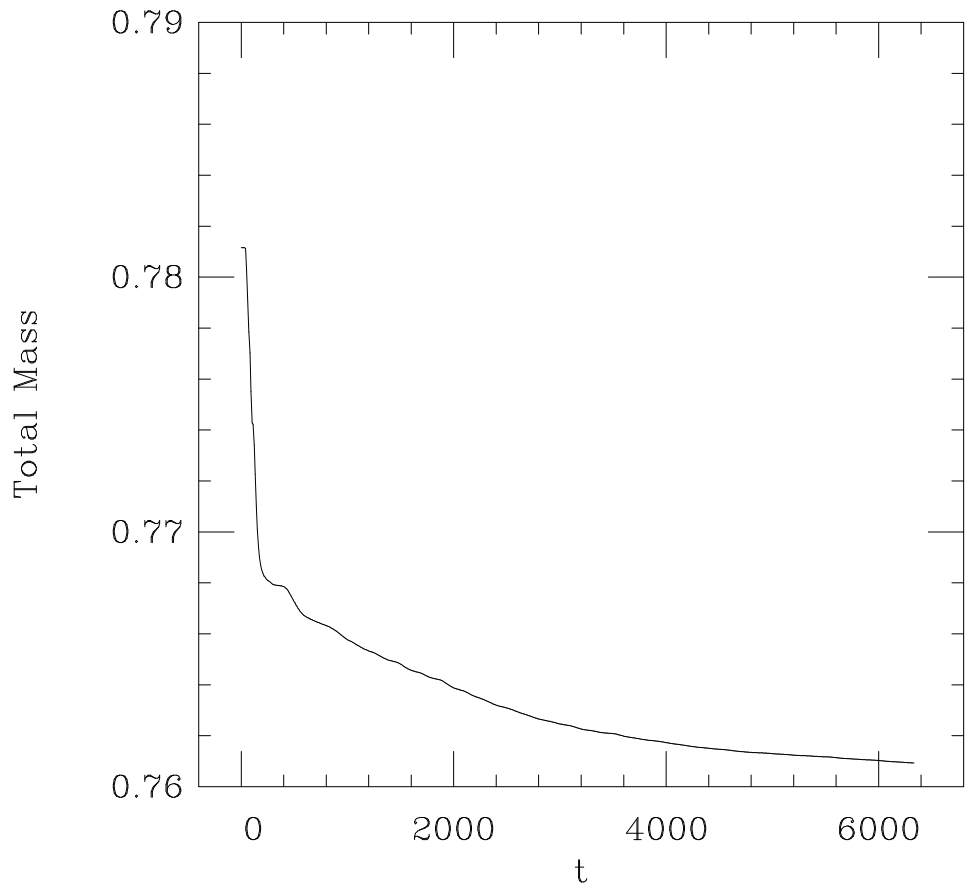}}
\end{center}
%%\begin{figure}
%%\hspace{-36pt}
%%\vspace{-130pt}
%\epsfbox[0 0 450 500]{figs/masslama.ps}
%%\epsfbox[0 -150 350 500]{figs/masslamae.ps}
\caption{
 The total mass of the
star is plotted as a function of time. The mass loss through scalar
radiation decreases in time as the
oscillations start damping out.}
\label{fig_lamgrramass}
\end{figure}

Fig.~\ref{fig_lamgrraa} shows the radial metric as a function of distance for the same
configuration at various
times. The labels $ A,B,C,D$ correspond to times (in units of the inverse
of the underlying scalar field frequency)
$t=192,\,\,306,\,\,391,$ and $505$ respectively. The positions
of the peaks are labeled $R_0,\,\,R_A,\,\,R_B,\,\,R_C,$ and $R_D$, where $R_0$ is the
position of the initial unperturbed peak. Here $R_0=7.95,\,\,R_A=8.55,\,\,
R_B=6.6,\,\,R_C=8.2 $ and $R_D=6.65,$ where the
length scale is in terms of the inverse mass
of the boson. The oscillations are shown clearly in
Fig.~\ref{fig_lamgrrab} which is a plot of the maximum value
of the radial metric as a function of time. The point where this
function is a maximum corresponds
to the core of the star contracting
to its minimum size in a cycle. Similarly,
the maximum radial metric starts to decrease as the star expands.
At each expansion the star loses mass through scalar radiation. The
oscillations damp out in time as the star starts settling down to the new
configuration. A plot of mass vs. time is shown in Fig.~\ref{fig_lamgrramass}.
The
emission of scalar radiation decreases as the oscillations damp out,
as can be seen from the figure. The slope of the curve steadily decreases
as the star starts settling down to its new lower-mass
configuration. The mass is measured at the inner edge of the sponge
using (\ref{massaha}). Exact
details of the
curves have some dependence on the sponge parameters but the basic results are 
the same.

\begin{figure}
\hspace{-36pt}
\vspace{-130pt}
%\epsfbox[50 -50 450 550]{figs/lamqnma.ps}
\epsfbox[50 -50 450 550]{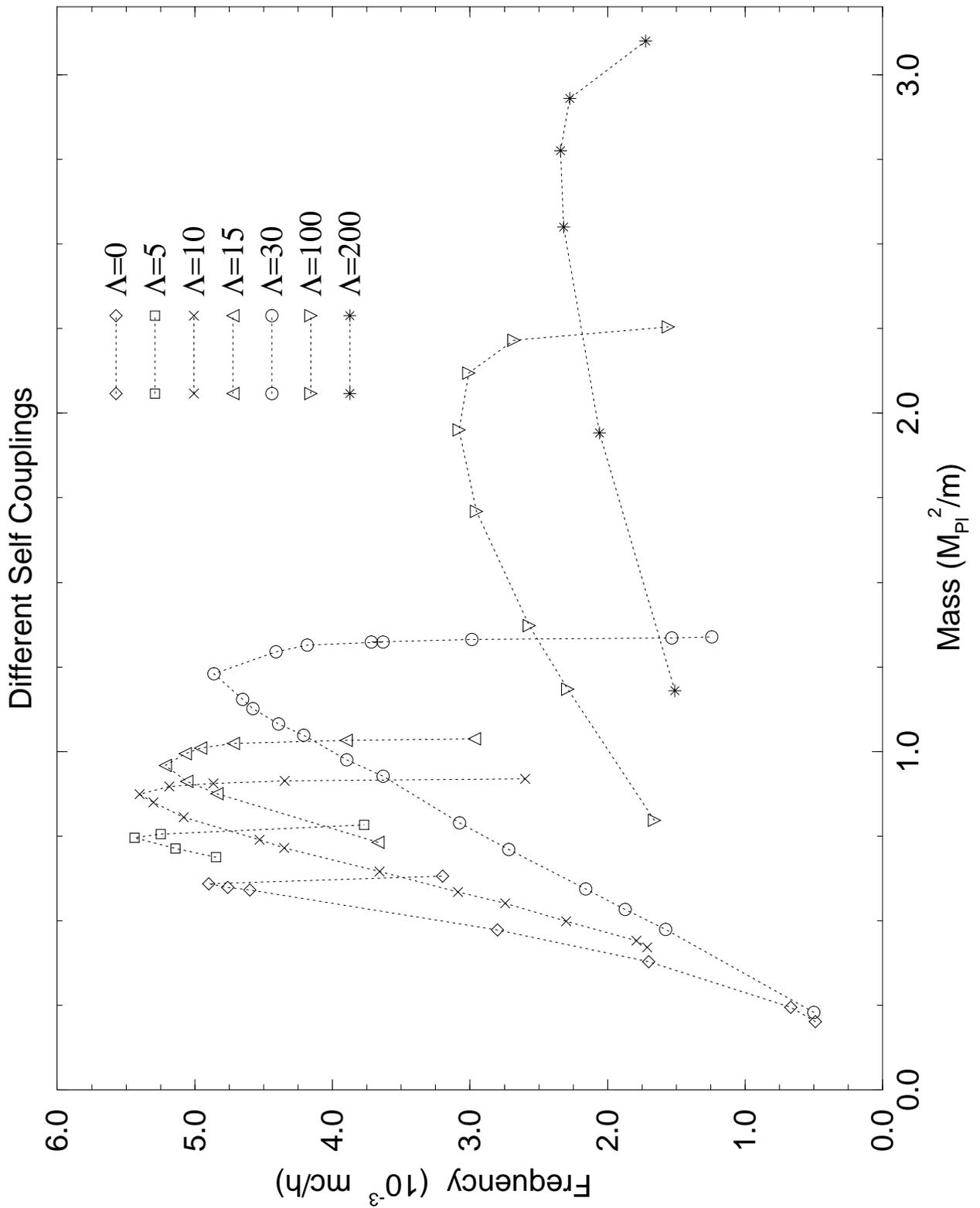}
\caption{
The oscillation frequencies of different ground state boson star
configurations are plotted as functions of mass, for
$\Lambda=0$, $5$, $10$, $15$, $30$, $100$, and $200$.
The curves
are obtained by slightly perturbing (perturbed mass within
$0.1\%$ of the unperturbed mass) $S$-branch stars.
They reach a peak and then drop down at the approach of the maximum mass
allowed for a given $\Lambda$, signalling a transition from stability to
instability.
}
\label{fig_lamqnma}
\end{figure}

A characteristic of the boson star that
might be important for its potential observation and identification
is its fundamental oscillation frequency which can be determined from
Fig.~\ref{fig_lamqnma}.
We found
$f=1/(199\,N(\infty)) =4.7\times 10^{-3}$. The oscillation frequencies for
a large number of S-branch stars have been compiled in this way.
Fig.~\ref{fig_lamqnma} shows a plot of the oscillation
frequency versus mass for many slightly perturbed configurations
(masses within 0.1\% of the unperturbed mass.) As the mass increases,
the frequency increases and then
drops down as the transition point
($dM/d\sigma(0)=0$)
is approached,
signalling the onset of
instability. This is seen for both the non self-interacting as well as the 
self-interacting case.
These quasinormal modes of oscillation characterize $S$-branch
stars. The point of transition from the $S$ to the $U$ branch
corresponds to a zero of the oscillation frequency.

For a given mass higher $\Lambda$ $ S $-branch stars have a lower
oscillation frequency than similar mass lower $\Lambda$ stars, unless
one is near the transition point of the lower $\Lambda$
configuration. This is not too surprising, since for a given mass the
radius of the star increases with increasing $\Lambda$. We have seen
this trend even for $\Lambda$ values as high as $1600$ (see discussion
on high $\Lambda$ stars in section (\ref{Appendix})). However since the maximum
mass of higher $\Lambda$ configurations is greater than that of lower
$\Lambda$ configurations their maximum oscillation frequency could be
greater than that for lower $\Lambda$ stars. (This can be understood
as a size effect: on the $S$-branch, higher mass stars are smaller and
have higher frequencies.) This can be seen in Fig.~\ref{fig_lamqnma} .
The maximum
oscillation frequencies of $\Lambda=5$ and $\Lambda=10$ configurations
are higher than that of the $\Lambda=0$ case. As the stars get much
larger, though, the highest frequency starts to decrease. For example
the maximum frequency for $\Lambda=30$ stars is less than that of
$\Lambda=15$ stars which is itself lower than that of the
$\Lambda=10$ case. A perturbation calculation for the high $\Lambda$
case is presented in section (\ref{Appendix}) to show the dependence of the quasinormal
mode frequencies on $\Lambda$ for high $\Lambda$ configurations. The
frequency is then proportional to the inverse of the square root of
$\Lambda$. Thus as the stars become extremely large they oscillate less and
less rapidly and it is no longer feasible to evolve them numerically in our
code. (The time step used in the numerical simulation cannot be increased
as it is determined by the intrinsic oscillation time scale of the
scalar field, which is many orders of magnitude shorter than the
oscillation time scale of the whole star for these cases.)

\begin{figure}[t]

\begin{center}
\leavevmode
\makebox{\epsfysize=15cm\epsfbox{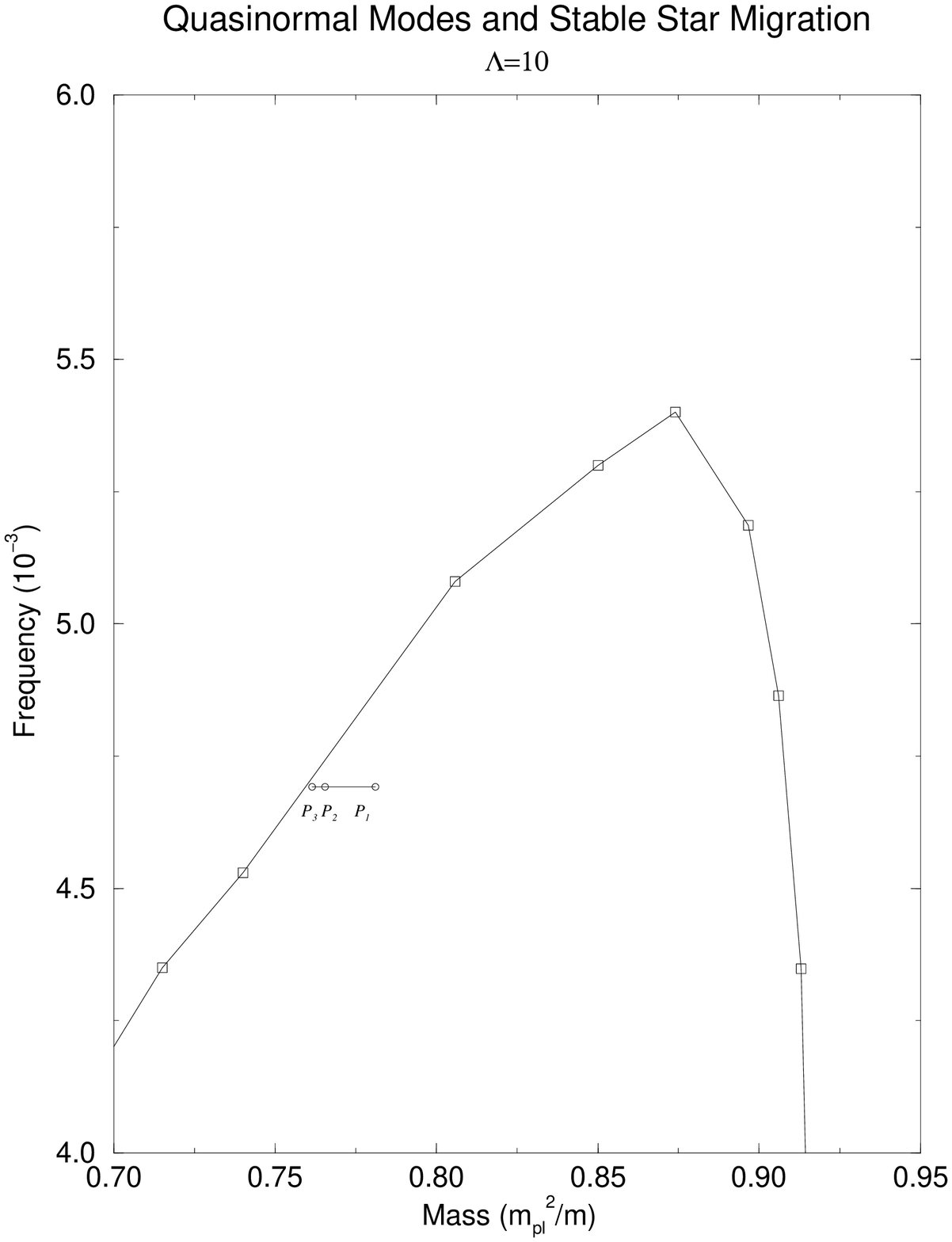}}
\end{center}
%%\begin{figure}
%%\hspace{-36pt}
%%\vspace{-130pt}
%\epsfbox[50 -50 450 650]{figs/lamqnmb.ps}
%%\epsfbox[0 -150 350 500]{figs/lamqnmbe.ps}
\caption{
The highly perturbed $\Lambda=10$ $S$-branch
star has an oscillation frequency below the
$\Lambda=10$ solid line. Its movement
towards the solid line is shown through points
$P1$, $P2$, and $P3$ corresponding to times 0, 1200,
and 4800 ($\hbar/mc$) respectively.}
\label{fig_lamqnmb}
\end{figure}

The quasinormal mode curves are also
useful in determining the evolution
of strongly perturbed $S$-branch stars and the final configurations they
could settle into. A
perturbed star loses
mass and settles to a final configuration corresponding to some position on the 
solid line in the figure.
In Fig.~\ref{fig_lamqnmb} we single out the $\Lambda=10$ curve and plot the evolution of the $
S$-branch star
discussed above. The
points $ P1$, $P2$, and  $P3 $ show the route to a new configuration. These
points correspond to
times $t=0$, $1200$, and $4800$ respectively. By extrapolating this line to 
where it meets
the $\Lambda=10$ curve one could expect a
final mass of about $0.76\,M_{Pl}^2/m$ .

%\begin{figure}
%\hspace{-36pt}
%\vspace{-130pt}
%\epsfbox[0 -250 450 500]{figs/masslam.ps}
%\epsfbox[-50 -150 350 500]{figs/masslame.ps}
%\caption{
%The mass profiles of ground state boson stars for different values
%of self-coupling constant $\Lambda$ are shown. The increase in mass with
%$\Lambda$ is clear. Note that the profiles are very similar.}
%\label{fig_masslam}
%\end{figure}

\begin{itemize}
\item{$U$-branch Stars}
\end{itemize}

Ground state $U$-branch stars (with or without self-coupling) can migrate to
the $S$ branch under certain perturbations. In general they are unstable
and cannot settle into new configurations on the same branch.
In ~\cite{kus}, with the aid of catastrophe theory, the instability of these unstable
stars is also discussed.
Their conclusion regarding unstable boson stars is that
they disperse or collapse. However we also demonstrate the presence
of migrations of these stars to configurations on the $S$ branch. This is
due to our perturbations being more
general, allowing for changes in mass and particle number, whereas they
keep them strictly constant.

\begin{figure}[t]

\begin{center}
\leavevmode
\makebox{\epsfysize=15cm\epsfbox{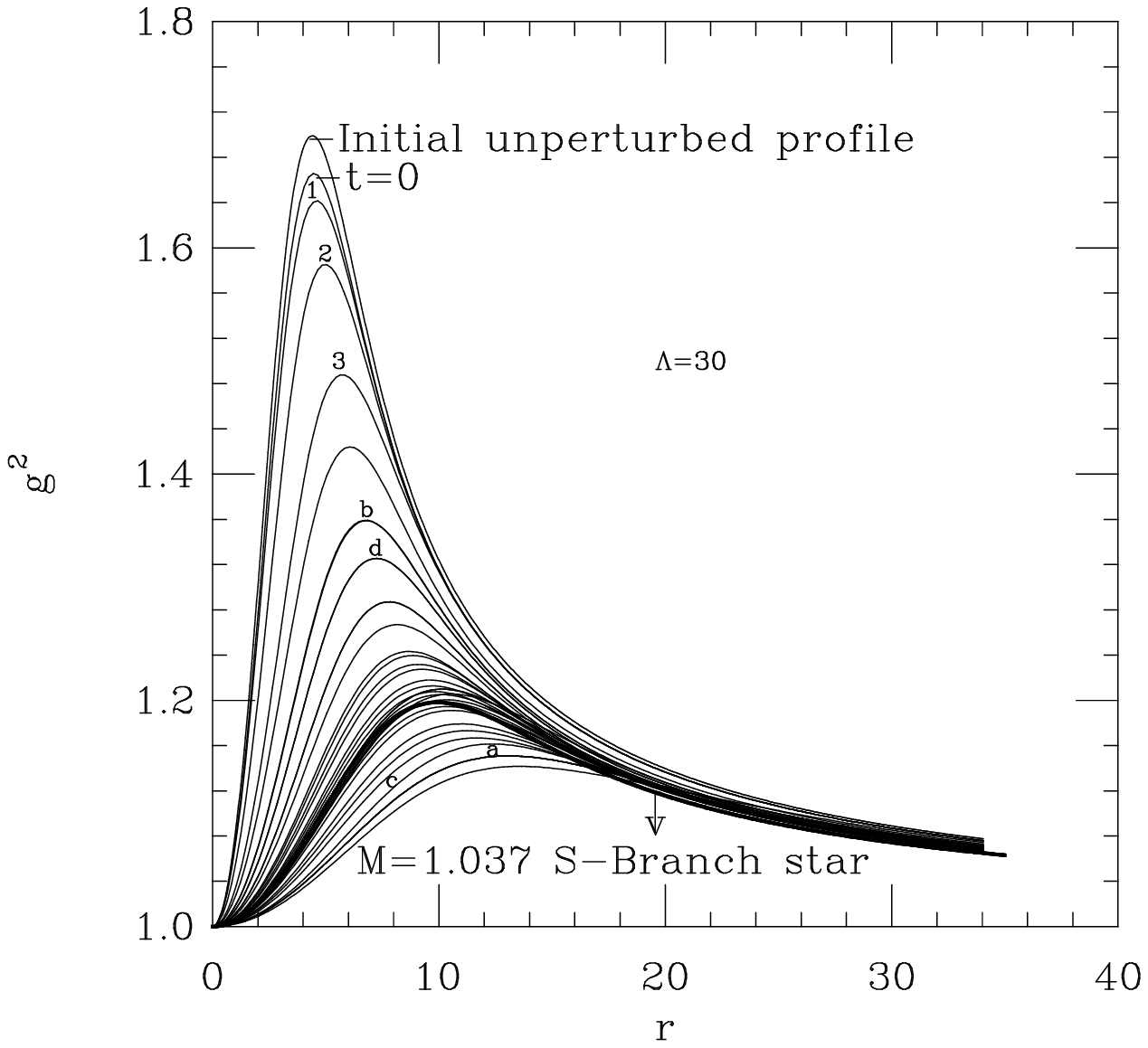}}
\end{center}
%%\begin{figure}
%%\hspace{-36pt}
%%\vspace{-130pt}
%\epsfbox[0 -250 450 500]{figs/lammiga.ps}
%%\epsfbox[0 -100 350 500]{figs/lammigae.ps}
\caption{
 The radial metric $g^2=g_{rr}$ of a perturbed $U$-branch ground
state star is shown at various times. The curves
$1\,\,,2,\,\,3,\,\,a,\,\,b,\,\,c$, and $d$ correspond to times $t=10,
\,\,20,\,\,30,\,\,340,\,\,440,\,\,540$, and $640$.
The unperturbed
star has a central density $\sigma(0)=0.23$ and a self coupling parameter
$\Lambda=30$. The initial equilibrium metric
configuration has also been shown.
The initial field  density of this star has been lowered by about
$10\%$. The $t=0$ curve corresponds to the initial perturbed
radial metric.
In the asymptotic
region $g^2$ is not oscillating but monotonically decreasing due to the mass
loss.}
\label{fig_lammiga}
\end{figure}

\begin{figure}[t]

\begin{center}
\leavevmode
\makebox{\epsfysize=15cm\epsfbox{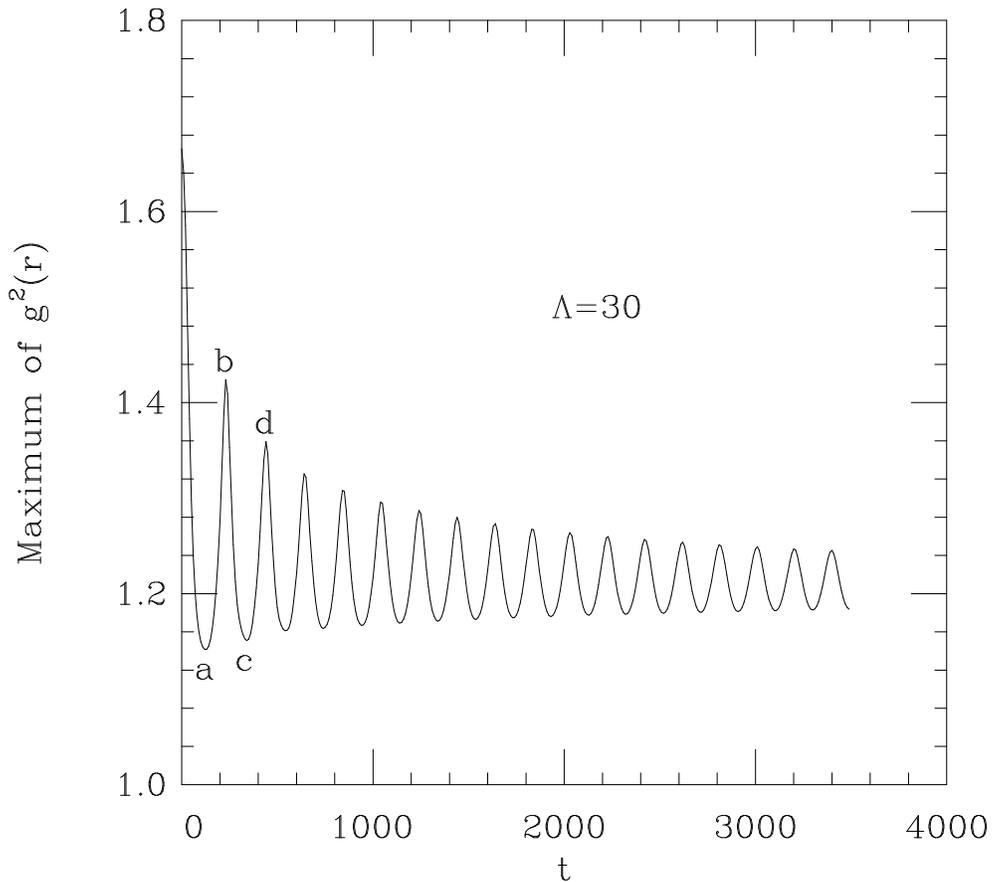}}
\end{center}
%%\begin{figure}
%%\hspace{-36pt}
%%\vspace{-130pt}
%\epsfbox[0 -250 450 500]{figs/lammigb.ps}
%%\epsfbox[0 -100 350 500]{figs/lammigbe.ps}
\caption{
 The maximum value of the radial metric is plotted as a function of
time. The initial sharp drop in the radial metric is due to the expansion of
the star as it proceeds to the stable branch. There it oscillates about
the new stable configuration that it will settle into. This corresponds
to a star of mass $M=1.037 \, M_{Pl}^2/m$ (whose
metric configuration has been shown in the previous figure as
a dark line).
The points $a$ and $c$ correspond to two minima in the peak
of $g_{rr}$ which occur when the core of the star reaches its local maximum
size. Likewise the maxima in the peak of $g_{rr}$ at $b$ and $d$ correspond
to the core of the star reaching its local minimum size.}
\label{fig_lammigb}
\end{figure}

\begin{figure}[t]

\begin{center}
\leavevmode
\makebox{\epsfysize=15cm\epsfbox{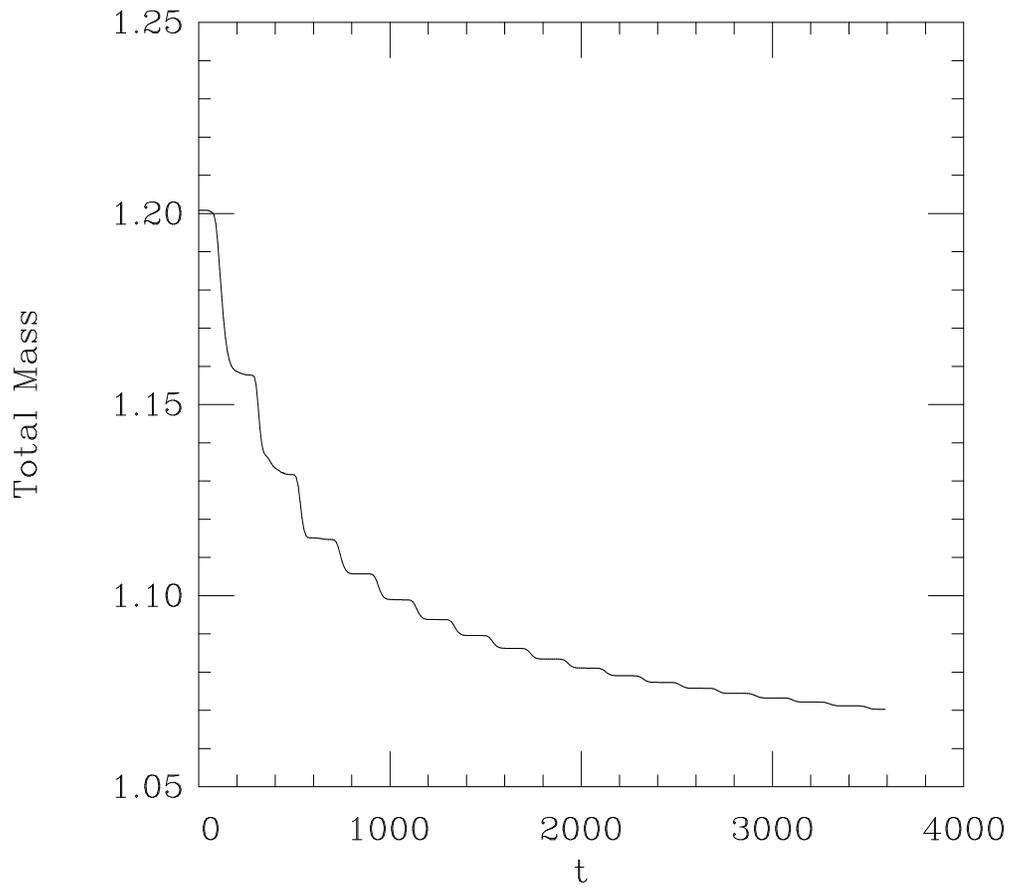}}
\end{center}
%%\begin{figure}
%%\hspace{-36pt}
%%\vspace{-130pt}
%\epsfbox[0 -250 450 500]{figs/lammigc.ps}
%%\epsfbox[0 -150 350 500]{figs/lammigce.ps}
\caption{
The mass is plotted
 against time.
The mass loss through scalar radiation at each expansion of the core
(corresponding to the maximum radial metric reaching a minimum) decreases
in time as the oscillations damp out.}
\label{fig_lammigc}
\end{figure}

Figs.~\ref{fig_lammiga}-\ref{fig_lammigc} show a migrating $\Lambda=30$ star whose unperturbed overall
field density $\sigma$
%$\sigma (0) =0.23$,
has been decreased by about 10\%. Fig.~\ref{fig_lammiga} shows the behavior
of the radial metric in time as a function of radius. It oscillates about the
final $S$-branch configuration
that it will settle into. This final state is shown on the plot as a dark line.
Fig.~\ref{fig_lammigb} shows the maximum radial metric as a function of time.
The star initially expands
rapidly as it moves to the $S$-branch. This can be seen from the sharp drop
in the radial metric. Once it moves to the stable branch, it oscillates about
the new configuration that it is going to settle into. Fig.~\ref{fig_lammigc} shows the mass of
the star as a
function of time. It loses mass at each expansion losing less and less mass at
each subsequent
expansion and the curve gets smoother and smoother as it settles to its final state.

\begin{figure}[t]

\begin{center}
\leavevmode
\makebox{\epsfysize=15cm\epsfbox{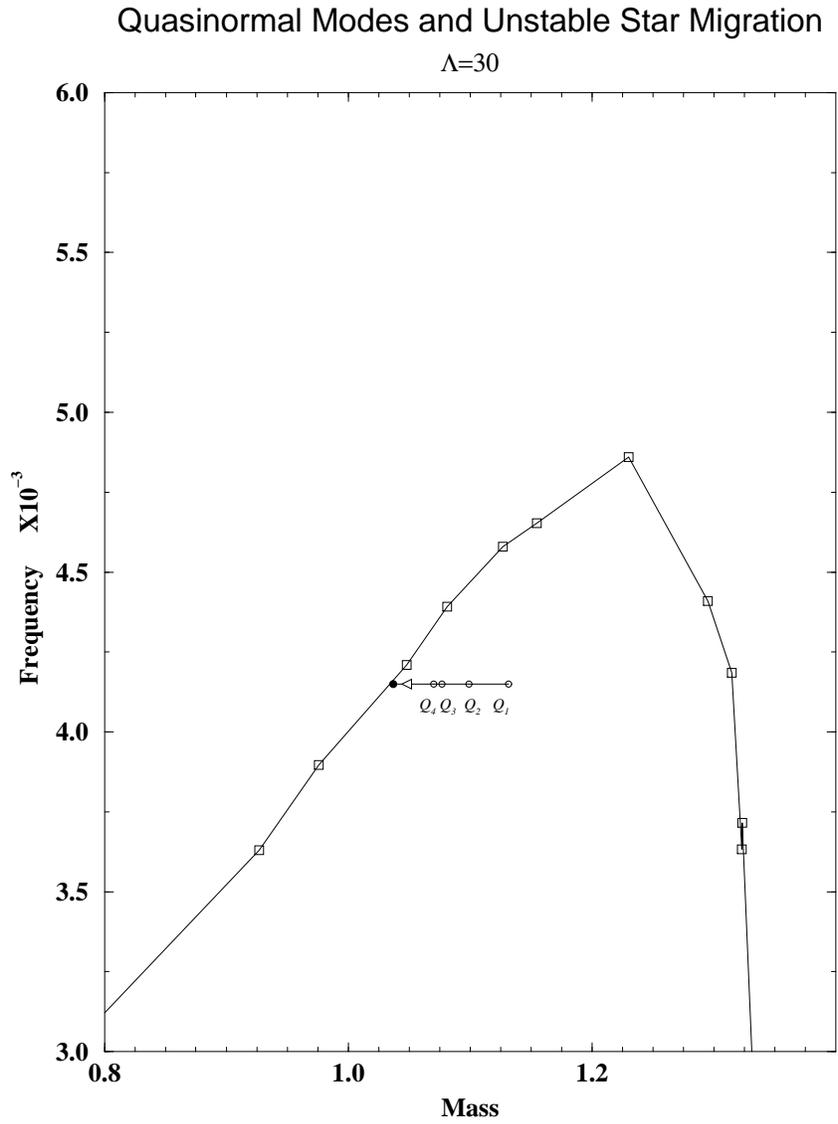}}
\end{center}
%%\begin{figure}
%%\hspace{-36pt}
%%\vspace{-130pt}
%\epsfbox[50 -50 450 650]{figs/lammigqnm.ps}
%%\epsfbox[0 -150 350 500]{figs/lammigqnme.ps}
\caption{The migration of the $U$-branch star considered in the previous figure
is shown after the star has moved to the $S$-branch. Points $Q1$, $Q2$, $Q3$, and
 $Q4$ correspond to
times 500, 1000, 2000,
and 3500 ($\hbar/mc$) respectively. Here too the mass loss decreases in time and the star
finally settles to a stable configuration.}
\label{fig_lammigqnm}
\end{figure}

Fig.~\ref{fig_lammigqnm}, which shows the oscillation frequency as a function of mass for $\Lambda=30$, can be
used to predict the end point of migration.
Points $Q1$, $Q2$, $Q3$, and $Q4$ show the migration of this star. These correspond to
 times of 500,
1000, 2000, and 3500. The oscillation is clearly damping out. The final
configuration it is expected to settle down to is shown as a dot and
corresponds to a stable star of central density
$\sigma(0)=0.0817$ with a mass of $1.037\, M_{Pl}^2/m$. This example is
typical of a number of simulations of $U$-branch ground state
configurations with self coupling.

For higher central density stars on the $U$-branch, the
mass versus central density curve has a second, gentler peak, similar to the
white dwarf neutron star situation. One might suspect that this corresponds
to another stable and unstable branch respectively.
However we find that configurations on both sides of the peak are
unstable. These configurations always disperse upon perturbation,
consistent with the fact that they have
$M\,>\,N\,m$, where $M$ is the mass of the star and $N\,m$ is the
number of bosons multiplied by the mass of an individual boson.

\subsection{Excited States}

\begin{figure}[t]

\begin{center}
\leavevmode
\makebox{\epsfysize=15cm\epsfbox{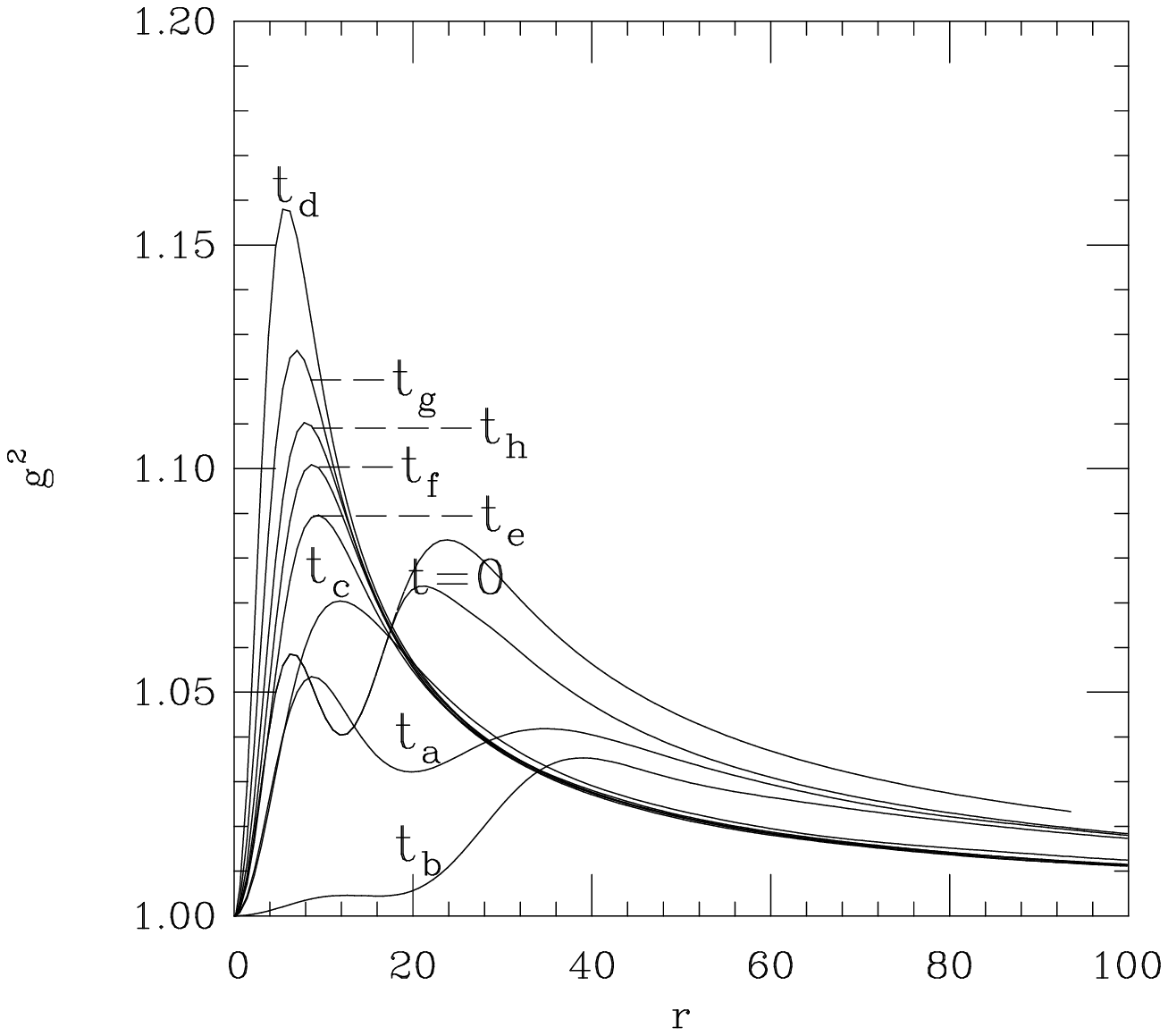}}
\end{center}

%%\begin{figure}
%%\hspace{-36pt}
%%\vspace{-130pt}
%\epsfbox[0 -250 450 500]{figs/lamexctransa.ps}
%%\epsfbox[0 -80 350 500]{figs/lamexctransae.ps}
\caption{
The transition of a first excited state star to the ground state is
shown. Here the radial metric is plotted against radius for various times
starting from $t=0$ and then for
$t_a=250$, $t_b=500$, $t_c=1245$, $t_d=4000$,
$t_e=5000$, $t_f=5505$, $t_g=6000$, and $t_h=6370$.
The initial
unperturbed and perturbed configurations
are shown. (The perturbed configuration
has the lower second peak.)
The initial mass of the star was $M=0.901 \, M_{Pl}^2/m$ after
perturbation. The radial metric
initially has two peaks indicative of a one-node configuration.
As the star evolves and goes to the
ground state one peak disappears.
This can be seen in the curves from $t_c$--$t_h$. Once on
the ground--state branch it oscillates and
finally settles into a stable ground--state configuration.}
\label{fig_lamexctransa}
\end{figure}

\begin{figure}[t]

\begin{center}
\leavevmode
\makebox{\epsfysize=15cm\epsfbox{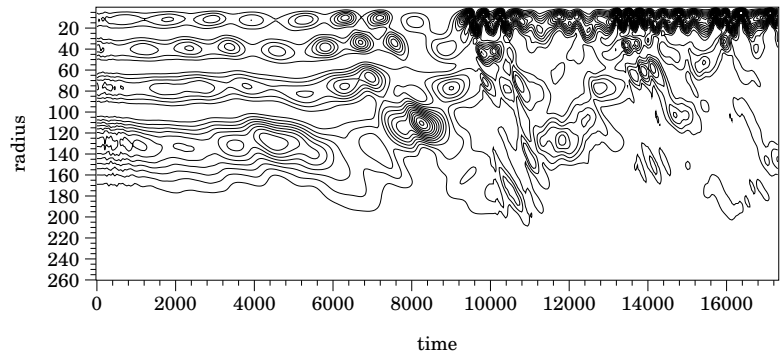}}
\end{center}

%%\begin{figure}
%%\hspace{-36pt}
%%\vspace{-130pt}
%\epsfbox[0 -250 450 500]{figs/lamexctransb.ps}
%%\epsfbox[0 -150 350 500]{figs/lamexctransbe.ps}
\caption{
The
transition of a three-node configuration of
mass $0.91 \, M_{Pl}^2/m$. This star loses enough
mass during the course of its evolution to move to the
ground state.
}
\label{fig_lamexctransb}
\end{figure}

The similarity of mass profiles of 
excited boson stars to their ground state counterparts, 
(Fig.~\ref{fig_lamexcmass})
might lead one to expect
 stable and unstable configurations to the
left and right of the maximum mass respectively in analogy with ground state 
configurations.
However, our numerical studies show that the
excited boson star configurations on {\em both} sides of
the peak are inherently unstable except that the time scales for
instability are different.
If they cannot lose enough mass
to go to the ground state they either become black holes or totally disperse. This 
occurs even if no
explicit perturbations are put in the numerical evolution other
than those introduced by the finite differencing error in the
numerical integration. We have also carried out perturbations that correspond
to scalar particles falling onto the star, or those that
decrease the scalar field strength
at the center point (corresponding to scalar particles decaying through
some channel). The instability shows up in all cases
studied. We note that this instability is $NOT$ in contradiction with the
result of \cite{jet}, which concluded that $S$ branch excited states are stable under
infinitesimal perturbations that strictly conserve $M$ and $N$,
(where $M$ is the
mass of the star, $N$ the number of bosons and $m$ the boson mass.)
Our result of instability under generic perturbation is consistent with the
studies of $\Lambda=0$
stars under infinitesimal perturbation \cite{leepert}.
The presence of a self-coupling term
increases the time scale of instability
but the essential pattern remains the same.
Fig.~\ref{fig_lamexctransa} shows a perturbed 1-node star (whose
mass has been reduced by about $8\%$ to $0.9\,M_{Pl}^2/m$ by a perturbation)
making a transition to the ground
state although the mass is greater than the maximum ground state mass
of $0.633 M_{Pl}^2/m$.  A substantial
amount of scalar radiation is
emitted in the dynamical evolution, which brings
the mass below the critical value.
The evolution of the radial metric function is
shown.  
That there are two peaks at $t=0$
is indicative of a first excited state. One of the peaks disappears
gradually as the star goes to the ground state branch. The star then
oscillates about the ground state configuration that it will finally
settle into. 
In Fig.~\ref{fig_lamexctransb} a three-node configuration with a total mass
of of $0.92 M_{Pl}^2/m$ is shown going to the ground state branch,
after radiation by scalar waves carries off the excess mass and kinetic
energy. The plot shows the density function against the radius and time of
evolution.
By the density function we mean density $\rho$ multiplied by $r^2$,
that is, the mass per $dr$ at radius $r$. The
density function, $\rho\,r^2$, has $n+1$
maxima for an $n$ node star and hence here we have four sets of lines initially.
At the end of the simulation, we see that it settles down to a ground
state configuration with small oscillations with ever decreasing
amplitude.  For these simulations of low central density stars
we put in an explicit perturbation to the equilibrium configuration
since otherwise the instability time scales are extremely long.

\begin{figure}[t]

\begin{center}
\leavevmode
\makebox{\epsfysize=15cm\epsfbox{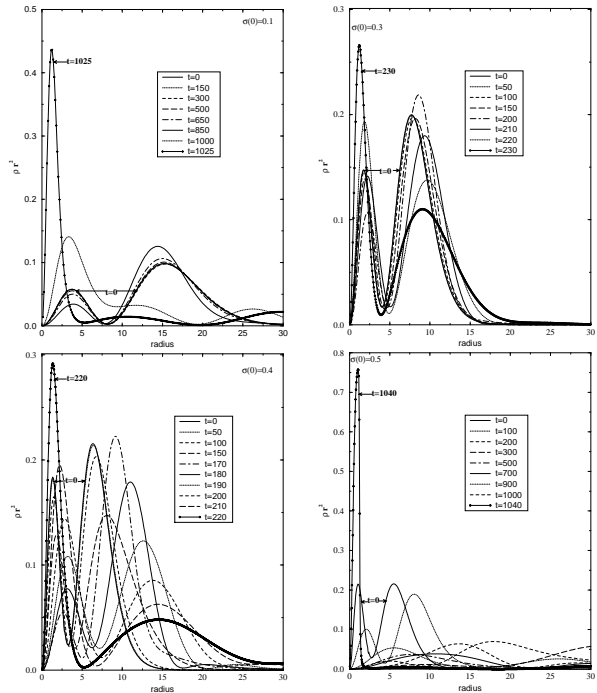}}
\end{center}
%%\begin{figure}
%%\hspace{-36pt}
%%\vspace{-130pt}
%\epsfbox[0 -250 450 500]{figs/lamdecaya.ps} 
%%\epsfbox[0 -80 350 500]{figs/lamdecayae.ps} 
\caption{
A comparison of the manner of black hole formation of four excited state
configurations.
The first frame is an $S$ branch star of central density $\sigma(0)=0.1$, that
tries
to go to the ground state but fails to. As the central density increases,
decays to black holes occur on a shorter time scale. A plot of the collapse of
a $U$-branch star of central density $\sigma(0)=0.3$ is shown in
the next frame. Stars get more diffuse as one moves farther along the
$U$ branch. A star of central density $0.4$ (shown in
the third frame) has a decay time close to
the previous one. Decay times then start to increase. A star of
central density $0.5$ (close to the point $M=N\,m$ with $\sigma(0)\sim 0.541$)
disperses to over ten times its radius before collapsing to a black hole.}
\label{fig_lamdecaya}
\end{figure}

\begin{figure}[t]

\begin{center}
\leavevmode
\makebox{\epsfysize=15cm\epsfbox{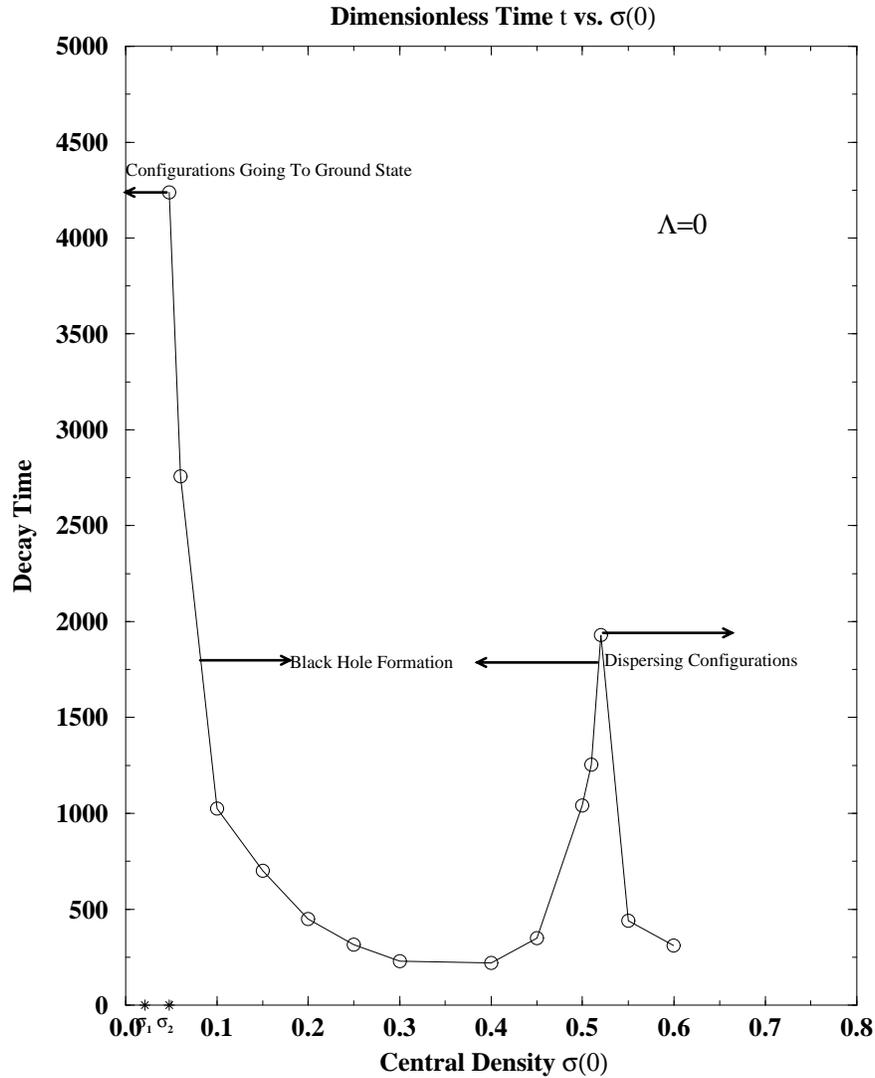}}
\end{center}
%%\begin{figure}
%%\hspace{-36pt}
%%\vspace{-130pt}
%\epsfbox[0 -250 450 500]{figs/lamdecayb.ps} 
%%\epsfbox[0 -30 350 500]{figs/lamdecaybe.ps} 
\caption{
The decay time to black holes is plotted as a function of
central density, for one-node configurations (first excited states)
of boson stars without self coupling.
Perturbations are due only to the finite differencing effects of
the numerical scheme. To make the comparisons meaningful  the
$95\%$ mass radius (which is our definition for radius of a star)
of every
configuration considered was covered by the same number of grid points.
Configurations for which the central
density $\sigma(0) < \sigma_2$ move to the ground state. The central density
$\sigma_1$ corresponds to a mass $M=0.633 \,M_{Pl}^2/m$, which is the
maximum mass of a ground state boson star.  The decay time
decreases with increasing $\sigma(0)$ and this continues even for
$\sigma(0) > 0.25$ which is the transition point from the
$S$-branch to the $U$-branch.
The decay time then starts to increase as
one approaches the $M=N\,m$ point corresponding to a central density
$\sigma(0)=0.541$, beyond which
the stars
disperse to infinity rather than becoming black holes.
The dispersion times of two such stars to
ten times their original radius ($95\%$ mass radius)
are also shown on the figure.
}
\label{fig_lamdecayb}
\end{figure}

For stars with higher central density, there is
a critical density above which the stars cannot lose enough mass to go to
the ground state but collapse to black holes.
In our numerical simulation for one node stars this critical density is
$\sigma(0)=\sigma_2=0.048$. As the central density is increased the
kinetic energy of the highly compressed initial equilibrium configuration
is increased and the star first expands before the eventual collapse
to a black hole.  As the central density  further increases towards the
$M=N\,m$ point
the expansion phase becomes longer.
In Fig.~\ref{fig_lamdecaya}
we show the density function ($\rho r^2$) against radius
at various times for 4 configurations (an $S$-branch configuration with
$\sigma(0)=0.1$, and three $U$-branch configurations of central densities
$\sigma(0) = 0.3$, $\sigma(0)=0.4$, and $\sigma(0)=0.5$).
The initial configurations are the equilibrium ones without any
explicit perturbation (except those introduced by the discretization
used in the numerical simulations).
The first frame shows
the $S$-branch star radiating a  little as it almost makes a
transition to the ground state.
However it cannot sustain this state for long and it collapses to
a black hole.
The decay time decreases in the case of a $U$-branch stars of
central density $0.3$ and $0.4$. However for the $\sigma(0)=0.5$ star, the
star is clearly getting more diffuse than the previous ones. It
goes through an expansion phase initially although
it finally collapses to a black hole.
Configurations with $\sigma(0) > 0.54$ have
$M>N\,m$. They do not collapse to black holes but disperse to
infinity.
Next we turn to
a study of instability time scales.
In simulations where a black hole will form, the imminent development of
an apparent horizon leads to a rapid collapse
of the lapse due to the polar slicing
used in the evolution.
We take the time of collapse of the lapse at
the origin
to $\sim 10^{-6}$ of its initial central value
to be the
approximate time for formation of the black hole.
Fig.~\ref{fig_lamdecayb} shows this time scale for a 1-node
star without self coupling.
We plot the decay time scales of first excited state stars as a
function of central field density.  Again no explicit perturbation is
applied in these evolutions other than the discretization error
in the simulation.
In order to make a fair comparison of the time scale due to
such a perturbation we cover the
radius of the star in all cases by the same number of grid
points.
The maximum ground state
mass for stars without self coupling is around 0.633\,$M_{Pl}^2/m$. This
corresponds to a central density of $\sigma_1=0.021$ for a one-node star.
We mentioned earlier that
stars with central densities below $\sigma_2=0.048$ (mass of $0.91
M_{Pl}^2/m$) lose enough mass and move to the ground state. Beyond
that and up to the point $M=N\,m$ they collapse to black holes,
while stars with $M > Nm$, corresponding to a central field
density of $\sigma(0)>0.541$, disperse to infinity. The time by which
collapse takes place to a black hole
decreases with increasing central density along
the $S$ branch. This trend continues for a while into the $U$ branch (starting
at $\sigma=0.25$ until $\sigma\sim 0.4$),
but as one approaches the $M=N\,m$ point
the stars
lose a significant amount of  matter to infinity
before they collapse to black holes and evolve on a longer time scale.
For example,
a star of central density $0.5$ has an initial
radius of $r\sim9$.
(Remember that the radius of the star is defined as the radius which
contains $95\%$ of the mass of the star.)
Its radius increases to as much as 115, more than an order of magnitude,
before it starts collapsing. The dispersion time scales
of a couple of typical examples of stars for $M>Nm$ which disperse to
infinity (instead of collapsing to
a black hole) are also shown on the figure. To give
a sense of the instability time scales of these
configurations we take the time scale to be
the time by
which these stars disperse to ten times their original radii.
This time drops drastically for stars with $M\gg Nm$.

\begin{figure}[t]

\begin{center}
\leavevmode
\makebox{\epsfysize=15cm\epsfbox{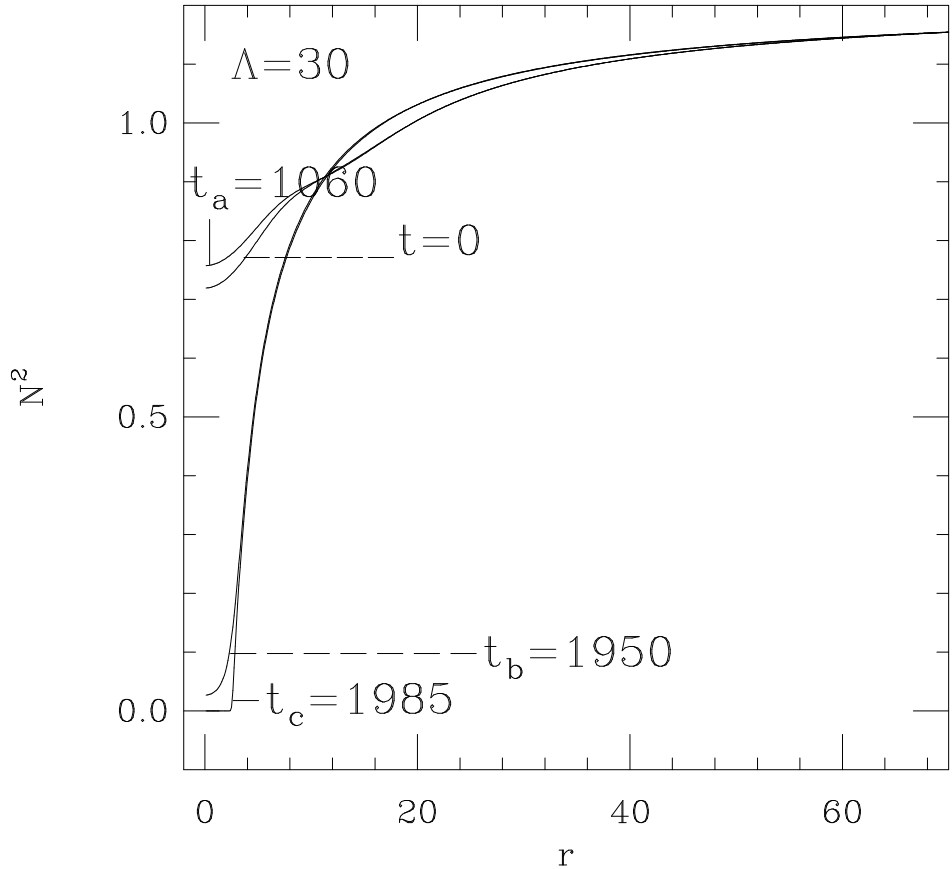}}
\end{center}
%%\begin{figure}
%%\hspace{-36pt}
%%\vspace{-130pt}
%\epsfbox[0 -250 450 500]{figs/lamexcbha.ps}
%%\epsfbox[0 -120 350 500]{figs/lamexcbhae.ps}
\caption{
The evolution of the metric function $N^2=-g_{tt}$ for a
$\Lambda=30$, $\sigma(0)=0.1$ boson star in the first excited state, without
any explicit added perturbation, is shown. The configuration lies on the
$S$-branch and has an initial mass $M=1.743 \, M_{Pl}^2/m$. The various time slices
correspond to times $t=0$, $t_a=1060$, $t_b=1950$, and $t_c=1985$.
The lapse function collapses
as an apparent horizon is approached, signaling the formation of the black
hole. (This is indicative of an inherent instability of the excited states).}
\label{fig_lamexcbha}
\end{figure}

\begin{figure}[t]

\begin{center}
\leavevmode
\makebox{\epsfysize=15cm\epsfbox{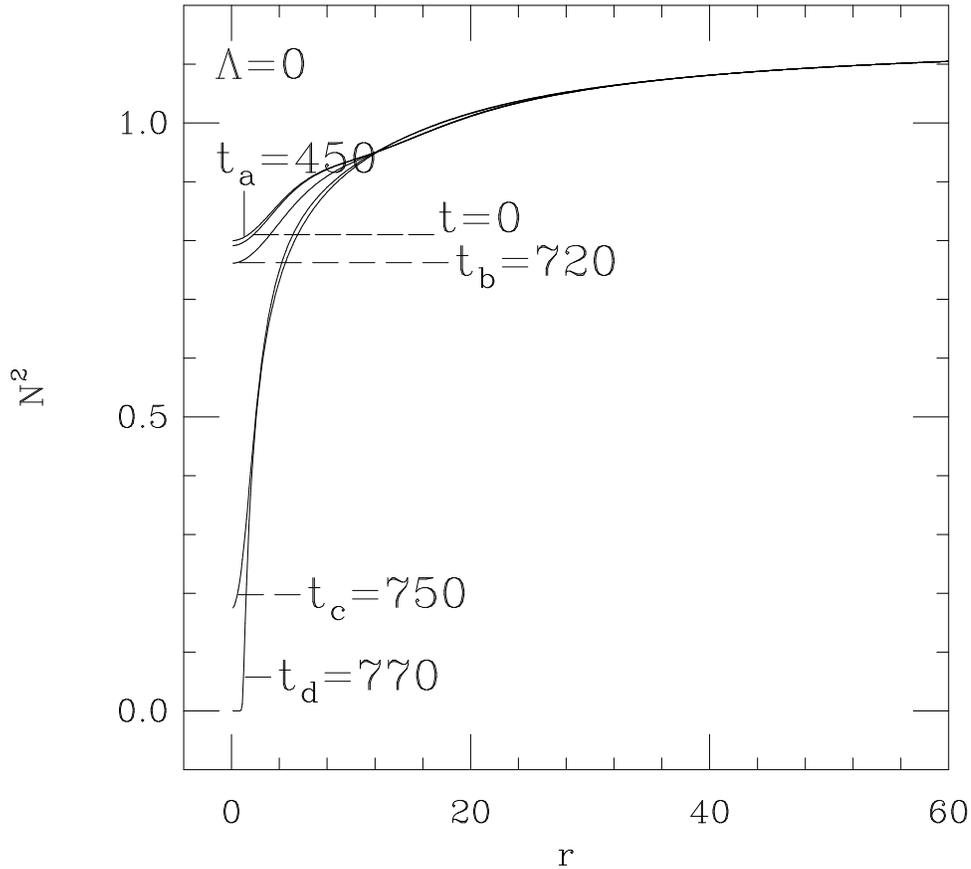}}
\end{center}
%%\begin{figure}
%%\hspace{-36pt}
%%\vspace{-130pt}
%\epsfbox[0 -250 450 500]{figs/lamexcbhb.ps}
%%\epsfbox[0 -100 350 500]{figs/lamexcbhbe.ps}
\caption{
The evolution of the metric function $N^2=-g_{tt}$ for a
$\Lambda=0$, $\sigma(0)=0.122$ boson star in the first excited state, without
any explicit perturbation added, is shown. The configuration lies on the
$S$-branch and has an initial mass $M=1.23 \,M_{Pl}^2/m$. The various time slices
correspond to times $t=0$, $t_a=450$, $t_b=720$, $t_c=750$, and $t_d=770$.
This star has a radius of
$20.7$ which is about the same as that of the configuration shown in the
previous figure. Again,
the lapse function collapses when an apparent horizon is approached, as a
black hole is being formed. The time scale of collapse is much less than for
the $\Lambda=30$ case in the previous figure.
}
\label{fig_lamexcbhb}
\end{figure}

Next we turn to the case of $\Lambda \ne 0$.
Fig.~\ref{fig_lamexcbha}
demonstrates black hole formation for a $\Lambda=30$ star in the first
excited state, with a central density $\sigma(0)= 0.1$. This star had an initial
radius of about 20.7 where the radius is again
defined as that containing $95\%$ of the
mass. The lapse finally collapses towards zero indicating that an
apparent horizon is about to form.
The time scale of collapse
to the black hole was around 1985 compared to a
time scale of less than 800 for a $ \sigma(0)=0.122 $ star of similar radius
without self-coupling which is shown in Fig.~\ref{fig_lamexcbhb}.
This is to be expected as the
$\Lambda$ term represents a repulsive force.
The collapse time
is again defined by the lapse collapsing to $10^{-6}$ of its
original value at $r=0$.

\begin{figure}[t]

\begin{center}
\leavevmode
\makebox{\epsfysize=15cm\epsfbox{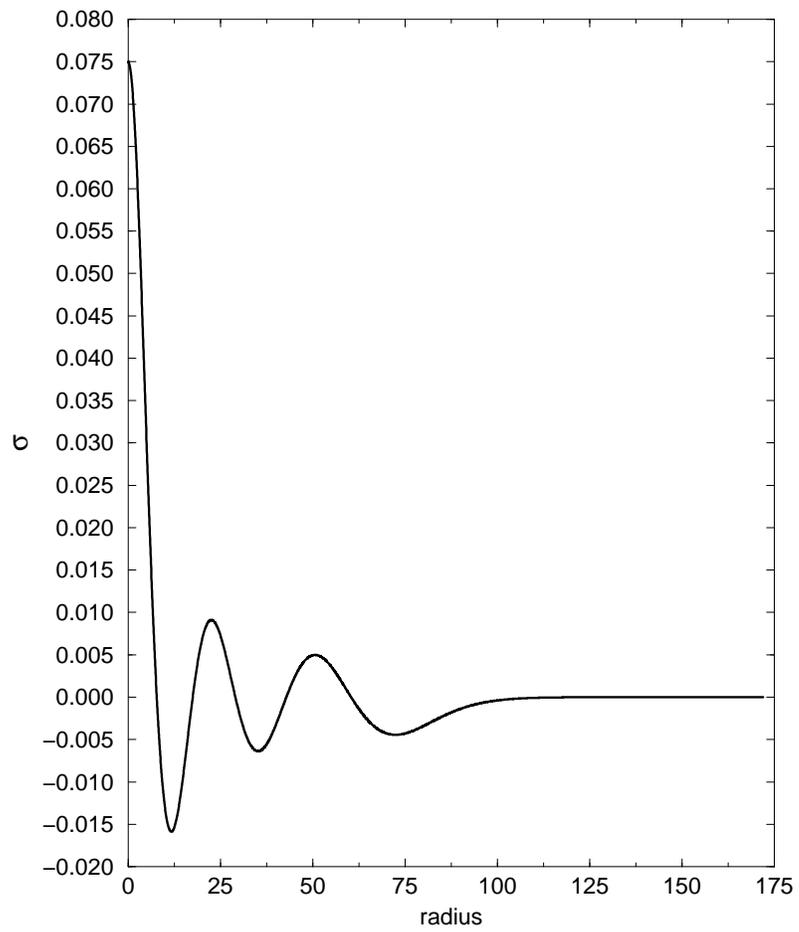}}
\end{center}
%%\begin{figure}
%%\hspace{-36pt}
%%\vspace{-130pt}
%\epsfbox[0 0 450 500]{figs/lamcasca.ps}
%%\epsfbox[0 -150 350 500]{figs/lamcascae.ps}
\caption{
The initial field configuration of a five-node star.
The field has five nodes or extrema. The absolute value of each extremum is
clearly smaller than the one preceding it.
%appears as a bold line while the unperturbed one as a dotted one.
}
\label{fig_lamcasca}
\end{figure}

\begin{figure}[t]

\begin{center}
\leavevmode
\makebox{\epsfysize=15cm\epsfbox{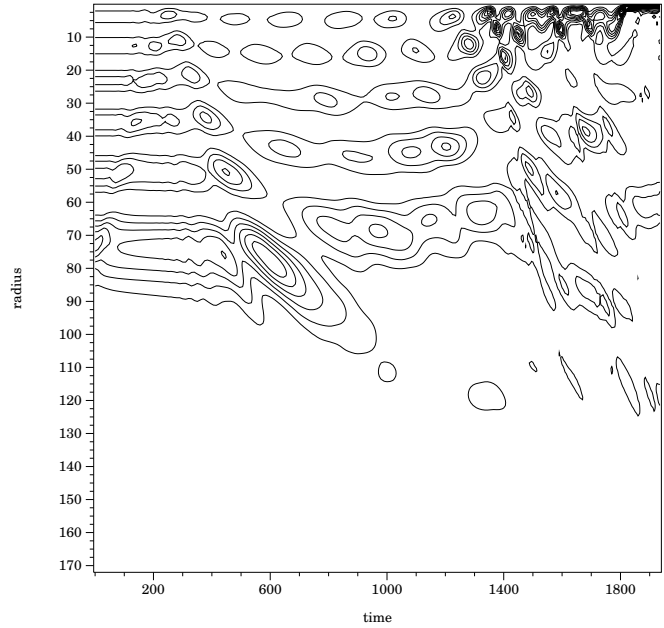}}
\end{center}

%%\begin{figure}
%%\hspace{-36pt}
%%\vspace{-130pt}
%\epsfbox[0 -250 450 500]{figs/lamcascb.ps}
%%\epsfbox[0 -50 350 500]{figs/lamcascbe.ps}
\caption{
A contour plot of a perturbed five-node star that decays into a
black hole showing $\rho\; r^2$
as a function of distance (vertical axis) and of time (horizontal axis) is
shown. The density $\rho$ is highest at the origin and has five other local
maxima, each smaller than the previous one. The value of the maxima at the
end are very small compared to the earlier ones, and to enhance the features
$\rho \; r^2$ rather than just $\rho$ is shown in the plots. Each set
of lines represents the maxima of $\rho\; r^2$ and the number of
lines in a set gives an indication of the height of the maximum.
This particular configuration has a mass $M=3.07\, M_{Pl}^2/m$. This star
cascades through an intermediate four-node state (around $t=1000$)
before proceeding to form a black hole. Cascades are characteristic
of excited boson star decays, similar
to atoms in excited states going through
intermediate states when decaying to the ground
state.}
\label{fig_lamcascb}
\end{figure}

We now turn to the evolution of a highly excited state. In Fig.~\ref{fig_lamcasca}
we show the initial field
configuration of a star containing five nodes.
For a five-node star the density has a central maximum, and then five local
maxima, each subsequent one smaller than the one
preceding it. This star is then evolved without any perturbation
except those introduced by the discretization
error of the numerical evolution.
In Fig.~\ref{fig_lamcascb}
we show a
contour plot of the evolution of the density function in time:
$\rho\,r^2$ has $n+1$ maxima for an $n$ node star and hence here we initially
have six sets of lines
centered at dimensionless $r=5,\,\,13,\,\,35,\,\,52$, and $75$ respectively.
This star has central density $\sigma(0)=0.075$ and it collapses to
a black hole after a long evolution. In the process we see intermediate
states with fewer numbers of nodes.

\begin{figure}[t]

\begin{center}
\leavevmode
\makebox{\epsfysize=15cm\epsfbox{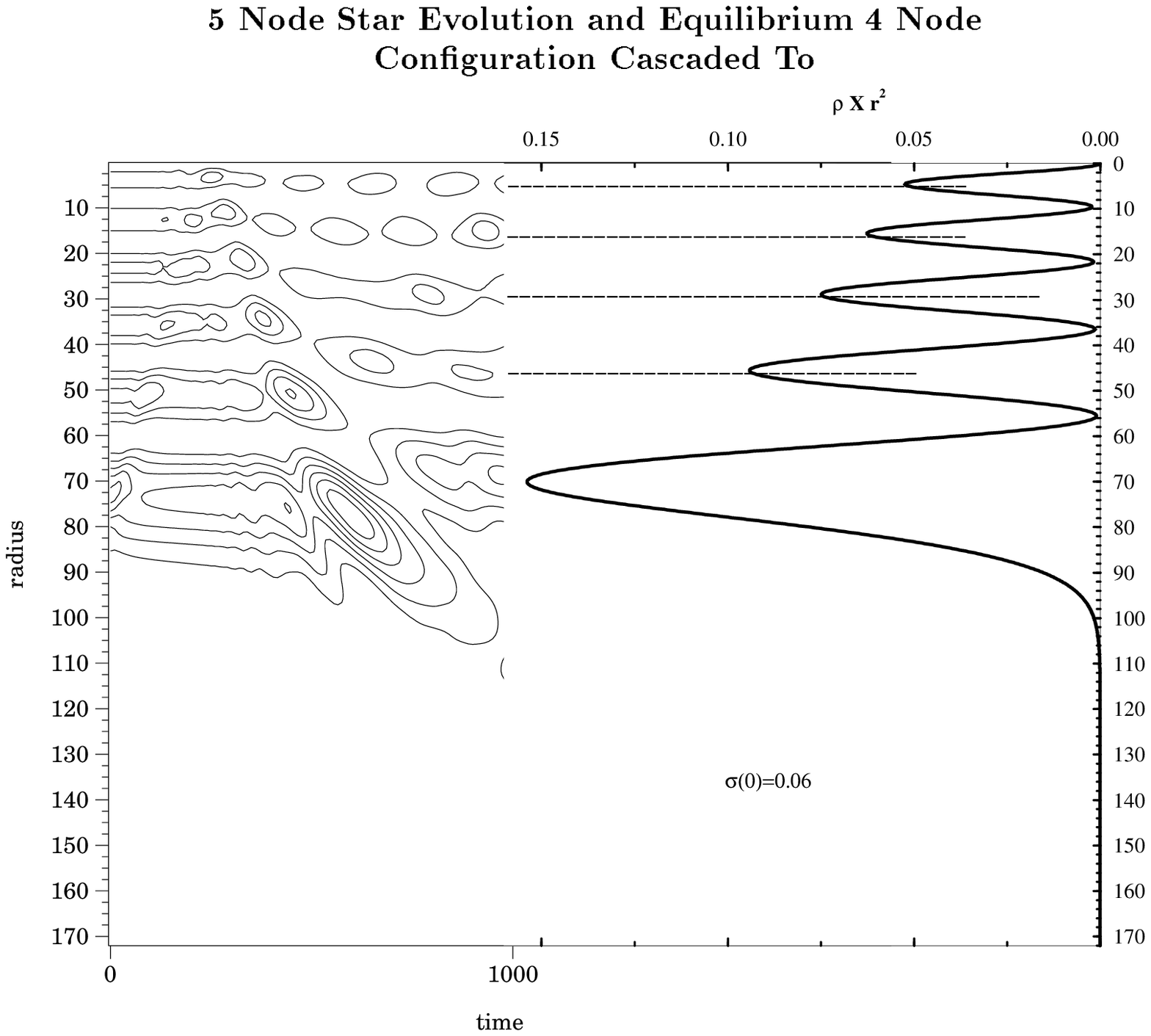}}
\end{center}
%%\begin{figure}
%%\hspace{-36pt}
%%\vspace{-130pt}
%\epsfbox[0 -250 450 500]{figs/lamcascc.ps}
%%\epsfbox[0 -140 350 500]{figs/lamcascce.ps}
\caption{
The equilibrium density function of a
four-node star of central density $\sigma=0.06$ (right frame) is
placed alongside the
contour plot of the five-node star described in the previous figure
%\ref{fig_lamcascb}
up to a time of $t=1000$ when it has 
decayed into a four-node state (left frame).
This plot shows that the transition of the
five-node star is to a perturbed four-node state close to the one shown
before it continues its evolution to a black hole.
}
\label{fig_lamcascc}
\end{figure}

For comparison, in Fig.~\ref{fig_lamcascc} we
show the contour plot
for the five-node
star
up to a time of $t=1000$,
at which time it has decayed into a four-node
state, against the equilibrium density function
of a four-node star with central density $\sigma = 0.06$.
Very clearly the maxima are at similar radii (although the size
of the peaks are somewhat different).

This feature, that of
nodes disappearing and the star cascading through lower excited states,
is characteristic of the decay of
higher excited states of boson stars. Although
in this particular case and at this particular time the decaying star is close
to a specific lower excited state, in general the
decaying star is (roughly speaking) a combination of
lower excited configurations. This is similar
to the decay of atoms in excited states.
However, we note that for the ``gravitational atom''\cite{Lid} there
is no exact superposition, due to the intrinsic non-linearity of the system.

Whenever we report black hole formation, we are of course talking about
imminent formation since our highly singularity-avoiding polar
slicing condition causes the lapse to collapse as soon as
an apparent horizon forms. However taking the data just before such
formation, and evolving it with a 3D code with maximal-slicing has allowed us
to confirm that a black hole indeed forms.

\subsection{Very High $\Lambda$ Configurations}
\label{Appendix}

While calculating the eigenvalue frequency for the equilibrium boson star
it is found that it gets increasingly difficult 
to calculate the eigenvalue as the value of $\Lambda$ gets large, because
the equations become very stiff. There are two scales to the
problem: a scale of slow variation of the field inside a certain radius
(related to $\Lambda$) followed by rapid 
decay outside it. This produces an effective surface layer to the star making
it similar to neutron stars.
It turns out that the large $\Lambda$ limit can be treated using a
set of approximate equations that are exact in the $\Lambda=\infty$ limit~[6]. By making the
following change of variables:
\begin{equation}
{\bar r} =  \frac{r}{\sqrt{\Lambda}}, \quad \quad {\bar \sigma} = \sqrt{\Lambda}\,\sigma
\label{A1}
\end{equation}
the equilibrium equations reduce to 
\begin{equation}
\frac{1}{\Lambda}\, {\bar \sigma}''= -\frac{1}{\Lambda}\,\left[\frac{1}{{\bar r}}+\frac{g^2}{{\bar r}}-
{\bar r}\, g^2 \, {\bar \sigma_0}^2\right]\,{\bar\sigma'} -
\frac{1}{\sqrt{\Lambda}}\,\left[\frac{1}{N^2}-1\right]\,{\bar \sigma_0}\,g^2+
\frac{1}{\sqrt{\Lambda}} \, ( g^2 \, {\bar\sigma_0^3}),
\label{A2}
\end{equation}
\begin{equation}
g'=\frac{1}{2} \, \left[\frac{g}{{\bar r}}-\frac{g^3}{{\bar r}}
+{\bar \sigma_0}^2 \, {\bar r} \, g^3\, \left[1+\frac{1}{N^2}
\right] + \frac{1}{\sqrt {\Lambda}} \, {\bar r} \, g 
\, {{\bar \sigma_0'^2} }
+\frac{1}{2} \, (g^3  \, {\bar r} \, {\bar \sigma_0^4})\right]
\label{A3}
\end{equation}
\begin{equation}
N' =\frac{1}{2} \, \left[
-\frac{N}{{\bar r}}
+\frac{N \, g^2}{{\bar r}}
+\frac{{\bar r} \, g^2 \, {\bar \sigma_0}^2}{N} \,
(1-N^2)
+\frac{1}{\sqrt{\Lambda}}\, {\bar r} \, N \, {\bar \sigma_0'^2}-\frac{1}{2}\, g^2 \,  N \, {\bar r} \, {\bar \sigma_0}^4\right].
\label{A4}
\end{equation}
where the primes refer to differentiation with respect to ${\bar r}$. In the limit of $\Lambda =\infty$,
one can keep terms to leading order in $\frac{1}{\sqrt \Lambda}$ in (\ref{A2})
--(\ref{A4}).
In particular (\ref{A2}) reduces to an algebraic constraint
\begin{equation}
N=({\bar \sigma}^2 + 1)^{-\frac{1}{2}}.
\label{A5}
\end{equation}
However, we note that this is valid only
for a ground state configuration. For a state with nodes it would not be
reasonable to neglect derivative terms compared to terms proportional
to ${\bar \sigma}$, because it is zero at every node.

\begin{figure}[t]

\begin{center}
\leavevmode
\makebox{\epsfysize=15cm\epsfbox{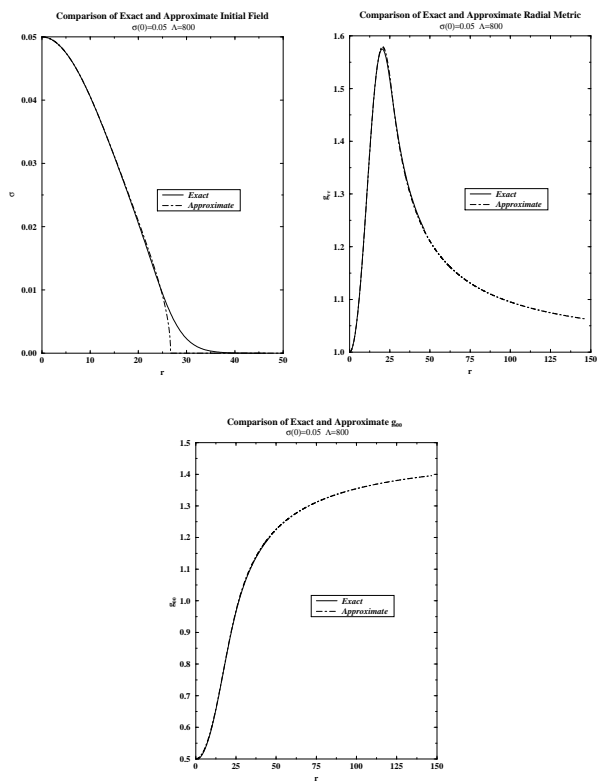}}
\end{center}

%%\begin{figure}
%%\hspace{-36pt}
%%\vspace{-130pt}
%\epsfbox[0 -250 450 500]{figs/lamhighequil.ps}
%%\epsfbox[0 -140 350 500]{figs/lamhighequile.ps}
\caption{ The equilibrium profiles of a $\Lambda=800$ star with central
density
$\sigma=0.05$ as derived
from the high-$\Lambda$ approximate equations and the exact equations are
compared. A Schwarzschild exterior is attached to the approximate solution
after the field vanishes.  The three plots show the
 field $\sigma$, $g_{rr}$, and $g_{00}$ respectively. Clearly, the
 approximation matches the exact solution very well.}
 \label{12old}
 \end{figure}

\begin{figure}[t]

\begin{center}
\leavevmode
\makebox{\epsfysize=15cm\epsfbox{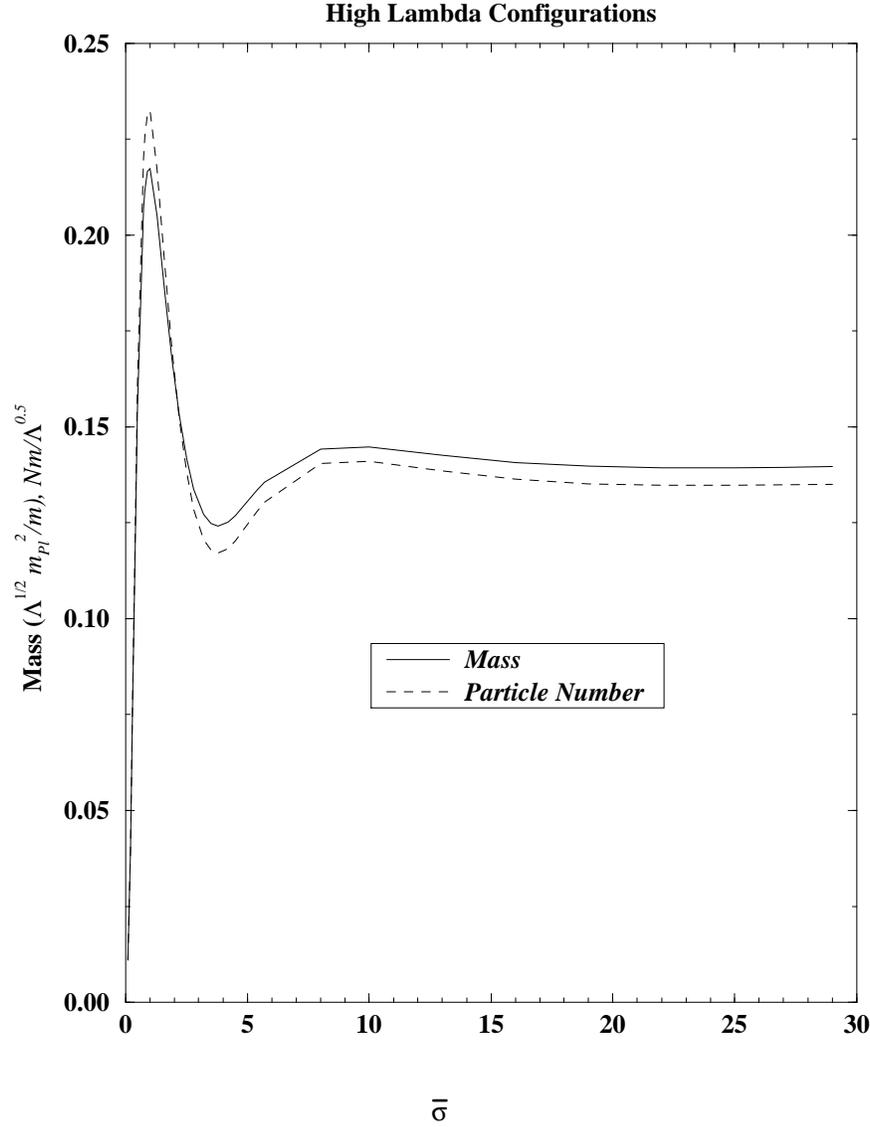}}
\end{center}
%%\begin{figure}
%%\hspace{-36pt}
%%\vspace{-130pt}
%\epsfbox[0 -250 450 500]{figs/lamhighmass.ps}
%%\epsfbox[0 -120 350 500]{figs/lamhighmasse.ps}
\caption{ The mass of a high $\Lambda$ star generated from the approximate 
equations is plotted as a function of $\bar{\sigma}$ 
($\sigma/\Lambda^{\frac{1}{2}}$). It
shows the same basic structure as the profiles generated for low $\Lambda$
using the exact equations. The peak is at about $0.22 \Lambda  ^{1/2}M_{Pl}^{2}/m $,
which means that
to achieve $ 0.1 \msun $ we would have to take $\Lambda$ 
of order $10^{38}$. This would be a very large star
to evolve numerically. Also plotted is $N\, m$ ($N$ is the particle number
and $m$ the mass of a boson). The crossing point 
of the two curves represents
transition from negative to positive binding energy.}
\label{13old}
\end{figure}

To get an estimate of the accuracy of the high $\Lambda$ approximation we
compare the solution using the approximate equation for high $\Lambda$ to the
brute force numerical solution of the complete set of equations, for a
$\Lambda =800$, $\sigma=0.05$
boson star in the ground state (see Fig.~\ref{12old}).
The agreement of the fields is quite good
until the outer region where the approximate equations cause the field to abruptly fall to zero.
Comparisons of the radial metric and the lapse are also shown.
In the $\Lambda=\infty$ limit the approximate equations are exact and the 
star really has an outer surface reminiscent of a neutron star. In fact, an effective equation of state can be
written~[6]. In Fig.~\ref{13old} the mass and particle number 
versus central density $\bar {\sigma}$ for high $\Lambda$ stars is shown. One expects (as in
the case of other ground state configurations) that the configurations with
$M > N_pm$ disperse when perturbed, while those on the $U$ branch with $M < N_pm$ would be unstable and,
if unable to migrate to the $S$ branch under perturbations, would form black holes. (Here we use the symbol $N_p$ for the particle number and not $N$ so
as not to confuse it with the lapse as both these functions figure prominently
in the analysis that follows.)
The particle number is calculated from the current $J^0$ and is given by
\begin{equation}
N_p = 4\pi\int\,r^2\frac{g}{N}\,\omega\, \sigma ^2dr.
\label{A6}
\end{equation}
Here $\dot \sigma$ has been replaced by 
$\omega\,\sigma$ and $d^3 r$ by $4\pi r^2dr$ for the spherically
symmetric case. 
In terms of the $bar$ coordinates we see that ${\bar N_p} = \sqrt{\Lambda} N_p$.

Next we turn to the determination of the quasi-normal frequency (QNM) of the
high $\Lambda$ stars. In principle one could determine 
the QNM using the dynamical studies as performed 
for the $\Lambda=0$ case. However the procedure is
extremely computationally expensive. 
The scalar field has an inherent oscillation period of about $2\pi$ and 
the evolution time steps must be small enough to resolve it. However the code must run long enough
to see a few metric oscillations in order to determine the quasinormal mode. As 
$\Lambda$ gets large, the sizes of the stars also get
large leading to a lower frequency of oscillation. In order to determine the QNM we use instead the
following perturbation analysis based on~[8,11] but using our own notation
for lapse, fields, time, radius and self-coupling as defined in section II.

We write the perturbed fields as:
\begin{equation}
\sigma =  (\sigma_1 + i \, \sigma_2) \, e^{i \omega t},  \qquad g=g_0+\delta g, \qquad N=N_0+ \delta N
\label{A7}
\end{equation}
where
\begin{equation}
\sigma_1=\sigma_0(r) \, (1+\delta\sigma_1(r,t)),\qquad \sigma_2=\sigma_0(r)
\, \delta\sigma_2(r,t).
\label{A8}
\end{equation}
the Klein--Gordon equation can be written as
\begin{eqnarray}
\sigma_1'' &+&\left(\frac{2}{r} + \frac{N'}{N} -\frac{g'}{g}\right) \, \sigma_1' +g^2
\, \left(\frac{1}{N^2} 
- 1-\frac{1}{2} \, \Lambda \, {\sigma_1}^2\right) \, \sigma_1
-\frac{g^2}{N^2} \, \ddot\sigma_1  \label{A9}\\  \nonumber
&+&\frac{g^2}{N^2}\, \left(
\frac{\dot N}{N} 
-\frac{\dot g}{g}\right)\, \left(\dot \sigma_1+ \sigma_2\right)
-2 \, \frac{g^2}{N^2} \, \dot\sigma_2.
\end{eqnarray}
The field $\sigma_0$ satisfies the equilibrium equation 
\begin{equation}
\sigma_0'' +\left(\frac{2}{r} + \frac{N_0'}{N} -\frac{g_0'}{g}\right)\,\sigma_0' +{g_0}^2
\left(\frac{1}{2\, {N_0}^2} 
- 1-\frac{\Lambda}{2}\, {\sigma_1}^2\right) \, \sigma_0
-\frac{{g_0}^2}{{N_0}^2} \, \sigma_0 = 0.
\label{A10}
\end{equation}
Expanding the perturbations to first order using (\ref{A9}) and (\ref{A10}) we get

\begin{eqnarray}
\delta{\sigma_1}''&+&\left(\frac{2}{r}+\frac{N_0'}{N} -\frac{g_0'}{g} + 2\,
\frac{{\sigma_0}'}{\sigma_0}\right)\,
\delta{\sigma_1}'
- \frac{{\sigma_0}'}{\sigma_0}\,
\left(\frac{g_0\, {\delta g}'-g_0'\, \delta g}{{g_0}^2}
\right.
\label{A11}
\\ \nonumber
&-&\left.\frac{N_0\, {\delta N}'-N_0' \, \delta N}{{N_0}^2}\right) 
- \frac{{g_0}^2}{{N_0}^2}\, \left(2\, \delta \dot \sigma_2 +\delta
\ddot \sigma_1\right) +  \frac{2}{{N_0}^2}\,
\left(g_0 \, \delta g - \frac{\delta N}{{N_0}}\right)\\ \nonumber
&-& 2 \, g_0\left(1 + \frac{1}{2}\, \Lambda \, {\sigma_0}^2\right)\, \delta g 
-{g_0}^2 \, \Lambda \, {\sigma_0}^2 \, \delta \sigma_1=0.
\end{eqnarray}
From 
\begin{equation}
{\bf R_{\mu\nu}} -\frac{1}{2} \, g_{\mu\nu} \, {\bf R} = -8\, \pi \, G \,{\bf T_{\mu\nu}}
\label{A12}
\end{equation}
and
\begin{equation}
T_{\mu\nu} = \sigma_{,\mu}^{\ast} \, \sigma_{,\nu} + c.c. -g_{\mu\nu}
\, \left[g^{\alpha\beta} \, \sigma_{,\alpha}^{\ast}\,
\sigma_{,\beta} -(1+\frac{1}{4} \, \Lambda \, {\Vert\sigma\Vert}^2)
\, {\Vert \sigma\Vert}^2\right]
\label{A13}
\end{equation}
(where  $T_{\mu\nu} = 4\pi G {\bf T_{\mu\nu}}$) we get
we get the equations
\begin{equation}
\delta T^{0}_{0}=\frac{1}{2\, r^2}\, \left(\frac{2\, r\, \delta g}{{g_0}^3}\right)',
\label{A14}
\end{equation}

\begin{equation}
\delta T^{1}_{1}=
\frac{\delta g}{{g_0}^3}\, \left(\frac{1}{r^2}-2\, \frac{{N}_{0}''}{N_0}\right)-\frac{{\delta N}'}{r\, {g_0}^2 \, N_0}
-\frac{{N_0}' \, \delta N}{r \, {g_0}^2 \, N^2},
\label{A15}
\end{equation}
and
\begin{eqnarray}
\delta T^{2}_{2}&=&\frac{1}{2} \, \Bigg[
\frac{2 \, \delta g}{g_0}
\left\{\left(\frac{{N_0}'}{{N_0}}\right)^2 
\right.
%\left.\left.
+\frac{N_0 \, {N_0}''-{{N_0}'}^2}{{N_0}^2}-
\frac{{N_0}' \, {g_0}'}{N_0 \, g_0}
%\right.
%\right.
\label{A16}
\\ \nonumber
&+&\frac{1}{r} \, \left(\frac{{N_0}'}{N_0}
\left.
-\frac{{g_0}'}{g_0}\right)
\right\}
+\frac{1}{{N_0}^2\, g_0} \delta \ddot g 
- \frac{1}{{g_0}^2}\, \Bigg(\frac{{\delta N}''}{N_0}
- 2 \, \frac{{N_0}'}{{N_0}^2} \, {\delta N}'
+\frac{{N_0}''}{{N_0}^2} \, \delta N
%\right.
\\ \nonumber
&-& \frac{{N_0}'}{N_0 \, g_0}
\left(\delta g' - \frac{{g_0}'}{g_0} \, \delta g\right)
- \frac{{g_0}'}{N_0 \, g_0}\left({\delta N}'-
\frac{{N_0}'}{N_0} \, \delta N\right)  
\\ \nonumber
&+&
2 \, \frac{{N_0}'}{{N_0}^2}\left({\delta N}' - \frac{{N_0}'}{N_0}
\, \delta N\right)
+\frac{1}{r} \, \left\{\frac{{\delta N}'}{N_0}-\frac{{N_0}'}{{N_0}^2}\,\delta N 
-\frac{{\delta g}'}{g_0}+\frac{{g_0}'}{{g_0}^2} \, \delta g \right\}
\Bigg)
\Bigg],
\end{eqnarray}
and the equations
\begin{eqnarray}
\delta T^{0}_{0}=-\frac{2 \, \sigma_0^2}{{N_0}^3} \, \delta N
-\frac{2}{{g_0^3}} \, \sigma'{_0}^{2} \, \delta g 
&-& \delta \dot \sigma_2 \, \frac{2}{{N_0}^2} \, {\sigma_0}^2 
+ \frac{2}{{g_0}^2} \, \sigma_0 \, {\sigma_0}' \, \delta{\sigma_1}'
\label{A17}
\\ \nonumber
&+&\delta\sigma_1 \, \left(\frac{2}{{N_0}^2} \, {\sigma_0}^2 
+\frac{2}{{g_0}^2} \, {{\sigma'}_0}^2 
+2 \, (1+\frac{1}{2} \, \Lambda \, {\sigma_0}^2) \, {\sigma_0}^2\right),
\end{eqnarray}
\begin{equation}
\delta T^{1}_{1}=-\delta T^{0}_{0} + 4 \, {\sigma_0}^2 \, \delta\sigma_1,
\label{A18}
\end{equation}
and
\begin{eqnarray}
\delta T^{2}_{2}=-\frac{2}{{N_0}^3} \, \sigma_0^2 \, \delta N
-\frac{2}{{g_0^3}} \, \sigma'{_0}^{2} \, \delta g 
&+& \delta \dot \sigma_2 \, \frac{2}{{N_0}^2} \, {\sigma_0}^2 
+ \frac{2}{{g_0}^2} \, \sigma_0 \, {\sigma_0}' \, \delta{\sigma_1}'
\label{A19}
\\ \nonumber
&-&\delta\sigma_1 \, 
\left(\frac{2}{{N_0}^2} \,  {\sigma_0}^2 -\frac{2}{{g_0}^2} \, {{\sigma'}_0}^2 
-2\,(1+\frac{1}{2} \, \Lambda \, {\sigma_0}^2) \, {\sigma_0}^2\right),
\end{eqnarray}
respectively.
Adding $\delta T^{0}_{0}$ to $\delta T^{1}_{1}$ we get
\begin{eqnarray}
\left(\frac{{\delta N}'}{N_0}-\frac{{N_0}' \, \delta N}{{N_0}^2} - 
\frac{{\delta g}'}{g_0}+\frac{g' \, \delta g}{{g_0}^2} \right)=
\frac{2}{{g_0}}&& \left(\frac{1}{r}+\left(\frac{{N_0}'}{N_0} 
- \frac{{g_0}'}{g_0}\right)\right) \, \delta g
\label{A20}
\\ \nonumber
&-&4 \, r \, 
{\sigma_0}^2 \, \left(1+\Lambda \, {\sigma_0}^2\right)\,\delta \sigma_1.
\end{eqnarray}

Calculating $\delta \dot \sigma_2$ from (\ref{A16}) and (\ref{A19}),
and substituting into (\ref{A11}) gives, along with
equation (\ref{A20}),
\begin{eqnarray}
\delta {\sigma_1}'' + \delta {\sigma_1}'\left(\frac{2}{r}
+\frac{{N_0}'}{N_0}-\frac{{g_0}'}{g_0}\right) &+&
\frac{1}{{g_0}^2 \, r \, {\sigma_0}^2} \,\left(g_0 \, \delta g' -{g_0}' \, \delta g\right)
\label{A21}
\\ \nonumber
-\frac{{g_0}^2}{{N_0}^2} \, \delta \ddot \sigma_1 
+ \frac{2\, \delta g}{g_0}\, \Bigg[{\frac{{\sigma_0}'}{\sigma_0}}^2 
&+& \frac{{g_0}^2}{{N_0}^2} 
+ \frac{1-2\, r \, \frac{{g_0}'}{g_0}}{2\, r^2 \, {\sigma_0}^2} -
{g_0}^2 \, \left(1+ \frac{\Lambda \, {\sigma_0}^2}{2}\right)
\\
\nonumber
+ \frac{{\sigma_0}'}{\sigma_0} \,
\left(\frac{1}{r}+\frac{{N_0}'}{N_0}-\frac{{g_0}'}{g_0}\right)\Bigg]
&-& {g_0}^2 \, \delta \sigma_1 \Bigg[
\frac{1}{{N_0}^2}+\frac{1}{{g_0}^2}\,
 \left(\frac{{\sigma_0}'}{\sigma_0}\right)^2 
 \\ \nonumber
&+&\left(1+ \Lambda \, \sigma_{0}^{2}\right) 
+ 2 \, r \, \sigma_{0}'\, \sigma_0 \left(1 +\frac{\Lambda \,  \sigma_{0}^2}{2}
\right)\Bigg]=0.
\end{eqnarray}
Adding $\delta T_{0}^{0}$ to $\delta T_{2}^{2}$, and substituting in equation 
(\ref{A20}) and its derivative, we get
\begin{eqnarray}
\frac{2}{g_0} \, {\delta g}'' &-& \frac{2\, g_0}{{N_0}^2} \, \ddot{\delta g} 
+ 8 \, \left( 2\, \sigma_0 \, {\sigma_0}' 
-r \, \left(1+\frac{1}{2} \, \Lambda \, {\sigma_0}^2\right)\, 
{\sigma_0}^2 \, {g_0}^2\right)\,
\delta {\sigma_1}'
\label{A22}
\\
\nonumber
&+& 8 \, \left[
2 \,  {{\sigma_0}'}^2 -r \,{\sigma_0}^2 \, {g_0}^2 \,
\left(1+\frac{1}{2} \, \Lambda \, {\sigma_0}^2\right)\,
\left(\frac{2 \, {\sigma_0}'}{\sigma_0}+\left\{\frac{2\,{N_0}'}{N_0}+\frac{{g_0}'}{g_0}\right\}\right)
\right] \, \delta \sigma_1\\ \nonumber
&-&\left[\frac{4 \,{g_0}'}{{g_0}^2}+\frac{6}{g_0} \,\left(\frac{{N_0}'}{N_0}-\frac{{g_0}'}{g_0}\right)\right]\,
\delta g' + 
\frac{2 \, \delta g}{g_0} \, \Bigg[\frac{{g_0}''}{g_0}+2\,
\left(\frac{{g_0}'}{g_0}\right)^2
\\ \nonumber
&&-3 \, \frac{{g_0}'}{g_0} \, \left(\frac{{N_0}'}{N_0}-\frac{{g_0}'}{g_0}\right)
-8 \, {{\sigma_0}'}^2-\frac{2}{r^2} - \frac{2}{{g_0}^2}
\,\left(g_0 \, {g_0}''-{{g_0}'}^2\right)
\\ \nonumber
&&+2\left(\frac{{N_0}'}{N_0}-\frac{{g_0}'}{g_0}\right)^2+\frac{2}{r}\,
\left(\frac{2 \, {N_0}'}{N_0}+\frac{{g_0}'}{g_0}\right)
\Bigg]=0.
\end{eqnarray}
Using the expression for the particle number $N_p=\int_0^{\infty} d^3x J^{0}\sqrt{g}$ where
$J^{0}=i g^{00}\left(\phi_{,0}\phi^{\ast} -c.c. \right)$ we get
\begin{eqnarray}
\delta N_p =+4\, \pi \, \int^{\infty}_{0}\, &dr& \; {r}^2 \, \frac{g_0}{N_0}{\sigma_0}^2\, {\bf X} \left\{ 
 \frac{1}{{g_0}^2 \, r \, {\sigma_0}^2}
 \, \left(g_0 \, \delta g' -{g_0}' \, \delta g\right)
\right. 
\label{A23}
\\ \nonumber
&+& 
\frac{2 \, \delta g}{g_0}\left({\frac{{\sigma_0}'}{\sigma_0}}^2 
+ \frac{1-2 \, r \, \frac{{g_0}'}{g_0}}{2\,
{r}^2 \, {\sigma_0}^2} 
+ \frac{{g_0}^2}{{N_0}^2}\right) +
\frac{{\sigma_0}'}{\sigma_0} \, \delta \sigma '
\\ \nonumber
&-& \left. 
{g_0}^2 \, \delta  \sigma_1 \left(\frac{1}{{N_0}^2}
 + \frac{1}{{g_0}^2} \, \left(\frac{{\sigma_0}'}{\sigma_0}\right)^2 +
\left(1+\frac{\Lambda}{2}\,
 \sigma_{0}^{2}
\right) \right)
\right\}.
\end{eqnarray}

In terms of bar coordinates, defined in the beginning of this section, equation
(\ref{A21}) becomes
\begin{eqnarray}
\frac{1}{\Lambda ^{1.5}}\, \delta {\bar \sigma_1}'' 
&+& \frac{1}{\Lambda^{1.5}} \, \delta {\bar \sigma_1}' \,
\left(\frac{2}{\bar r}
+\frac{{N_0}'}{N_0}-\frac{{g_0}'}{g_0}\right) + 
\frac{1}{{g_0}^2 \, {\bar r} \, {\bar \sigma_0}^2}
\,\left(g_0 \, \delta g' -{g_0}' \, \delta g\right)
\label{A24}
\\ \nonumber
&-&\frac{1}{\Lambda^{0.5}} \, \frac{{g_0}^2}{{N_0}^2} 
\, \delta \ddot {\bar \sigma_1}
+ 
\frac{2 \, \delta g}{g_0} \, \left(\frac{1}{\Lambda}\,
\left(
{\frac{{\bar \sigma_0}'}{\bar \sigma_0}}\right)^2 +
 \frac{{g_0}^2}{{N_0}^2} +
   \frac{1-2 \, {\bar r} \, \frac{{g_0}'}{g_0}}{2\, {\bar r^2}
   \,{\bar \sigma_0}^2} -
{g_0}^2 \, \left(1+ \frac{ {\bar \sigma_0}^2}{2}\right)
\right.
\\ \nonumber
&+&\left. \frac{1}{\Lambda} \, \frac{{\bar \sigma_0}'}{\bar \sigma_0} 
\, \left(\frac{1}{\bar r}+\frac{{N_0}'}{N_0}-\frac{{g_0}'}{g_0}\right)\right)
-\frac{1}{\Lambda^{0.5}} \, {g_0}^2 \, \delta \bar \sigma_1
\Bigg(
\\ \nonumber
&&
\frac{1}{{N_0}^2}+
\frac{1}{\Lambda \, g_0^2}\,
\left ({\frac{{\bar \sigma_0}'}{\bar \sigma_0}}\right)^2 +
\left(1+  \bar \sigma_{0}^{2}\right) 
+ \frac{1}{\Lambda} \, 2  \, {\bar r} \, \bar \sigma_{0}'
\, \bar \sigma_0 \, \left(1 +\frac{ \bar \sigma_{0}^2}{2}
\right) \Bigg)=0,
\end{eqnarray}
and equation (\ref{A22}) becomes
\begin{eqnarray}
&&\frac{1}{\Lambda} \, \frac{2}{g_0} \, {\delta g}'' - \frac{2\,g_0}{N^2}\,
\ddot{\delta g} 
+\frac{ 8}{\Lambda^{1.5}} \, \left(
\frac{2}{\Lambda} \, \bar  \sigma_0 \, {\bar \sigma_0}' -r \, 
\left(1+\frac{1}{2} \, {\bar \sigma_0}^2\right)
\,{\bar \sigma_0}^2 \,{g_0}^2\right)\,
\delta {\bar \sigma_1}'
\label{A25}
\\ \nonumber
&+& \frac{8}{\Lambda^{1.5}} \, \left[
\frac{2}{\Lambda} \, {\bar {{\sigma_0}'}^2} 
-r\, \left(1+\frac{1}{2} \, {\bar \sigma_0}^2\right)\,
{\bar \sigma_0}^2 \, {g_0}^2\,
\left(\frac{2\,{\bar \sigma_0}'}{\bar \sigma_0}+\left(\frac{2\, {N_0}'}{N_0}+\frac{{g_0}'}{g_0}\right)\right)
\right] \, \delta \bar \sigma_1\\ \nonumber
&-&\frac{1}{\Lambda}\left[\frac{4 \, {g_0}'}{{g_0}^2}+\frac{6}{g_0}
\, \left(\frac{{N_0}'}{N_0}-\frac{{g_0}'}{g_0}\right)\right]\,
\delta g' + \frac{2}{\Lambda \, g_0}
\, \left[\frac{{g_0}''}{g_0}+2 \,(\frac{{g_0}'}{g_0})^2
-3 \, \frac{{g_0}'}{g_0}
\,\left(\frac{{N_0}'}{N_0}-\frac{{g_0}'}{g_0}\right)
\right.\\ \nonumber
&-&\left.\frac{8}{\Lambda} \,
{\bar {{\sigma_0}'}^2}-\frac{2}{r^2} - \frac{2}{{g_0}^2}
\, \left(g_0 \, {g_0}''-{{g_0}'}^2\right)
+2 \, \left(\frac{{N_0}'}{N_0}-\frac{{g_0}'}{g_0}\right)^2+\frac{2}{r}\,
\left(\frac{2\, {N_0}'}{N_0}+\frac{{g_0}'}{g_0}\right)
\right] \, \delta g=0.
\end{eqnarray}

Similarly equation (\ref{A23}) becomes
\begin{eqnarray}
\delta N_p =4\pi\int^{\infty}_{0} && dr\, r^2\, \frac{g_0}{N_0}\,{\bar \sigma_0}^2\, {\bf X} \left\{ 
 \frac{1}{{g_0}^2 \, \bar r \, {\bar \sigma_0}^2}
 \, \left(g_0\, \delta g' -{g_0}' \, \delta g\right)
 \right.
\label{A26}
\\ \nonumber
&+&
\frac{2\,\delta g}{g_0}\left(\frac{1}{\Lambda}
\, ({\frac{{\bar \sigma_0}'}{\bar \sigma_0}})^2 
+ \frac{1-2 \, \bar r \, \frac{{g_0}'}{g_0}}{2\,
{\bar r}^2 \, {\bar \sigma_0}^2} \,
\frac{{g_0}^2}{{N_0}^2}\right) +
\frac{1}{\Lambda^{1.5}}
\, \frac{{\bar \sigma_0}'}{\bar \sigma_0} \, \delta \bar \sigma '
\\ \nonumber
&-& \left. 
\frac{{g_0}^2}{\Lambda^{0.5}}\, \delta  \bar \sigma_1 \left(\frac{1}{{N_0}^2} 
+\frac{1}{{g_0}^2 \, \Lambda} \,\left(\frac{{\bar \sigma_0}'}{\bar \sigma_0}
\right)^2 +
\left(1+ \frac{1}{2}\,
 \bar \sigma
_{0}^{2}\right) \right)
\right\}.
\end{eqnarray}
Now using $\delta N_p =0$ (charge conservation), which is appropriate
for large $\Lambda$ where $\delta N_p$ is given by (\ref{A26}),
we get on putting this in (\ref{A24}):
\begin{eqnarray}
\frac{1}{\Lambda ^{1.5}}
\, \delta {\bar \sigma_1}'' &+& \frac{1}{\Lambda^{1.5}}
\, \delta {\bar \sigma_1}' \, \left(\frac{2}{\bar r}
+\frac{{N_0}'}{N_0}-\frac{{g_0}'}{g_0}\right)
-\frac{1}{\Lambda^{0.5}}\, \frac{{g_0}^2}{{N_0}^2} \,
\delta \ddot {\bar \sigma_1}
\label{A27}
\\ \nonumber
&+& 
\frac{2 \, \delta g}{g_0} \, \left(
-{g_0}^2 \, \left(1+ \frac{ {\bar \sigma_0}^2}{2}\right)
+ \frac{1}{\Lambda}\,
\frac{{\bar \sigma_0}'}{\bar \sigma_0} \,\left(\frac{1}{\bar r}+\frac{{N_0}'}{N_0}-\frac{{g_0}'}{g_0}\right)\right)
\\
\nonumber
&-&\frac{1}{\Lambda^{0.5}}\,
{g_0}^2 \, \delta \bar \sigma_1  \,
\left(
\frac{ \bar \sigma_{0}^{2}}{2} 
+ \frac{1}{\Lambda}\,2 \, {\bar r} \, \bar \sigma_{0}' \,\bar \sigma_0 
\, \left(1 +\frac{ \bar \sigma_{0}^2}{2}
\right)\right) -\frac{1}{\Lambda^{1.5}}\,
\frac{\bar \sigma_0'}{\bar \sigma_0}
\, \delta \bar \sigma=0.
\end{eqnarray}
This equation suggests that $\delta g$ goes like $1/\sqrt\Lambda$. 
Writing $\ddot \delta g$
as $-\chi^2 \delta g$ and $\ddot \delta {\bar \sigma}$ 
as $-\chi^2 \delta \sigma$ we then see from (\ref{A25}) and (\ref{A27}) that the quasinormal mode frequency $\chi$ for
high $\Lambda$ configurations must go like $1/\sqrt\Lambda$ for a given $\bar \sigma$. 
We have numerically evolved stars with the same $\bar \sigma$ and compared 
the QNM frequencies obtained
to the inverse ratios of the square root of their $\Lambda$ values for $\Lambda=800$, $\Lambda=1200$,
and $\Lambda=1600$ confirming the above analysis.
In Table I we show the comparison between the perturbation analysis and the
numerical result. We see that the perturbation result gets more accurate for
increasing $\Lambda$.
We note that 
configurations that have the
size of a neutron star would have to have
$\Lambda$ of order $10^{38}$ with QNM frequency of order
$\chi = 10^{-19}$.
 
%\begin{itemize}
%\item{Excited States}
%\end{itemize}

\footnotesize
\begin{center}
\begin{tabular}{|c|c|c|c|c|}
\multicolumn{5}{c}{Table 1.}   \\
\hline

$\Lambda$ &
$1/f$ & $f_{1600}/f$ & $ \sqrt{\frac{\Lambda}{1600}} $
&
 \% error
 \rule[-0.1in]{0.0in}{0.3in}\\
  \hline
  1600&       1220&         1&                   1& --
  \rule[-0.1in]{0.0in}{0.3in}\\
   \hline
   1200&       1070&        0.877&                  0.866&
   1.25
   \rule[-0.1in]{0.0in}{0.3in}\\
    \hline
    800 &        880&        0.7213&                 0.707     &
    1.9
    \rule[-0.1in]{0.0in}{0.3in}\\
     \hline
     600 &        770&        0.6311&                 0.612 &
      3
      \rule[-0.1in]{0.0in}{0.3in}\\
      \hline
      \end{tabular}

      \end{center}

\normalsize

%\normalsize

\begin{table}
\vspace{0.4in}
\caption{The ratio of the QNM frequency for $\Lambda=1600$ 
to the QNM frequency for a given $\Lambda$
is compared to 
$ \Lambda^{0.5}/ 40$ 
(which is the predicted ratio for large $\Lambda$)
for $\Lambda = 1200$, $800$, and $1600$. As expected,
the higher $\Lambda$ values match better.
The initial central density is $\bar \sigma (0) =0.4.$
}
\label{tbh}
\end{table}

\section{Boson Stars: Formation and Stability in {\em BD} Theory}

We now describe our work on the study of boson stars in the {\em BD} theory~\cite{balabd}. We hope to extend this work to general Scalar--Tensor theories
in the future. 

The equations and set up have already been described earlier.
We use the evolution code described in the {\em GR} section
of this chapter to study the dynamical evolution of this system.
The equilibrium configurations are described in~\cite{CS97}. A modified
version of the equilibrium code used in {\em GR} was used to find
the equilibrium configurations. We then studied the time
evolution of these configurations using our evolution code.

\subsection{Boundary Conditions}

For the boundary conditions, regularity dictates that the
radial metric is equal to 1 at the origin. The boson field and the
{\em BD} field are both specified at the origin. The boson field goes
to zero at $\infty$ and the {\em BD} field goes to a constant which is
fixed during the evolution. This constant does not enter into any
of the evolution equations as all the terms in the set of equations
are derivatives in the {\em BD} field.

The inner boundary at the origin requires that
the derivatives of the metric and all the fields vanish
at this point.  This is implemented
by extending the range of $r$ to include some negative values.
The metric components $g$, $N$,
the boson field and the {\em BD} field are required to be symmetric
about $r=0$. While the boson field falls off exponentially with $r$ in
the asymptotic region, the {\em BD} field is more slowly varying with a $1/r$
dependence.

For the {\em BD} field, $\varphi$, the wave equation can be
written as
\begin{equation}
\varphi(r,t) = \varphi_\infty + {1\over r}\, F\left(t-{r\over c}\right) \label{e:ccfi}
\end{equation}
which is an exact condition for 1-D waves. Differentiating, one gets
\begin{equation}
\left .
{1\over c} \, {\partial \varphi \over \partial t} + {\partial \varphi \over
\partial r} + {(\varphi-\varphi_\infty) \over r} \,
\right |_{\text{outer edge}} = 0.
\end{equation}
Note that this technique used by Novak~\cite{novak} in a study of stellar
collapse.

For the boson field, $\Psi$, an outgoing boundary condition to order $1/r$
is
\begin{equation}
\beta^2 \, k^2=\omega_E^2-N^2 \, m^2 \, e^{2a}.
\end{equation}
Therefore at the outermost grid
point we require that
\begin{equation}
\partial_t \partial_t \tilde{\Psi} =
-\beta \, \partial_t \partial_r \tilde{\Psi} -\frac{N^2}{2}\,
\tilde{\Psi} \, e^{2a},
\end{equation}
where $\tilde{\Psi} =r\,\Psi$ is the field variable whose evolution
equations are set up in the code.

We also use a sponge to remove second order reflections. It reduces the
momentum of the boson field
artificially, and is irrelevant for the massless {\em BD} field.
It is a damping term that effectively simulates a
potential term in the wave equation. This potential is large for incoming waves
 (proportional to $k+\omega_E$) and small for outgoing waves
 (proportional to $k-\omega_E$).
 Therefore,
 we add an additional term in the evolution equation for $\Pi_\Psi$
 (\ref{dyn-4})
 \begin{equation}
  \frac{V(\Psi\Psi^\dagger)}{e^{a} \, N}(\Pi_\Psi+\partial_r \tilde{\Psi})
  \end{equation}
  for $r_{end}-D\le  r \le r_{end} $
  where $r_{end}$ is the $r$ value of the outermost grid point and
  $D$ is an adjustable parameter representing the width of the sponge.
  $D$ is typically chosen to be a few times the wavelength
  of the scalar radiation moving out.
  
  The equilibrium sequences have similar profiles to {\em GR}~\cite{CS97} with
of course the additional {\em BD} profile.
The perturbations are described in the {\em GR} subsection. While the mass loss
in {\em GR} must solely be due to scalar field radiation there is an additional
possibility here of gravitational field radiation through oscillations
of the {\em BD} field even though we have spherical symmetry.

\subsection{Ground  State Results}

\begin{itemize}
\item{$S$-branch evolutions}
\end{itemize}
After confirming that our code was stable and the equilibrium profile of
the star was undisturbed for several thousand time steps, we let the
system evolve under infinitesimal perturbations (less than $0.1 \%$ mass
change from equilibrium). The system started oscillating at its fundamental
quasinormal mode frequency. The {\em BD} field also acquired this oscillation
frequency. 

In GR,
a QNM frequency increases as $\Phi_c$ become larger
 (radius become smaller) up to a point before starting to decrease
rapidly
to zero as it approaches the density corresponding to maximum
mass signalling the onset of instability~\cite{seinum,balalam}. We found the same feature in {\em BD}.

\begin{figure}[h]
%\vspace*{1.5cm}
\setlength{\unitlength}{1in}
\begin{picture}(7.5,5.0)
\put(0.75,2.0){\epsfxsize=5.0in \epsfysize=2.0in
\epsffile{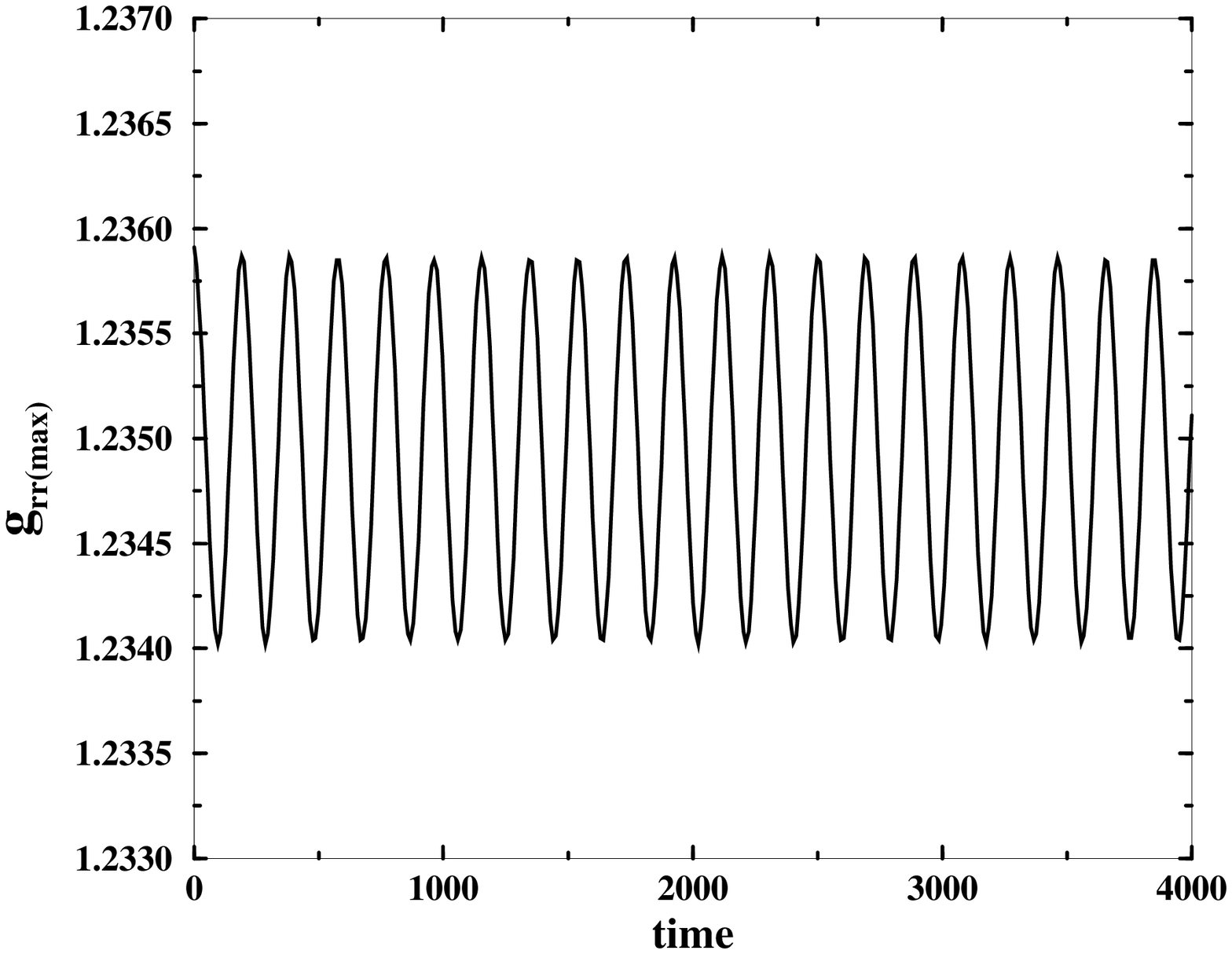} }
\put(0.75,-1.0){\epsfxsize=5.0in \epsfysize=2.0in
\epsffile{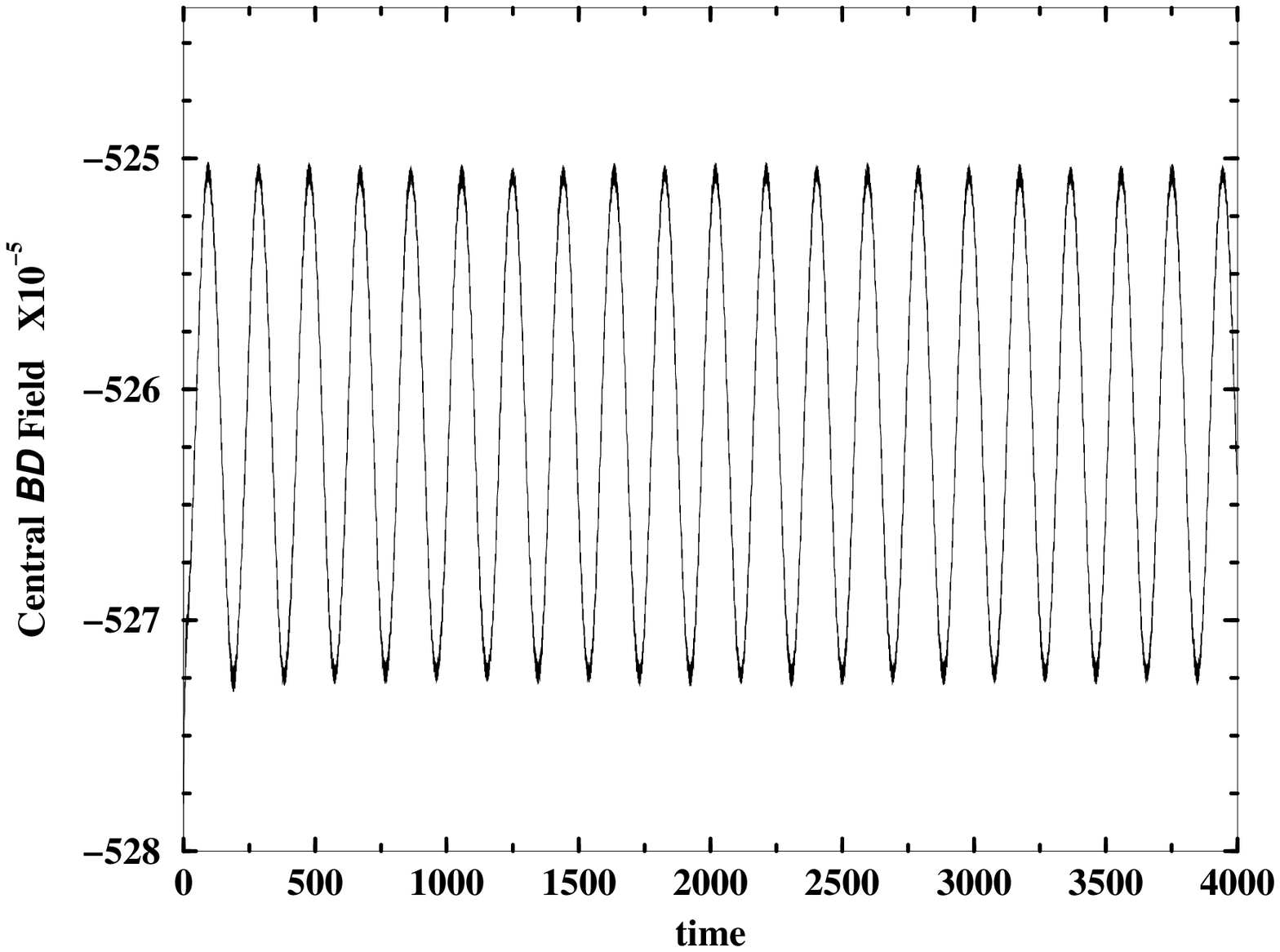} }
\end{picture}
\caption[Quasinormal Modes of a Stable Star in {\em BD} Theory]{
\linebreak
\linebreak
\linebreak
\hspace{0.3 in} Quasinormal mode oscillation of a stable boson
star.
The maximum value of the metric $g_{rr}$,
and the central
Brans-Dicke field $\varphi(r=0)$
are plotted as functions of time. Both of them
take on the  QNM
frequency of the star. The oscillation is
virtually undamped for a long period of time.
}
\label{stable_grr}
\end{figure}

\begin{figure}[h]
%\vspace*{1.5cm}
\setlength{\unitlength}{1in}
\begin{picture}(7.5,2.5)
\put(0.75,0.0){\epsfxsize=5.0in \epsfysize=2.0in
\epsffile{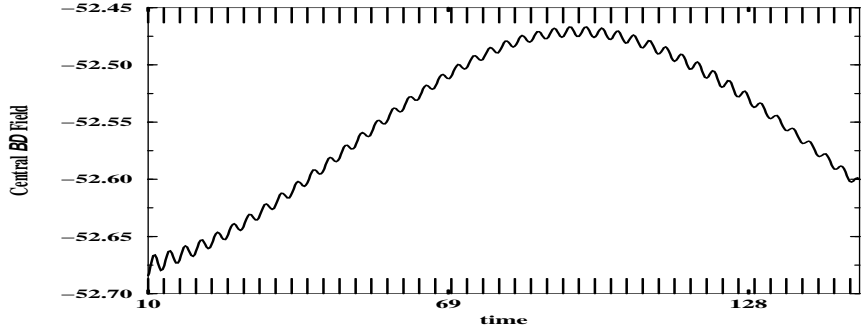} }
\end{picture}
\caption[Underlying Oscillation of the Star]{
The magnification of the second figure of Fig.~\ref{stable_grr} is shown.
 In the dimensionless units we have adopted, the time interval between
 the ticks on the horizontal axis is $\pi$.
 We see the Brans-Dicke field oscillates with period
 $\pi$. Thus it has twice the frequency of the
 boson field, which oscillates with period $2\;\pi$. 
 }
 \label{stable_pi}
 \end{figure}

In addition, the system takes on the proper underlying frequency,
which originates from the time dependence of the boson field $\Psi
\sim e^{i t}$. 
This frequency
corresponds to period $2\pi$ in $t$ in our units,
and leads to metric and {\em BD} field
 oscillations with period $\pi$,
from the structure of the equations.
To show this feature we
have enhanced the {\em BD} field oscillation (Fig.~\ref{stable_grr})
in Fig.~\ref{stable_pi}. The time interval between
the ticks on the horizontal axis is $\pi$.

Under large perturbations,  a stable boson star in GR expands
and contracts, losing mass at each expansion.  The
oscillations damp out in time
and the system finally settles down
into a new configuration on the $S$-branch. These features
are now also observed in {\em BD} theory.

We show the effects of a large perturbation on a stable
{\em BD} boson star described above.
The example we present is the case of initial data
with $\Phi_c=0.2$, mass $M=0.540G_*/m$ after perturbation
(about $13\%$ lower in mass compared to unperturbed equilibrium
configuration of $M=0.622G_*/m$).

\begin{figure}[h]
%\vspace*{1.5cm}
\setlength{\unitlength}{1in}
\begin{picture}(7.5,3.25)
\put(1.0,0.0){\epsfxy \epsffile{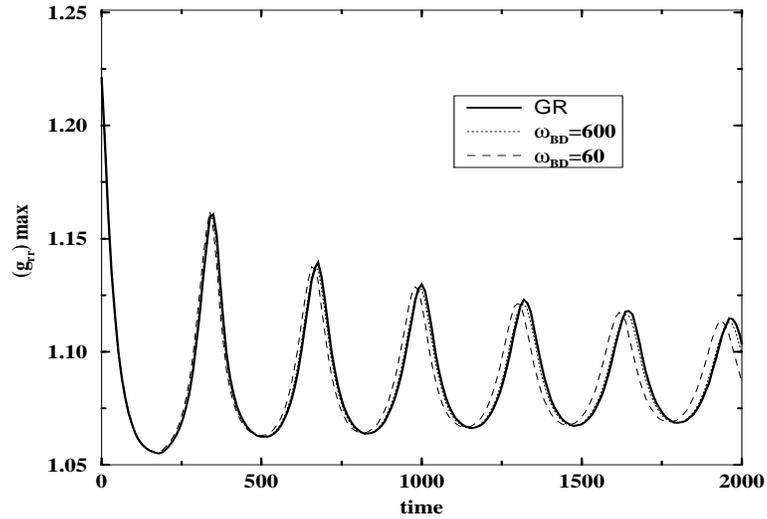} }
\end{picture}
\caption[Oscillation of the Metric Under Finite Perturbations of the Stable
Star in {\em BD} Theory]{
Finite perturbation of an $S$-branch boson star.
Maximum metric $g_{rr}$ is plotted.
The metric is damped in time as the star settles to a new
configuration.
}
\label{stBS_grr}
\end{figure}

\begin{figure}[h]
%\vspace*{1.5cm}
\setlength{\unitlength}{1in}
\begin{picture}(7.5,4.0)
\put(1.0,0.0){\epsfxy
              \epsffile{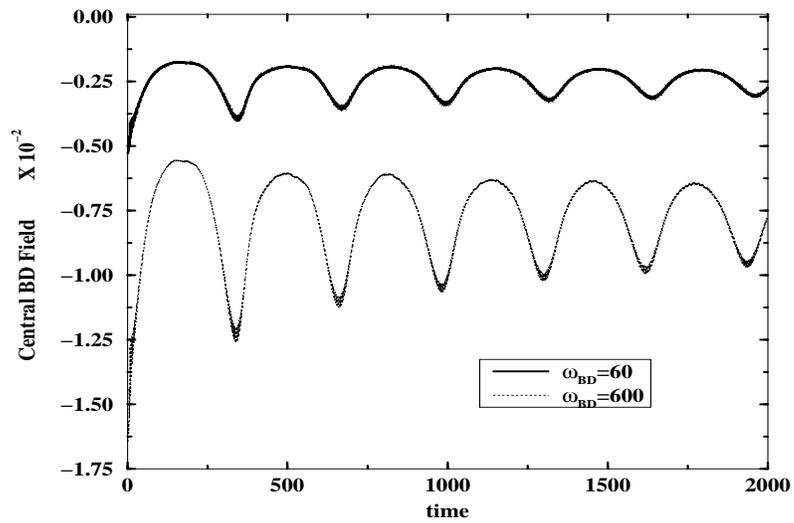} }
\end{picture}
\caption[Oscillations of the {\em BD} Field for a Perturbed Stable Star]{
The same model as Fig.~\ref{stBS_grr}.
 The central Brans-Dicke field $\varphi(r=0)$ is plotted.
We see the oscillations damp out as the star settles down,
indicating again
transition to a new stable boson star configuration.
}
\label{stBS_vphi}
\end{figure}

The maximum radial metric $g_{rr}$ and the central {\em BD} field
as a function of time are shown in Fig.~\ref{stBS_grr} and
Fig.~\ref{stBS_vphi}, respectively. In both figures, we have plotted
the case of $\omega_{BD}=600$, and $\omega_{BD}=60$ as well as GR.
The increase of the  maximum  $g_{rr}$ indicates
the star is contracting,  reaching
its maximum value at the end of the contraction in a cycle.
Then, as the star expands,  the maximum  $g_{rr}$ decreases,
reaching its
minimum at the end of the expansion.
These processes repeat themselves
with the oscillations damping out in time as the star settles
to a new stable
configuration with maximum radial metric of smaller
value than it started with (lower mass). The lower value
of the {\em BD} parameter shows a phase shift in comparison to {\em GR},
which might be suggestive of a different rate of approach
to the final configuration.
We see the same dynamical behavior in the {\em BD} field (Fig.~\ref{stBS_vphi}).
It has the same oscillation frequency as that of
the metric. The oscillations damp out in time and the {\em BD} field settles
to a value closer to zero than it started at (lower final mass).

The system loses mass through
radiation during its evolution.
A comparison
of the mass as a function of time for the {\em BD} case ($\omega_{BD}=
600$ and $60$) as well as {\em GR} shows little
 difference, indicating
that the radiation is mostly scalar field radiation from the boson field
and not scalar gravitational radiation due to the {\em BD} field.
This is despite the {\em BD} field
oscillating in the {\em BD} case and being zero in the {\em GR} case.

\begin{figure}[h]
%\vspace*{1.5cm}
\setlength{\unitlength}{1in}
\begin{picture}(7.5,3.25)
\put(1.0,0.0){\epsfxy
              \epsffile{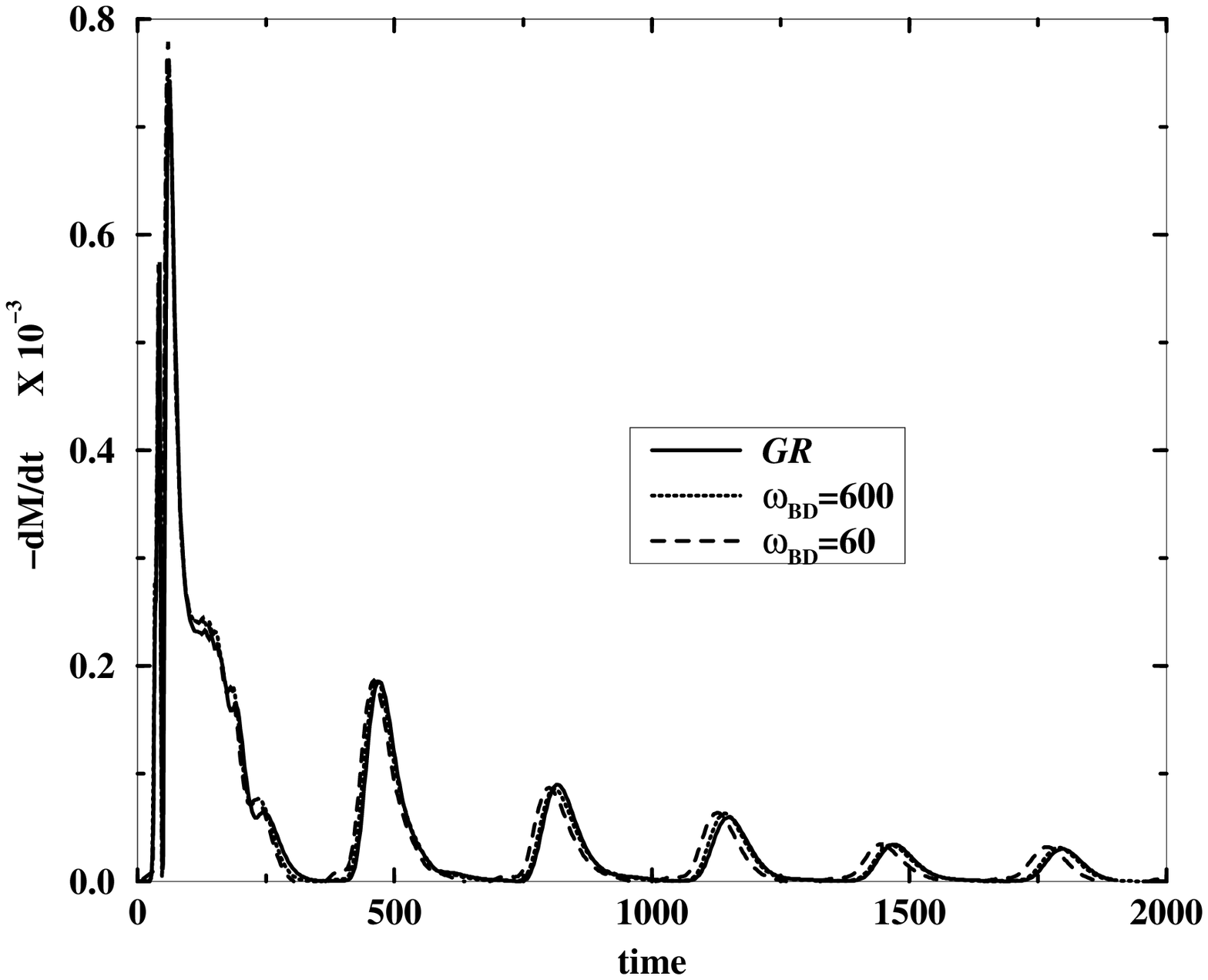} }
\end{picture}
\caption[Stable Branch: Mass Loss Under Perturbations in {\em BD} Theory]{
The same model as the previous figures.
`Luminosity' $L=-dM/dt$  is plotted versus time.
Clearly the radiation is decreasing in time.
}
\label{stBS_dmdt}
\end{figure}

The amount of mass radiated progressively decreases, as
can be seen in the luminosity profile
($-{dM}/{dt}$ versus time) shown in~Fig.~\ref{stBS_dmdt}.
Here again a comparison between {\em GR} and the
two {\em BD} parameters is shown. Again we see the phase shift for the
lower {\em BD} parameter (further from {\em GR}).
The system finally settles to a new state of smaller mass.

 %******************************     \ref{stable_pi}  <<<<<.

%****************************** Fig.\ref{stBS_grr}  >>>>>.
%******************************     \ref{stBS_grr}  <<<<<.

%****************************** Fig.\ref{stBS_vphi}  >>>>>.
%******************************     \ref{stBS_vphi}  <<<<<.
%******************************     \ref{stBS_dmdt}  <<<<<.
\begin{itemize}
\item{Evolution of $U$-branch stars and excited stars}
\end{itemize}

Boson stars on the $U$ branch and excited states
are inherently unstable in {\em GR} \cite{seinum,balalam}.
Under perturbations that reduce the mass, boson stars on the
$U$ branch can migrate to the stable branch.
In this section, we will see that these features also exist in {\em BD} theory.
Configurations can be perturbed in such a way so as to decrease
their mass enough that they migrate to new configurations on the $S$-branch.
On the other hand, in {\em GR},
if they do not lose mass and migrate, then
the boson stars of configurations
with $M<N_p\,m$
%(where $M$ is the total mass of the star,
%$N_p$ the
%particle number and $m$ the mass of a boson)
collapse to black holes.
Stars with $M> N_p\,m$ disperse and radiate out to infinity.
We also see these features in {\em BD} theory.

\begin{figure}[h]
%\vspace*{1.5cm}
\setlength{\unitlength}{1in}
\begin{picture}(7.5,4.0)
\put(1.0,0.0){\epsfxy
              \epsffile{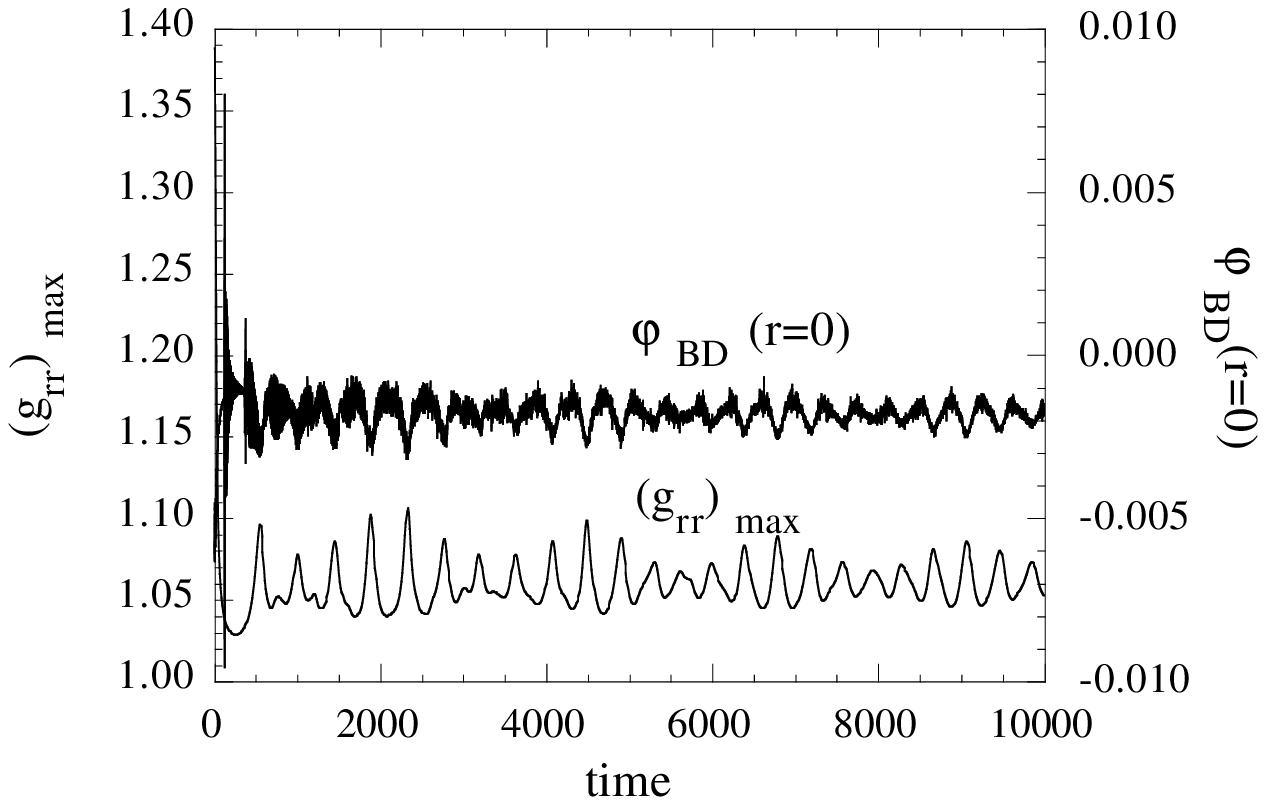} }
\end{picture}
\caption[Radial Metric Oscillations of a Migrating Unstable {\em BD} Star]{
Migration of an unstable boson star to a stable configuration;
central Brans-Dicke field $\varphi(r=0)$ and maximum of
$g_{rr}$ is plotted.
There is a sharp initial drop in the radial metric as the
star moves to the stable branch.
The oscillations damp out in time as the star settles down.
}
\label{trBSgrrvphi}
\end{figure}

Our first dynamical example from $U$-branch boson stars is
a migration process.
As in {\em GR}, we have also seen migrations of these stars to the stable
branch when we remove enough scalar field smoothly from some
region of the star so as to decrease the mass by about $10\%$.
In particular, we show the migration of a star of central boson field
 $\Phi_c=0.35$ with unperturbed mass 0.625 $G_*/ m$.
After perturbation, its mass is reduced to 0.558 $G_*/ m$.
% Fig grr and vphi ---
In Fig.~\ref{trBSgrrvphi}, we show the maximum value of metric $g_{rr}$
and central value of {\em BD} field $\varphi(r=0)$ versus time.
The initial sharp drops in both  lines occur as the star
rapidly expands and moves to the stable branch.
After that it oscillates and
finds a new configuration to settle into.  The damping of oscillations
as it settles down is clearly seen in the figure.
We see also  the BD field oscillations
damping out as the star gets closer and closer
to its final state.

\begin{figure}[h]
%\vspace*{1.5cm}
\setlength{\unitlength}{1in}
\begin{picture}(7.5,3.25)
\put(1.0,0.0){\epsfxy
              \epsffile{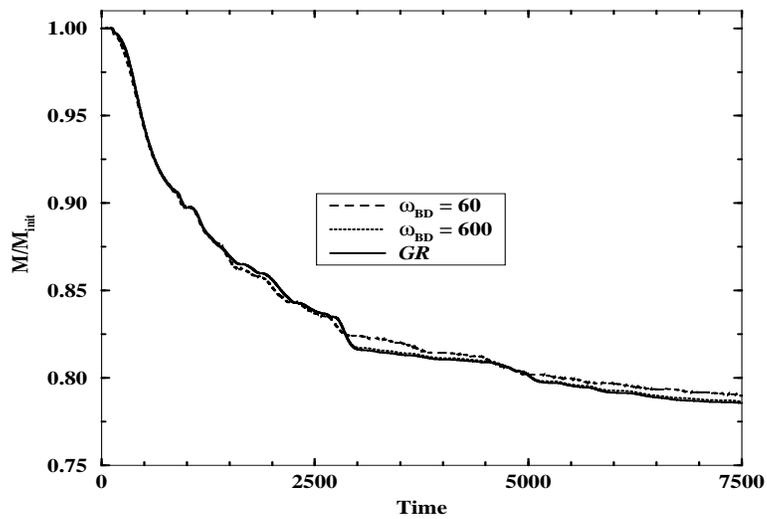} }
\end{picture}
\caption[Mass Loss for a Migrating Star: Comparisons of {\em BD} and {\em GR}]{
Comparisons of total mass of the system $M$ rescaled by its initial
mass $M_{init}$
 during a migration process from $U$-branch
star.  Three lines are plotted. Although the mass-loss is
similar in the three cases, the curve corresponding to
the higher Brans-Dicke parameter is clearly
closer to GR (as it should be).
}
\label{trBS_mloss}
\end{figure}

The ratio of mass at time $t$ to the
initial mass for {\em BD} with parameter $\omega_{BD}=60$, $600$
and the {\em GR} case is shown in Fig.~\ref{trBS_mloss}.  The flattening
of the curve at later time is indicative of the star settling down
to a new configuration.
Although convergence towards GR with increasing
$\omega_{BD}$ is clearly
indicated,  there is no significant difference between the
three cases.  The amount of total mass extraction from the system
is slightly suppressed if we evolve in the BD theory.
By the time of 7500 shown in the plot, we see that the mass of the
star is
about 0.045 $G_{\ast}/m$, which corresponds to an equilibrium
configuration with
$\Phi_c=0.06$,
while our central density $\Phi_c$ is about  0.061,
 meaning that the star is quite
close to its final configuration.

\begin{figure}[h]
%\vspace*{1.5cm}
\setlength{\unitlength}{1in}
\begin{picture}(7.5,3.25)
\put(1.0,0.0){\epsfxy
              \epsffile{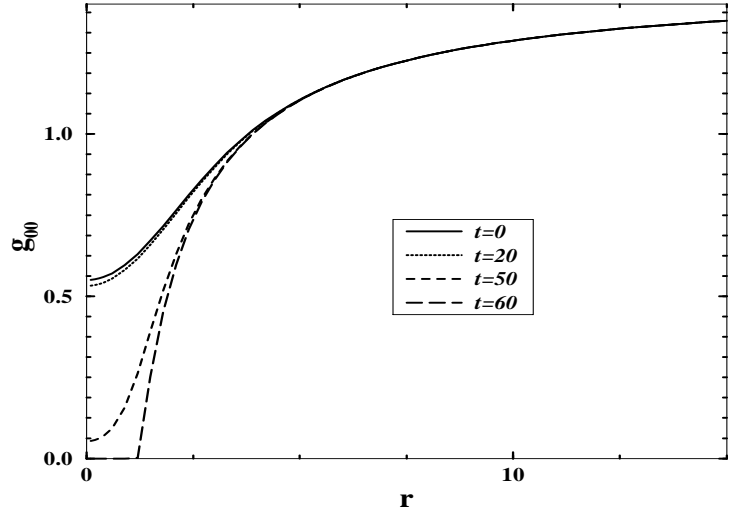} }
\end{picture}
\caption[Black Hole Formation of an Unstable Star in {\em BD} Theory]{
Dynamical transition from $U$-branch star to a black hole. The metric $g_{00}$ is plotted. The collapse of the lapse function
is indicative of imminent black hole formation.
}
\label{trBH_g00}
\end{figure}

In contrast to the previous example, if we add a small mass to
$U$-branch stars, we can see the formation of a black hole in its
evolution.
In Fig.~\ref{trBH_g00}, we plotted an example of such an evolution,
suggesting formation of a black hole.
The initial data of this plot is boson star of central boson field
$\Phi_c=0.35$ (the same value with the previous migration case) and
the system is perturbed very slightly so that
the perturbed mass is less than $0.5\%$ greater than its initial mass.
The sudden collapse of the lapse function is
indicative of the imminent formation of an apparent horizon.
(Due to the polar-slicing condition in our code~\cite{bard} we will never see
the horizon actually form).
In addition to this the radial metric starts to grow rapidly and the
code is no longer capable of handling the resulting sharp gradients.
As an indicator of the
suddenness of the process, we see that in the configuration
shown the lapse has fallen to a value
of about $0.003$ by a time of 60 after being at $0.230$ at a time of $55$
and $0.5$ at a time of $50$ (the latter two points are not shown in the
plot).  %This is in terms of the dimensionless time of the code.

There is almost no loss in mass in this system and the time of
collapse is quite close to that in {\em GR}.
In the {\em GR} case, we confirmed the
formation of black hole~\cite{BSS_3D} very shortly after this point
($\approx 3\,M$ where $M$ is the mass of the system) by switching this data
into 3-dimensional code and evolving the system.
Therefore we expect almost the same behavior in the {\em BD} case.

The black hole formation in the {\em BD} theory has been investigated by
Scheel, Shapiro and Teukolsky for dust collapse~\cite{SST}.
They found that the dynamical behavior of the apparent horizon
is quite different in the physical
Brans-Dicke frame, but the same as {\em GR} in the Einstein frame.
Their results therefore also support our discussion of
the formation of black hole in the {\em BD} theory.

\subsection{Excited States}

\begin{figure}[h]
%\vspace*{1.5cm}
\setlength{\unitlength}{1in}
\begin{picture}(7.5,3.25)
\put(1.0,0.0){\epsfxy
              \epsffile{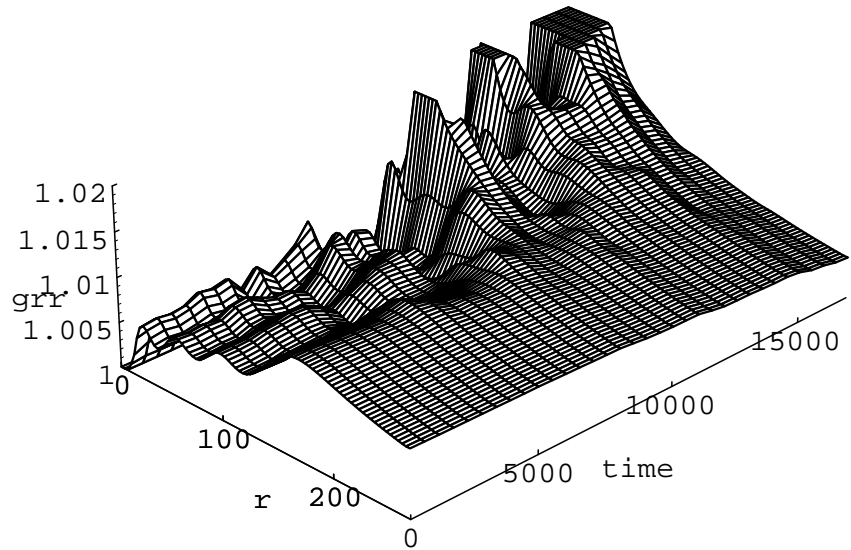} }
\end{picture}
\caption[Cascade of an Excited Star to the Ground State in {\em BD} Theory]{
Dynamical transition from an excited state to a ground state
 boson star configuration.  The metric  $g_{rr}$ is plotted.
The initial four peaks indicative of a three-node star cascades to
the ground-state branch after a long time evolution.
}
\label{node3grr}
\end{figure}

\begin{figure}[h]
%\vspace*{1.5cm}
\setlength{\unitlength}{1in}
\begin{picture}(7.5,3.25)
\put(1.0,0.0){\epsfxy
              \epsffile{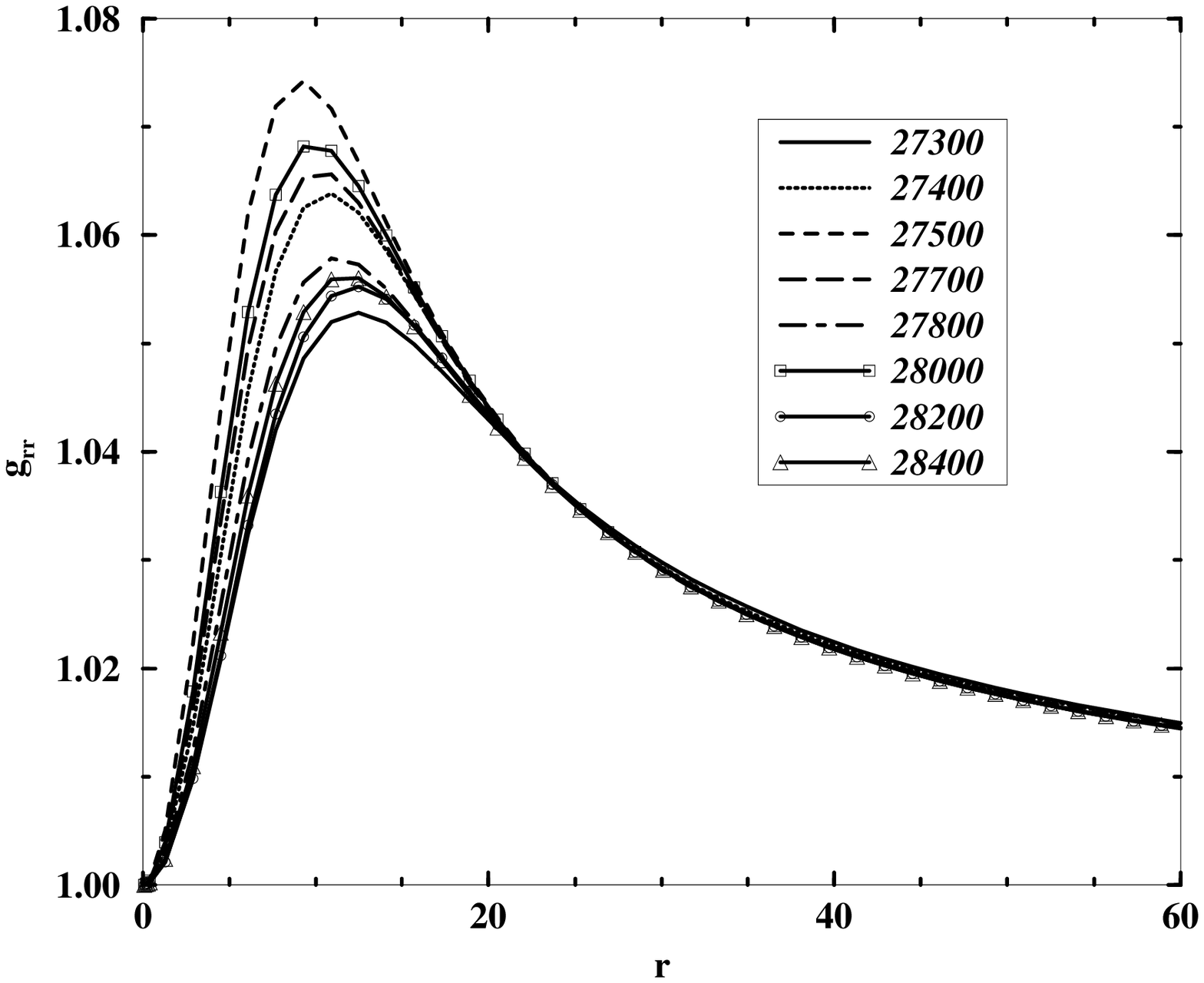} }
\end{picture}
\caption[Transition from Excited state to Ground State of {\em BD} Star: Late
Time]{
Dynamical transition from an excited state to a ground state
boson star configuration.  The metric  $g_{rr}$ is plotted
at later times to show its oscillations after the star reaches
a ground state.
}
\label{node3grrend}
\end{figure}

Excited states of boson stars in general are not stable in {\em GR}.
They form black holes if they
cannot lose enough mass to go to the ground state. We confirm the
same features for the {\em BD} case.
In Fig.~\ref{node3grr}, we plot the metric $g_{rr}$ of
dynamical transition from an excited state with 3 nodes to a ground state
boson star configuration. The initial configuration has four metric maxima and the
final has one.
After it goes to the ground-state branch, it oscillates and compactifies
to form a new configuration.
We show the oscillations of the star
from a time of $27300$ to $28400$ in Fig.~\ref{node3grrend}.
The $95\%$ mass radius
at this stage is about $100$ and the star has still to reach its final
state.

%---------------------------------------------------------------

%****************************** Fig.\ref{trBSgrrvphi}  >>>>>.
% HS Kaleidagraph eps
%******************************     \ref{trBSgrrvphi}  <<<<<.

%****************************** Fig.\ref{trBS_mloss}  >>>>>.
% original name unstabmigr.eps
%******************************     \ref{trBS_mloss}  <<<<<.

%****************************** Fig.\ref{trBH_g00}  >>>>>.
%******************************     \ref{trBH_g00}  <<<<<.

%****************************** Fig.\ref{node3grr}  >>>>>.
%******************************     \ref{node3grrend}  <<<<<.
%\begin{itemize}
%\item{Formation of Boson Stars in Brans-Dicke theory}
%\end{itemize}

\subsection{Formation of Boson Stars}

In the previous sections, we have analyzed boson stars in {\em BD} theory, starting
with equilibrium or perturbed equilibrium configurations.
However, we have not discussed whether or how such an equilibrium
configuration actually forms from generic initial data in BD theory.
In this section, we answer this question by demonstrating the
formation of boson stars in BD theory.
The formation of boson stars in GR has been discussed by
Seidel and Suen~\cite{SS94}, for some set of initial data of non-zero
measure.

We start our evolution with initial data given by a
Gaussian packet in the bosonic field $\Phi$. This represents a local
accumulation of matter field:
\begin{equation}
\Phi= a \exp (-b x^2),
\end{equation}
where $a$ and $b$ are free parameters.
We set the BD field $\varphi$ to be flat at the
initial stage, so as to see if local inhomogeneity of the matter will
form a boson star in BD theory.
We reintegrate
the lapse equation and Hamiltonian
constraint equation to provide metric
functons on the initial slice. We then observe its evolution.

\begin{figure}[h]
%\vspace*{1.5cm}
\setlength{\unitlength}{1in}
\begin{picture}(7.5,3.5)
\put(1.0 , -0.25){\epsfxy \epsffile{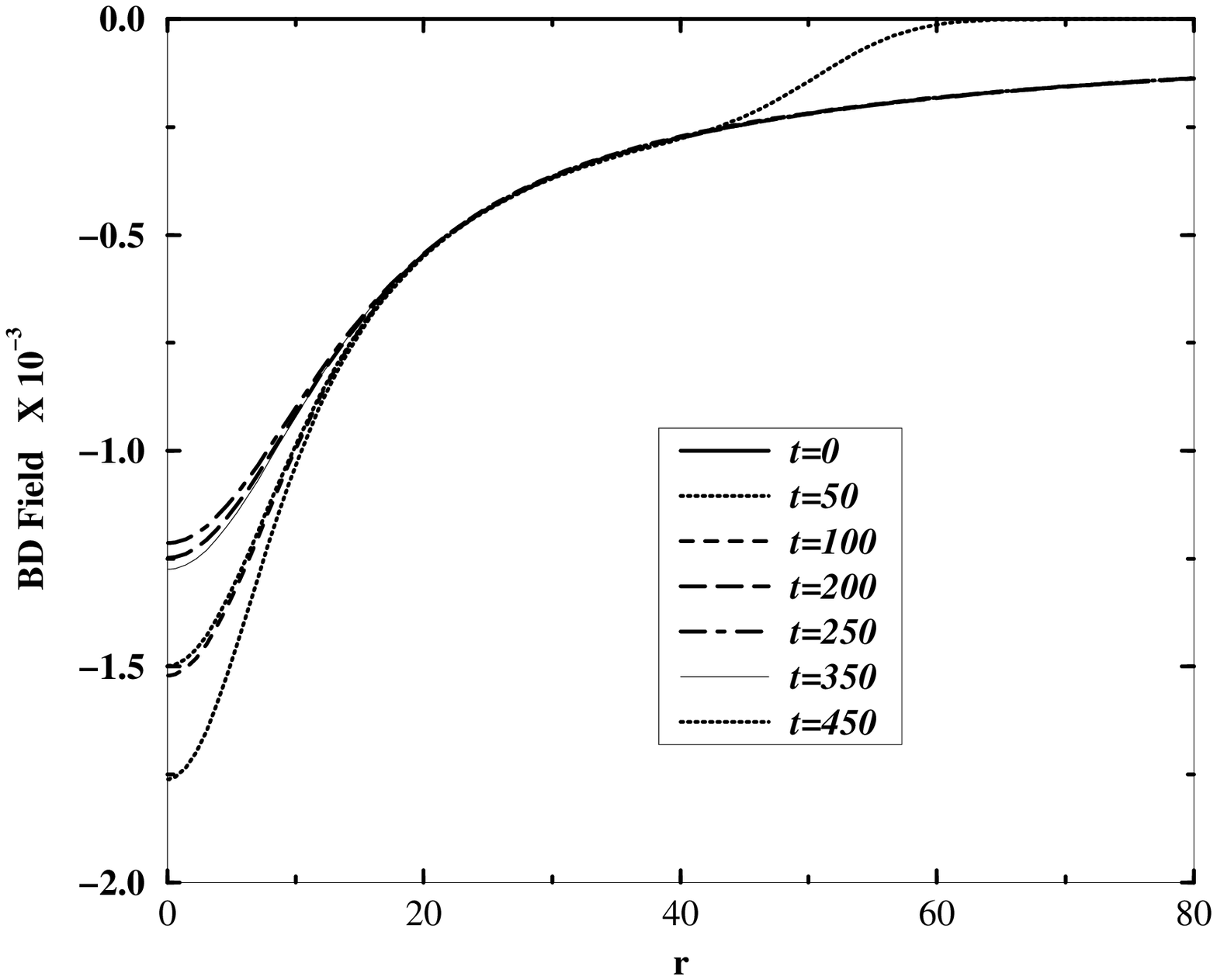}}
\end{picture}
\caption[Formation of {\em BD} Star: Early Time Behavior of {\em BD} Field]{
An example of formation of boson star in Brans-Dicke theory.
Snapshots of $\varphi_{BD} (r)$ are plotted for initial stage of
evolution.
}
\label{fig_formC_1}
\end{figure}

\begin{figure}[h]
%\vspace*{1.5cm}
\setlength{\unitlength}{1in}
\begin{picture}(7.5,3.5)
\put(1.0 , -3.2){\epsfxsize=5.25in \epsffile{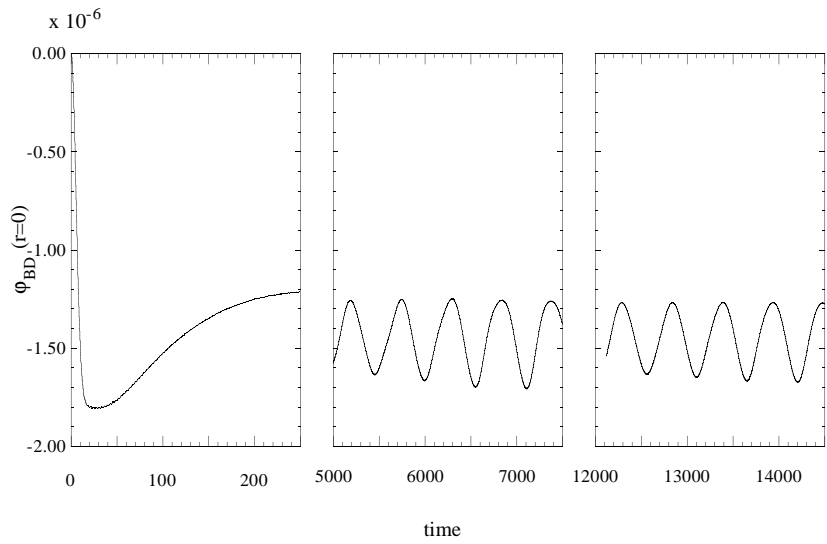}}
\end{picture}
\caption[Formation of {\em BD} Star: Comparison of Behavior of {\em BD} Field at
Various Times]{
An example of formation of boson star in Brans-Dicke theory.
Dynamical behavior of the
Brans-Dicke scalar field $\varphi_{BD} (x=0)$ is
 plotted for three evolution regions:
early time, intermediate time, and late time. We can see the field
settling down to an equilibrium configuration (periodic oscillation).
}
\label{fig_formC_2}
\end{figure}

We find that, with particular parameters $a$ and $b$, this
system forms a stable equilibrium configuration, which
can be recognized as the formation of a boson star.
As a demonstration, we here show an evolution with parameters $a=0.1$
and $b=0.025$. The BD parameter $\omega_{BD}$ is taken to be $600$.
In Fig.~\ref{fig_formC_1}, we show the BD field $\varphi$ as a function
of radial coordinate at various earlier times of evolution.
We see that the BD field becomes negative quickly
and begins oscillating around a particular value.
In the figure one can see that the BD field $\varphi$ between 
$r=40$ and $r=50$, and at around $t=50$ is still not 
near its final configuration. The long time evolution of BD field is shown in
Fig.~\ref{fig_formC_2}, in which we show the
BD scalar field $\varphi_{BD}$ at the center
for early times, intermediate times, and late times.
We can see the field
settling down to a periodic oscillation in final phase, similar to
the migration and transition cases in the previous section.
The initial mass of this configuration in units of
$G_{\ast}$ is $0.39$ and the final mass (at the end of our simulation)
about $0.384$.  At this stage the magnitude of the central
boson field is oscillating between 0.032 and 0.048. The BD field
oscillates
between $-0.00126$ to $-0.00166$. This range of boson oscillations
corresponds to masses between 0.342 and 0.410 respectively while
the BD field oscillations give
a mass between 0.355 and 0.405 respectively.
Given that the mass at this
stage is $0.384$ (consistent with the above) we expect that
the final mass will be between $0.355$ and $0.384$.

\begin{figure}
%\vspace*{1.5cm}
\setlength{\unitlength}{1in}
\begin{picture}(7.5,3.5)
\put(1.0 , -3.2){\epsfxsize=5.25in \epsffile{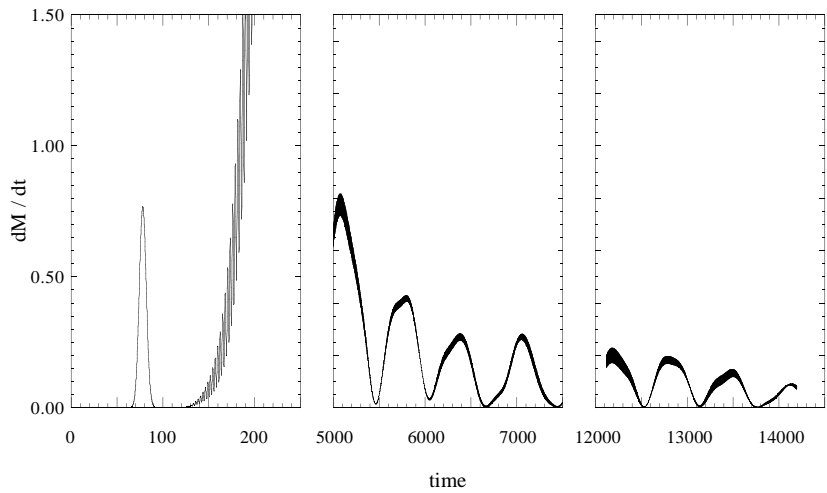} }
\end{picture}
\caption[Mass Loss Rates During Formation of {\em BD} Boson Star]{
The emitted luminosity $L=-dM/dt$ is plotted as a function of time.
In the first stage of evolution the luminosity data takes on an
underlying oscillation of period $\pi$.
The frequency of this oscillation is twice the frequency
of the underlying boson field.
This occurs after the emission of one Gaussian-shaped scalar pulse,
related to the initial field configuration.
The amount of mass loss decreases in time as
the star forms and settles down.
}
\label{fig_formC_3}
\end{figure}

We show the luminosity $L (-{dM}/{dt})$ versus time $t$ curve in
Fig.~\ref{fig_formC_3} during the same periods as in
Fig.~\ref{fig_formC_2}.
In the early stage, we see one pulse is emitted from the system.
This is related to the outgoing pulse from our initial boson field
configuration.
After this initial pulse, the system starts oscillating
with a period $\pi$ (twice the frequency of the boson field oscillation)
and the star begins forming.
After that,
we see the luminosity $L$ begin damped oscillations as a function of $t$.
(We cut out the initial large amplitude luminosity around $t=200$).
Even though the system's evolution is followed for a long time, the
accuracy of the calculation is quite good with the Hamiltonian
constraint satisfied to
order $10^{-7}$ or better.

We also note that certain parameters $a$ and $b$
(mentioned at the beginning of this section)
will result in star formation but others will not.
If we choose large amplitude $a$ and small $b$, then
the initial configuration has too large a mass and is not
dispersive enough
resulting in black hole formation during evolution.
In the opposite limit if we have a very narrow localized wave
packet it has a tendency to be dispersive (as per the wave equation).
So if $b$ is too large
no boson star forms.
Intermediate between these two are configurations that form stable stars.
For example, if $a=0.1$,
then at $b=0.01$ a black hole is formed, while at $b=0.025$ a boson star
forms as we
have shown, and at $b=0.035$ the configuration
disperses away to flat space at the end of the evolution.
On the other hand for $b=0.01$ and $a=0.05$ a boson star forms.
This demonstrates
that the boson star is a realizable object even in the BD
theory, and opens windows to study them and other
similar nontopological solitonic objects
in astrophysical roles.

%****************************** Fig.\ref{fig_formC_1}  >>>>>.

%****************************** Fig.\ref{fig_formC_2}  >>>>>.
%******************************     \ref{fig_formC_2}  <<<<<.
%****************************** Fig.\ref{fig_formC_3}  >>>>>.

%******************************     \ref{fig_formC_3}  <<<<<.

\section{Boson Halos}

In the last chapter we described the nature of the equilibrium configurations of Boson Halos, made
of massless scalar particles.
To study the formation of these objects we use the evolution equations
($BD$ evolution equations with $a=m=0$) and use the same code described in
the first section of this chapter.
Earlier studies, both numerical and analytical,
have confirmed that
self-gravitating objects with massless scalar fields cannot be compact
\cite{sei2,chr}. Thus taking a Gaussian type
localized distribution of scalar matter and evolving the system using the
evolution equations
described above, results in dispersal of the scalar matter without forming
any self-gravitating
object. Even sinusoidal functions that were damped by exponential decays met
with the same fate.
These were functions like $\exp(-r)\sin(r)/r$ and $\exp(-r)\cos(r)$. 
In Fig.~\ref{msls2}, a plot of the
density versus the radius for different times is shown for
$\sigma = 0.001 \cos (r) \exp(-r)$.
The star dissipates very quickly.
A Gaussian
distribution represents a local accumulation of matter and is obviously
the most likely happening. It seems though that only specialized
distributions can give rise to these halos.

The initial configurations for the scalar field that yield stable
configurations were those
characterized by a $1/r$ times a sinusoidal dependence at
large $r$ as well as an energy density
$\rho$ that had points of inflection. These were typically functions like
%$\cos(r)/(1+r)$, $\cos(r) (1-\exp(-1/r))$, and
%$\sin(x) (1+1/r) \log[1+1/(1+r)]$.
\ben
&&\frac{\cos(r)}{1+r}, \\ \nonumber
\cos(r) &&
\left [ 1-\exp \left ( -\frac{1}{r} \right ) \right ] ,
\quad \mbox{and}
\\ \nonumber
\sin(r) && \left ( 1+ \frac{1}{r} \right )\,
\log \left [1+ \frac{1}{1+r} \right ]
.
\een
We show the density and radial metric evolutions
for a field of the form $0.003\cos(r) (1-\exp(-1/r))$. This settles
into a self-gravitating
object after some time. Fig.~\ref{msls3}a
shows the density as a function of
$r$. The central density $\rho(0)$ increases during the evolution
from its initial value at $t=0$. Fig.~\ref{msls3}b
reveals the radial metric as it
evolves in time as a
function of radius. This too displays the settling to a stable
configuration.
In Fig.~\ref{msls3}c
we show the mass loss for the system as it finds itself
a configuration. The mass at fixed radial values for these times is
shown in Table {\ref{mslstbl1}.
The amount of radiation for this system is relatively small and decreases in
time.

On the other hand, functions like $1/(r+1)$ , $\cos (r)/{(1+r)}^2$ and
$\sin (r^{1.01})/r^{1.01})$ failed
to settle to a bound state and just dispersed away. Scalar fields represented
by these functions did not result in $\rho$ having points of inflection.

\subsection{Analytical Proof of Newtonian Stability}

To investigate the stability of Newtonian solutions under
small perturbations using the perturbative method
\cite{leepert,J89,GW89}, we write the linear scalar field perturbation in
the form
\be
\Phi (r,t):=\sigma(r) e^{-i\omega t} + \delta \sigma(r) e^{i k_n t} \; ,
\ee
where $k_n$ are the frequencies of the quasinormal modes. The perturbed
scalar field equation
\be
\delta \sigma''+2 \delta \sigma'/x+k_n^2 \delta \sigma = 0 \; ,  \label{deltap}
\ee
together with the boundary conditions
\be
\delta \sigma'(0)=\delta \sigma (R)=0,
\ee
(for regularity of $\delta \sigma$ at the origin and at the radius $R$ of the
boson halo) defines a Sturm-Liouville eigenvalue problem with real
eigenvalues
\be
k_1^2 < k_2^2 < \ldots  \; .
\ee
Eigenfunctions of this particular differential equation (\ref{deltap})
are
\be
\delta \sigma = \frac{\sin(k_n x)}{x} \; ,
\ee
where the eigenvalues are real,
\be
k_n = n \frac{\pi }{R} \; ;
\ee
$n$ is an integer and $R$ the radius of the solution. Thus
the eigenvalues $k_n^2$ are positive.
This means that all modes are stable. 

\subsection{Numerical Study of Stability}

We study the evolution of dilute configurations using
the same numerical code we used for studying boson stars.
We show a perturbed Newtonian configuration as it settles to
a new Newtonian configuration.
The perturbation that is used in this case mimics an annihilation of
particles. A Gaussian bump
of field is removed from a part of the star near the origin.
Fig.~\ref{msls5}a is a plot of the unperturbed versus the perturbed density at
$t=0$. The perturbed configuration
is evolved and settles to a new configuration. The scalar radiation moves
out as shown in Fig.~\ref{msls5}b.
In Fig.~\ref{msls5}c the density profile is shown after the system settles down.
The system is very
clearly in a new stable configuration. In Fig.~\ref{msls5}d
the mass is plotted
as a function of radius
for various times. Again one can see that the mass loss is decreasing by
the end of the run showing
that the system is settling down to a new configuration. The mass as a
function of time is presented in Table \ref{mslstbl2}
for different radii.

We have so far only been successful in evolution studies
of dilute configurations.
The reason for this is that denser configurations
need much better resolution in order for their evolution to be
studied in our code. This is coupled with
the difficulty that we still
need the boundary to be very far away, so that the density has significantly
fallen off, for an outgoing boundary
condition to work. One possibility is to match our configurations
with some other kind of matter at the boundary.
For example, we could have a massless scalar field
configuration surrounded by a massive field configuration like
a boson star. Unfortunately, while we can find such configurations
with smooth mappings of the metric and field density, the scalar field
itself does not match smoothly for such configurations. 
So far, we are using either an outgoing wave condition
or exact boundary conditions where the latter one simulates a vacuum
energy. Further investigation is needed before we
can decide whether these
non-Newtonian configurations are inherently unstable.

\subsection{Radially Oscillating Solutions: Massive Field Case}

A boson star consists of scalar massive particles, hence in the simplest
model one has a potential $U=m^2 |\Phi|^2$. Exponentially decreasing
solutions exist for special eigenvalues of the scalar
field $\omega < m$,
so that the star has a finite mass.
In the case of $\omega > m$, oscillating scalar field solutions can
be found for all values of $\omega $. The energy density reveals
minima and maxima as opposed to the points
of inflection seen for the massless case.
Fig.~\ref{msls6}a
shows a comparison of the density profile for the two cases.
In Fig.~\ref{msls6}b
we show the mass profile and the particle number as functions
of central density. The binding energy is always positive making the
system unstable. Numerically we find the system to completely disperse
(as it should),
making these parameters unfavourable as halo models.

\begin{figure}
\centering
\leavevmode\epsfysize=7.5cm \epsfbox{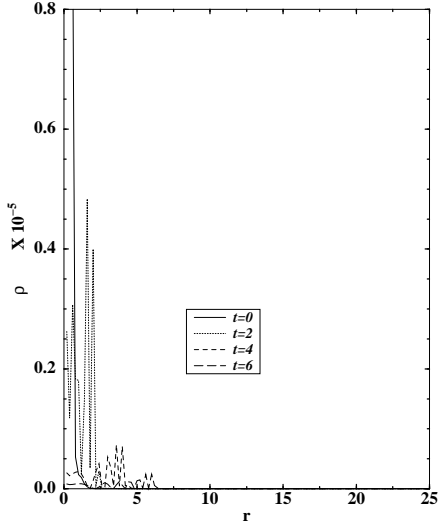}\\
\caption[Dispersion of Gaussian Packet]
{The inability of massless scalar field configurations to form
compact self-gravitating objects as
discussed by Christodoulou and others \cite{sei2,chr} is verified
numerically.
An initial field configuration of the
form $\sigma = 0.001 \cos(r) \exp(-r)$ is seen dispersing in the plot.
The dispersal takes place very quickly.}
\label{msls2}
\end{figure}

\begin{figure}
\centering
\leavevmode\epsfysize=7.5cm \epsfbox{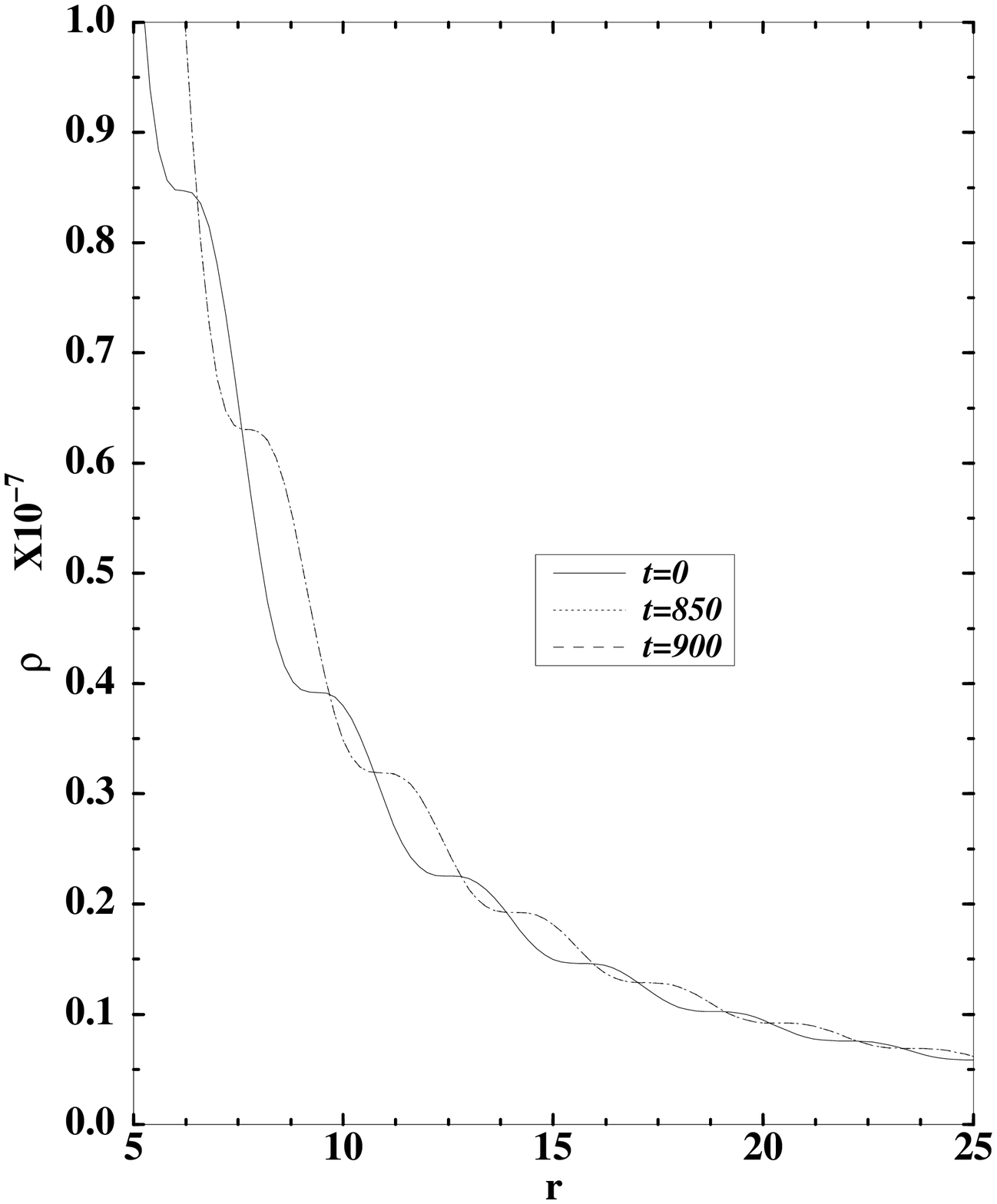}\hskip0.5cm
\leavevmode\epsfysize=7.5cm \epsfbox{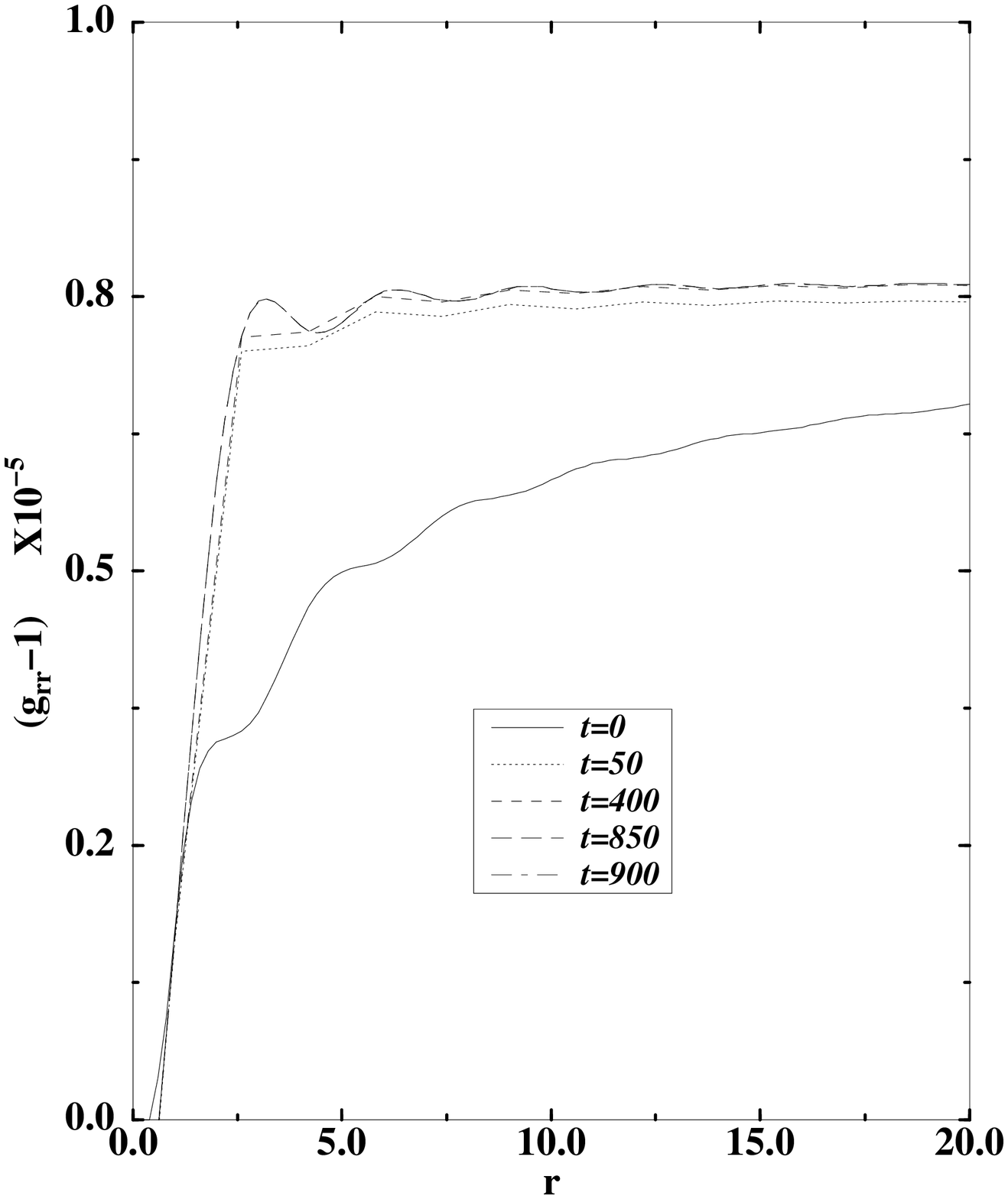}\hskip0.5cm
\leavevmode\epsfysize=7.5cm \epsfbox{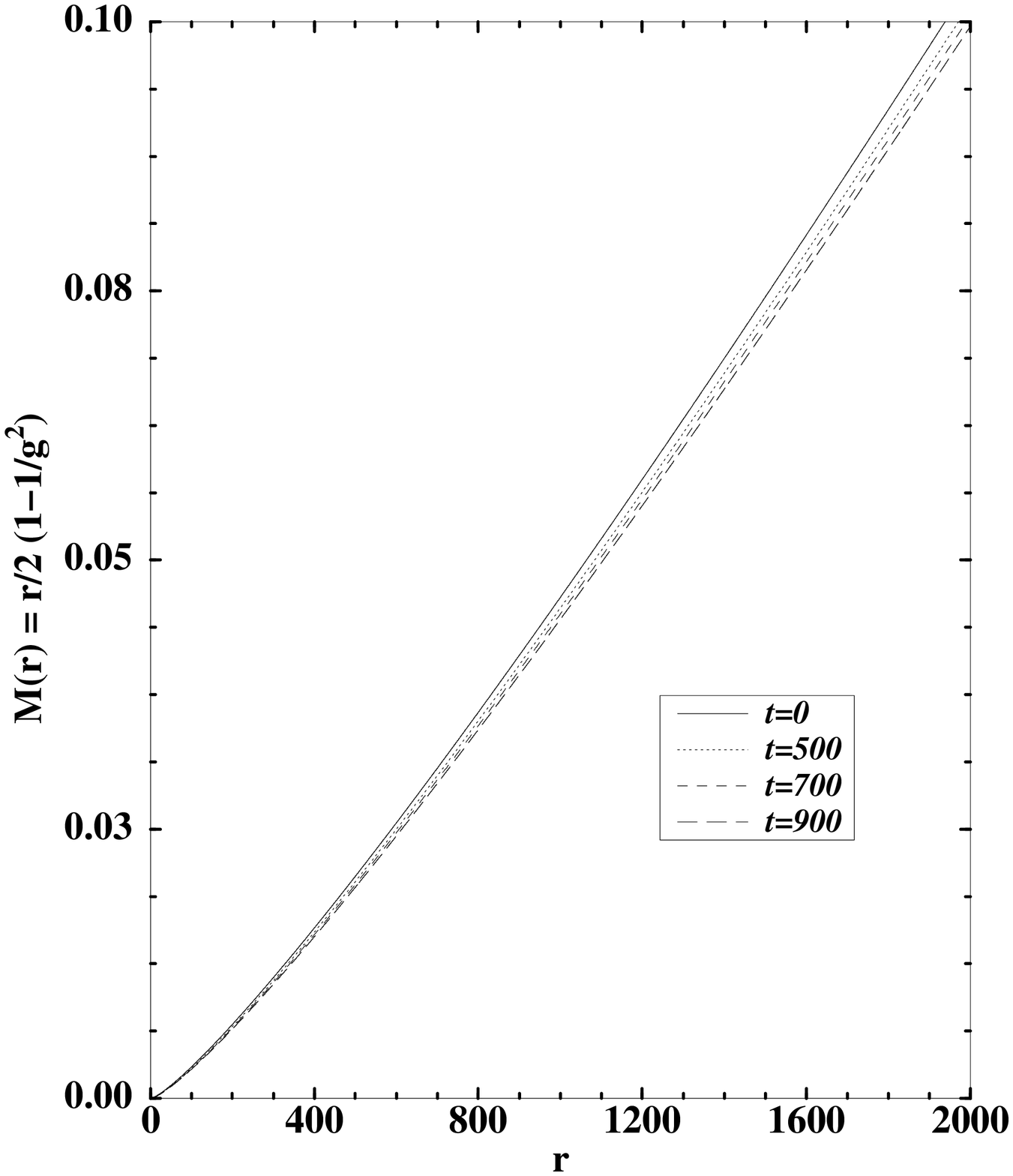}\\
\caption[Formation of Halo]
{(a) Top left: The formation of a self-gravitating massless scalar field object
is
shown here for
an initial field configuration of the form $0.003 \cos (r) (1-\exp(-1/r))$.
The density in this
case increases from its initial value as the configuration settles down.
(b) Top right: The radial metric for this configuration is shown evolving
to a stable final configuration.
(c) Bottom: The mass is plotted as a function of radius at different times.
As the configuration evolves it loses less and less mass as it settles
down.}
\label{msls3}
\end{figure}
\begin{figure}
\centering
\leavevmode\epsfysize=8.5cm \epsfbox{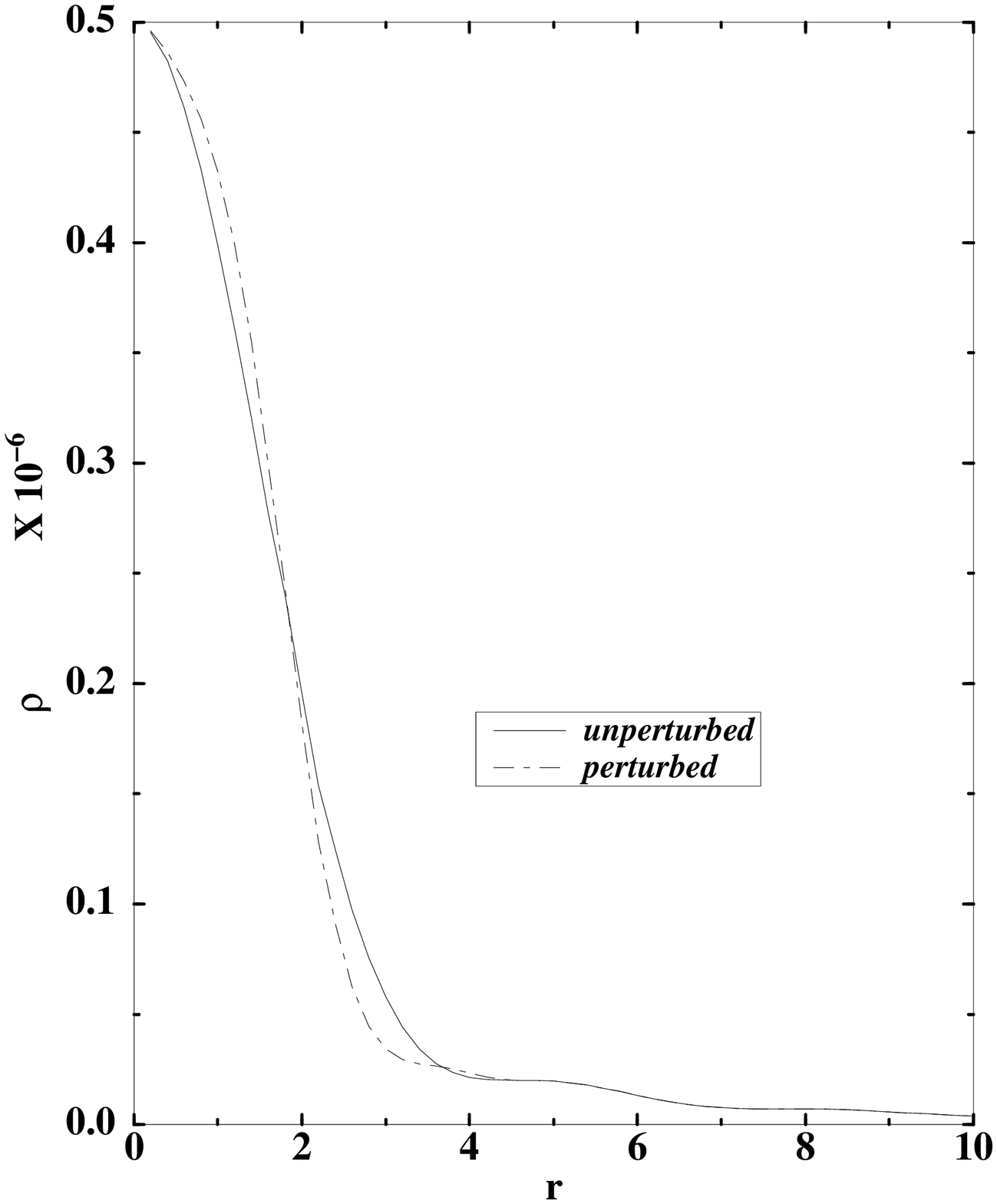}\hskip0.5cm
\leavevmode\epsfysize=8.5cm \epsfbox{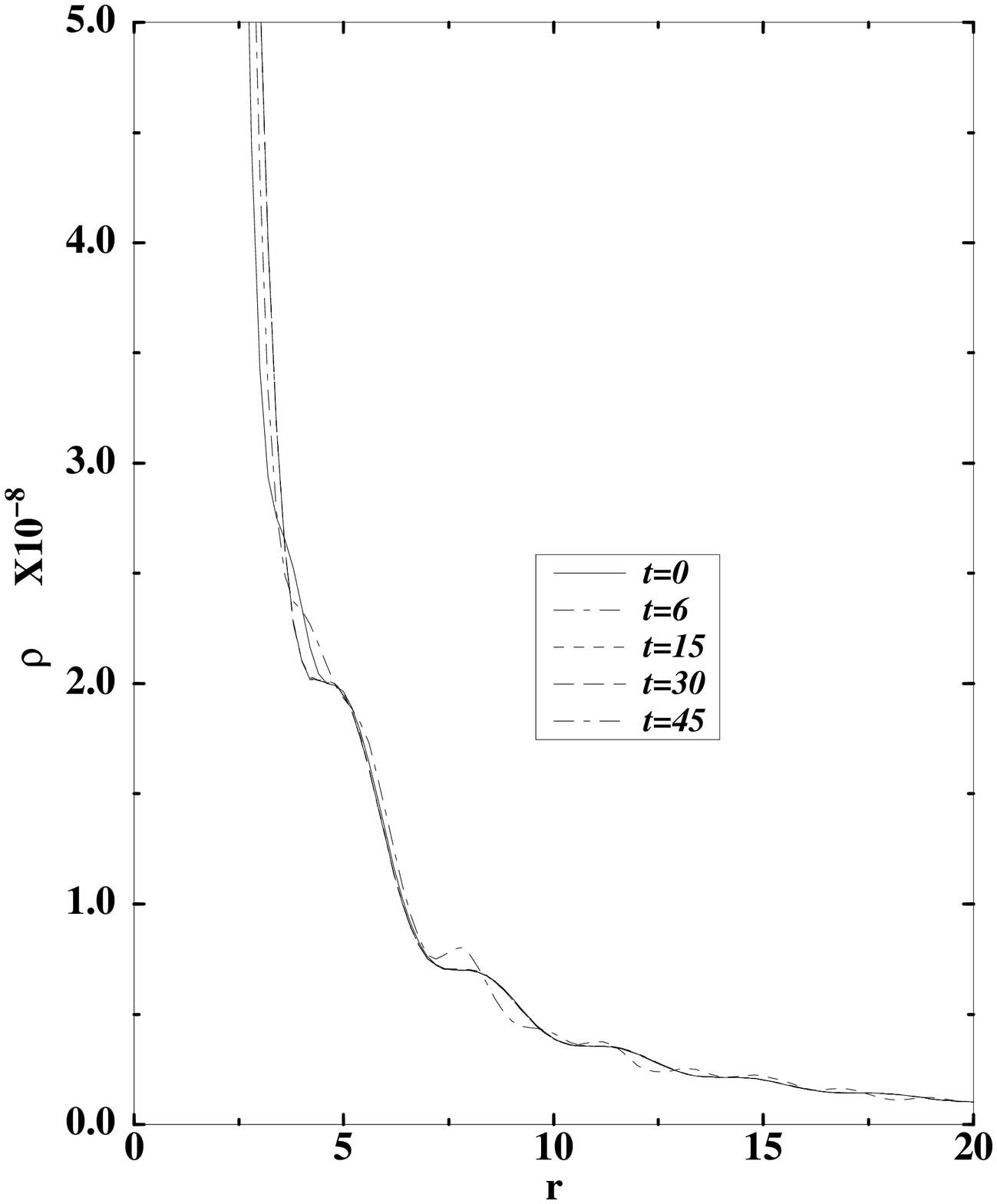}\\
\leavevmode\epsfysize=8.5cm \epsfbox{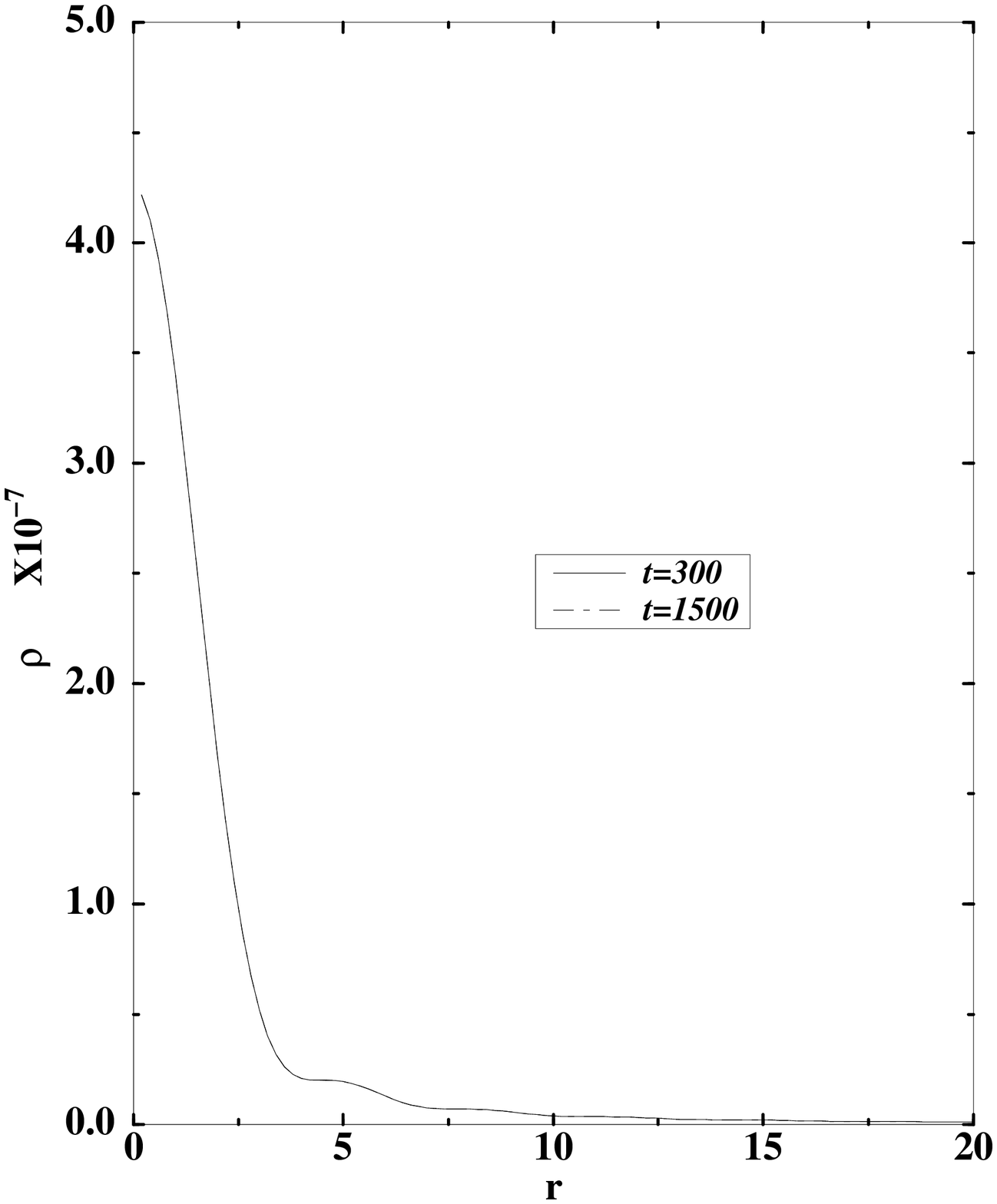}\hskip0.5cm
\leavevmode\epsfysize=8.5cm \epsfbox{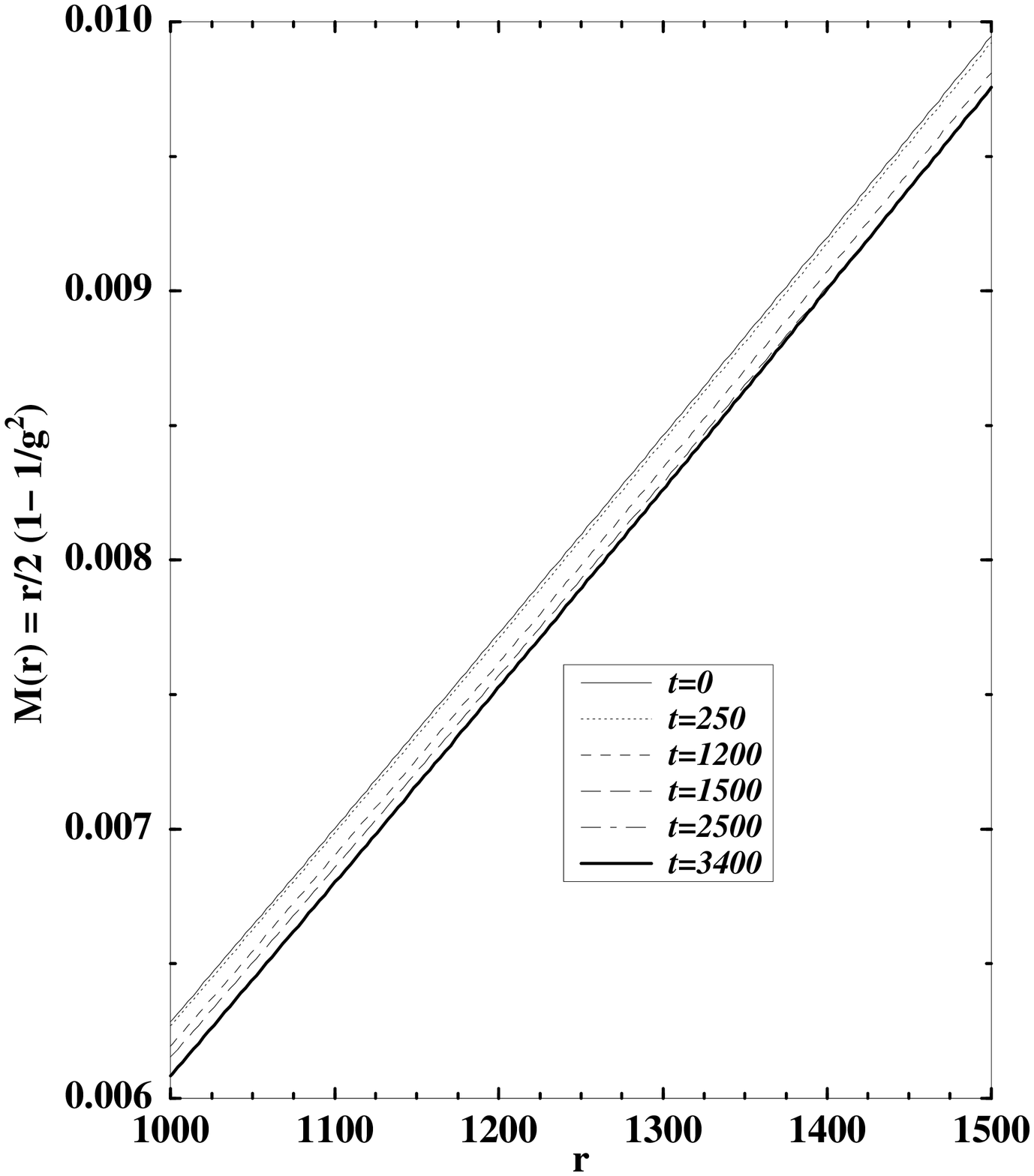}\\
\caption[Behavior of Perturbed Halo Configuration]
{(a) Top left: An exact Newtonian configuration is perturbed and evolved
using the GR code.
The initial unperturbed and perturbed densities
are shown. A Gaussian profile of scalar field is removed from the
equilibrium configuration, and the
constraint equations are reintegrated to give new metrics before
beginning evolution.
(b) Top right: The outgoing scalar radiation can be seen in a plot of the
density at early times.
(c) Bottom left: The density profile is shown after it starts to settle
down.
Between times of 300 and 1500
very little has changed in the density profile.
(d) Bottom right: The mass is plotted as a function of radius for
different times. In order to enhance the features
only a small part of the radial region is shown. The amount of mass
loss is clearly lessening in time.}
\label{msls5}
\end{figure}

\begin{figure}
\centering
\leavevmode\epsfysize=7.5cm \epsfbox{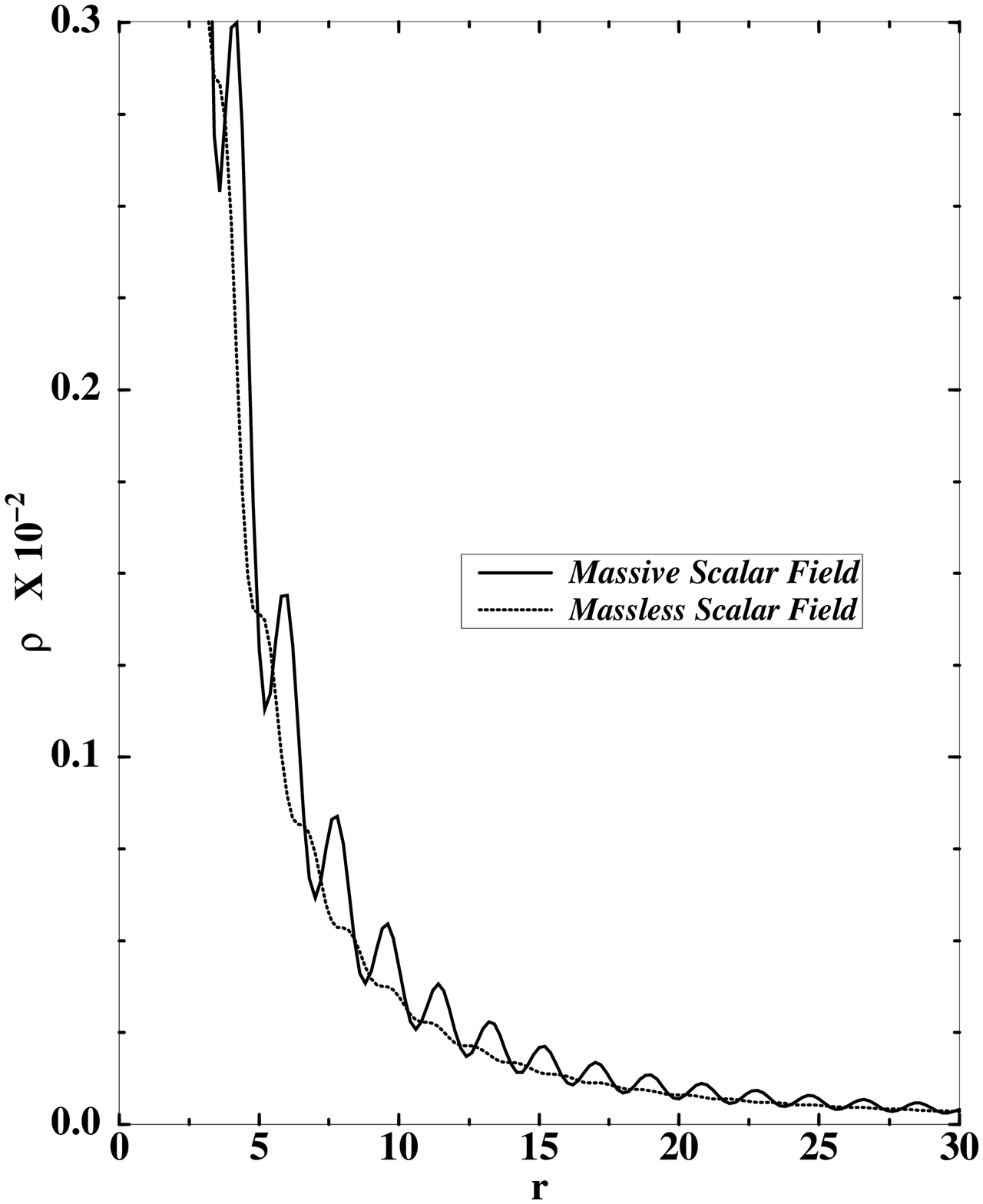}\hskip0.5cm
\leavevmode\epsfysize=7.5cm \epsfbox{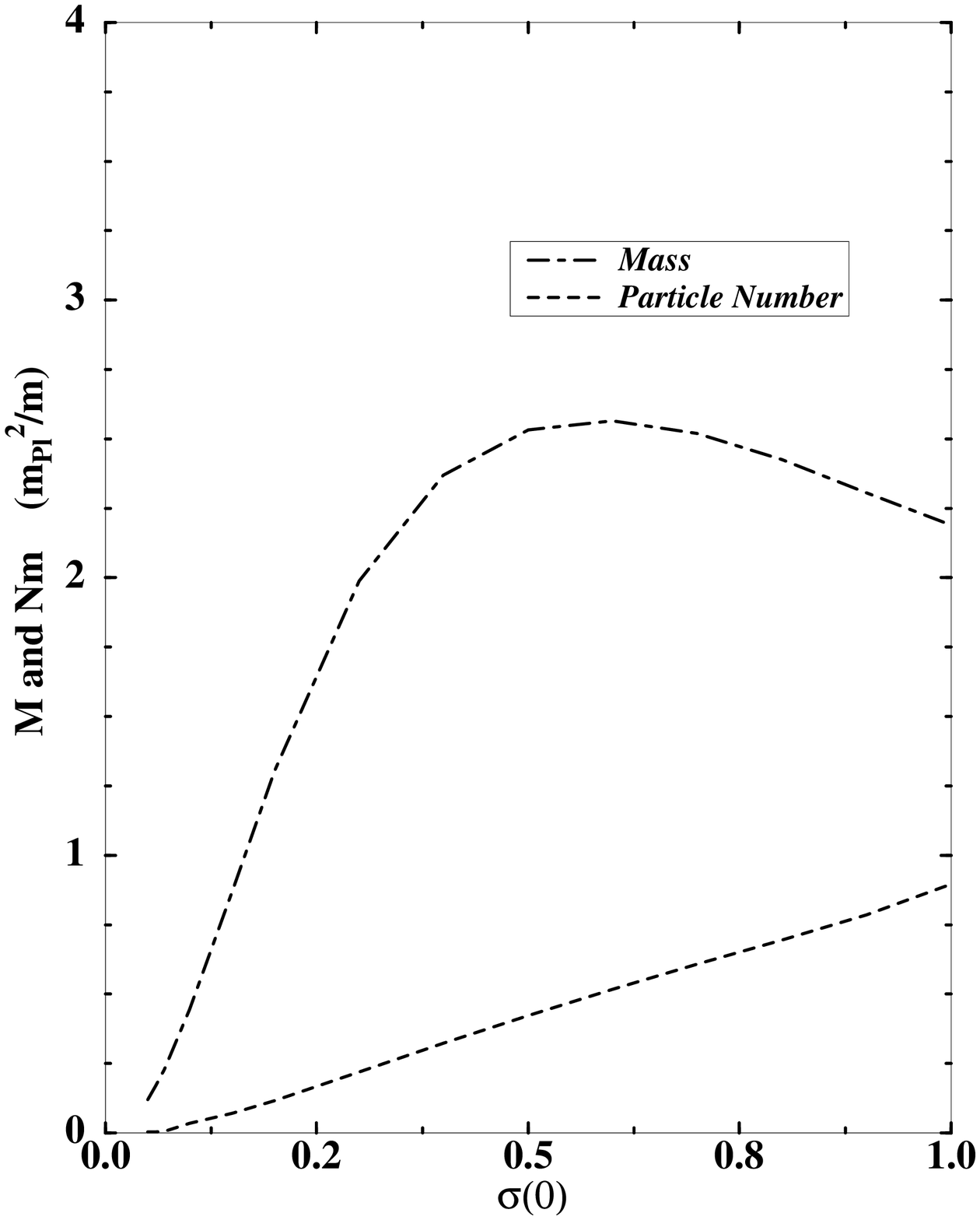}\\
\caption[Radially Oscillating Solutions for Massive Field]
{(a) Left: The density profiles of massive scalar field configurations with
$\omega > m$ is compared to that of
a massless field. The former has maxima and minima while the latter
has points of inflection. This
structure may be related to stability issues.
(b) Right: The mass and particle number are plotted against the central
density
for boson star oscillators. We find in all cases that the mass is always
greater than the
particle number explaining their instability against a collective
dispersion.}
\label{msls6}
\end{figure}

\begin{table}
\caption[]{Mass values at different radii and different times for the
evolution in Fig.~{\ref{msls3}}. One recognizes that the change in the mass at
fixed radius decreases in time, showing that the system
is settling down to a stable
configuration.}
\begin{tabular}{|c|c|c|c|c|}
\multicolumn{5}{c}{\label{mslstbl1}}   \\
\hline
 & $t=0$ &
 $t=500$ & $t=700$& $t=900$
 \rule[-0.1in]{0.0in}{0.3in}\\
  \hline
  $r=800$ &$0.03589375$&
  $0.0350625$ & $0.0346250$& $0.034234375 $
  \rule[-0.1in]{0.0in}{0.3in}\\
   \hline
   $r=1200$ & $0.05750000$&
   $0.0563125$ & $0.0556250$& $0.055062500 $
   \rule[-0.1in]{0.0in}{0.3in}\\
    \hline
    $r=1600$ & $0.08018750$&
    $0.0786250$ & $0.0776875$& $0.076937500 $
    \rule[-0.1in]{0.0in}{0.3in}\\
    \hline
    \end{tabular}
    \end{table}

    \begin{table}
    \caption[]{Mass values at different radii and different times for the
    evolution in Fig.~{\ref{msls5}}. One recognizes that the change in the mass at
    fixed radius decreases, hence it proves the settling to a stable
    configuration.}
    \begin{tabular}{|c|c|c|c|c|}
    \multicolumn{5}{c}{\label{mslstbl2}}   \\
    \hline
     & $t=0$ &
     $t=1500$ & $t=2500$& $t=3400 $
     \rule[-0.1in]{0.0in}{0.3in}\\
      \hline
      $r=500$ &$0.00284700$&
      $0.00278000$ & $0.00268750$& $0.00268375 $
      \rule[-0.1in]{0.0in}{0.3in}\\
       \hline
       $r=1000$ & $0.00628625$&
       $0.00619375$ & $0.00608495$& $0.00608335 $
       \rule[-0.1in]{0.0in}{0.3in}\\
	\hline
	\end{tabular}
	\end{table}

\chapter{Boson Stars in $3D$ Numerical Relativity}

We now describe our work on Boson stars in $3D$. One of the
major drawbacks in going from a $1D$ code to a $3D$ code was
the paucity of grid points. Computer memory and time constraints
limited grid resolution. There was no simple outgoing wave condition (the boundary
was no longer one point) and there was no reasonable way to
implement a sponge. A sponge, while
it prevents reflections, also uses up grid points.
Besides, the length of the sponge had to be of the
order of the wavelength of the outgoing radiation. Gravitational waves
have longer wavelengths than scalar waves and this further inhibited
the use of a sponge in a $3D$ code.

The $3D$ code that had been developed 
to deal with vacuum spacetimes was extended
to include matter terms. We used our
model to test the code by comparing with established results in
$1D$, as well as to results of $3D$ perturbation studies. Once stability was
achieved, we then used the code itself to get physical results of Boson stars
under nonspherical perturbations. The main advantage
with Boson stars as the matter source was that there were
ordinarily no spacetime singularities as in black hole evolutions,
as well as none of the surface problems that people dealing with neutron star
configurations have to routinely face. The Boson field
exponentially decays to zero at spatial $\infty$. Nevertheless, the small amount
of matter in the outer regions that prevents surface problems
could, for the purposes of wave extraction, be perfectly well
regarded as a vacuum. The code used a leapfrog scheme with the
extrinsic curvature
(related to the time derivative of the metric) and
time derivatives of the Boson field on 
half integral time steps ($n+1/2$) while metric fields and Boson fields were
on the integral timesteps ($n$). To prevent grid point to
grid point instabilities due to mesh drifting we routinely used a small
amount of diffusion in the center of the star, as described in the
last section of the introduction of this thesis.

In an appendix we describe how we scaled the variables in our $3D$
code. This appendix shows how to relate a unit of computer time to
physical time in terms of the mass of the
configurations.

\section{Stress Energy Tensor, Energy Density, Momentum Density, Sources}

The metric components in the $ADM$ formalism are given in (\ref{downmet}) and (\ref{upmet}). The
lapse is $N$ and the shift components are $N^i$.
In the introduction to the thesis we wrote down the matter Lagrangian for a
Boson star in scalar-tensor theory and showed how
to get the stress-energy tensor from it. In $GR$ we get the equations
with the parameter $a$ set to zero. The stress-energy tensor is then
\be
T_{\mu\nu} = \bosa_{,\mu} \, \bosa_{,\nu} - \frac{1}{2}\,  g_{\mu\nu} \,
[\bosa_{,\alpha} \, \bosa^{,\alpha} + m^2 \, \bosa^2]
+ (\bosa\rightarrow \bosb).
\ee
where the subscripts $1$ and $2$ refer to the two real components
of the complex Boson field.
Since we rescale quantities to dimensionless ones no factor of $m$ finally
appears in the 
Einstein equations. (An appendix attached to this thesis
explains the scaling in the $3D$ code). 
Thus we can write
\ben
T_{\mu\nu} = 
\bosa_{,\mu} \,   \bosa_{,\nu} &-& \frac{1}{2}\, g_{\mu\nu}\Bigg[
\\ \nonumber
&&g^{00} \, (\partial_t \bosa)^2 + \bosa_x\,
( g^{xx} \, \bosa_x + g^{xy} \, \bosa_y +
g^{xz} \, \bosa_z)
\\ \nonumber
&+&
\bosa_y \, ( g^{yy} \, \bosa_y + g^{xy} \, \bosa_x +
g^{yz} \, \bosa_z)
\\ \nonumber
&+& \bosa_z \, ( g^{zz} \, \bosa_z + g^{zy} \, \bosa_y +
g^{xz} \, \bosa_x)
+\bosa^2 \Bigg] 
\\ \nonumber
&&+ (\bosa\rightarrow \bosb).
\een
Here the subscripts $x$, $y$, and $z$ refer to a partial derivatives with respect
to those variables.
The form of the
density is given by $n^\mu n^\nu T_{\mu \nu}$ where $n^0 =1/N$
and $n^i = -N^i/N$.
Thus
\be
\rho = \frac{1}{N^2} \, T_{0 0} -\frac{N^i}{N^2} \, T_{0i} + 
2 \, \frac{N^iN^j}{N^2} \, T_{ij}.
\ee

The scalar field equation is the gravitational Klein--Gordon equation:
\ben
0&=&\dal \psi -\psi = (g^{\alpha\beta} \, \psi_{,\beta})_{;\alpha}-\psi
\label{psiform}
\\ \nonumber
&=& \psi^{,\alpha} {}_{,\alpha} + \Gamma^{\alpha}
{}_{\alpha \beta} \, \psi^{,\beta}-\psi
\\ \nonumber
&=&\psi^{,\alpha} {}_{,\alpha} + \frac{1}{2}\,g^{\alpha \nu} \,g_{\nu \alpha,\beta}
\, \psi^{,\beta}-\psi
\\ \nonumber
&=& \psi^{,\alpha} {}_{,\alpha} + \frac{1}{\sqrt {-g}} \, (\sqrt {-g})_{,\beta} 
\psi^{,\beta}
-\psi
\\ \nonumber
&=& \frac{1}{\sqrt {-g}} (\sqrt{-g} \, \psi^{,\alpha})_{,\alpha}-\psi.
\een
where the factor $g$ represents the determinant of the 4 metric. Thus
\be
\frac{1}{\sqrt {(N^2-N^i \, N_i)}\,
\sqrt \gamma}\,\left (\sqrt{(N^2-N^i\,N_i)} \, \sqrt \gamma \,
g^{\alpha\beta} \, \psi_{,\beta} \right)_{,\alpha}-\psi=0.
\ee
These second order (in time) equations ($\alpha=\beta=t$) can be conveniently
converted to first order by defining a new
variable  $\pi= \sqrt{(N^2-N^i \, N_i)} \, (\sqrt \gamma \, g^{tt} \,
\dot \psi)$ or, since we have two fields
\be
\pi_{a} = \frac{\sqrt{(N^2-N^i \, N_i)}}{N^2}\left(\sqrt \gamma \,
\dot {\psi_a} \right)\quad
a=1,2.
\ee
This gives the evolution of $\pi$ to be of the form
\ben
\dot \pi =& -&\sqrt {(N^2-N^i \, N_i)}\, \sqrt \gamma \, \psi 
\label{pievolution}
\\ \nonumber
&+&
\partial_i\left({N^i} \, \pi\right) + \partial_t\left(\sqrt{(N^2-N^i \, N_i)}\,
\sqrt \gamma \, \frac{N^i}{N^2} \, \partial_i{\psi}\right) 
\\ \nonumber
&+& \partial_i\left(\sqrt{(N^2-N^i\, N_i)} \, \sqrt \gamma \,
\gamma^{ij} \, \partial_j \psi \right).
\een

\section{Testing the Scalar Evolver}

We computed the derivatives of the fluxes $\gamma^{ij} \sqrt{(N^2-N^iN_i)}
\sqrt \gamma$ using a second-order scheme. We also
adopted a second-order
scheme for the first and second-order derivatives $\partial_i\psi_a$
and $\partial_j \psi_a$. (The subscript $a$ runs from $1$ to $2$ representing
the two real components of the boson field). We used a staggered grid
(the origin was not one of the points). In order to be better able to
resolve the star we observed evolutions on an octant rather than the full
grid so our perturbations also had to retain this octant symmetry. The
first grid point was actually staggered half a spacing step away from the origin
in one direction and the second grid point was half a step in the other
direction. 

The first stage of the analysis was to get initial data. 
Spherically symmetric equilibrium data was obtained from the $1D$ code, described in the last chapter. This data was then interpolated
onto a $3D$ grid. This was done by identifying the positions
$r(i,j,k)$ on the $3D$ grid, finding out which point
it was closest to on the $1D$ grid, and using the $1D$ data before that point, at that
point and the point after that, to get the value at $r(i,j,k)$ with a three-point interpolator. This $3D$ data for the boson fields and for the metrics was then converted from $r,\theta,\phi$ coordinates to $x,y,z$ using
\ben
x^2 &+& y^2+z^2 =r^2,
\\ \nonumber
x &=& r\, \sin\theta \, \cos\phi,
\\ \nonumber
y &=& r \, \sin\theta \, \sin\phi 
\\ \nonumber
z &=& r\cos\theta, \,\,\text{and}
\\ \nonumber
g_{ij} &=& \frac{\partial i'}{\partial i}\frac{\partial j'}{\partial j}g_{i'j'},
\een
where the $i$ and $j$ coordinates refer to $x,y,z$ and the prime coordinates
to $r, \,\theta,\, \phi$.

To test the scalar evolver we then froze the metric evolution and just allowed
scalar field evolutions. (This is physically reasonable
since an equilibrium boson star is characterized by a static metric.)
When, after thousands of time steps, the boson field continued oscillating
with a period of $2\pi$ (since the boson fields have time dependence of the
form $\cos(t_{code})$ and $\sin(t_{code})$ where $t_{code} =\omega_Et$), we were
convinced that our scalar evolver was doing well. 

The boundary conditions used were reflection symmetry at the origin
and flat boundary conditions for the fields at the outer boundary.
We present a plot of the scalar evolver in Fig.~\ref{fig_scalaevol}.

\begin{figure}
\hspace{-36pt}
\vspace{-130pt}
\epsfbox[0 0 450 500]{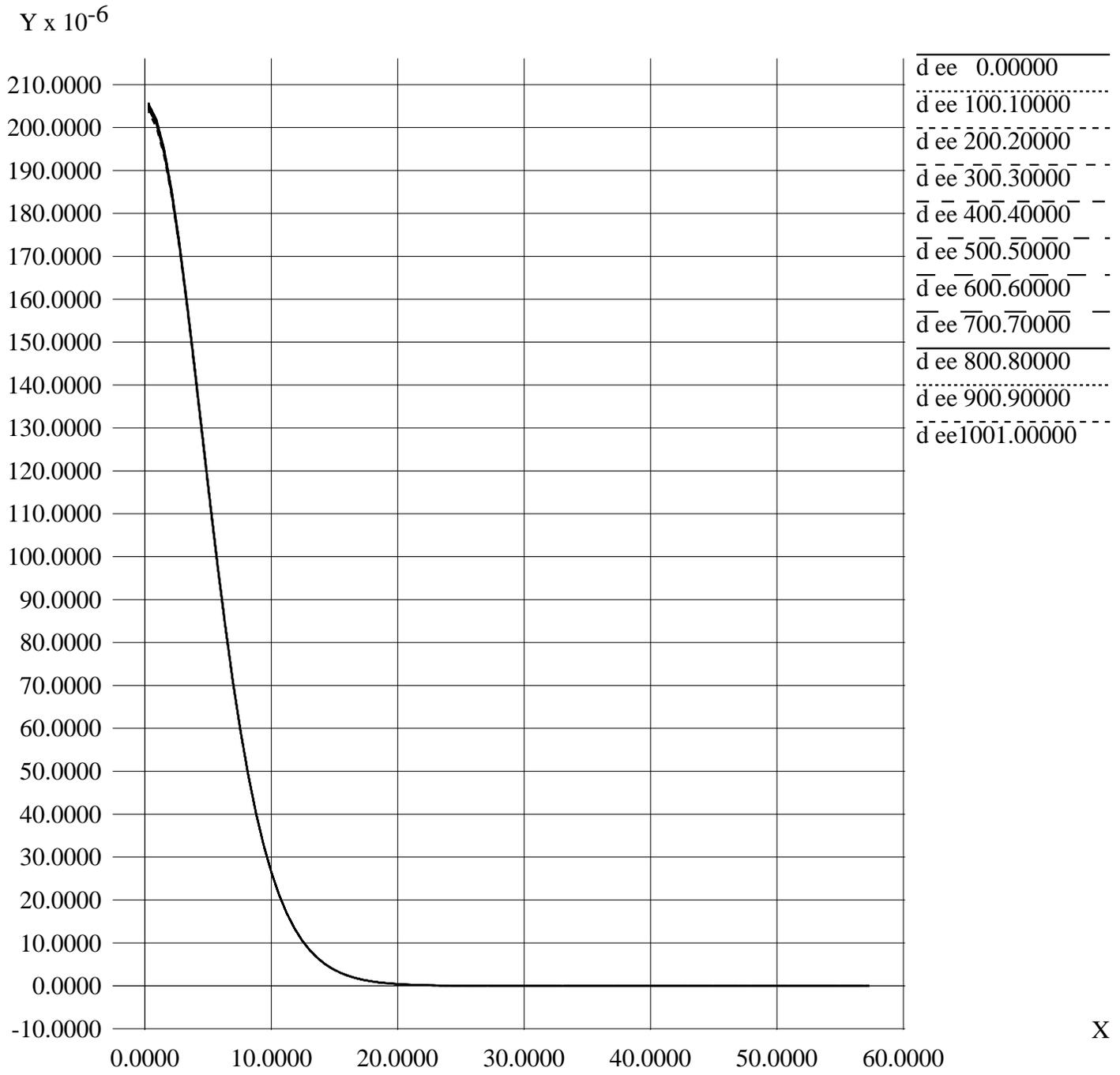}
%\epsfbox[0 -150 350 500]{figs/scalaevole.ps}
\vspace{0.75 in}
\caption{A plot of the density $\rho$ as a function of $r$ until a time
$t=1000$ for a boson star of central field density $\sigma=0.05$ 
(mass $0.416\,M_{Pl}^2/m$), without any
metric evolution. The drift in density is insignificant over this
time showing the scalar evolver to be functioning well. (That is: all the
curves lie on top of each other and cannot be visually distinguished.)
}
\label{fig_scalaevol}
\end{figure}

When calculating $(f \psi')'$ (where $f$ is the flux and $\psi$
is the boson field) in the evolution of
$\pi$ (equation (\ref{pievolution})), it was numerically
more advantageous to first write
\begin{equation}
(f\,\psi')' = f \,\psi'' + f' \,\psi',
\end{equation}
and then to calculate the two terms on the RHS separately.
This avoids the numerical instabilities typically generated by
finite differencing a finite difference.

\section{Equilibrium Boson Star}

The first step in testing the efficacy of the code was to take the
known results of the $1D$ code and reproduce them in the full $GR$ code.
Once the scalar evolver was found to be producing accurate
results we wanted to test the code with full metric evolutions.
Initially, we used equilibrium data obtained from our $1D$ spherical code.

The equations for full evolution have been developed in the introductory
part of this thesis (\ref{subsection:ADM}). The four gauge degrees of freedom
were used
to set the three shift components to zero and then a maximal slicing
condition was used for the lapse.

\subsection{Maximal Slicing}

The condition for maximal slicing~\cite{York79,piran}
is $Tr K =0$. Thus, on replacing for ${}^3R$, from the 
Hamiltonian constraint equation (\ref{hamilform}) (with $K=0$), into
(\ref{trkevolv}), we get
\be
\frac{d Tr\,K}{dt} =0  = -N^{|i}{}_{|i} + N\left[K_{ij}K^{ij} + \fpi(\rho
+Tr\,S)\right].
\label{maximalform}
\ee
Here $S = S^i_i$ where $S_{ij}=T_{ij}$.
The reason for writing it in this form was due to the presence
of second derivatives in
the Ricci tensor. The code had originally been developed for black hole
spacetimes. Late time peak development in the metric functions
rendered computation of second derivatives difficult. The
above form was therefore numerically more advantageous~\cite{ani2bh}. This was important
in the boson star collapse problem as well. 

One of the advantages of the maximal slicing condition was that it
provided a linear equation for the lapse. The
disadvantage was that an elliptic equation had to be solved on
each time slice. Asymptotic flatness required the lapse to go to
some constant in the limit $r\rightarrow \infty$. In our code
the value of this constant was influenced by the underlying frequency of the
scalar field and the lapse value at $\infty$ was the value of that frequency
(see chapter 2). 

The main advantage of maximal slicing was its
singularity avoiding ability. One of the advantages of a good coordinate
system is avoidance of physical singularities by slowing down of evolution
in the spatial region near such a singularity. The
maximal lapse collapsed, vanishing exponentially, near the singularity and froze the evolution
there (near $r=0$, the physical singularity for a Schwarzschild blackhole
or for a Kerr metric), allowing the maximal slices to span the rest of the
future development of the initial data. This was important in the boson
star collapse problem. Coordinate volumes usually shrink in the
neighborhood of singularities. This is avoided by maximal slicing
since the $Tr K =0$ condition translates in the zero
shift case to the volume $\sim \gamma$ being constant.
Another advantage of the
maximal slicing condition is that asymptotically it turns to a natural radiation
gauge in which the gravitational radiation has a wavelike propagation.

\subsection{The Elliptic Solver}

In equation (\ref{maximalform}) there is a term of the form ${}^3\dal N$ 
and, as
we did for the Klein--Gordon case (\ref{psiform}), we write the maximal slicing
lapse equation:
\be
\partial_i(\sqrt{\gamma} \, \gamma^{ij} \,\partial_j N)
=N\, \sqrt{\gamma}\, \left(K^{ij}\, K_{ij} + \fpi \, (\rho +S)\right)
\ee

The code used a conjugate matrix solver developed at the $NCSA$ to which we
added the matter terms. The basic technique was to first finite difference the
above equation to second order. Because we were not dealing with
flat space there were
mixed derivative operators $\partial_i\partial_j$ with $i\ne j$
as well. Thus the second order finite differencing did not just
connect points ($i+1,j,k$), ($i,j,k$), ($i-1,j,k$), ($i,j+1,k$), 
($i,j-1,k$), ($i,j,k+1$), and ($i,j,k-1$),
a seven-point stencil. It also connected 12 more points
($i+1,j\pm 1,k$), ($i-1,j\pm1,k$),
($i+1,j,k\pm 1$), ($i-1,j,k\pm 1$), ($i,j+1,k\pm 1 $), and
($i,j-1, k\pm1$) where $i$ $j$ and $k$ are the $x$, $y$ and $z$
coordinate positions respectively. Thus we were dealing with a 19-point
stencil. For ease of visualization we
show the $2D$ case in Fig.~\ref{fig_twodstencil} and the $3D$ case
in Fig.~\ref{fig_threedstencil}.
(In $2D$ we have  a five-point stencil when there
are no mixed derivatives and a nine-point stencil
when mixed derivatives are present.)

%%\begin{figure}[t]

%%\begin{center}
%%\leavevmode
%%\makebox{\epsfysize=15cm\epsfbox{storedir/grrmax3D.ps}}
%%\end{center}
%\makebox{\epsfysize=15cm\epsfbox{storedir/linefig44.ps}}
\begin{figure}
\hspace{-36pt}
\vspace{-130pt}
%%   \vspace{0.5 in}
%\epsfbox[0 -150 350 500]{2dstencil.ps}
\epsfbox[0 30 350 500]{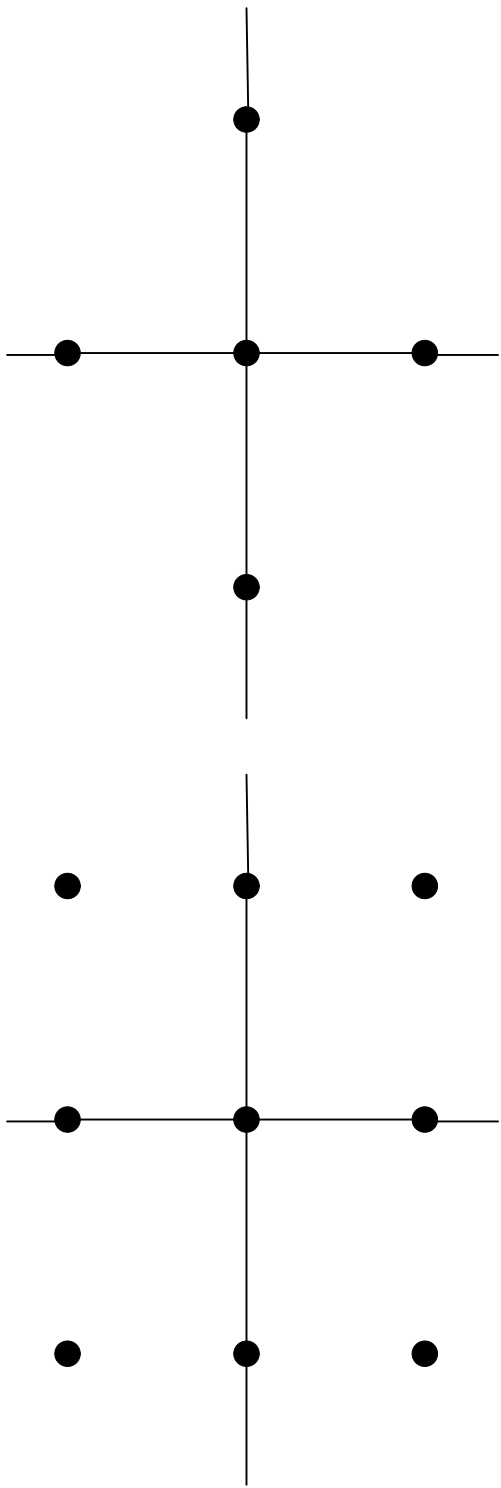}
%\epsfbox[0 30 350 500]{storedir/grrmax3D.ps}
\vspace{0.5 in}
\caption{a) Top: The five-point stencil in $2D$ in flat space is shown.
The absence of mixed second-order derivatives in the $x$ and $y$ directions
ensures that no points along the diagonal (except the center) are
required. b) Bottom: In curved space the second-order derivatives
include mixed derivatives ($\frac{\partial^2}{\partial x\partial y}$).
We now have a nine-point stencil that includes the corner
points.
}
\label{fig_twodstencil}
\end{figure}

%\begin{figure}[t]

%\begin{center}
%\leavevmode
%\makebox{\epsfysize=15cm\epsfbox{storedir/cubefig45.ps}}
%\end{center}

\begin{figure}
\hspace{-36pt}
\vspace{-130pt}
%\epsfbox[0 -150 350 500]{3dstencil.ps}
%\epsfbox[0 -150 350 500]{storedir/cubefig45.ps}
\epsfbox[0 -150 350 500]{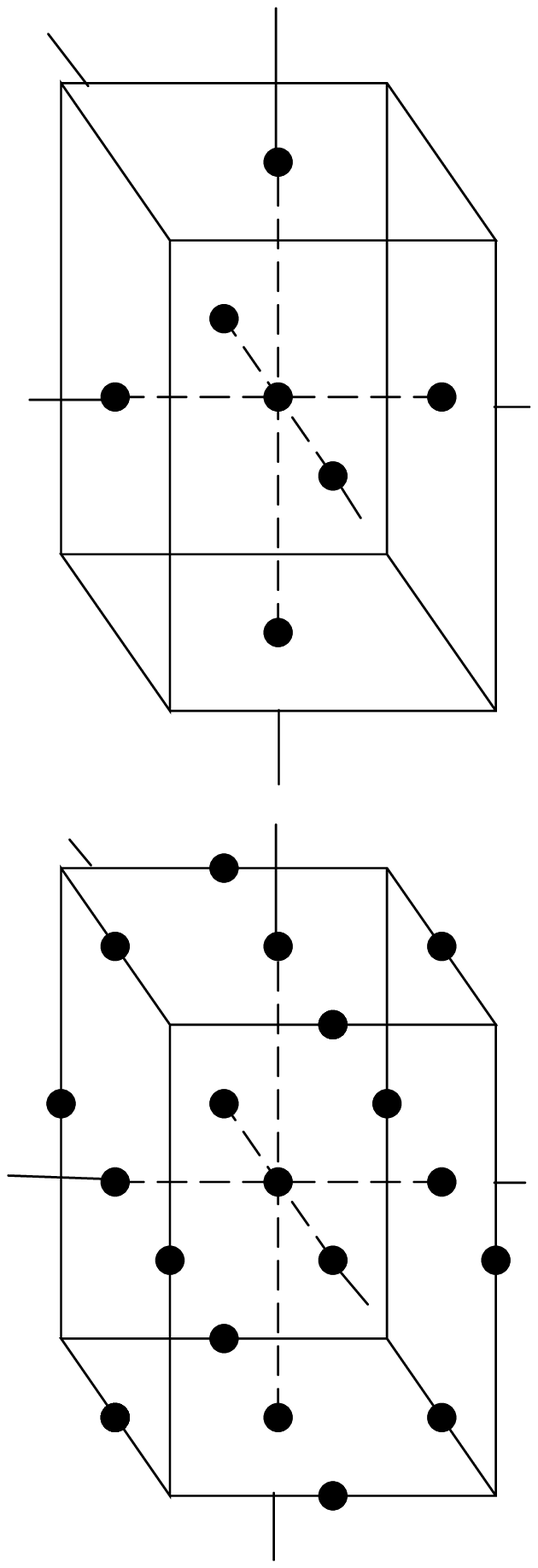}
\vspace{-1.5 in}
\caption{a) Top: An extension of the previous figure to $3D$ is shown.
The first figure shows a seven-point  stencil, for
the second-order derivatives in flat space.
b) Bottom: In curved space the second-order derivatives
include mixed derivatives ($\frac{\partial^2}{\partial x\partial y}$,
$\frac{\partial^2}{\partial x\partial z}$, and
$\frac{\partial^2}{\partial x\partial z}$, and their permutations).
We now have a 19-point stencil.
}
\label{fig_threedstencil}
\end{figure}

This gave an algebraic equation 
of the form
\be
\sum_{p,q,r=-1}^{p,q,r=+1}C_{i,j,k}^{p,q,r} \, N_{i+p,j+q,k+r} =0
\ee
where the sum was over non-repeating terms 
and had 19 coefficients $C$. 
It had to be solved at each grid zone except boundaries,

The conjugate gradient method involves solving the problem by
solving an analogous minimization problem.
Consider a set of difference equations
\be
{\bf A}f =b
\label{diffe}
\ee
where $f$ is the vector of unknowns (here our lapse value at each grid
zone) and ${\bf A}$ is an $N_g \times N_g$ square matrix ($N_g=$ number 
of grid zones) containing the finite differenced coefficients at
each grid zone. The right hand side is a vector of source terms.

Define the quadratic function
\be
V(f) = \frac{1}{2}\, f^{\dag} \, {\bf A} \,  f - b^{\dag} \, f.
\ee
Therefore 
\be
\frac{\partial V}{\partial f} = {\bf A} \, f - b ={\bf r}.
\ee
Thus solving (\ref{diffe}) is the same problem as minimizing the potential $V$.
Note that $A$ is the matrix
\be
A_{ij} = \frac{\partial^2\, V}{\partial f_i \partial f_j} ,
\ee
which is symmetric. So to apply this method
the finite difference equations have to be written in
symmetric form.

The minimization of the potential function is carried out by
generating a succession of search directions ${\bf s}$, with improved
minimizers $f_i$ at each stage. In a maximum of $N_g$ iterations
a solution is found. 

%FFig FIll in
However since we have a 19-point and not an $N_g$-point
stencil, a lot of the matrix elements will be zero.
(The matrix $A$ is a sparse matrix.) Since
the finite difference equations connect points immediately to either
side of a a given point, as one steps up the grid from point to
point, there is a definite structure to system. The matrix has
a banded diagonal structure with each diagonal band corresponding
to one of the 19 finite differenced coefficients. The coefficients
are stored as 19 three-dimensional arrays, meaning $19N_g$ not $N_g^2$ total
elements. This reduces the number of steps needed to minimize the function.
For details see~\cite{Press,ani2bh}.

\subsection{Results of Maximal Slicing for Equilibrium Boson Stars}

With the above machinery in place we ran the code with maximal slicing.
Our grid resolution was $dx=dy=dz=0.35$ and the grid dimension was $96\times 96\times 96$. The outer boundary condition on the boson fields was flat (equal
to the neighbor) and that
on the metrics was static.
In Fig. \ref{fig_grrmax3D} we show a plot of the radial metric as
a function of radius at various times. Clearly, this
metric was not the static metric expected of an equilibrium star.

\begin{figure}[t]

\begin{center}
\leavevmode
\makebox{\epsfysize=15cm\epsfbox{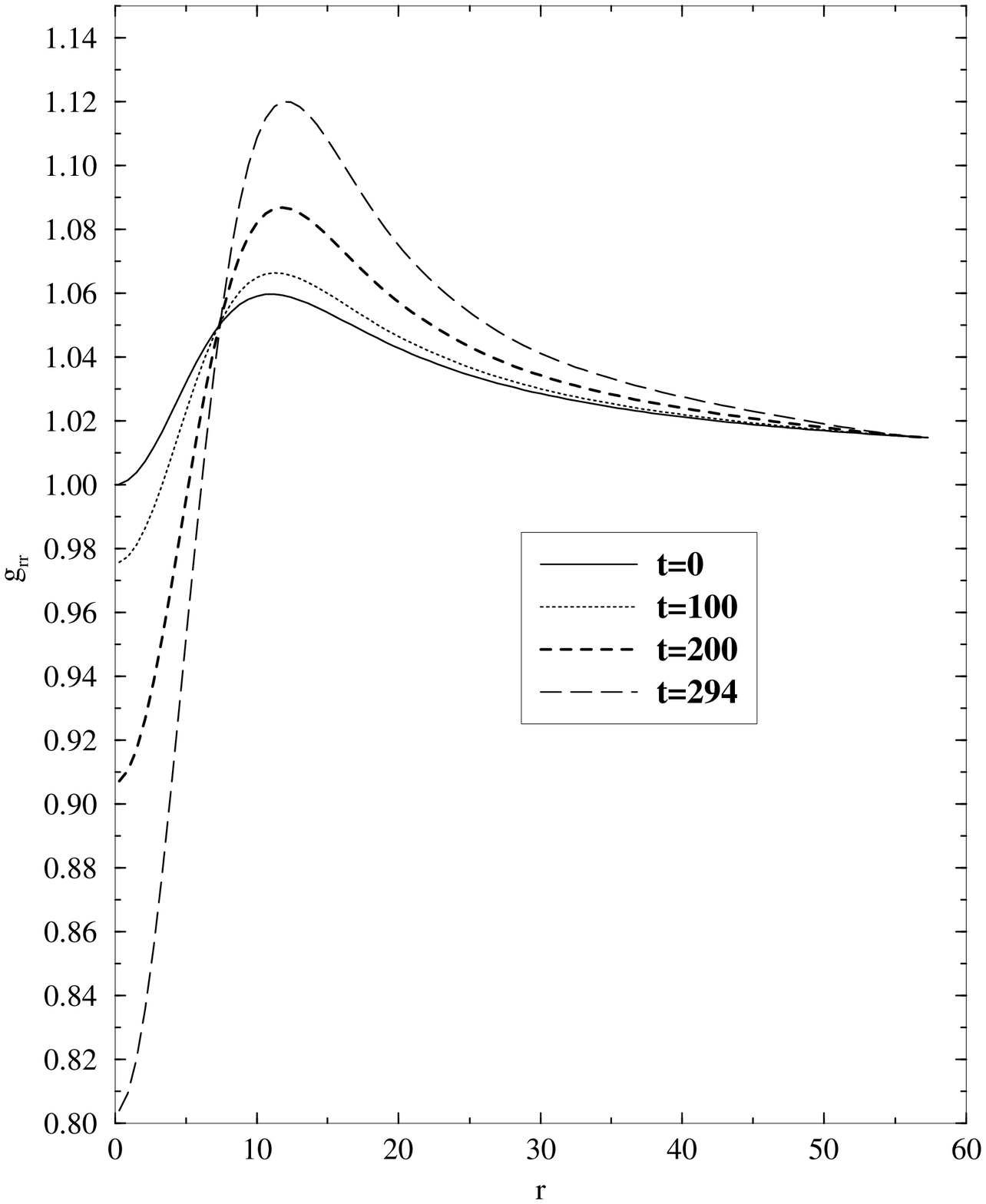}}
\end{center}

%%\begin{figure}
%%\hspace{-36pt}
%%\vspace{-130pt}
%%\epsfbox[0 0 450 500]{figs/grrmax3D.ps}
%%\epsfbox[0 -150 350 500]{figs/grrmax3De.ps}
\caption{ Radial metric drift, with maximal slicing, for an equilibrium boson
star of central field density $0.05$.
}
\label{fig_grrmax3D}
\end{figure}

The reason for this was a lack of gauge control. This is clear from
the metric function $g_{\theta\theta}$ which should really be $r^2$ in
a perfectly spherically symmetric run. For ease of visualization we
plot $g_{\theta\theta}/r^2$ as a function of $r$ at various times and
observe its drift from unity. This is shown in Fig. \ref{fig_gttmax3D}.

\begin{figure}[t]

\begin{center}
\leavevmode
\makebox{\epsfysize=15cm\epsfbox{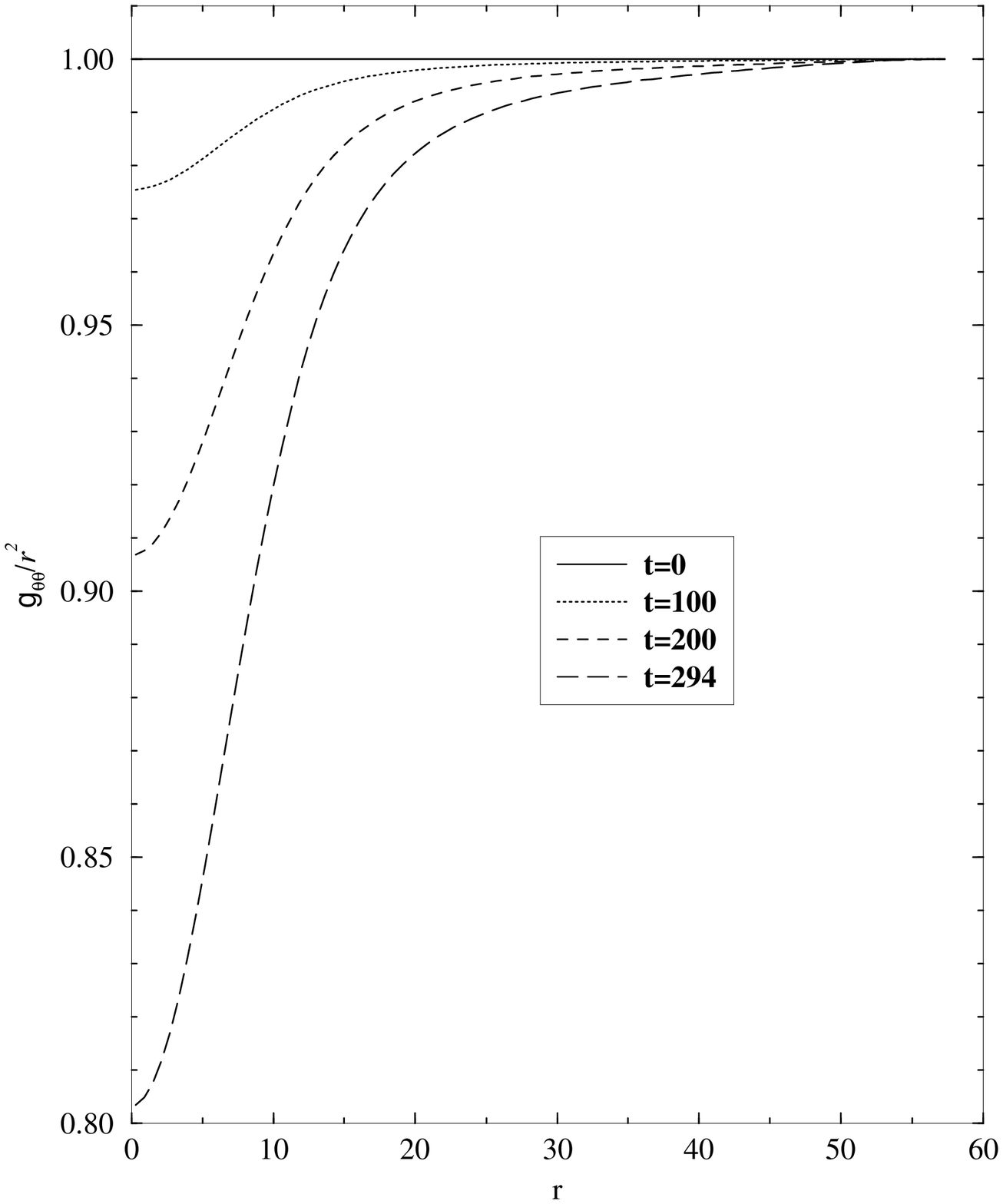}}
\end{center}
%%\begin{figure}
%%\hspace{-36pt}
%%\vspace{-130pt}
%\epsfbox[0 0 450 500]{figs/gttmax3D.ps}
%%\epsfbox[0 -150 350 500]{figs/gttmax3De.ps}
\caption{ The drift of $g_{\theta\theta}/r^2$ from unity is shown, as
an indicator of loss of gauge control, for a spherically symmetric
equilibrium boson star configuration.
}
\label{fig_gttmax3D}
\end{figure}
Attempts to regain gauge control are described in the next two subsections.

\subsection{Role of the Shift Vector}

As we saw in the description of $ADM$ in the introduction, we can
write $t^\mu$ as a sum of two terms, one along the normal to the
hypersurface and
the other the shift along the hypersurface.
So far we have been using normal coordinates
with shift vector equal to zero. Although this was the simplest
choice of shift, with this choice, we have squandered our
ability to implement gauge control by making specific choices of shift.
From the plots of $g_{\theta\theta}$ it is clear that there is
coordinate motion ($dr/dt$) due to numerical error. Maybe setting the
shift terms to zero as we have done left too much residual spatial coordinate
freedom? Might a shift term control this? The question now is how
to choose a shift. To start with, the obvious choice seemed to be
one that satisfied
\be
\frac{d\, g_{\theta\theta}}{dt}=0.
\ee
For simplicity we used a spherically symmetric form, with
\ben
K_{\theta\theta} &=&\frac{1}{2\,N}\left(\frac{d\, g_{\theta\theta}}{dt}
-2\,\Gamma^{i}{}_{\theta\theta}\,N_{i}\right)
\\ \nonumber
&=& \frac{1}{2\,N}\left(\frac{d\, g_{\theta\theta}}{dt}
+g^{rr}\,g_{\theta\theta,r}\,N_r\right).
\een
This yields, using the $\frac{d\, g_{\theta\theta}}{dt}=0$ condition,
\be
N^r = \frac{2\,N \,K_{\theta\theta}}{\frac{d}{dr}g_{\theta\theta}}.
\ee

Typically this shift was phased in over a period of time. That is, a
fractional amount of the
calculated value of the shift was used in the code. The fractional
amount was 
gradually increased from
0 to 1.
To circumvent dealing with boundary problems we isolated the points along the
diagonal $N^r(i,i,i)$ into a $1D$ array.
We interpolated points to
fill in values of $r$ along this $1D$ array that corresponded
to the $3D$ grid spacing (grid spacing along diagonal $=\sqrt{3}$ grid spacing
along $x$ direction if the $x$, $y$ and $z$ directions have the same number of
points). This was then extrapolated onto the $3D$ grid.
This meant we needed only
one boundary condition instead of worrying about all the boundary points
on a $3D$ grid.
Under our assumption of $N^\theta= N^\phi=0$ then,
\be
N^x = N^r \,\frac{x}{r}, \quad N^y  = N^r \, \frac{y}{r}
\quad \text{and} \quad N^z  = N^r \, \frac{z}{r}.
\ee

For the boundary condition we phased out the shift over the last 5 points
($N^r(nx-4)=0.8 \, N^r(nx-5)$, $N^r(nx-3) =0.6 \, N^r(nx-5)$ and so on till
$N^r(nx)=0$). Although the shift rose to a peak near the center of
the star, where $g_{\theta\theta}$ had drifted the most before, and
did have the intended effect of lowering the metric drift, it had a wavelike
structure outside the star. The wave moved out and on reaching the boundary invalidated the code.
A scheme of redefining the shift in the region where it first crossed the
axis was tried. We used an exponentially damped attachment of the form $N^r=A\,exp(-b\,r)$,
with A and b determined
at each time step by matching the actual shifts at points close to where it
crossed the axis with this expression.
This and other variations on the functional
form of the exponentially phased-out shift, lent a little additional
stability to the 
metric but the metric eventually developed a peak at the patching region of the
shift and invalidated the code. 
Nevertheless, $g_{\theta\theta}$ was fairly
well locked in place and so it was at least certain that the drift
had been due to coordinates drifting.

A kicking scheme was also tried.
The shift was fully phased in using the condition that best suited us at
that point.
A shift of the form $ae^{-br}$, matched just before
where the first shift peak crossed zero, was attached from
the matching point onwards. After fully phasing
in this shift, one discarded this method and tried to keep $g_{\theta\theta}$
locked at its value at full phase in, by taking a snapshot of it at that
time
and then measuring the deviation from that value at each subsequent time
step.
The new shift at a given time at any point was either increased or
decreased from its phased in value depending on whether $g_{\theta\theta}$
decreased or increased from its phased in value at that point.
Since $g_{\theta\theta} \sim r^2$ therefore $\frac{d}{dr} g_{\theta\theta} \sim 2 r $.
Thus coordinate drift could be controlled by a shift of the form:
\ben
N^r(t) &=& N^r (t_p)+ c \,\frac{dr}{dt} \\ \nonumber
&=& N^r(t_p) + \frac{c}{2\,r} \,
\frac{d g_{\theta\theta}}{dt},
\een
where $t_p$ was the phase in time and $c$ could be positive or negative
depending on which direction the
drift was heading. In principle $|c|$ was unity, but in
practice it tended to be a less and one had to tune this parameter. This
kicking scheme had previously been used to lock the horizon
in the black hole codes~\cite{anniho}.

Although we found definite coordinate control with these methods the
code always became unstable at the outer boundary. We present
comparisons of the radial metric function in simulations
with maximal slicing  with and without shift in Fig.~\ref{fig_kicknoshftgr}. In Fig.~\ref{fig_kicknoshftgt} we
present a comparison of $g_{\theta\theta}$ with and without a shift
phased in. Clearly the code was more stable with the shift phased in.
Unfortunately the code ran longer without a shift.

\begin{figure}[t]

\begin{center}
\leavevmode
\makebox{\epsfysize=15cm\epsfbox{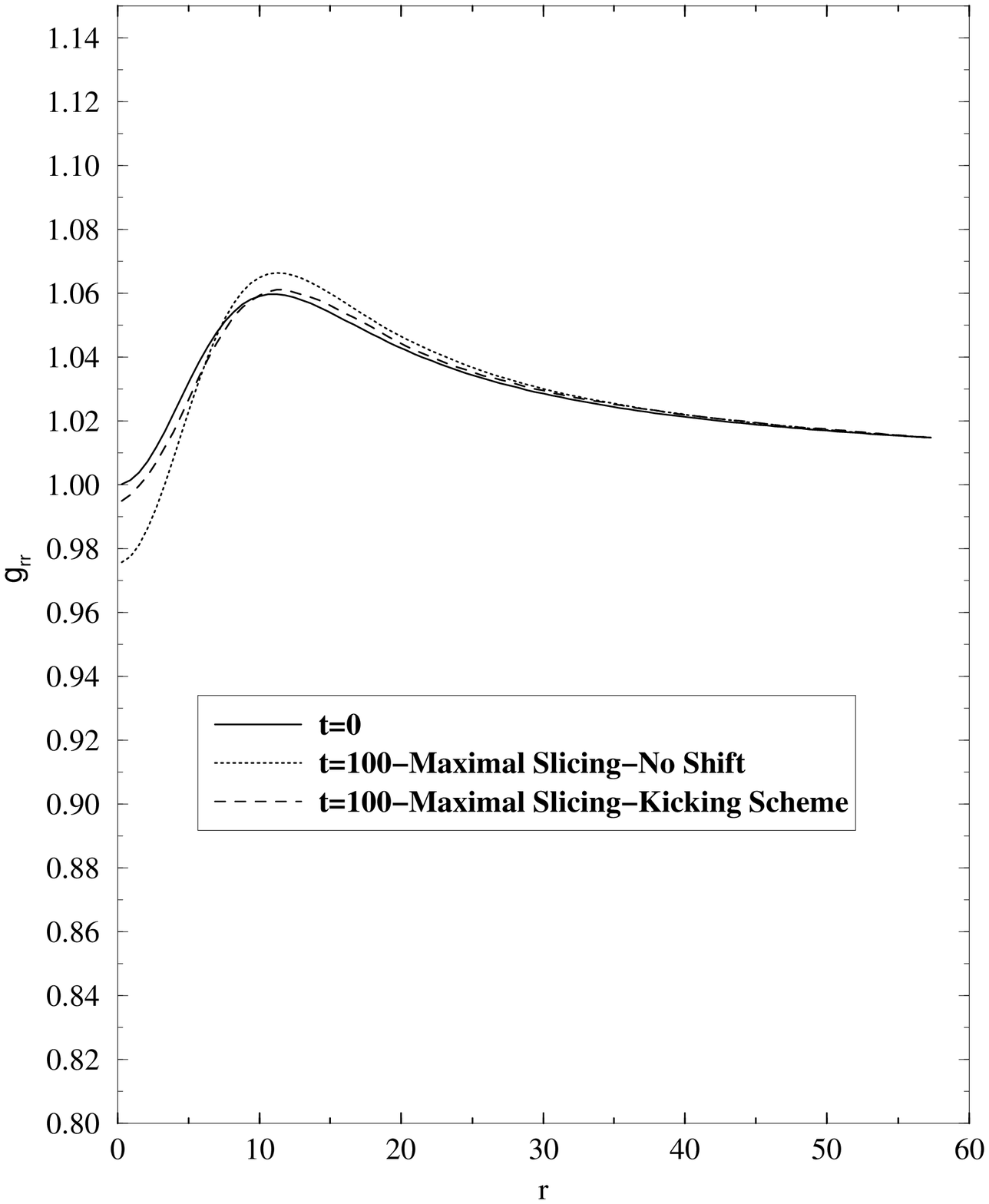}}
\end{center}
%%\begin{figure}
%%\hspace{-36pt}
%%\vspace{-130pt}
%\epsfbox[0 0 450 500]{figs/kicknoshftgr.ps}
%%\epsfbox[0 -150 350 500]{figs/kicknoshftgre.ps}
\caption{ The radial metric function, showing markedly more drift
at the same time under conditions of no shift, compared to a $g_{\theta\theta}$
freezing shift with a kicking scheme.
}
\label{fig_kicknoshftgr}
\end{figure}

\begin{figure}[t]

\begin{center}
\leavevmode
\makebox{\epsfysize=15cm\epsfbox{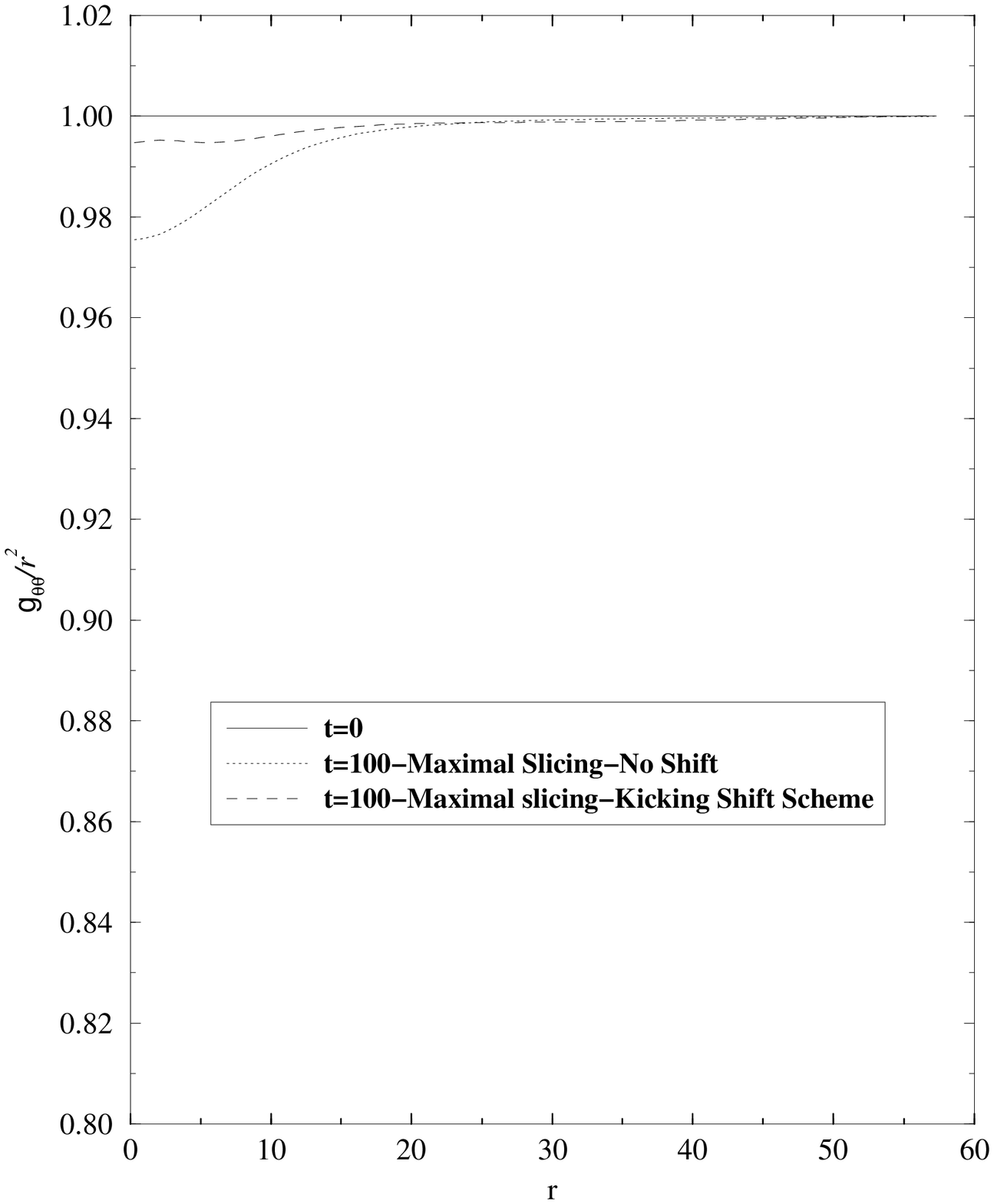}}
\end{center}
%%\begin{figure}
%%\hspace{-36pt}
%%\vspace{-130pt}
%\epsfbox[0 0 450 500]{figs/kicknoshftgt.ps}
%%\epsfbox[0 -150 350 500]{figs/kicknoshftgte.ps}
\caption{ A plot of $g_{\theta\theta}$  shows markedly more drift
at the same time under conditions of no shift compared to a
shift with a kicking scheme. Both are with maximal slicing.
}
\label{fig_kicknoshftgt}
\end{figure}

Even hybrid schemes were tried, with kicking done only in the outer region
and
the inner shift peak, that worked so well to lock $g_{\theta\theta}$, retained.
This led to grid point to grid point
instabilities that diffusion could not cure.

\subsection{The K--Driver}

With the shift schemes providing some control, but not to the degree that
was needed, it was timely that at this time the $K$--driver was
discovered. We were able to modify the $K$--driver for the boson star
problem~\cite{balacood}.

The whole point was that we needed long term gauge control, and
we had been using coordinate conditions that had been tried in
numerical relativity for the study of black hole evolutions. The
coordinate conditions themselves were becoming
unstable due to the accumulation of numerical errors during
long evolutions. Shift vectors, which shifted 
coordinate points so they did not move
normal to the constant time hypersurfaces, and time slicing conditions
determining the normal direction of motion of grid points, were
gauge freedoms 
satisfying equations that were themselves subject to numerical inconsistencies.

Consider the maximal slicing condition $K =\frac{\partial K }{\partial t}=0$
providing the elliptic equation for the lapse (\ref{maximalform}). Any
perturbation (numerical roundoff) in the lapse that made $K$ nonzero, could not be forced by
the slicing condition to get $K$ back to zero. The lapse might evolve
to its preferred value but errors made during its perturbation at
some time in the past could still create errors in the slicing
condition. The initial perturbation errors in the lapse might have been
caused by
numerical inaccuracies like errors due to
finite differencing of nonlinear equations, with low resolution.
The condition 
\be
\frac{\partial K}{\partial t} =0,
\ee
written with $K=0$ could evolve to something where $K\ne0$.
Instead one considered a ``K-driver'' slicing condition
\be
\frac{\partial K}{\partial t} +c\, K=0,
\ee
which without $K$ set to zero was
\be
c\,K  = N^{|i}{}_{|i} - N\,\left[K_{ij}\, K^{ij} - \fpi \, (\rho
+Tr\,S)\right] -N^i \, K_{,i}.
\ee
While (\ref{maximalform}) did not actively enforce the $K=0$
condition the above slicing condition drove
$K$ to zero when it was perturbed away from it.
The idea being that $K \sim e^{-ct}$
exponentially went to zero, in the long run, for a positive $c$.
One now could control the
stability of the code by fiddling with the value of $c$ that worked
best.

Finally, we actually used the K-driver for the boson star with
shift equal to zero to get our best results. We needed a parameter
$c=2$ near the center (typically about a third of
the grid) and then we let it smoothly damp away ($c e^{-
\frac{1}{3}(r-R/3)}$ with $R=r(nx,ny,nz)$ for all values of $r > R/3$).
This damping factor was needed to prevent boundary problems from
arising in the code. Since the drift
was typically observed where the star was concentrated, this condition
worked very well. We present a comparison of the zero shift maximal
slicing versus the zero shift K-driver in Fig.~\ref{fig_kdrivgrr}.

\begin{figure}[t]

\begin{center}
\leavevmode
\makebox{\epsfysize=15cm\epsfbox{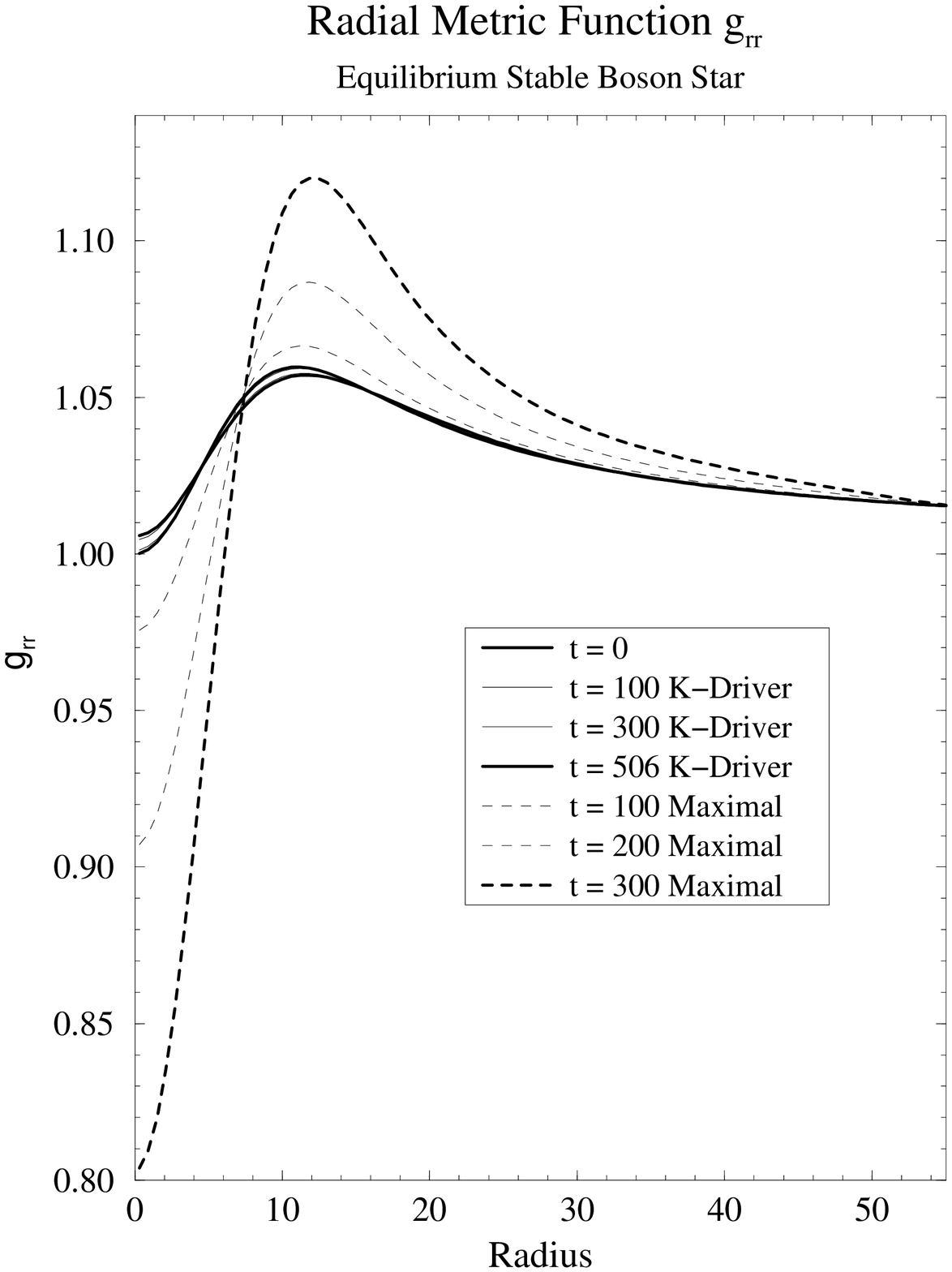}}
\end{center}
%%\begin{figure}
%%\hspace{-36pt}
%%\vspace{-130pt}
%\epsfbox[0 0 450 500]{figs/kdrivgrr.ps}
%%\epsfbox[0 -150 350 500]{figs/kdrivgrre.ps}
\caption{A comparison of the metric function drift in the time
evolution of an equilibrium boson star with maximal slicing and
K--driving condition is shown. The latter is clearly capable of
controlling metric drift.
}
\label{fig_kdrivgrr}
\end{figure}

\section{Spherically Symmetric Perturbations of a Boson Star.}

Once the $3D$ code had been stabilized to correctly display the
true evolutions of an equilibrium boson star, 
we went to the next step of our code tests. This was to take
a boson star and observe how the code displayed its evolutions under
spherical perturbations. A well tested 1-D code provided us with initial data.
The initial data was time symmetric and the momentum constraints
were satisfied by ensuring that the perturbation kept
$\psi_1=\dot \psi_2$ and $\psi_2=-\dot \psi_1 =0$ for the two boson fields.
We show an oscillation of the radial metric in figures
\ref{fig_oscgrrdown3D} and \ref{fig_oscgrrup3D} (the former showing the
downward part of the oscillation and the latter the upward part of it)
of a boson star of original mass $0.5916 \, M_{Pl}^2/m$ which has been reduced by about
$10\%$ after perturbation.
%|Ffig check this data.

\begin{figure}[t]

\begin{center}
\leavevmode
\makebox{\epsfysize=15cm\epsfbox{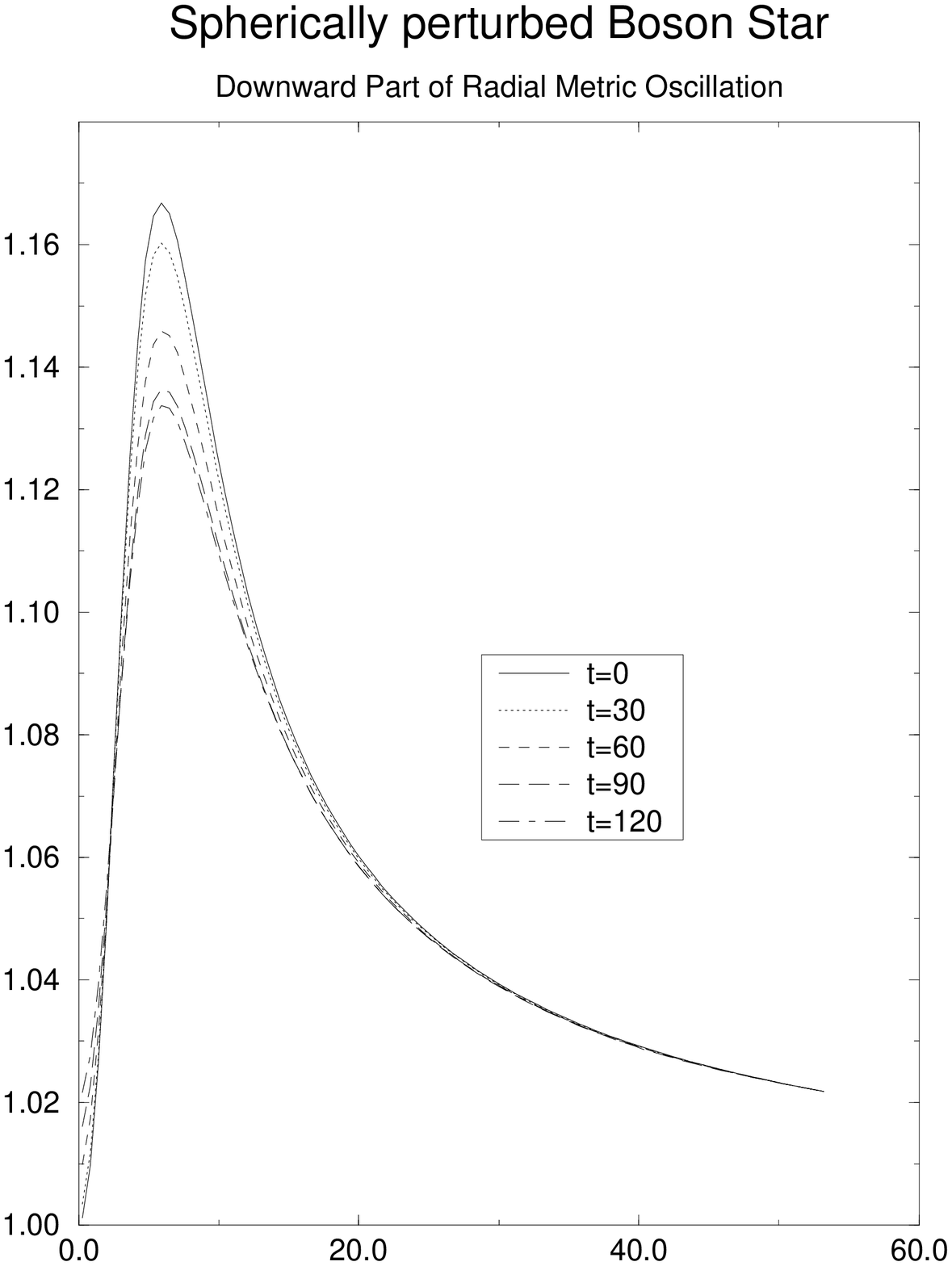}}
\end{center}
%%\begin{figure}
%%\hspace{-36pt}
%%\vspace{-130pt}
%\epsfbox[0 0 450 500]{figs/Sphpertgrrdow.ps}
%%\epsfbox[0 -150 350 500]{figs/Sphpertgrrdowe.ps}
\caption{We show the downward part of the radial metric oscillation of
a boson star that has been perturbed by a spherically symmetric
perturbation. As the star expands the radial metric descends.
}
\label{fig_oscgrrdown3D}
\end{figure}

\begin{figure}[t]

\begin{center}
\leavevmode
\makebox{\epsfysize=15cm\epsfbox{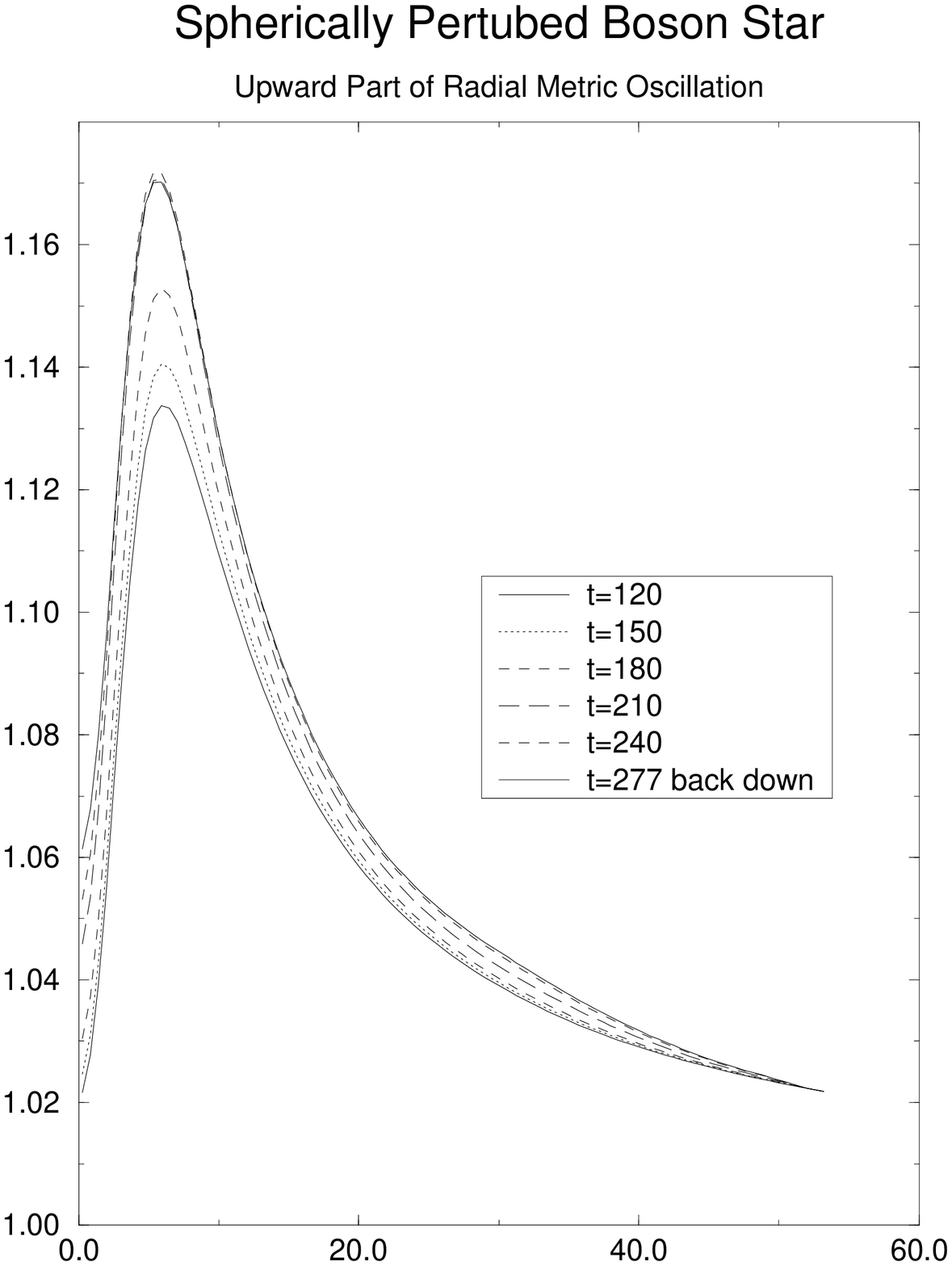}}
\end{center}
%%\begin{figure}
%%\hspace{-36pt}
%%\vspace{-130pt}
%\epsfbox[0 0 450 500]{figs/Sphpertgrrup.ps}
%%\epsfbox[0 -150 350 500]{figs/Sphpertgrrupe.ps}
\caption{ As the star starts to contract after its expansion in
reaction
to the perturbation, its radial metric starts to rise
again.
This part of its motion is shown in the figure. At the last step we start
to see it reverse its motion again. The central field density
of this star is $\psi_1(0) = 0.15$ in the dimensionless units discussed in
chapter 2. Its mass is $0.5916 \, M_{Pl}^2/m$.
}
\label{fig_oscgrrup3D}
\end{figure}

A successful code test we performed was in
calculating the mass loss of a boson star under spherical
perturbations and comparing it to the results we had in $1D$.
In the 1-D code we calculated the mass of the star using the
$ADM$ mass. Just before the sponge region (see chapter 2) we
calculated the mass of the star from the radial metric
\be
M= \frac{r}{2}\,\left(1-\frac{1}{g_{rr}}\right).
\ee
We could do this because in this region the field was small enough
for the spacetime to be considered to be Schwarzschild. This
was obviously not a good formula for the $3D$ code with static outer
boundary conditions. We needed a more global measure of the mass so that
dominant contributions from the interior of the star would not
be affected by outer boundary errors. In order to calculate
the mass loss we used the fact that $T^{tr}$ represents
the energy flux ($ergs/cm^2 sec$) in the radial direction.
Thus integrating $T^{tr}$ over time and elemental area provided a measure
of the mass loss at a given time
due to scalar radiation. We calculated
the mass loss in the $1D$ code by using the
$ADM$ mass formula as well as integrated $T^{tr}$ and
compared the results to the $3D$ case.
We ran the $1D$ code
with the same resolution as the $3D$ code and put
detectors at the same position. We found very good agreement
in all cases with different detectors and different types
of perturbations. We show two plots, one with a localized
spherical perturbation and the other with a global spherical perturbation,
in Fig.~\ref{fig_Elocspher} and Fig.~\ref{fig_Eglobalspher} respectively.
The latter plot also demonstrates the effect of resolution on the matching
of the energy of scalar radiation.

\begin{figure}[t]

\begin{center}
\leavevmode
\makebox{\epsfysize=15cm\epsfbox{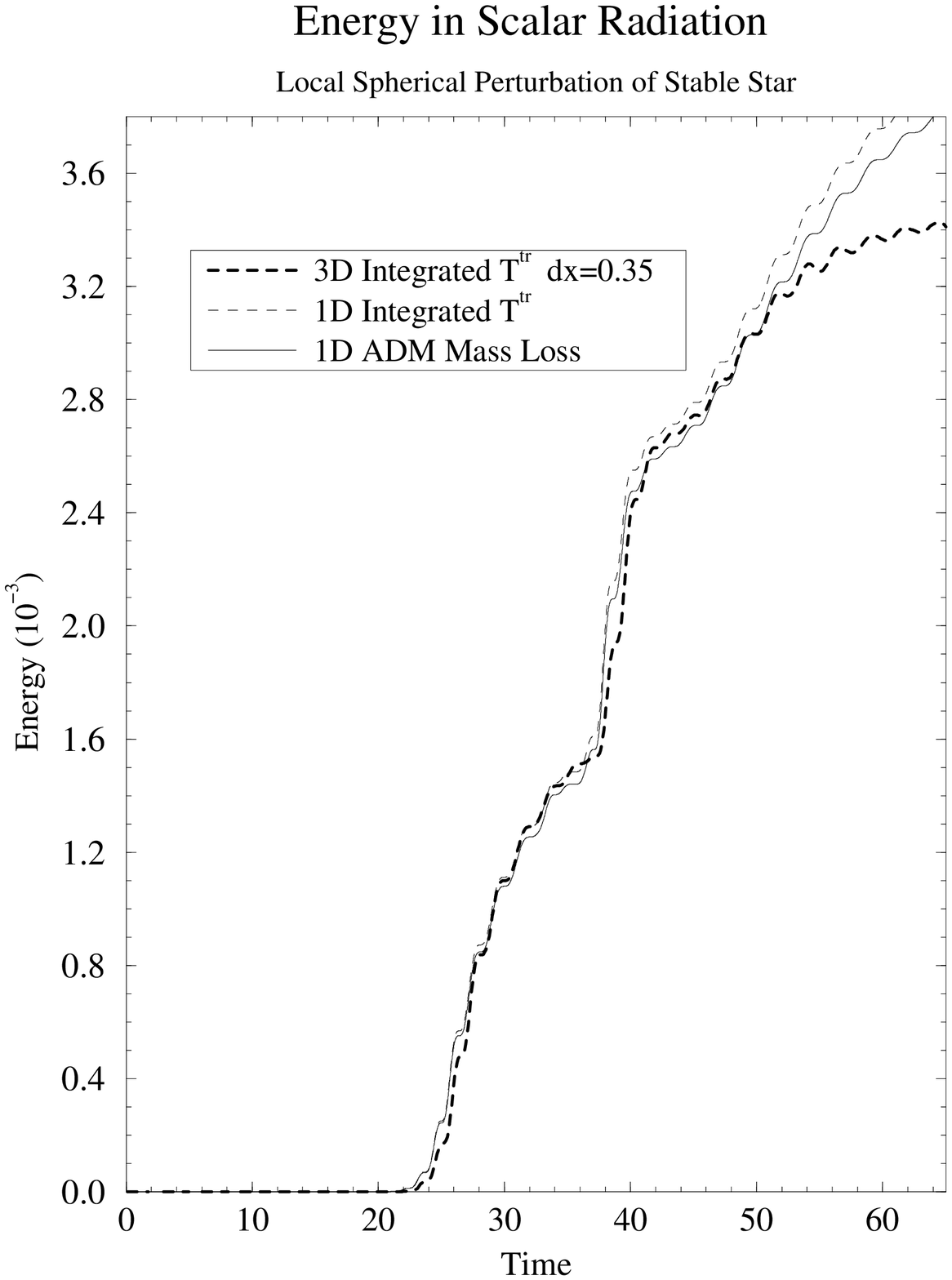}}
\end{center}
%%\begin{figure}
%%\hspace{-36pt}
%%\vspace{-130pt}
%\epsfbox[0 0 450 500]{figs/Elocspher.ps}
%%\epsfbox[0 -150 350 500]{figs/Elocsphere.ps}
\caption{The mass loss in $1D$ computed using the $ADM$ mass formula, and
integrated energy flux $T^{tr}$, is plotted along with the integrated energy
flux $T^{tr}$ using the $3D$ code for a local spherical perturbation.
The closeness of values is remarkable until the effects of
boundary reflections starts affecting the accuracy of the $3D$ code.
}
\label{fig_Elocspher}
\end{figure}

\begin{figure}[t]

\begin{center}
\leavevmode
\makebox{\epsfysize=15cm\epsfbox{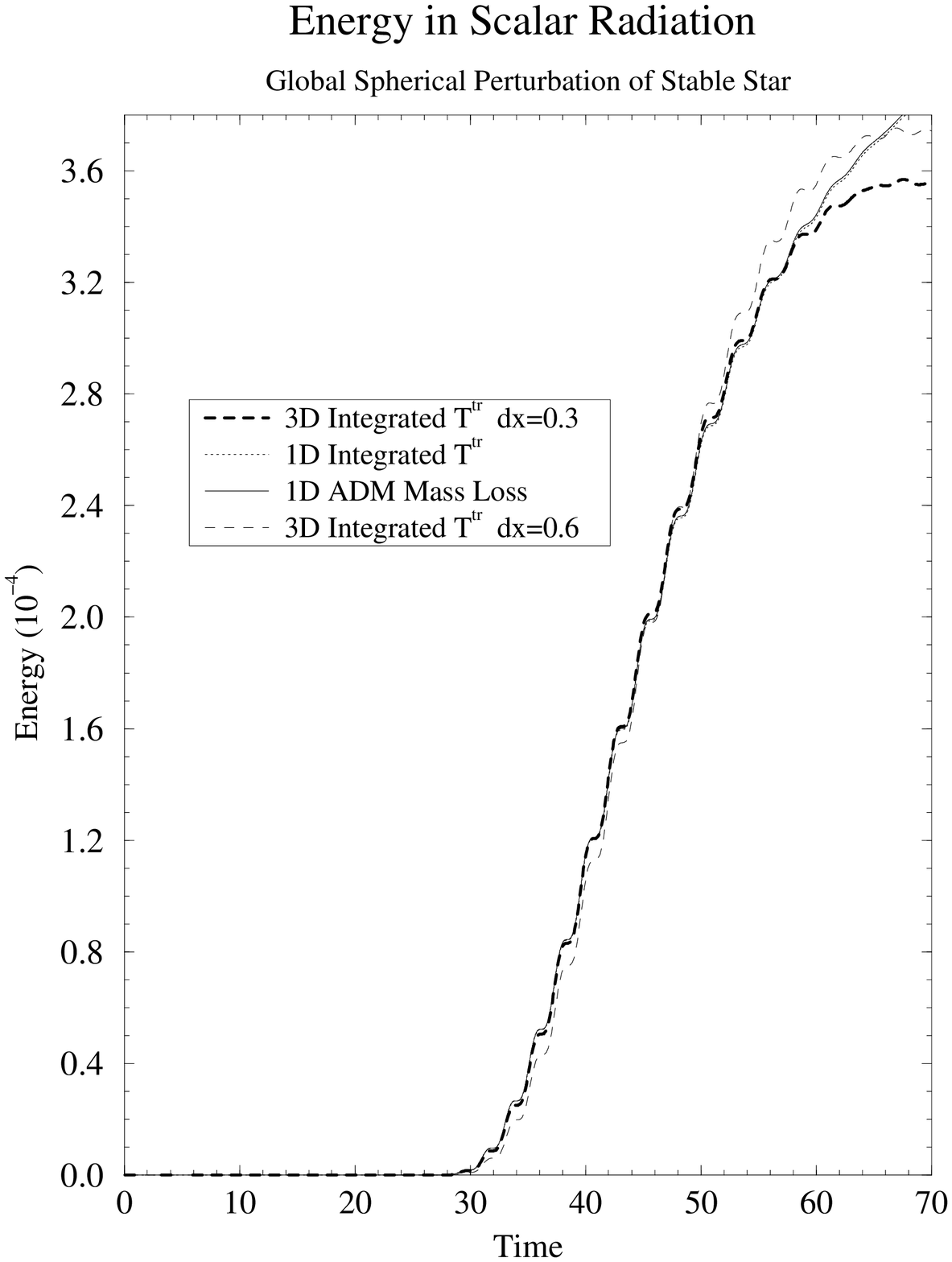}}
\end{center}
%%\begin{figure}
%%\hspace{-36pt}
%%\vspace{-130pt}
%\epsfbox[0 0 450 500]{figs/Eglobalspher.ps}
%%\epsfbox[0 -120 350 500]{figs/Eglobalsphere.ps}
\caption{The mass loss in $1D$ computed using the $ADM$ mass formula, and
integrated energy flux $T^{tr}$, is plotted along with the integrated energy
flux $T^{tr}$ using the $3D$ code for a global spherical perturbation.
The closeness of values is remarkable until the effects of
boundary reflections starts affecting the accuracy of the $3D$ code. In
addition the effect of grid resolution is shown by also plotting a lower
resolution $3D$ calculation which shows greater discrepancy from the
$1D$ result but which is still quite good.
}
\label{fig_Eglobalspher}
\end{figure}

This was indeed the key code test that showed that the
code was performing well given the computational resources available.

\section{Non-Spherical Perturbations of a Boson Star}

Finally we were ready to study the behavior of boson stars under
nonspherical perturbations. In the Introduction to this thesis
we have already stated how Einstein's equations under nonspherical
perturbations can give rise to a wave equation. These gravitational waves,
along with the scalar radiation due
to our scalar field, are the two big signatures of the boson star.
We have seen in the Introduction, that the {\em Zerilli functions,
Newman--Penrose Spin Coefficient} $\Psi_4$ and
the {\em Bel--Robinson vector} can provide information about the
nature of this radiation and the gravitational waveform.
One concern was of course whether one could extract the full waveform
since the outer boundary effects could cause inaccuracies
after a fairly short time as indicated by the results of
the last subsection. Since the waveform extractions had to
take place in the outer region of the star, which could effectively be
regarded as a vacuum region, reflections from the
boundary would cause problems earlier than if our detectors were placed
well within the star. Here was another advantage to dealing
with a boson star. As discussed in the introduction of
this thesis (\ref{subsection:bs}), in~\cite{yoshinonrad}
it was actually shown that the normal modes of boson stars
damp out on a short time scale.
The short time scale allowed us to measure all of the gravitational
wave signal that contributes significantly to the gravitational energy output.

We used a grid with one octant, making sure our perturbations were
axisymmetric. We used an equilibrium configuration determined
from the $1D$
code and imposed on it such a perturbation. These were typically
weak perturbations. Now the constraint equations had to be
solved on the initial time slice in order to provide consistent fields and metric components.
This was done using an initial value solver that used a relaxation
method to solve the elliptic equation.
\begin{itemize}
\item{Basics of the Initial Solver}
\end{itemize}

We used the initial solver to provide new metric components
by solving the Hamiltonian constraint equation (\ref{hamilform})
on the initial time slice. York's procedure was
followed~\cite{York79}. We specified the initial scalar
field configuration and an initial guess for the three-metric. The idea was to solve
for a conformal factor $\phi$ so as to get the real metric from the
guess through the relation
\be
\gamma_{ij} = \phi^4 \tilde \gamma_{ij}
\ee
where the trial metric was $\tilde \gamma_{ij}$. Assuming time-symmetric
data, $K_{ij}=0$, the Hamiltonian constraint equation takes the
form
\be
R = 16\pi\rho.
\ee
With the updated metrics (in terms of the guess and the conformal transformation)
we could calculate the connection coefficients in terms of
those corresponding to the guess and also get the Ricci scalar in
terms of the guess
\be
R =\tilde R \phi^{-4} - 8 \phi^{-5} {\tilde \partial_i}{\tilde \partial^i}\phi.
\ee
The field configuration was retained as initially specified. This,
along with
$\gamma_{ij} = \phi^4 \tilde \gamma_{ij}$, was used to calculate $\rho$. 
Then the Hamiltonian constraint $R=16\pi \rho$ was tested
to some tolerance. So the problem was one of
determining the conformal factor $\phi$ satisfying
\be
\tilde R \phi^{-4} - 8 \phi^{-5} {\tilde \partial_i}{\tilde \partial^i}\phi
= 16\pi \rho(\phi,\pi, \phi^4 \tilde \gamma_{ij}).
\label{conforma}
\ee
This could be solved using the $CMSTAB$ (conjugate gradient method)
by treating
$\phi^5 \times$ the right hand side ($RHS$)
of the above equation as
a constant for each iteration. The version we used in the
code uses a Jacobi type relaxation technique. This
was to write an equation for the conformal factor in
the form
\be
\partial _t \phi = \epsilon \, (LHS -RHS),
\ee
where $LHS$ and $RHS$ were the left and right hand sides of
the constraint (conformal) equation (\ref{conforma}). This gave the $n+1$th iteration
in terms of the $n$th one
\be
\phi^{n+1} =\phi^{n} + dt \, \epsilon \, (LHS-RHS),
\ee
so that one could relax to the solution and stop when $LHS$= $RHS$ (which
kept getting updated) was satisfied
within some tolerance. Typically $\epsilon  dt$ was chosen
to be a fraction of the square of the grid spacing.

Our perturbations were small enough that the initial guess for the metric
was close enough to the actual value that this was not extremely time
consuming.

\section{Results of Non-Spherical Perturbations}

Far from the source, $\Psi_4$ represents an outgoing wave. It is thus normal
to a two sphere of constant radius and the energy this wave carries
over the whole sphere per second must be its integrated intensity.
The Energy calculations from the Newman-Penrose scalar are then
based on the
formula
\be
E_{NP} = \frac{1}{4\pi}\int\,dt\int\,
\left[
\int_{0}^{t} \,dt' \Psi_4 
\right]^2 
r^2 d\Omega.
\ee

We first estimated the degree of asymmetry in the system when subjected to
spherical perturbations. These asymmetries could be due to numerical
errors induced by, for instance, 
always finite differencing second derivatives in
a particular order. For example taking the
$x$ derivative before the $y$ derivative whenever one needed to
find $\frac{\partial^2}{\partial x \,  \partial y}$ of any function.
In Fig.~\ref{fig_enercomppsi4} we show a
plot of the comparison of gravitational
energy measured at a detector for the case of a star that is unperturbed,
the same star under small and large spherical perturbations, and under
a small nonspherical perturbation.

\begin{figure}[t]

\begin{center}
\leavevmode
\makebox{\epsfysize=15cm\epsfbox{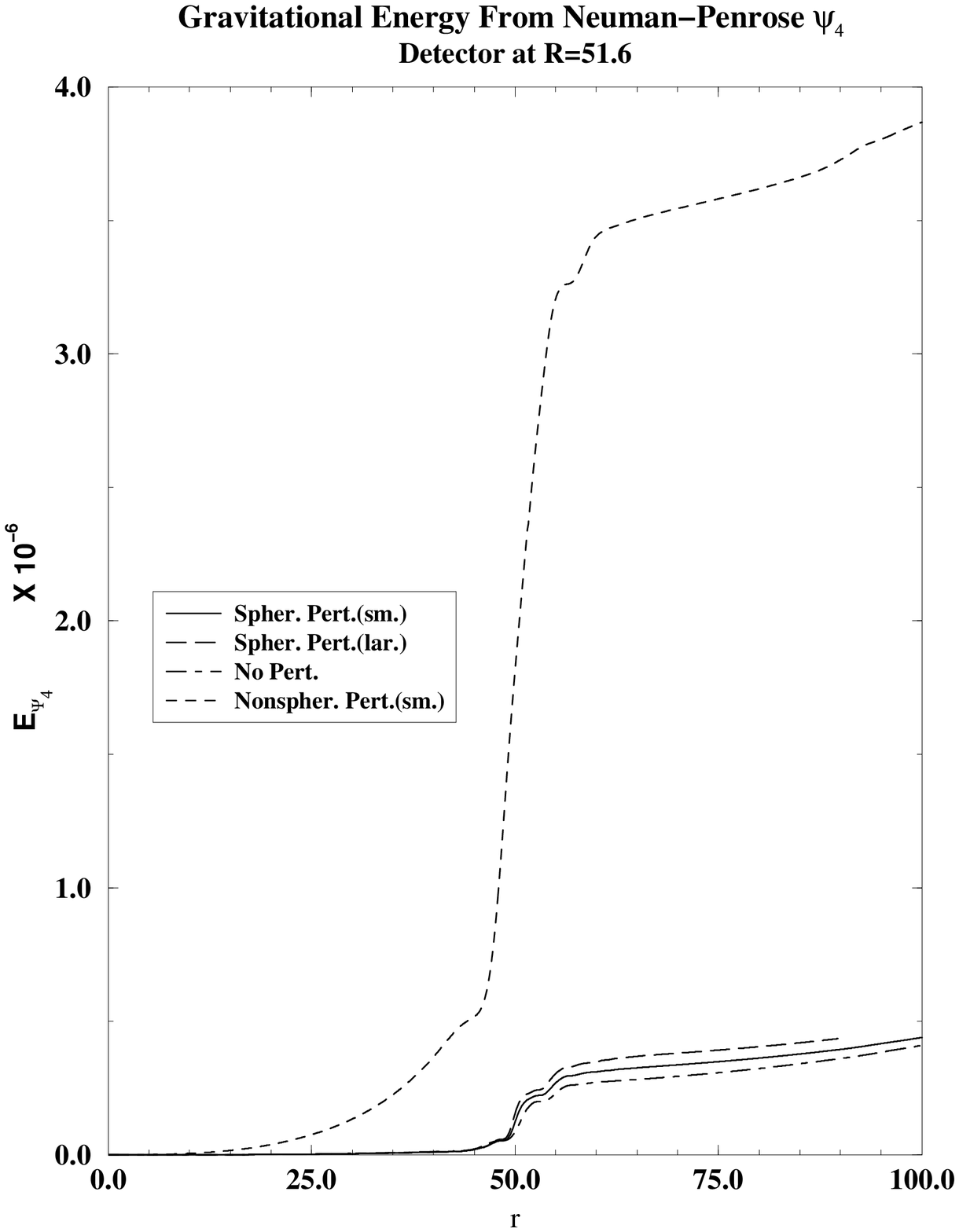}}
\end{center}
%%\begin{figure}
%%\hspace{-36pt}
%%\vspace{-130pt}
%\epsfbox[0 0 450 500]{figs/ThesEnercomppsi42.ps}
%%\epsfbox[0 -150 350 500]{figs/ThesEnercomppsi42e.ps}
\caption{ A comparison of gravitational energy output for a star
of unperturbed central field density $0.15$ for spherical and
nonspherical perturbations. 
While the level of this `background'
is not insignificant it is sufficiently small that the
results for the nonspherical perturbations may be trusted.
}
\label{fig_enercomppsi4}
\end{figure}

The energy was also calculated by other means, namely the Bel--Robinson
vector $p^i$:
\be
\frac{dE}{dt} =\frac{1}{4\pi}\int\,\left[\int_{0}^{t}\,dt'
\left(\pm\sqrt{|p^x x+p^y y +p^z z|/2}\right)\right]^2 r^2d\Omega,
\ee
as well as from the Zerilli functions $\Psi_Z$:
\be
\frac{dE}{dt} = \frac{1}{32\pi} \left(\frac{\partial \Psi_Z}{\partial
t}\right)^2.
\ee

We show a perturbed stable star with unperturbed central
field density $0.1414$ that had been given a very tiny non-spherical
perturbation so that we could obtain its quasinormal modes.
Fig.~\ref{fig_Enerps4BRzer} shows three plots of the gravitational energy
output as detected by three detectors placed at various points on the
grid. The first plot shows a Bel--Robinson calculation, the second a
Newman--Penrose calculation and the third a Zerilli calculation.
While the former two have consistent results the Zerilli plot shows
much more sensitivity to detector positioning. That is, boundary
effects as well as how close one is to being in a vacuum, seem to affect
the Zerilli calculation more than the others. In the case of the
Bel--Robinson and Newman--Penrose calculation the outermost detector
is most affected by the boundary 
errors, while the other
two give curves that flatten out. In Fig.~\ref{fig_Enercompsps4BRzer}
we show
three plots, each now an energy comparison by the three methods, at
a particular detector. The Bel--Robinson and Newman--Penrose
calculations are most sensitive to how close one is to being
in a vacuum while the Zerilli is most affected by boundary problems.
Taking the middle detector as the best compromise between boundary effects
and degree of vacuum, we see that the Bel--Robinson and Newman--Penrose
calculations match up very well for the most part until the point
where they level off, with about $12\%$ differences at levelling
off positions.

\begin{figure}[t]

\begin{center}
\leavevmode
\makebox{\epsfysize=15cm\epsfbox{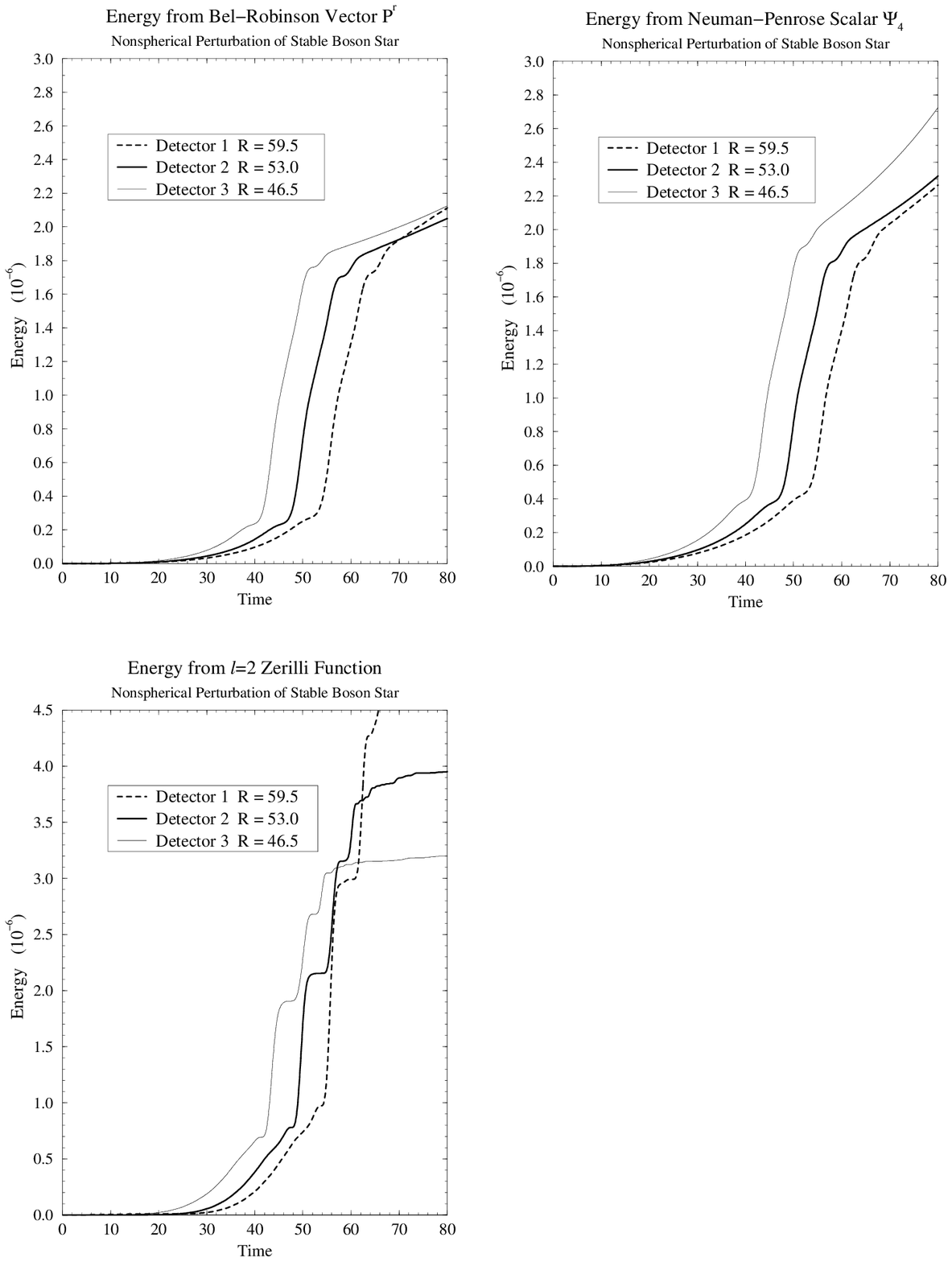}}
\end{center}
%%\begin{figure}
%%\hspace{-36pt}
%%\vspace{-130pt}
%\epsfbox[0 0 450 500]{figs/Enerps4BRzer.ps}
%%\epsfbox[0 -130 350 500]{figs/Enerps4BRzere.ps}
\caption{ 
The gravitational wave energy output of a non-spherically
perturbed stable boson star of central
field density $0.1414$ is shown as measured at three detectors at the
outer region of the grid. Three plots are shown, the first showing
measurements using the Bell--Robinson vector, the second using
the Newman--Penrose $\Psi_4$ parameter and the third using the
Zerilli functions. The most erratic are the Zerilli measurements.
}
\label{fig_Enerps4BRzer}
\end{figure}

\begin{figure}[t]

\begin{center}
\leavevmode
\makebox{\epsfysize=15cm\epsfbox{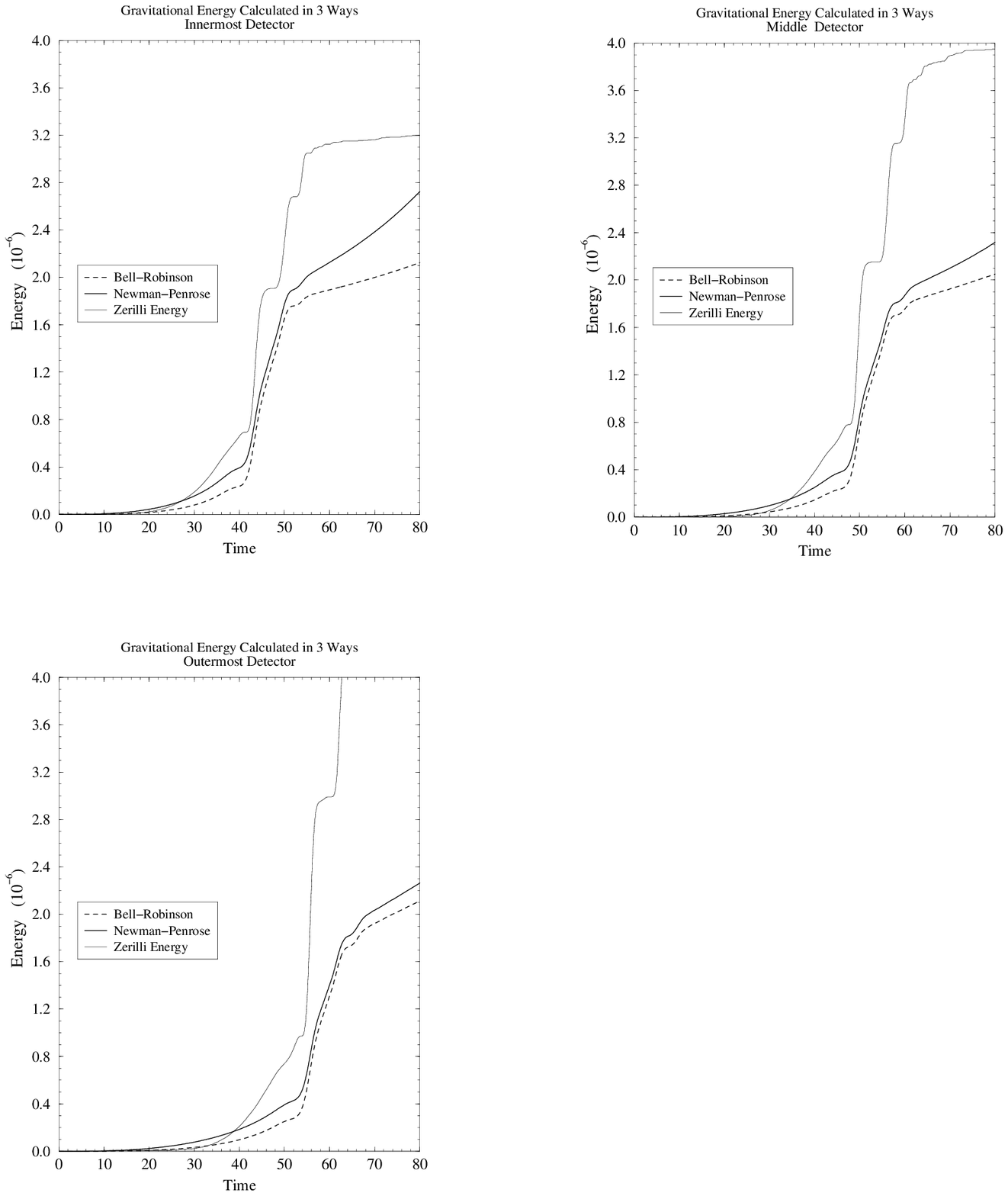}}
\end{center}
%%\begin{figure}
%%\hspace{-36pt}
%%\vspace{-130pt}
%\epsfbox[0 0 450 500]{figs/Enercompsps4BRzer.ps}
%%\epsfbox[0 -150 350 500]{figs/Enercompsps4BRzere.ps}
\caption{A comparison of the gravitational energy output using
the three measures defined in the previous figure is made
at three detectors for the non-spherically perturbed stable star described
in the previous figure. While the Bel--Robinson and Newman--Penrose 
$\Psi_4$ calculations are quite close 
the Zerilli
data is more erratic.
}
\label{fig_Enercompsps4BRzer}
\end{figure}

The scalar radiation still dominates the system's energy output as
seen in Fig.~\ref{fig_thesscalgrav} where we have plotted a tenth
of the scalar radiation and the total gravitational energy radiation
from Bel--Robinson calculations so as to be able to show them on the
same plot. The matter radiation clearly moves out slower. The star's
configuration after it gets rid of the gravitational radiation
becomes spherically symmetric, and it continues its evolution
through spherical oscillations.

\begin{figure}[t]

\begin{center}
\leavevmode
\makebox{\epsfysize=15cm\epsfbox{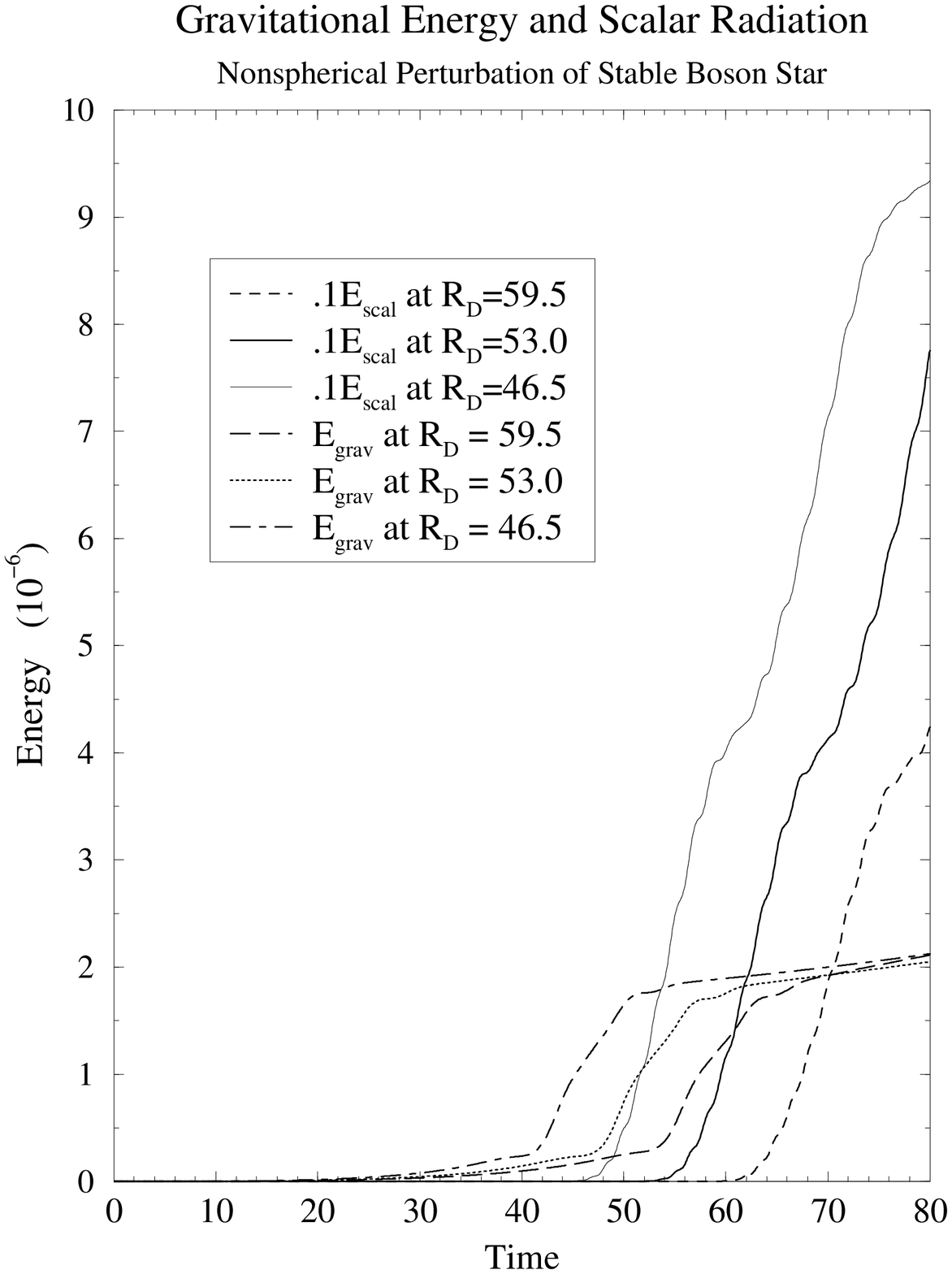}}
\end{center}
%%\begin{figure}
%%\hspace{-36pt}
%%\vspace{-130pt}
%\epsfbox[0 0 450 500]{figs/ThesEscalgrav.ps}
%%\epsfbox[0 -150 350 500]{figs/ThesEscalgrave.ps}
\caption{The stable star previously described, when subjected to
a nonspherical perturbation, emits gravitational and scalar radiation.
The gravitational radiation is far less in amount compared
to the dominant scalar energy radiated. The scalar radiation, due
to the bosonic matter, travels slower than the gravitational wave due
to curvature perturbations.
}
\label{fig_thesscalgrav}
\end{figure}

The
nonradial quasinormal modes for the configuration with central field
density $0.1414$ have been calculated in~\cite{yoshinonrad}
using perturbation theory. We have briefly
described this in the introduction of this thesis. We have looked
at our {\em Newman--Penrose} $\Psi_4$ function and performed a two-mode 
fit of their lowest two $l=2$ modes. Our results are shown in
fig.~\ref{fig_psi4fit}. Clearly we have a very close agreement
to their results. After the first burst of energy, the star
moves into the quasinormal mode and stays there till boundary effects
become significant.

\begin{figure}[t]

\begin{center}
\leavevmode
\makebox{\epsfysize=15cm\epsfbox{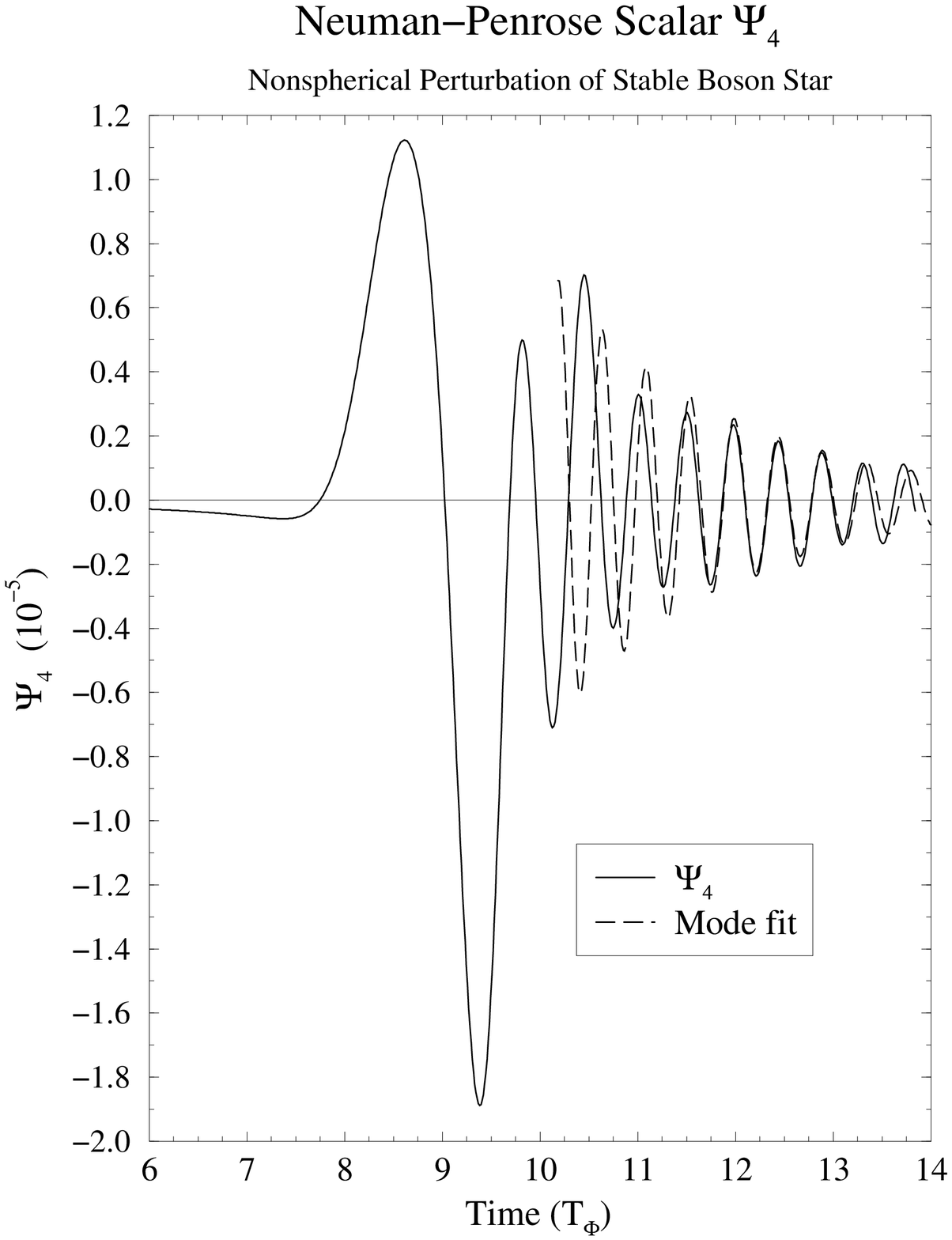}}
\end{center}
%%\begin{figure}
%%\hspace{-36pt}
%%\vspace{-130pt}
%\epsfbox[0 0 450 500]{figs/Thespsi4fit2.ps}
%%\epsfbox[0 -150 350 500]{figs/Thespsi4fit2e.ps}
\caption{The gravitational waveform $\Psi_4$ is shown, as is a quasinormal
mode fit for the stable boson star previously described. The star gradually starts oscillating in its quasinormal mode
and then at late times boundary effects start affecting the system.
The plot has time in units of the underlying oscillation of the
scalar field (time period of $2\,\pi$).}
\label{fig_psi4fit}
\end{figure}

We also compared two evolutions of a stable star, one with a small and the
other with a large nonspherical perturbation. We observed
that the energies of the gravitational waves and scalar radiation
were greater in the case of the large perturbation. An infinitesimal
perturbation caused the star to oscillate at the quasinormal mode of the original stable star
while a substantial perturbation caused it to move to a new
configuration, going through the modes of intermediate states. A
Fourier transform for a star of central field density $0.15$ with
a small and large perturbation exhibits the differences in frequency
in Fig.~\ref{fig_ftpsi4}.

\begin{figure}[t]

\begin{center}
\leavevmode
\makebox{\epsfysize=15cm\epsfbox{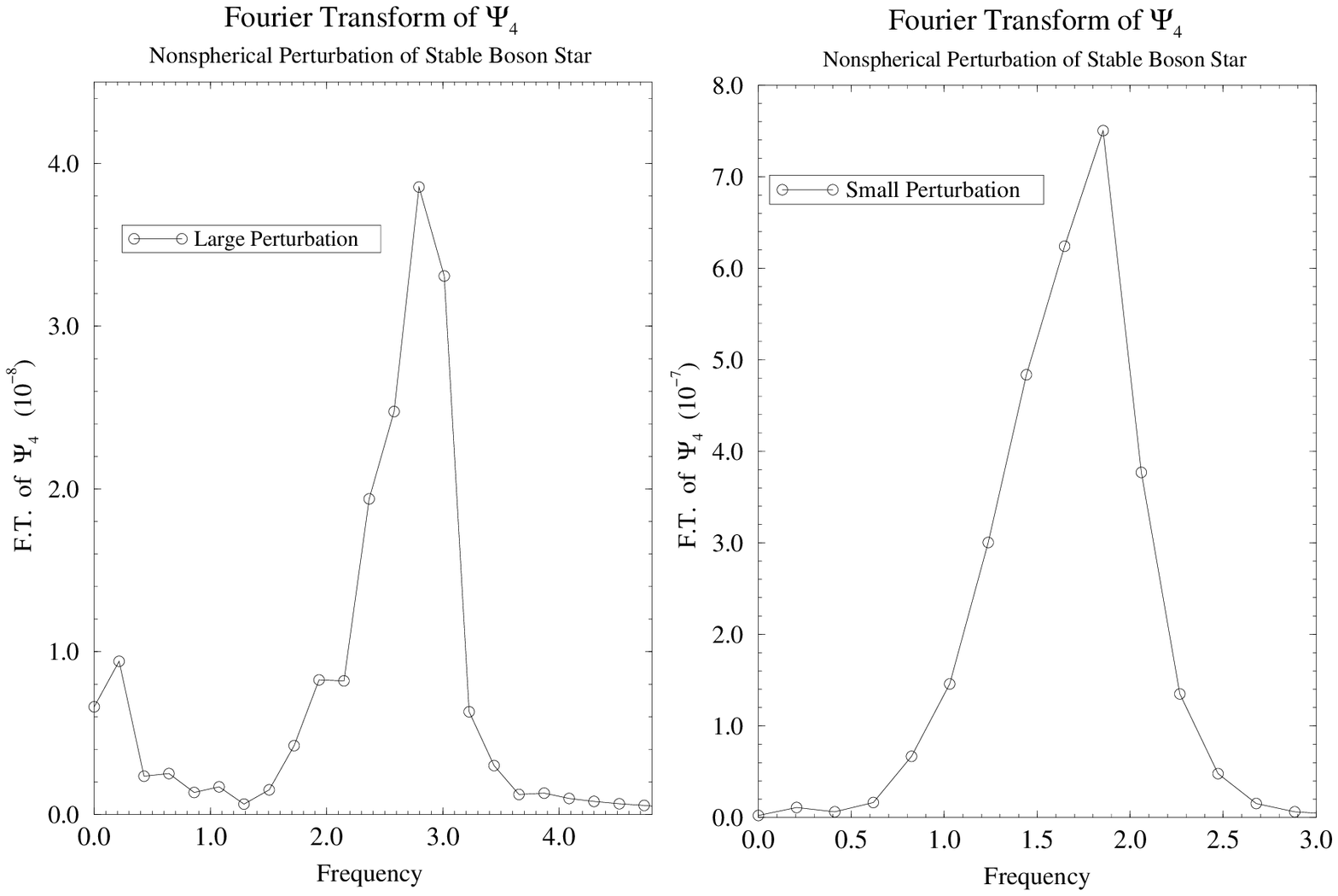}}
\end{center}
%%\begin{figure}
%%\hspace{-36pt}
%%\vspace{-130pt}
%%\epsfbox[0 0 450 500]{figs/ft_psi4a.ps}
\caption{
The nonradial modes of a strongly perturbed star are shown to be very
different from the nonradial modes of slightly perturbed one. The
former star will lose far more mass through scalar and gravitational radiation
and move to a new configuration.
The latter oscillates in the quasinormal mode of the original
configuration.
}
\label{fig_ftpsi4}
\end{figure}

Thus we have demonstrated that our code gives very consistent results and is
capable of providing a detailed description of the gravitational signatures
of a boson star.

\section{Colliding Two Boson Stars}

Another project, although still in its infancy, is that
of the collision of two boson stars. 
The $1D$ data of a single star was put on a $3D$
grid using reflection symmetry along an axis to simulate
the presence of two identical boson stars. In order to center
two stars on the $Z$ axis separated by some number of grid points
$2\,nzs$
the boson field  $\psi(i,j,nzs-k+1)$ was added to $\psi(i,j,nzs+k)$ 
to get the new field $\psi(i,j,k)$ for all $i,j,k$ on the grid.
As an initial guess metric
we just used the one star metric $g_{rr}(i,j,k)=g_{rr}(i,j,nzs-k+1)$
and $g_{00}(i,j,k) = g_{00}(i,j,nzs-k+1)$. We then used the
initial solver to get proper initial data assuming the momentum constraints
were satisfied (we had time-symmetric data and the $\pi$ fields
were defined appropriately). The main result that we have right now, is that
because the system produces much more gravitational radiation,
the errors in the Zerilli functions due to boundary problems is
much less significant. 
Typically, though, the first two detectors 
were in close agreement for the Zerilli function and the last two for the
Newman--Penrose calculation. This shows the Zerilli function is still somewhat
sensitive to outer boundary errors.
The disagreement between the the two measures
is markedly reduced compared to the earlier
cases with smaller signals. So far we have only observed
black hole formations from these collisions.
These black hole formations were accompanied
by negligible amounts of scalar radiation. Steep gradients
in the metrics invalidated the code as they could not be resolved as the
black hole formed.

In Fig.~\ref{fig_2bos} we show the gravitational energy radiated in the
black hole formation during a two boson star collision. The central
field density is $0.075$ for the individual stars. 
One reason to study the two boson star collision is that it
is a way to study the generalized two body problem. The amount
of energy radiated may not sensitively depend on the component
constituents of the collision. An order-of-magnitude calculation
of the energy output to mass ratio (of the two
boson star configuration) gives a value
$\sim 5\times 10^{-4}$ comparing well to the energy emitted 
($0.001$) by
two black hole collisions for black holes separated by large
distances~\cite{anhobi}. 

\begin{figure}[t]

\begin{center}
\leavevmode
\makebox{\epsfysize=15cm\epsfbox{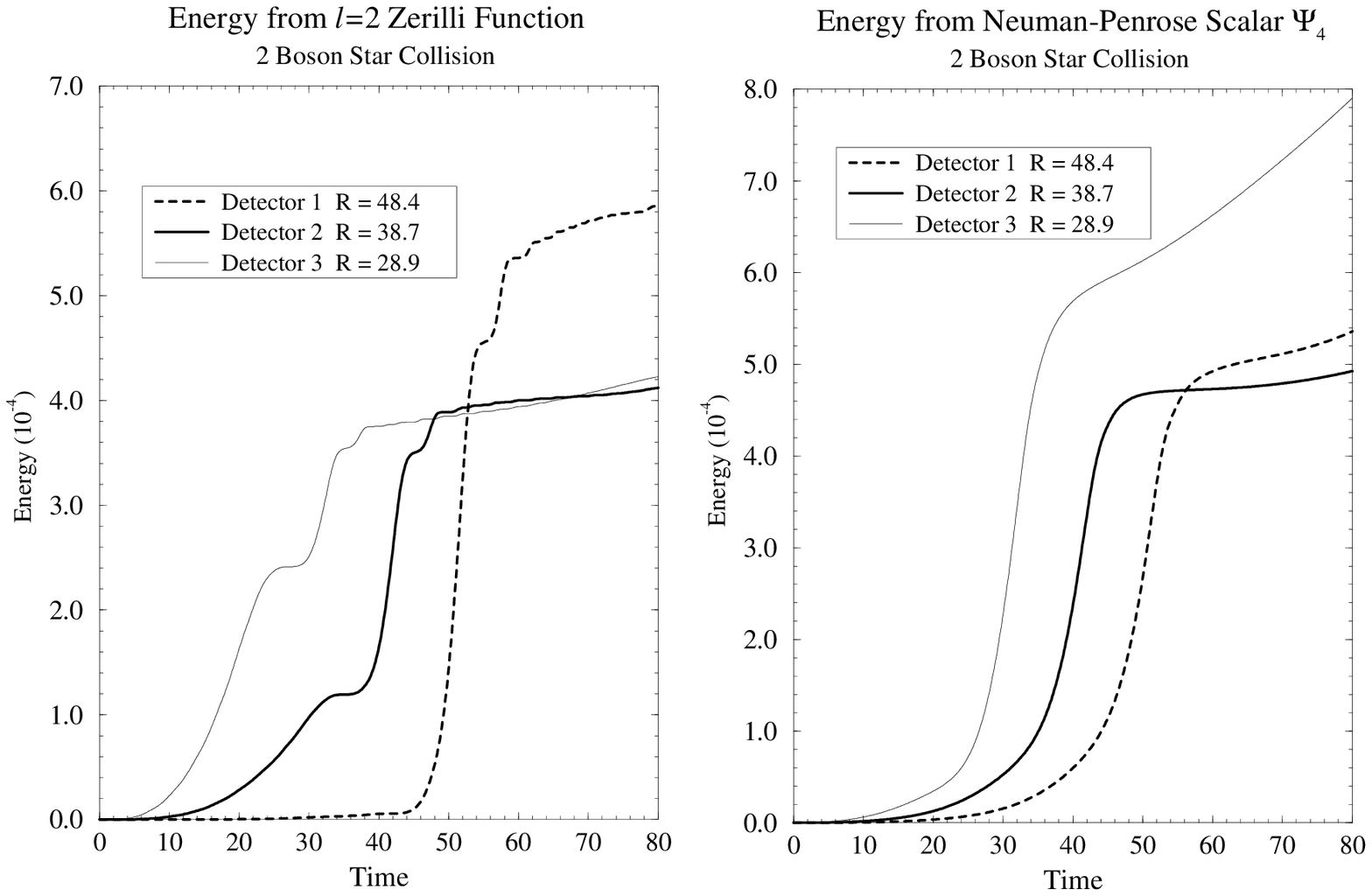}}
\end{center}
%%\begin{figure}
%%\hspace{-36pt}
%%\vspace{-130pt}
%%\epsfbox[0 0 450 500]{figs/cl2bose.ps}
\caption{
The energy carried by a gravitational wave as measured by three detectors,
is shown for a Zerilli calculation and a Newman-Penrose calculation.
The former shows very close agreement for the first two detectors
but outer boundary effects result in inaccuracies
in the third detector's signal. The latter on the other hand
shows a close agreement in the last two detectors' signals. 
}
\label{fig_2bos}
\end{figure}

\section{Boson Star: Black Hole Formation}

In the spherical collapse problem our $1D$ code with
the polar slicing condition caused the lapse to
collapse and radial metric to grow uncontrollably
at the approach of an apparent horizon. 
Our code was unable to handle these sharp
gradients and crashed (uncontrollable numerical instabilities arose). By using a $3D$ code with maximal slicing,
that was still singularity avoiding, but not as drastically as polar
slicing, one could confirm the formation of the black hole.
We took advantage of the $1D$ code's ability to resolve the star,
by monitoring the first part of its evolution in the $1D$ code.
Then after collapse was initiated, and before the gradients
got too large, we collected data and placed it on the $3D$ grid.
At this stage the star was smaller and the outer part of the grid
could be conveniently thrown away allowing us decent resolution in $3D$.

We used ``polar slicing'' in the $1D$ code, which
did not set $Tr K=0$ except on the initial time slice. As the system 
evolved in $1D$ it evolved to $Tr K \ne 0$. While maximal
slicing ($Tr K=0$) avoided problems like grid stretching, by keeping
the volume fixed in the absence of a shift, our initial
data did not have $Tr K =0$. Although the K-driver transformed the
slicing to $K=0$ after a short time the evolution exhibited severe
grid stretching and we could evolve with zero shift only to
a time of $22.6M$ as opposed to $t \sim 50M$ for vacuum Schwarzschild black
holes. We present a plot of 
$g_{rr}$ for the simulation of boson star collapse to
a black hole in Fig.~\ref{fig_kdrivbh}.

\begin{figure}[t]

\begin{center}
\leavevmode
\makebox{\epsfysize=15cm\epsfbox{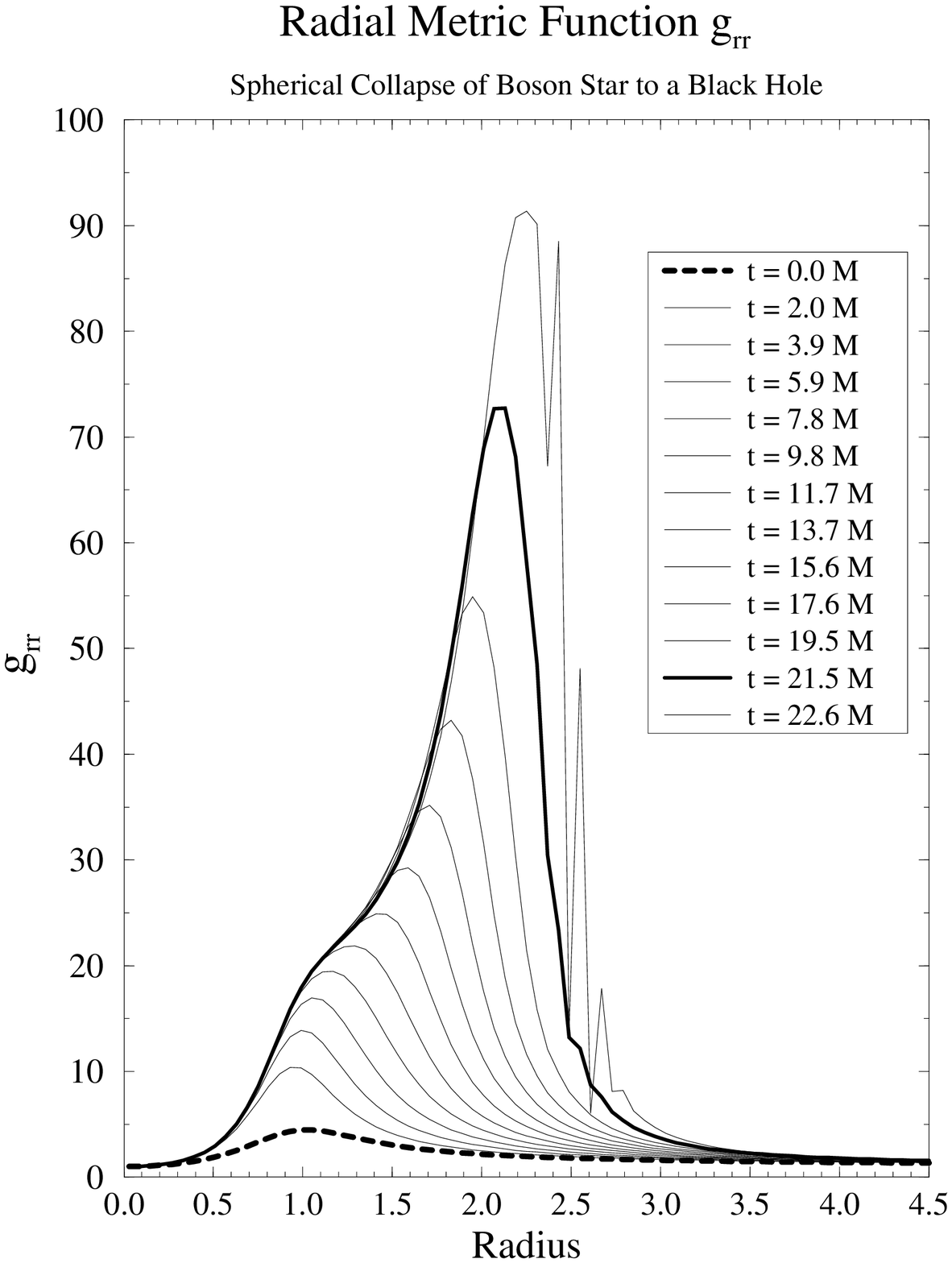}}
\end{center}
%%\begin{figure}
%%\hspace{-36pt}
%%\vspace{-130pt}
%%\epsfbox[0 0 450 500]{figs/kdriv_bhe.ps}
\caption{
This plot shows the evolution of the radial metric function
$ g_{rr} $ for the case of the spherical collapse of a boson star
to a black hole. In this simulation the shift is zero and
the lapse function is obtained by solving the maximal slicing
equation with a K-driver term  $\sim c K$.
}
\label{fig_kdrivbh}
\end{figure}

Singularity avoiding slicings offer a way to avoid evolution
in regions with singularities while continuing the evolution in
regions outside. However they only manage to delay the breakdown of a numerical
code. Pathological properties of the slicing, like sharp gradients near the
horizon, eventually invalidate the code. 

Since the region inside the black hole horizon cannot causally affect the region
outside, one can in principle cut away the interior singular
region and continue the evolution outside. An outside observer loses
no information by cutting away a causally disconnected region. This
requires an apparent horizon boundary condition ({\em AHBC}). 

For a collapsing star this can be implemented by evolving the star under
suitable gauge conditions for a while and then introducing
a shift vector after an apparent horizon is generated. This shift vector's
function would
be to maintain the apparent horizon at some fixed position~\cite{anniho}. This machinery had been developed for black holes~\cite{dauesphd} and was used
in our code to lock the horizon in the spherical collapse problem.
The simulation began without {\em AHBC} implemented and covered the $r=0$
region. After some evolution the apparent horizon formed and the
grid was cut. A B-locking shift as previously described was applied to
prevent coordinate drift. The lapse function was updated by a gamma-driver
slicing. The details of these conditions and the causal differencing
(using one-sided derivatives inside the apparent horizon)
can be found in~\cite{dauesphd} and~\cite{anniho}. With this machinery
our code ran twice as long. We show the radial metric evolution
in Fig.~\ref{fig_grrahbc}.

\begin{figure}
\hspace{-36pt}
\vspace{-130pt}
\epsfbox[0 0 450 500]{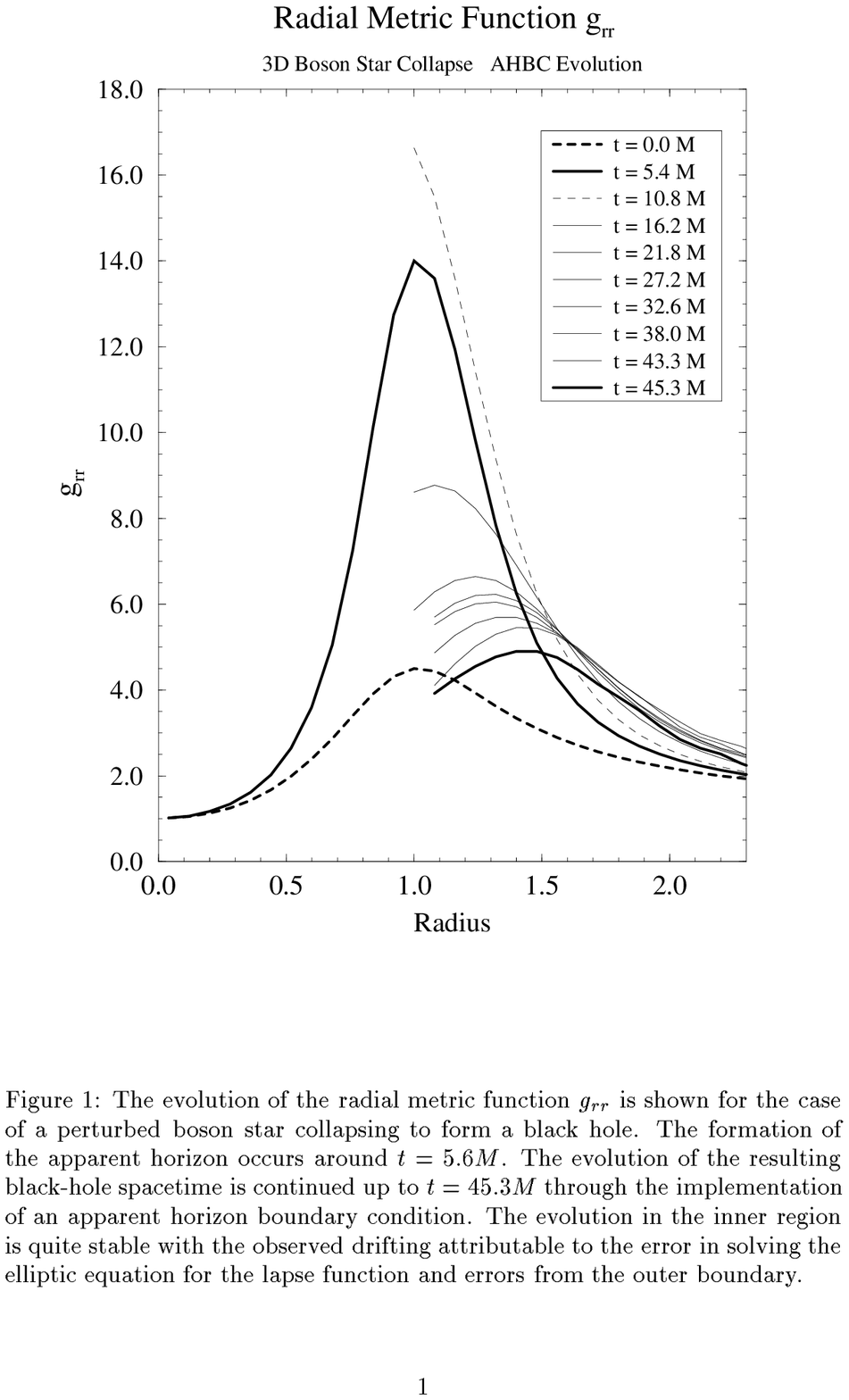}
%\epsfbox[0 0 450 500]{figs/psaahbc.ps}
%\caption{The evolution of the radial metric function $g_{rr}$ is shown for
%the case of a perturbed boson star collapsing to a black hole. The
%formation of the apparent horizon occurs around $t=5.6\,M$. The evolution of
%the resulting black-hole spacetime is continued up to $t=45.3\,M$ through the
%implementation of an apparent horizon boundary condition. The evolution in
%the inner region is quite stable with the observed drifting attributable to
%the
%error in solving the elliptic equation for the lapse function and errors
%propagating inwards from the outer boundary.
%}
\label{fig_grrahbc}
\end{figure}

\chapter{Conclusions}

Even the skeptic, who refuses to believe in the possibility of
the existence of self-gravitating bosons, would be forced to
admit that they are indeed very useful objects: That is to say,
they are rich in physical content, providing insights into the
nature of several physical systems. 

For instance the dark matter problem is real and there must be something to explain
it. We know that normal baryon densities are not enough
to close the universe, and we know that cold and hot dark
matter models fail to explain large and small scale structure
of the universe, respectively. Relativistic Bosons seem to offer 
a partial solution to these problems. Although they have particles
with high velocities to explain the
large scale power spectrum, they have more low velocity particles than
fermionic hot dark matter (because
of the Bose-Einstein distribution as opposed to the
Fermi-Dirac distribution). Thus while small scale power 
is erased for hot dark matter due
to free streaming, enough survives on the smallest scales with this
kind of bosonic hot dark matter~\cite{madpower}.

Also, many particle physics and cosmological models rely on some form of scalar
particles. These unseen scalars could very well be
dark matter constituents. From the primordial soup of
scalar particles self-gravitating objects can form by the Jeans instability
mechanism. Perturbations larger than the Jeans' length
on a homogeneous background could cause the compactification of
matter~\cite{scottrem}.

Self-gravitating objects made from complex scalar fields are called
Boson stars. These solutions to the Klein--Gordon--Einstein equations
are held together by the balance between attractive forces of gravity
and the dispersive forces of the wave equation. In the literature
they are referred to as mini-soliton stars~\cite{leepert} because of
their small sizes. The addition of a small coupling parameter can
however increase their size to neutron star dimensions, making
them astrophysically interesting. 

The masses of these stars increases with central density up to a peak and 
then decreases.
We have
investigated the stability of ground and excited state boson stars
with self-coupling \cite{balalam}. 
The branch to the left of the peak is called an $S$ or stable
branch while the branch to the right is called the $U$-branch. 
Configurations on the $S$ branch have very specific quasinormal modes
of oscillation under small perturbations. Under finite perturbations they
settle down to new configurations on the same branch.
When perturbed, configurations
that lie on the unstable or $U$-branch either migrate to new configurations
on the $S$-branch or collapse to black holes. These
characteristics are shared by boson stars with or without
self-coupling.

Excited states are configurations with nodes.
The field of an $ n^{\rm{th}} $
excited state star has $n$ nodes and its radial metric has $n+1$ peaks.
Their mass profiles are similar to the profiles of boson stars in the
ground state, which makes it appear as if they have a stable and an unstable
branch of configurations. However, irrespective of which branch they lie on,
excited boson stars are unstable with different
instability timescales. Low density excited stars
having masses close to ground state configurations
will form ground state boson stars after evolution. Denser configurations
form black holes with the decay time decreasing with increasing central
density till one approaches the density corresponding to zero
binding energy. As the central density approaches this critical central
density the kinetic energy of the star starts to increase as it
becomes more dispersive. It still collapses to a black hole
but on a larger time scale. Beyond this point, for densities corresponding
to positive binding energy, the stars disperse to infinity.

An interesting feature in the collapse of excited state boson stars is that,
during this process, they cascade through intermediate states,
rather like atoms transiting from excited states to the ground state,
suggesting that boson stars behave in some ways like gravitational atoms~\cite{Lid}.
However, for boson stars, an investigation of the possible decay channels (selection
rules) seems much more difficult (if at all possible or meaningful)
due to the intrinsic nonlinearity of the theory.

We have also studied the stability and formation of boson stars in
Brans-Dicke theory~\cite{balabd}. The arbitrariness of units of
length, time, and
masses, and the observed coincidences in the values of dimensionless constants,
resulted in alternative theories of gravity, like {\em BD} theory
and scalar-tensor
theories, being propounded. While it has been established that the
weak-field differences between these theories and gravity is
insignificant at the present times, the question is whether a
boson star model can be a source of strong-field differences. We
have worked on a {\em BD} model, with the view of carrying
on the calculation to more general {\em ST} theories in the future.

We find the basic features of $U$-branch and $S$-branch stability
and excited state transitions to be similar to {GR}. We do see
the {\em BD} field, which is static for an equilibrium configuration,
begin to oscillate when perturbed,
as do the metric components. However these oscillations of
the gravitational field are not accompanied by any significant
gravitational radiation.

Using a Gaussian initial boson scalar field configuration, to
represent the local accumulation of matter, we were able to
demonstrate the formation of boson stars in $BD$ theory. This demonstrates that
a boson star is a physically realizable object in $BD$ theory.

The flatness of galactic rotation curves indicates the presence of dark
halos extending beyond the last visible point, with masses proportional
to distance and densities inversely proportional
to distance squared. While boson stars in the ground state have
too fast a mass fall off to fit galactic rotation curves,
third and fourth excited state stars with and without self-coupling
have been used to fit them quite well \cite{sin}, \cite{jaewonleehalo}.
These states are unstable but their energy differences from lower
excited states are large enough to warrant large instability
time scales.
As we have seen in our own
simulations, a repulsive self interaction term further
increases this instability time scale.

Another model that can explain galactic rotation curves involves
massless scalar fields \cite{sch1halo}, \cite{sch2halo}. The
dispersive effects of the scalar field prevents collapse of
these halos to black holes. These configurations exist
for different values of the scalar field oscillation frequency
($\phi\sim e^{i \omega t}$) and for each value of central density. While the value
of the central density is related to the orbital velocity, the scalar
field oscillation frequency determines the mass and energy density.
The Newtonian configurations had already been described and we studied the
stability and formation of these objects in a relativistic code\cite{balamsls}.
The mass and energy densities still had the desirable properties
for flattening out rotation curves. 

We have also shown that massless complex scalar fields are able to settle down
to a stable configuration. It seems that the formation process needs
a special profile for the energy density. The appearance of
points of inflexion within the decreasing density supports the `birth' of the boson
halos.
In contrast, extremal values for the density cannot be compensated
and lead to the destruction of the initial configuration.
Stable configurations cannot be formed from initial Gaussian distributions. This
could mean that the formation of boson halos could be difficult.
However under perturbations a known stable solution settles to a new one.

The thing that makes these models undesirable though is that they are
not asymptotically flat. The cut-off radius of these configurations
is quite arbitrary. It could be that the size of the halo
depends on the value of the cosmological constant. We are working
on boundary conditions to make this model more viable.
In the process of studying these solutions we have found similar
solutions for the massive scalar field. These solutions
have energy density characterized by a series of maxima and minima rather
than saddle points. They have positive binding energy and as expected
disperse in our numerical simulations.

We have furthermore investigated boson stars in a general three-dimensional
code demonstrating the key role that boson stars play in the
field of numerical relativity.
By exploiting the difference between boson stars and neutron stars,
we have managed to deal with them numerically. There are no surface
problems in a boson star system. Nevertheless the similarities in
their structures has allowed us to study stellar dynamics using them
as a model. 

On the numerical front we have developed a fully relativistic code
with matter sources. By comparisons to known results, we have
been convinced that the code is indeed stable within the
computational resources available to us. We have tested our
scalar evolver by itself and seen that it is stable for
thousands of timesteps. We have seen metric components remain
static for equilibrium boson stars, for several time steps before
outer boundary problems invalidate the code. The amount of scalar
radiation under spherical perturbations matches very well with
our $1D$ code until boundary reflections become significant.
We have observed one full metric oscillation of the quasinormal
mode under spherical perturbations. 

We have then moved a step further to study real physics with this
code by studying the behavior of boson stars under non-spherical
perturbations. We have measured the gravitational wave output
using different techniques with comparable results. We have
calculated the quasinormal modes of the star under infinitesimal
perturbations. Our results are consistent with those of~\cite{yoshinonrad}. The nonradial modes
for boson star configurations are strongly damped. This
allows us to get the full gravitational wave output for these
stars in a relatively short evolution timescale. After
the emission of gravitational waves the
star becomes spherical and then oscillates in its spherical quasinormal
mode. The amount of scalar radiation is much more than the gravitational
radiation for these configurations. The scalar radiation
decouples from the gravitational radiation.
While the latter propagates at the speed of light the scalar radiation is
slower and arrives later at our detectors.
Under larger non-spherical perturbations 
the star emits more radiation and does not oscillate in the quasinormal
mode of the initial configurations. By comparing the amount of
gravitational radiation for unperturbed, spherically perturbed, and
non-radially perturbed configurations, we have convinced ourselves that
numerically induced asymmetries are negligible. 

We have collided two boson stars and formed black holes. This is with
the view of studying the generalized two-body problem in $GR$. The
amount of scalar radiation in these black hole collapses is negligible
and the gravitational radiation dominates the system. The ratio of
energy radiated to initial mass of the system is of order
$0.0005$ as opposed to $0.001$ for two 
black hole collisions initially largely seperated. These numbers
are comparable and might suggest that the two body problem may be independent
of the nature of the constituent objects. Thus a boson system might,
with its absence of singularities and surface problems, be ideal
to study such systems.

We have also seen the collapse to black holes of unstable boson stars.
This is of importance on two accounts. When we performed
these simulations in $1D$ we used a polar-slicing condition which
caused the lapse to collapse and radial metric to blow
up at the approach of an apparent horizon. Thus we could
not carry the simulation through to actual black hole formation . By
using a maximally-sliced $3D$ code we can carry our simulation
further and confirm black hole formation. 
In addition by running the $1D$ code with very fine resolution
until gradients start getting
steep and collapse has been initiated, we could throw away the
outer part of the star (its radius has been reduced due to collapse).
It could then be placed on a $3D$ grid with finer resolution
than if we started the simulation from the beginning in $3D$.
After this we could turn on the $AHBC$ and evolve the resulting
black hole with matter present in the space-time. This added
a new dimension to the black hole problem.

Thus with a simple system of bosons, we have touched
upon numerous physical and numerical issues. 
We now have stable $1D$ and $3D$ codes that are capable
of handling different system configurations. There
are several models whose equilibrium
configurations have been described but whose evolutions have
not been studied. 

For example we have considered boson stars
which are minimally coupled to gravity. If one considers the
relevance of boson stars in the early universe, it would be worth
investigating scalar fields non-minimally coupled to gravity.
Since the Pecci-Quinn symmetry \cite {pecciquinn}
is not exact, QCD effects place the coupling
strength for axions at $\xi\approx 10^{-20}$ (not 0). This
concept might extend to complex scalar fields. In~\cite{bijminimal} a
nonminimal coupling
of the form $\xi\,|\phi|^2R$ is chosen and the equilibrium configurations
are described. For a coupling strength  $\xi>4.0$ the mass is always less
than the particle number. One expects the branch to the right of
the maximum mass to be unstable but if the particle number is greater
than the mass this instability must always
manifest itself in gravitational
collapse and not dispersion. These things can only be confirmed in
an evolution code. 

There are also descriptions of the equilibrium sequences
of mixed fermion-boson stars~\cite{ing2,hen1}.
Pure boson stars are qualitatively quite
similar to pure fermion stars. A mixture
of bosons and fermions interacting only by gravitational forces is of
interest in
field theory as a study of the self-confinement of two quantum fields~\cite{ing2}. Also
bosonic axion fields could be captured by neutron stars and influence their
stability. Similarly nuclear matter in high density neutron cores could be subject
to phase transitions into the pion condensate. Again the stabilities
and behavior of the system can be studied only in an evolution code.

One interesting project would be to confirm whether real scalar field
configurations could exhibit long term stability under non-spherical
perturbations. While these configurations
had no static equilibrium solutions~\cite{rosen1,rosen2,jetzreal},
periodic solutions have found~\cite{sei1}. Evolution
studies in a $1D$ code showed that
there were oscillating configurations
that oscillated about some configuration for a long time, without
observable diminishing mass. This study could be taken to the next step
by studying the behavior of the system under more realistic
non-spherical perturbations. Since axions are described by real fields
this would be of particular importance. 

The list is endless and one hopes that the tools
described and developed in this thesis will make their contribution in
furthering knowledge in physics and numerical relativity.

%\include{chap6}
%\section*{scaling}
\appendix
\addcontentsline{toc}{chapter}{Appendix}{}
    \section*{Appendix: Scaling}
    In transferring initial $1D$ data from the $1D$ boson star code, to the $3D$
``G'' code, 
where it was interpolated onto a $3D$
grid and evolved, one had to make sure all coordinates were rescaled appropriately. Essentially this meant going from a code where $\hbar =1$ and $c=1$ (the $1D$ code) to a code where $G=1$ and $c=1$.
In terms of coordinates this translates to the spatial coordinates going from units of inverse
mass to units of
mass. Consider for example the $1D$ Hamiltonian constraint.

\begin{equation}
g'=\frac{1}{2}\left[\frac{g}{{\bf r}}-\frac{g^3}{{\bf r}}+4\pi G\left[m^2\phi^2
{\bf r}g^3+\dot{\phi}^2{\bf r}g^3/N^2
 + {\bf r} g {\phi'}^2\right]\right].
\end{equation}
In dimensionless coordinates ($r$, $t$, $\sigma$, and $N$)
\begin{equation}
r= m{\bf r}, \quad t=\omega_0{\bf t},\quad \sigma=\sqrt{4\pi G}\phi, \quad 
N= {\bf N}\frac{m}{\omega_0},
\label{dimenappendix}
\end{equation}
where $m$ refers to the mass of a boson  and $\omega_0$ the oscillation frequency of the
boson field, this becomes
\begin{equation}
g'=\frac{1}{2}\left[\frac{g}{r}-\frac{g^3}{r}+rg^3\left[\sigma^2+\frac{\dot {\sigma^2}}{N^2}
\right] + r g {\sigma'}^2\right].
\label{gappendix}
\end{equation}
Since the field is very small at the edge of the grid the metric components can be taken to
be Schwarzschild there. So $g^2(\infty)\sim {1}/(1-{2MG}/{\bf r})$.
This means the mass of the star is given by
\begin{equation}
M=\frac{{\bf r}}{2G}\left(1-\frac{1}{g^2}\right),
\end{equation}
calculated at the edge of the grid.

Using $\frac{c\hbar }{G}=M_{Pl}^2$ (for the $1D$ code therefore $G=1/M_{Pl}^2$) and
 ${\bf r}= r/m$, where $r$ is the dimensionless radius read off from the output of the $1D$ code, we get
\begin{equation}
M=\frac{r}{2}\left(1-\frac{1}{g^2}\right ) \, \frac{M_{Pl}^2}{m}.
\end {equation}

Going back to equation (\ref{dimenappendix}), one sees that when the code runs for a certain
dimensionless time $t_{\mathrm code}$ it corresponds to physical
time ${\bf t} = 1/\omega_0\, t_{\mathrm code} =\frac{N}{{\bf N }m}\, t_{\mathrm code}$, where $\omega_0$ has been written in terms of the lapse. For a boson star
of mass $0.633 \, M_{Pl}^2/m$ (maximum mass of a ground state star)
we can compute the physical time 
\begin{eqnarray}
{\bf t} &=& 
\frac{1}{0.633}\frac{N}{{\bf N }}\, G\,M\,t_{\mathrm code},
\\ \nonumber 
&=& 
\frac{1}{0.633}\, \left(\frac{m\,c^2}
 {{\hbar\,\omega_0 }}\right)\, \left(\frac{G\,M}{c^3}\right)\,t_{\mathrm code},
\end{eqnarray}
where we have written the mass of the boson in terms of the mass of the
star (Schwarzschild radius of the star $\sim 1/m$). Computing $\frac{N}{{\bf N}}$ at the edge of the grid, where 
${\bf N} \sim \sqrt{1-2M/r}$, we find this fraction to be of order 1.
Thus 1 unit of code time in terms of the mass of the
star is about $1.6 \,G M/c^3$.

In the $3D$ code however
G=1 but $\hbar $ is not. So consider the Hamiltonian  constraint
equation again. There is no special reason for this but it suffices to
use it to figure out the scaling parameters. Every derivative must be 
multiplied by $\hbar $, 
which had been set equal to 1 in the $1D$ case.
Now $G$ is set equal to 1. So one has for the Hamiltonian constraint
equation
\begin{equation}
g'=\frac{1}{2}\left[\frac{g}{{\bf r}}-\frac{g^3}{{\bf r}}+4\pi \left[m^2\phi^2
{\bf r}g^3+\hbar ^2 \dot{\phi}^2{\bf r}\frac{g^3}{N^2}
 + {\bf r} g \hbar ^2 {\phi'}^2\right]\right],
\end{equation}
substituting dimensionless coordinates
\begin{equation}
r= m{\bf r}/\hbar, \quad t=\omega_0{\bf t},\quad \sigma=\sqrt{4\pi}\hbar\phi, 
\quad 
N= {\bf N}\frac{m}{\hbar\omega_0},
\end{equation}
we get back equation (\ref{gappendix}).
Now to interpret our results, which have thus far been stated as raw 
dimensionless numbers:
Without metric evolution the code runs to a time of 2000. The more complete 
problem must include
metric evolutions. For this case, when we have in our reports been
stating that the code runs to a time of 140, we
actually mean that the code runs to a time of $140/\omega_0$ with
$1/\omega_0=\hbar/m\frac{N(\infty)}{{\bf N}(\infty)}$.
In $G=1$ units we have
\begin{equation}
M=\frac{{\bf r}}{2} \, \left(1-\frac{1}{g^2} \right),
\end{equation}
calculated at the edge of the grid. This gives
\begin{equation}
M=\frac{ r}{2} \, \left(1-\frac{1}{g^2} \right)\,\frac{\hbar}{m}.
\end{equation}
Using the fact that $G=1$ and $c=1$ implies $\hbar=M_{Pl}^2$ one gets the same
result as before. That is 
\begin{equation}
M=\frac{r}{2} \, \left(1-\frac{1}{g^2} \right)\, \frac{M_{Pl}^2}{m},
\end {equation}
which for the problem in consideration is
$M=0.05 \, M_{Pl}^2/m$.
Replacing for $1/m$ in the equation for $1/\omega_0$
gives  $1/\omega_0=\frac{N(\infty)}{{\bf N}(\infty)}
\frac{M}{0.05}$ where the $\hbar$ cancels the factor $M_{Pl}^2$. 
$N$ is $>1$ and ${\bf N} <1$ with the ratio slightly greater than 1.
So an order of magnitude for 1 time step in terms of mass for the
maximum mass boson star is $1/0.633\sim 1.6$ times the mass of the star.

For our equilibrium configurations with full
evolutions we use star masses of order $0.5 \, 
M_{Pl}^2/m$ so one time step is slightly more than 
two times the mass of a star. The code runs to a dimensionless time of over $500$ or
over thousand times the mass of the star ($t_{\mathrm physical}> 1000\,GM/c^3$).

For non-spherical perturbations the mass was of order $0.584 \, M_{Pl}^2/m$.
One time step was roughly 1.87 times the mass of the star.

\addcontentsline{toc}{chapter}{Bibliography}{}
\bibliographystyle{unsrt}
\bibliography{thesis}
%\addcontentsline{toc}{chapter}{Bibliography}

\end{document}